\documentclass{elsart1p}
\usepackage{amsmath}
\usepackage{psfrag}
\usepackage[dvips]{graphicx}

\def \ba {\begin{eqnarray}}
\def \ea {\end{eqnarray}}
\def \nn {{\nonumber}}

\def \AA {{P}}
\def \AAb {\mbox{\boldmath${\cal A}$}}
\def \al {{(\alpha)}}

\def \alpm {{(\alpha)\pm}}
\def \albe {{(\alpha\beta)}}
\def \be {{(\beta)}}
\def \bnu {\mbox{\boldmath$\nu$}}
\def \bxi {\mbox{\boldmath$\xi$}}
\def \bxic {{\acute {\mbox{\boldmath$\xi$}}}}
\def \bzeta {\acute{\bf k}}
\def \cos {{\rm cos~}}

\def \D {{\bf D}}
\def \DD {{M}}
\def \e {{\bf e}}

\def \EEm {{N^-}}
\def \EEp {{N^+}}
\def \ex {{\rm e}}
\def \exp {{\rm exp~}}
\def \i {{\rm i}}
\def \Im{{\rm Im}~}

\def \g {{\bf g}}

\def \k {{\bf k}}
\def \kbc {{\acute {\bf k}}}

\def \kc {{\acute k}}
\def \n {{\bf n}}
\def \nbc {\acute{\bf n}}

\def \O {{\bf O}}
\def \p {{\bf p}}
\def \Re{{\rm Re}~}
\def \rr {{\acute r}_\perp}
\def \sin {{\rm sin~}}
\def \sgn {{\rm sgn}}
\def \t {{\bf t}}
\def \T {{\bf T}}
\def \TT {\mbox{\boldmath${\cal T}$}}
\def \tt{\mbox{\boldmath${\hat {\cal \tau}}$}}

\def \thec {{\acute \theta}}
\def \u {{\bf u}}
\def \U {{\bf U}}
\def \UU {\mbox{\boldmath${\cal U}$}}
\def \v {{\bf v}}
\def \V {{\bf V}}

\def \varrhoc {{\acute \varrho}}
\def \varthec {\acute\vartheta}
\def \varphic {\acute\varphi}
\def \x {{\bf x}}
\def \xbc {\acute{\bf x}}
\def \xc {{\acute x}}
\def \xic {{\acute \xi}}

\def \ybc {\acute{\bf y}}
\def \yc {{\acute y}}

\begin{document}
\begin{frontmatter}

\title{Diffraction coefficients of a semi-infinite
planar crack embedded in a transversely-isotropic
space}
\author[Iowa]{A. Gautesen},
\author[South]{V. Zernov\corauthref{cor}}
\corauth[cor]{Corresponding author.}\ead{zernovv@lsbu.ac.uk}, and
\author[South]{L. Fradkin}

\address[Iowa]{Dept. of Mathematics, Iowa
State University and Ames Laboratory, Ames,
 IA 50011, U.S.A.}

\address[South]{Waves and Fields Research
Group, Department of Electrical, Computer and Communication
Engineering, Faculty of Engineering, Science and Built Environment,
London South Bank University, SE1 0AA, U.K.}

\begin{abstract}
We have considered a semi-infinite crack embedded in a transversely
isotropic medium and studied two special cases, one, in which the
axis of symmetry is normal to the crack face and the wave incidence
is arbitrary and another, in which the axis lies in the crack plane
normal to the edge and the incident wave vector is also normal to
the edge. The problem is of interest in  Non-Destructive Evaluation,
because austenitic steels that are found in claddings and other
welds in the nuclear reactors are often modeled as transversely
isotropic. In both of cases, we have expressed the scattered field
in a closed form and computed the corresponding diffraction
coefficients.
\end{abstract}
\begin{keyword}
diffraction coefficients \sep transversely isotropic medium \sep
elastic waves \PACS 43.20.El \sep 43.20.Gp
\end{keyword}
\end{frontmatter}

\section{Introduction}
 \label{introduction}

The main aim of this article is to evaluate---in two special
cases---diffraction coefficients of a semi-infinite planar crack
embedded into a TI (transversely isotropic) solid. TI is an
anisotropic medium which is  invariant  under  the rotation around a
symmetry axis, that is, a TI medium exhibits isotropy in the plane
perpendicular to that  axis. The symmetry of this kind is ubiquitous
in nature and manmade materials and TI solids have been widely
studied e.g.~in crystallography, seismology and NDE (Non-Destructive
Evaluation).

The diffraction coefficients relate the
 far field  amplitudes of the corresponding waves
diffracted by a straight crack edge to the amplitude of an incident
plane wave.  The rays, along which energy propagates and the  wave
fronts, which separate the disturbed regions in space from
undisturbed, are
  the basic concepts of the so-called ray theory. In the far field (high-frequency) approximation, this provides a
  convenient description of the wave phenomena
such as propagation,  reflection and refraction ~(see
e.g.~\cite{duff60}).

   It is well known that in an isotropic solid,  the fronts of waves radiated by a point source
comprise three concentric spheres. At each moment in time, the
sphere of the largest radius is the region covered by the
disturbance due to P (compressional or longitudinal) mode of
propagation while the remaining two spheres are coincident and
describe the region disturbed by S (shear or transverse) waves. The
latter degeneracy is usually resolved by identifying two orthogonal
components of polarization, which are named the SV and SH modes,
with V standing for vertical and H for horizontal. The rays are
normal to wave fronts. The nature of wave propagation in anisotropic
elastic solids is significantly different, since they support three
distinct types of elastic waves, qP, qSV and qSH, with q standing
for quasi, none of which is in general purely compressional or
shear. The rays are not normal to the wave fronts either
\cite{pay83}

A well known extension of the ray theory, GTD (the Geometrical
Theory of Diffraction), suitable for description of diffraction in
the far-field approximation was first introduced  by Keller
\cite{kel57}--\cite{kel78}.
 He considered
diffraction of the plane scalar wave by an edge embedded into an
isotropic medium whose radius of curvature is much larger than the
wave length. According to GTD, a plane wave incident on such an edge
produces a cone of diffracted rays. The apex of the cone coincides
with the point of diffraction; it is centered on the straight line
which is tangent to the edge at this point, and the cone's solid
angle  is determined by the Snell's law. Moreover, according to
Keller, GTD applies even when the incident wave is not plane, but at
the point where it strikes the edge, the  radii of curvature of the
wave front  are   much larger than the wave length. Special cases
which can be treated using the high-frequency approximation but for
which GTD fails can be described with the so-called uniform GTD
which lies outside the scope of this article.

In \cite{ach82} Keller's GTD was extended to elastodynamics and
derived formulae for the diffraction coefficients for a
semi-infinite planar crack in an isotropic solid. They achieved this
by reformulating the original problem as a Wiener-Hopf matrix
functional equation in a complex Fourier variable. The problem was
solved by the Wiener-Hopf factorization technique~(see
e.g.~\cite{nob60}) and explicit formulae for the diffraction
coefficients were derived using the method of steepest descent. The
latter provides an explicit construction of the diffraction cones as
well as complex amplitudes along the rays.

For a general anisotropic solid, a semi-analytical approach to the
problem was previously developed in \cite{lew96}. The authors have
reduced the problem to a Wiener-Hopf functional equation which, in
general, has no known analytical solution and have used a numerical
scheme to factorize the underlying Wiener-Hopf matrix. In
\cite{lew00}, they gave a  description  of numerics and showed the
dependence of the magnitude of the backscatter $qP-qP$ diffraction
coefficient on the observation (incidence)  angle.

Here we aim to produce a procedure for calculating  some diffraction
coefficients for semi-infinite planar cracks in austenitic steels.
The problem is of interest in  NDE, because austenitic steels are
found in claddings and other welds in the nuclear reactors.  It is
well known that the austenitic steel can be modeled as  a TI
material (see e.g.~\cite{char89}).   A simple case of normal
incidence in a TI material that supports three convex slowness
surfaces  (see below) has been considered before \cite{nor84}. We
address a more challenging oblique incidence case and deal with an
extra complication due to the fact that in the austenitic steel, one
of the slowness surfaces has inflections.

The article is organized as follows: in Section \ref{sec:problem} we
state the problem in terms of partial differential equations and
boundary conditions. In Section \ref{sec:tt} we discuss the
corresponding transfer tensor, that is the free-space Green's tensor
in the Fourier domain. In Section \ref{sec:intform} we reformulate
the problem in terms of an integral equation (Green's formula). In
Section \ref{sec:functional} we reformulate it again as a 3D
functional  equation and solve this equation analytically in Section
\ref{sec:case_per} for the case of the symmetry axis perpendicular
to the crack plane and in Section \ref{sec:case_par} for the case of
the symmetry axis lying in the crack plane perpendicular to the
crack edge. In Section \ref{sec:dc} we calculate the corresponding
diffraction coefficients.

\section{The problem statement} \label{sec:problem}
Let the medium be a homogeneous elastic solid governed by the
Hooke's law
\begin{eqnarray}
\label{eq:hooke_law} \sigma^d_{ij}(\x^d,t) =
c^d_{ijkl}\epsilon_{kl}(\x^d,t),
\end{eqnarray}
where the superscript ${}^d$ is used to denote the dimensional
quantities (whenever the non-dimensional versions are also used),
$\x^d$ is an arbitrary point in space, $t$ is time and everywhere
$\sigma^d(\x^d,t)$ and $\epsilon(\x^d,t)$ are respectively, stress
and strain tensor of the second order while $i,j,k,l=1,2,3$ are
indices of the tensor component corresponding to any three
dimensional Cartesian coordinate system. Here and everywhere below,
unless otherwise stated, we employ the summation convention over the
repeated index.

As already mentioned above, a TI material has one axis of symmetry.
It is well known that the corresponding stiffness tensor $c^d$
involves five unknowns ~\cite{pay83}.  Also, the stress tensor
$\sigma^d(\x^d,t)$ and strain tensor $\epsilon(\x^d,t)$ are both
symmetric, which allows us to reduce their order by using  the
so-called Voigt notations: Introducing the medium Cartesian
coordinate system $\{\e_1,\e_2,\e_3\}$, with the $\e_3$ axis running
along the symmetry axis, we replace the tensor $\sigma^d(\x^d,t)$ by
the six dimensional vector whose the components
$\sigma^d_{11}(\x^d,t),\sigma^d_{22}(\x^d,t),\sigma^d_{33}(\x^d,t)$
and
$\sigma^d_{23}(\x^d,t),\sigma^d_{31}(\x^d,t),\sigma^d_{12}(\x^d,t)$
are called the normal and shear stresses, respectively. Then the
Hooke's law~(\ref{eq:hooke_law}) takes the form
\begin{eqnarray}
\label{eq:hooke_tiso} \left(
\begin{array}{c}
\sigma^d_{11}(\x^d,t) \\ \sigma^d_{22}(\x^d,t) \\ \sigma^d_{33}(\x^d,t) \\ \sigma^d_{23}(\x^d,t) \\
\sigma^d_{31}(\x^d,t) \\ \sigma^d_{12}(\x^d,t)
\end{array}
\right) = \left(
\begin{array}{cccccc}
A^d_{11} & A^d_{12} & A^d_{13} & 0 & 0 & 0 \\
A^d_{12} & A^d_{11} & A^d_{13} & 0 & 0 & 0 \\
A^d_{13} & A^d_{13} & A^d_{33} & 0 & 0 & 0 \\
0 & 0 & 0 & 2B^d_1 & 0 & 0 \\
0 & 0 & 0 & 0 & 2B^d_1 & 0 \\
0 & 0 & 0 & 0 & 0 & 2B^d_3
\end{array}
\right) \cdot \left(
\begin{array}{c}
\epsilon_{11}(\x^d,t) \\ \epsilon_{22}(\x^d,t) \\ \epsilon_{33}(\x^d,t) \\ \epsilon_{23}(\x^d,t) \\
\epsilon_{31}(\x^d,t) \\ \epsilon_{12}(\x^d,t)
\end{array}
\right),
\end{eqnarray}
where $B^d_3= (A^d_{11}-A^d_{12})/2$. It follows that any TI medium
is characterized by five independent elastic moduli $A^d_{11},
A^d_{12}, A^d_{13}, A^d_{33}$ and $B^d_1$ as well as density $\rho$.

To continue,   in the absence of body forces, the  elastodynamic
equation  can be written as
\begin{eqnarray}
\label{eq:forceequilibrium} \nabla^d\cdot\sigma^d(\x,t)  = \rho
\partial_{t}^2 \u^d(\x,t),
\end{eqnarray}
where the nabla operator
$\nabla^d=(\partial^d_1,\partial^d_2,\partial^d_3)$, $\partial_t$ is
the partial derivative with the respect to time and $\u^d(\x, t)$ is
the displacement. Let us further assume that the medium is
irradiated with an incident wave $\u^{d (in)}(\x,t)$, which is
harmonic, plane and has amplitude $U^d_0$. The corresponding stress,
strain and displacement fields can be written as $\sigma^{d
(sc)}(\x,t)=\sigma^{d( sc)}_m(\x)\exp(-\i\omega t)$, $\epsilon^{
(sc)}(\x,t)=\epsilon^{( sc)}_m(\x)\exp(-\i\omega t)$ and $\u^{d
(sc)}(\x,t)=\u^{d( sc)}_m(\x)\exp(-\i\omega t)$, respectively, where
the subscript ${}_m$ denotes functions whose vector arguments are
expressed in the medium coordinates. Below we simplify the
presentation by dropping the factor ${\rm exp}(-\i\omega t)$
everywhere. A further simplification can be achieved by
non-dimesionalizing all physical variables, except $\omega$ and $t$
which do not feature below,  using the material density $\rho$, the
S wave speed along the symmetry axis $c_0$ (where a degeneracy takes
place and there is no distinction between the $qSV$ and $qSH$
modes), reference wave number $k_0=\omega/c_0$ and the amplitude of
the incident wave $U^d_0$, that is by introducing non-dimensional
variables
\begin{eqnarray}
&& A_{ij}=\frac{A^d_{ij}}{\rho c_0^2},  \,\,\,B_i=\frac{B^d_i}{\rho c_0^2}, \,\,\,\sigma=\frac{\sigma^d}{\rho c_0^2},\nn \\
&&  \k=k_0^{-1}\k^d, \,\,\, \u=(U^d_0)^{-1} \u^d,   \nn\\
&&\x=k_0 \x^d, \,\,\,\nabla=k_0^{-1} \nabla^d.
\end{eqnarray}
The dimensionless form of the reduced elastodynamic equation is \ba
\nabla\cdot\sigma +\u =0, \label{ede} \ea where
$\sigma=\sigma_m(\x)$ and $\u=\u_m(\x)$ . Since for small
deformations, the components of strain tensor
$\epsilon=\epsilon_m(\x)$ are defined by
\begin{eqnarray}
\epsilon_{ij} = \frac{1}{2}\left( \partial_j u_i + \partial_i
u_j\right),
\end{eqnarray}
the dimensionless form of the Hooke's law~(\ref{eq:hooke_law})
allows us to relate the stress tensor to the displacement using the
formula
\begin{eqnarray}
\sigma_{ij}=\Sigma^{(i)}_{jk}(\nabla)u_k, \label{eq:sigmaS}
\end{eqnarray}
where  $\Sigma(\nabla)$ is a  differential operator with the
elements $\Sigma^{(1)}, \, \Sigma^{(2)}$ and $\Sigma^{(3)}$, which
are the 3x3 matrices
\begin{eqnarray}
&&\Sigma^{(1)}(\nabla)= \left[
\begin{array}{ccc}
A_{11}\partial_1 & A_{12}\partial_2 & A_{13}\partial_3 \\
B_3\partial_2 &  B_3\partial_1 & 0 \\
B_1\partial_3 & 0 &  B_1\partial_1
\end{array}
\right],\nn\\
&&\Sigma^{(2)}(\nabla)= \left[
\begin{array}{ccc}
B_3\partial_2 & B_3\partial_1 & 0  \\
A_{12}\partial_1 & A_{11}\partial_2 & A_{13}\partial_3 \\
0 & B_1 \partial_3 & B_1 \partial_2,
\end{array}
\right],\nn\\
&&\Sigma^{(3)}(\nabla)= \left[
\begin{array}{ccc}
B_1\partial_3 & 0 & B_1\partial_1 \\
0 & B_1\partial_3 &  B_1\partial_2 \\
A_{13}\partial_1 &  A_{13}\partial_2 & A_{33}\partial_3
\end{array}
\right]. \label{def:Sigma} \ea

 Note that the stress tensor is symmetric, $
\sigma_{ij}=\sigma_{ji},
$
and therefore, the divergence of the stress tensor can be written as
\begin{eqnarray}
\nabla \cdot \sigma = L(\nabla)\u, \label{def:L}
\end{eqnarray}
where in the medium coordinate system the operator $L=L(\nabla )$ is
given by
\begin{eqnarray}
L_{11}(\nabla) &=& A_{11}\partial^2_1 + B_{3}\partial^2_2 +
B_{1}\partial^2_3,
\nn \\
L_{22}(\nabla) &=& B_{3}\partial^2_1 + A_{11}\partial^2_2  +
B_{1}\partial^2_3,
\nn \\
L_{33}(\nabla) &=& B_{1}\partial^2_1 + B_{1}\partial^2_2 +
A_{33}\partial^2_3,
\nn \\
L_{12} (\nabla) &=& L_{21} (\nabla) =
(A_{11}-B_{3})\partial_1\partial_2, \nn \\
L_{j3} (\nabla) &=& L_{3j} (\nabla) =
(A_{13}+B_{1})\partial_j\partial_3, \mbox{\ \ \ \ \ \ \ \ \ \ \ \
j=1,2.}
\end{eqnarray}
It follows that   elastodynamic equation~(\ref{ede}) can be
rewritten as
\begin{eqnarray}
L \u + \u = {\bf 0}. \label{eq:elmotion}
\end{eqnarray}

Let us now assume that the medium contains a semi-infinite planar
crack with non-contacting faces  (see Fig.~1).  Then  the problem~
(\ref{eq:elmotion}) should be supplemented with the boundary
conditions: Firstly, we assume both crack faces $\Gamma^+$ and
$\Gamma^-$ to be traction-free (see Fig. 1), so that we have \ba
\t\Big|_{\Gamma^\pm}={\bf 0}, \label{bc} \ea
 where the traction $\t=\t(\x )$ is a vector defined by
\begin{eqnarray}
\t=\bnu\cdot \sigma
\end{eqnarray}
with $\bnu$---the inner normal to the upper crack face (pointing
into the solid). Note that using~(\ref{eq:sigmaS}), we can write
\begin{eqnarray}
\t=S_m(\nabla)\u, \label{def:transferoperator}
\end{eqnarray}
so that $\t$ is related to $\u$ via the displacement-traction
transfer operator
\begin{eqnarray}
(S_m)_{jk}(\nabla)=\nu_i\Sigma^{(i)}_{jk}(\nabla),\,\,\,\,\,j,k=1,2,3.\label{toinF}
\end{eqnarray}
 Secondly, at
infinity we assume the radiation boundary condition in the form of
the limiting absorbing principle~(see e.g.~\cite{har01}). This
implies that once a small positive imaginary part is introduced into
a wave number, all scattered waves decay at infinity. Thirdly, at
the tip of the crack, we impose the so-called tip condition,
requiring the  elastic energy of the total field to be bounded~(see
e.g.~\cite{kar62a}, \cite{kar62b}). Combined
with~(\ref{eq:elmotion}) this reduces to the requirement
\begin{eqnarray}
\t= \O(r^{-1/2}_{tip}),\label{tip_cond}
\end{eqnarray}
where $r_{tip}$ denotes the dimensionless distance from the point of
observation to the tip of the crack.
\begin{figure}[ht]
\psfrag{e1p}{$\e'_1$} \psfrag{e2p}{$\e'_2$} \psfrag{e3p}{$\e'_3$}
\psfrag{e1}{$\e_1$} \psfrag{e2}{$\e_2$} \psfrag{e3}{$\e_3$}
\psfrag{t0}{$\theta_0$} \psfrag{p0}{$\phi_0$}
\psfrag{Sp}{$\Gamma^+$}\psfrag{Sm}{$\Gamma^-$} \hfil
\includegraphics[height=80mm,width=100mm]{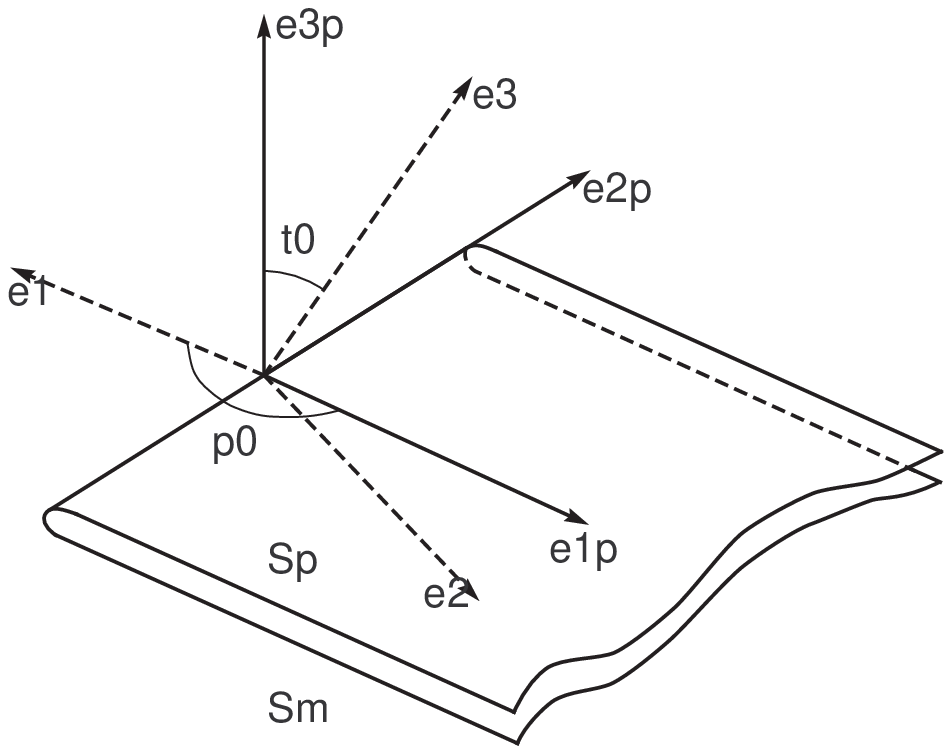}
\hfil
  \caption{The geometry of the problem: the half-plane crack with
non-contacting faces $\Gamma^+$ and $\Gamma^-$, the medium
coordinate system $\{\e_1,\e_2,\e_3\}$ and the crack coordinate
system $\{{\acute \e}_1, {\acute \e}_2, {\acute \e}_3\}$.}
\end{figure}

\section{Transfer Tensor} \label{sec:tt}
Let us introduce the Green's tensor, $u^G=u^G_m(\x)$, the solution
of the system of elastodynamic equations
\begin{eqnarray}
\label{eq:greene2} L_{ij}(\nabla) u^G_{jk}(\x) +  u^G_{ik}(\x)= -
\delta (\x) \delta_{ik},
\end{eqnarray}
 where $\delta (\x)$ is the delta function and $\delta_{ik}$ is the Kronecker
delta.
Taking the triple Fourier transform, which is denoted everywhere by
the hat $~{\widehat{}}~$, Eq.~(\ref{eq:greene2}) gives us
\begin{eqnarray}
\label{eq:algssysgreen} {\widehat L}_m(\bxi) {\widehat u}_m ^G(\bxi)
- {\widehat u}_m ^G(\bxi)=  I,
\end{eqnarray}
where $I$ is the $3\times3$ identity matrix and as above, the
subscript $m$ is used to denote a function whose vector argument is
expressed in medium coordinates. The solution
of~(\ref{eq:algssysgreen}) is called the transfer tensor.

To continue, the determinant of the so-called Kelvin-Cristoffel
matrix ${\widehat L}_m(\bxi)-\lambda_m(\bxi) I$ can be expressed in
the form \ba |{\widehat L}_m (\bxi) - \lambda_m(\bxi) I|\ =
\Delta_{1-2}(\bxi) \cdot \Delta_3(\bxi), \ea where we employ the
notations \ba & & \Delta_{1-2}(\bxi) = \ [A_{11} \xi_\bot^2+B_1
\xi_3^2 - \lambda_m(\bxi) ][B_1 \xi^2_\bot + A_{33} \xi_3^2 -
\lambda_m(\bxi) ] - (A_{13} + B_1)^2 \xi_\bot^2 \xi_3^2,
\nn\\
& & \Delta_3(\bxi) = B_3 \xi_\bot^2+B_1 \xi_3^2 - \lambda_m(\bxi),
\label{eq:delta_quad}
\end{eqnarray}
with $\xi^2_\bot=\xi_1^2+\xi_2^2$. Thus, eigenvalues of ${\widehat
L}_m(\bxi)$, that is zeros of $\Delta_{1-2}$ and $\Delta_3$ are
\begin{eqnarray}
& &\lambda_m^\be (\bxi) =  \nn
\\
& &\frac12 \Bigl[(B_1+A_{11})\xi_\bot^2 + (B_1+A_{33})\xi_3^2 - \nn \\
& & (-1)^\beta\left\{[(A_{11}-B_1)\xi_\bot^2 +
(B_1-A_{33})\xi_3^2]^2 +
4(B_1+A_{13})^2\xi_\bot^2\xi_3^2\right\}^{\frac12}
\Bigr],\,\,\,\,\beta=1,2,\nn\\
& & \lambda_m^{(3)}(\bxi) = B_3 \xi_\bot^2 + B_1
\xi_3^2,\label{quad}
\end{eqnarray}
and the corresponding unit eigenvectors are \ba  {\widehat
\p}_m^\be(\bxi) = \frac{\V_m^\be(\bxi)}{V_m
^\be(\bxi)},\label{ue}\,\,\,\,\beta=1,2,3 \ea where  no summation is
implied   and we use \ba &&\V_m ^\be(\bxi)=\left(\begin{array}{c}
(B_1+A_{13})\xi_1\xi_3\\(B_1+A_{13})\xi_2\xi_3 \\
\lambda_m^\be(\bxi)-A_{11}\xi^2_\bot -B_1\xi_3^2
\end{array}\right),
\V_m ^{(3)}(\bxi)= \left(\begin{array}{c} -\xi_2\\\xi_1\\0
\end{array}\right).
\ea Note that if $\xi_2=0$, as $\xi_1 \rightarrow 0,$ we have \ba
{\widehat \p}_m ^{(1)}(\bxi)\rightarrow\e_3, \,\,{\widehat \p}_m
^{(2)}(\bxi)\rightarrow\e_1, \,\,{\widehat \p}_m
^{(3)}(\bxi)\rightarrow\e_2. \ea The unit eigenvectors of
$L_m(\bxi)$ form an orthonormal basis, which can be used to  expand
any function, that is, we can write
\begin{eqnarray}
{\widehat u}_m ^G(\bxi) = {\bf a}^\be {\widehat \p}_m ^\be(\bxi),
\label{def:ubasis}
\end{eqnarray}
 where the outer product ${\bf a}{\bf b}$ is a
tensor with components $({\bf a}{\bf b})_{ik}=a_ib_k.$  Here and
everywhere below, unless stated otherwise,  we imply summation over
the repeated superscript $\beta=1,~2,~3$.

Substituting ~(\ref{def:ubasis}) into ~(\ref{eq:algssysgreen}) and
using the definition of eigenvectors, we find that the vectors ${\bf
a}^\be$ satisfy the matrix equation
\begin{eqnarray}
[\lambda_m^\be(\bxi)-1]{\bf a}^\be{\widehat \p}^\be_m(\bxi)=I.
\label{eq:systemc}
\end{eqnarray}
 Dot-multiplying~(\ref{eq:systemc}) by ${\widehat \p}^\be_m,\,\beta=1,2,3$
from the right, we obtain the transfer tensor
\begin{eqnarray}
\widehat u_m ^G (\bxi) = \frac { {\widehat \p}^\be_m (\bxi)
{\widehat \p}^\be_m(\bxi) } {\lambda_m^\be(\bxi) - 1 }.
\label{def:ueigen}
\end{eqnarray}

\subsection{Slowness surfaces and wavefronts}
In view of the above, the TI media are conveniently described by the
so-called slowness surfaces generated by the slowness vectors $[c_m
^\be(\n)]^{-1}\n$ , where $c$ is the wave speed and the unit vector
$\n$ indicates the direction of wave propagation \cite{pay83}. The
slowness curves, cross-sections of these surfaces with $\xi_2=0$ are
shown in Fig.~2~(a). The slowness surfaces can be used to construct
\begin{figure}[ht]
{\bf~a}$\quad\quad\quad\quad\quad\quad\quad\quad\quad\quad\quad\quad\quad\quad\quad\quad\quad\quad\quad${\bf~b}\hfil\break
\hfil \psfrag{xi1}{$\xic_1$}
\psfrag{xi3}{$\xic_3$} \psfrag{a}{}
\includegraphics[height=50mm,width=60mm]{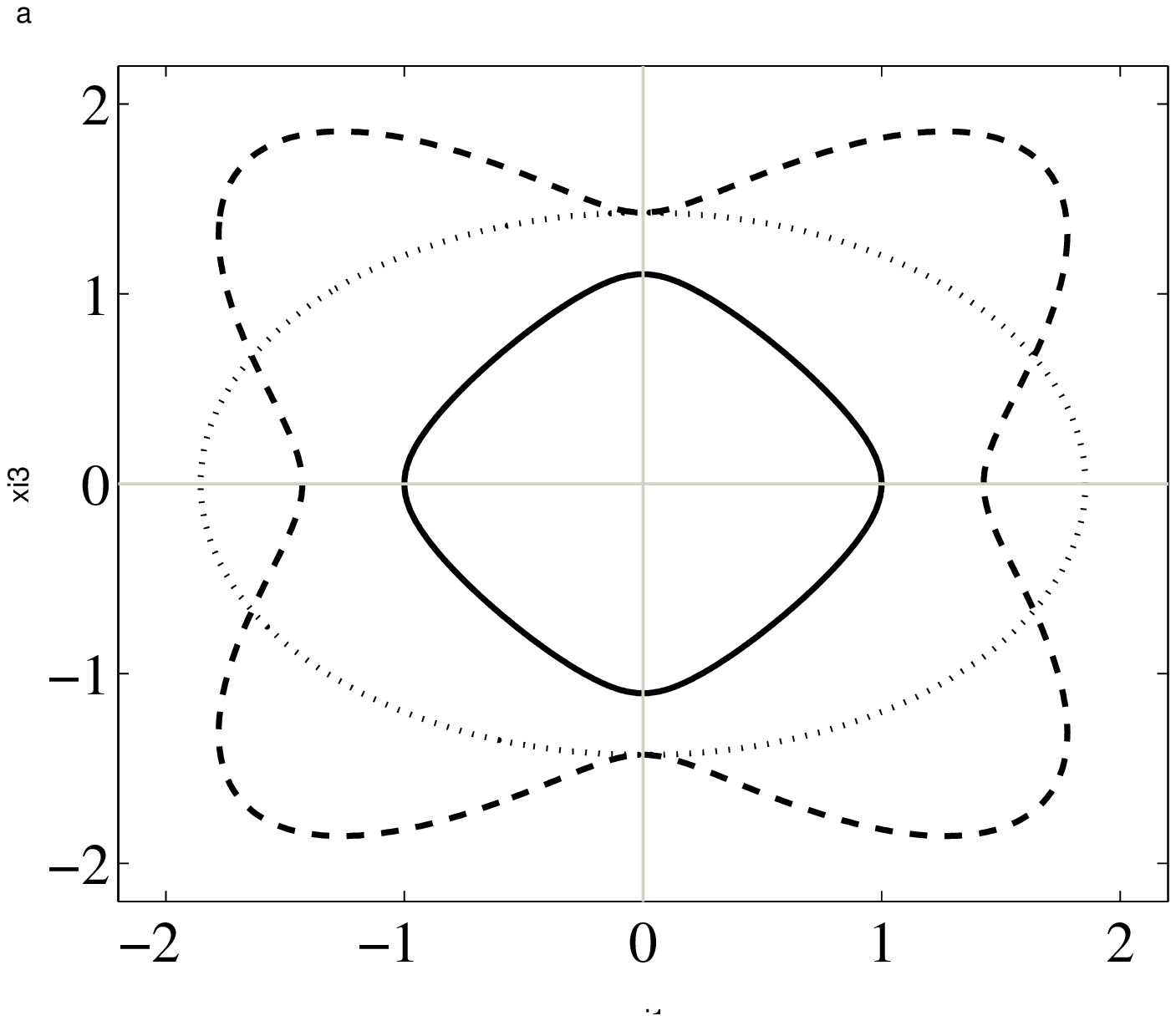}\hfil
\psfrag{x1}{$x_1$} \psfrag{x3}{$x_3$}
\includegraphics[height=50mm,width=60mm]{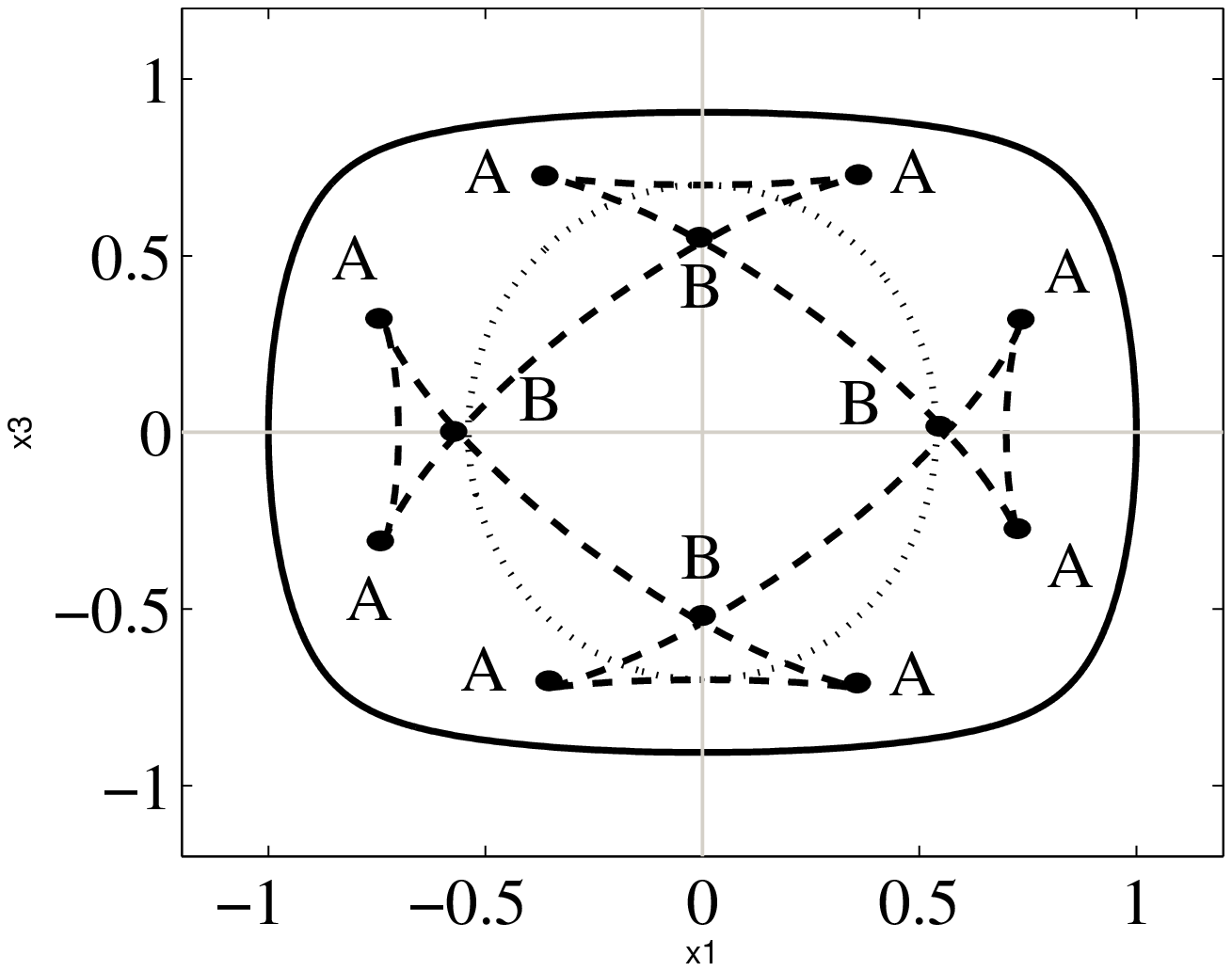}
\hfil \caption{(a) The slowness curves and (b) wave curves for
$\xi_2=0$.  When the wave curve is rotated around the $x_3$ axis,
points A circumscribe the so-called cuspidal edges. Points B are
called the conical points. Key: $qP$ (solid line), $qSV$ (dashed
line) and $qSH$ (dotted line).}
\end{figure}
the wavefronts of waves radiated by a point source at the origin.
Indeed,  the front of each infinitesimal plane wave of type $\beta$
radiated by this source in the direction of a unit vector $\n$ moves
to, after a unit of time, the location $\x$ such that we have\ba
\n\cdot\x=c_m ^\be(\n). \label{wavefront}\ea The envelope of the
above plane fronts is obtained by differentiating the  wavefront
equation (\ref{wavefront}) with respect to $\n$. Therefore, the
front of the wave radiated by the point source is generated by the
rays \ba \x=\frac{\partial c_m ^\be(\n)}{\partial \n}.
\label{ray}\ea  The wave curves, the cross-sections of the above
wave fronts with $\xi_2=0$ are shown in see Fig.~2~(b). Note that
the rays indicate the direction of the energy propagation and are
perpendicular to the slowness surfaces and not the wavefronts (see
Fig.~3~(a)).

\section{The reciprocity theorem (Green's formula)}
\label{sec:intform}

Let us introduce the Cartesian system $\{{\acute \e}_1,{\acute
\e}_2,{\acute \e}_3\}$ associated with the crack, such that its
${\acute \e}_3$ axis is perpendicular to the crack plane, ${\acute
\e}_2$ runs along the crack edge and ${\acute \e}_1$ is the inner
normal to the edge. Note that there exists a matrix of coordinate
rotation $Q$, such that $Q^{- 1}=Q^T$, where the superscript ${}^T$
denotes a transpose, and the medium coordinates $v_i$ of any vector
$\v=\k,\,\n,\,\x, \bxi\,etc.$ are related to its crack coordinates
${\acute v}_j$ via \ba v_i=Q^T_{ij}{\acute v}_j,\,\,i=1,2,3. \ea
 Below,  the bold symbol ${\acute \v}$ is used to
denote the triplet of coordinates of vector $\v$ in the primed
(crack) coordinate system. When working in the unprimed (medium)
coordinate system the symbol $\v$ is retained to denote the triplet
of the corresponding coordinates.  Since the medium is isotropic in
the $\e_1\e_2$-plane, we can always rotate the medium coordinate
system around the axis of symmetry $\e_3$ so that ${\acute \e}_3$,
the normal to the crack plane, lies in the $\e_2\e_3$-plane and
therefore, $\e_1$ lies in  the crack plane ${\acute \e}_1{\acute
\e}_2$. Then without loss of generality, the matrix of coordinate
rotation is \ba Q=\left[
\begin{array}{ccc}
\cos \phi_0 & \sin\phi_0\cos\theta_0 & \sin\phi_0\sin\theta_0 \\
-\sin\phi_0& \cos\phi_0\cos\theta_0 &  \cos\phi_0\sin\theta_0 \\
0 & -\sin\theta_0 & \cos \theta_0
\end{array}
\right], \ea where $\theta_0$ is the angle between the symmetry axis
and normal to the crack, and $\phi_0$ is the angle between $\e_1$
and ${\acute \e}_1$.

We can now use the reciprocity theorem~(see e.g \cite{ach82}) to
reformulate the problem stated in Section \ref{sec:problem} as an
integral equation for the scattered field (otherwise known as the
Green's formula). We start by decomposing the total fields $\u$ and
$\sigma$ as
\begin{eqnarray}
\u=\u^{in} + \u^{sc}, \\
\sigma=\sigma^{in} + \sigma^{sc},
\end{eqnarray}
with $\u^{in}$ and $\sigma^{in}$ representing the respective
incident displacement and stress fields.  The incident displacement
field can be written as
\begin{eqnarray}
\u^{in} (\xbc) = {\widehat \p}^\be(\kbc^{in}) \ex^{\i
\kbc^{in}\cdot\xbc}, \label{def:incwaven}
\end{eqnarray}
where we use the notation
\begin{eqnarray}
\u^{in} (\xbc)=\u^{in}_m (\x)\Bigr|_{\x=Q^T\xbc};\label{u_conv}
\end{eqnarray}
the polarization vector ${\widehat \p}^\be={\widehat \p}^\be(\bxic)$
is
\begin{eqnarray}
{\widehat \p}^\be(\bxic)={\widehat
\p}^\be_m(\bxi)\Bigr|_{\bxi=Q^T\bxic};\label{p_conv}
\end{eqnarray}
the wave vector is $\kbc^{in}=k^\be\nbc^{in}$, where $\nbc^{in}$ is
the propagation direction of the incident wave, given in the primed
(crack) coordinates,
 and the wave number $k^\be=\omega/c^\be(\nbc^{in})$ is
\begin{eqnarray}
k^{in}=[\lambda^\be(\nbc^{in})]^{-1/2},\label{kbe}
\end{eqnarray}
where similarly to (\ref{u_conv}) and (\ref{p_conv}), we use the
notation
\begin{eqnarray}
\lambda^\be(\bxic)=\lambda^\be_m(\bxi)\Bigr|_{\bxi=Q^T\bxic}.
\end{eqnarray}

We note that the incident stress field gives rise to the incident
tractions
\begin{eqnarray}
\t^{in}(\xbc)={\widehat
\t}^\be(\kbc^{in})\ex^{i\kbc^{in}\cdot\xbc},\label{tinc_def}
\end{eqnarray}
and the  boundary condition (\ref{bc}) can be rewritten as \ba
\t^{sc}(\xbc)\Big|_{\Gamma^\pm }= -\t^{in}(\xbc)\Big|_{\Gamma^\pm}.
\label{bc_scat} \ea The amplitude of the incident tractions
${\widehat \t}^\be(\kbc^{in})$ is related to the amplitude of the
incident displacement field ${\widehat \p}^\be$ via
\begin{eqnarray}
{\widehat \t}^\be(\bxic)=\i S(\bxic) {\widehat
\p}^\be(\bxic).\label{t_t}
\end{eqnarray}
The latter relationship is obtained by substituting
(\ref{def:incwaven}) into (\ref{def:transferoperator}), applying the
triple Fourier Transform to the result, changing from $\bxi$ to
$\bxic$ and finally using (\ref{toinF}) and the notation
\begin{eqnarray}
S(\bxic )=S_m(\bxi)\Bigr|_{\bxi=Q^T\bxic}.\label{d_tt}
\end{eqnarray}
As before, the presence (absence) of subscript $m$ means that the
function has the vector argument expressed in the medium (crack)
coordinates.

By taking into account that both crack faces are traction-free, we
then write the reciprocity theorem for the scattered field $\u^{sc}$
as
\begin{eqnarray}
\u^{sc}(\xbc)=  -\int_0^\infty d\yc_1 \int_{-\infty}^\infty d\yc_2\,
t^G(\xbc-\ybc){\bf \Delta} \u^{sc}(\ybc) , \label{integralformU1}
\end{eqnarray}
where   $t^G=\bnu\cdot \sigma^G$ is the Green's traction tensor,
$\bnu={\acute \e}_3$,
$\sigma^G_{ijk}=\Sigma^{(i)}_{j\ell}(\nabla)u^G_{\ell k}$ is the
Green's stress tensor and $\Delta \u^{sc} (\xc_1,\xc_2)$ is the
so-called COD (Crack Opening Displacement) (see e.g \cite{ach82}),
which is defined as
\begin{eqnarray}
\Delta \u^{sc} (\xc_1,\xc_2) = \u^{sc} (\xc_1,\xc_2,0+) - \u^{sc}
(\xc _1,\xc_2,0-).
\end{eqnarray}
Above, $0+$ refers to the upper face of the crack and $0-$ to the
lower.

Let us note that the scattered field is invariant with respect to
translations along the edge of the crack. Therefore, all the fields
$\u, \t, \sigma$ and $\Delta \u^{sc}$ have a common factor ${\rm
exp}(-\i\xic_2 \xc_2)$, where we have \ba \xic_2=-\kc^{in}_2,
\label{xic2} \ea and all can be factorized as follows
\begin{eqnarray}
{\bf v} (\xbc) = {\bf V} (\xc_1,\xc_3;\xic_2) \ex^{-\i \xic_2 \xc
_2}. \label{def:uscfactor}
\end{eqnarray}
Here and everywhere  the argument $\xic_2$ is separated from other
arguments by a semi-column to emphasize the fact that in the problem
under consideration it is just a fixed parameter. Also, the bold
capitals are used everywhere to denote the preexponential factors of
the quantities denoted by the corresponding lower case letters. Then
the integral equation (\ref{integralformU1}) can be rewritten as \ba
&&\U^{sc} (\xc_1,\xc_3;\xic_2) = \nn\\&&-\int_0^\infty d\yc_1
\int_{-\infty}^\infty d\yc_2\, t^G (\xc _1 - \yc_1,\xc
_2-\yc_2,\xc_3) {\bf\Delta}\U^{sc} (\yc_1;\xic_2) \ex^{\i \xic_2
(\xc_2-\yc_2)}, \label{disp:intform} \ea  where the components of
vector functions $\U^{sc},~\Delta \U^{sc}$ and tensor $t^G$ are all
given in medium coordinates while the crack coordinates are used to
represent the components of vector arguments.  This choice leads to
simpler formulas.
\section{The 3D functional equation} \label{sec:functional}
Noting that the $\yc_2-$integral in (\ref{disp:intform}) is actually
a single Fourier transform,  which is denoted everywhere by the bar
$~{\bar {}}~$, and applying the convolution theorem to the
$\yc_1-$integral, the single Fourier transform of
(\ref{disp:intform}) in $\xc_1$ gives us the  vector functional
equation
\begin{eqnarray}
\overline {\U}^{sc} (\xic_1,\xc_3;\xic_2)=  -\widetilde
t^G(\xic_1,\xc_3; \xic_2) \overline{\Delta \U}^{sc} (\xic_1;\xic_2).
\label{eq:last}
\end{eqnarray}
Above and everywhere below, the tilde $~{\widetilde {}}~$ denotes
the double Fourier transform The tensor ${\widetilde
t}^G(\xic_1,\xc_3;\xic_2)$ is the inverse Fourier transform in
$\xic_3$ of the traction tensor ${\widehat t}^G(\bxic)$, which is
given by
\begin{eqnarray}
\widehat {t}^G (\bxic) = \frac { {\widehat \p}^\be (\bxic) {\widehat
\t}^\be(\bxic)} {1-\lambda^\be(\bxic)  }. \label{def:teigen}
\end{eqnarray}
This means that we can write \ba {\widetilde
t}^G(\xic_1,\xc_3;\xic_2)= \frac1{2\pi} \int_{-\infty}^\infty
{\widehat t}^G(\bxic) \ex^{-\i\xic_3 \xc_3} d\xic_3. \label{int} \ea
Note that as $\xic_3\rightarrow\infty$, ${\widehat
t}^G(\bxic)=O(\xic_3^{-1})$. Therefore, above we can apply the
Jordan Lemma if for $\xc_3
>0+$ the contour of integration  is closed in the lower
$\xic_3$-plane and for $\xc_3<0-$, in the upper half-plane. In both
cases considered below, for each $\xic_1$, inside the chosen
contour, the Transfer Tensor has two poles
$\xic^{\alpm}_3(\xic_1;\xic_2)\,\,\alpha=1,\,2$, which are roots of
the quartic expression in (\ref{eq:delta_quad}) with the
corresponding $\lambda^\al(\bxic)=1$, and  one pole
$\xic^{(3)\pm}_3(\xic_1;\xic_2)$, which is a root of the quadratic
expression in (\ref{eq:delta_quad}) with $\lambda^{(3)}(\bxic)=1$.
Above, the top sign is chosen when $\xc_3
>0+$ and bottom when $\xc_3<0-$; and for the cases under consideration,
$\xic^\alpm_3(\xic_1;\xic_2)=\mp\xic^\al_3(\xic_1;\xic_2)$; where
the functions $\xic^{\al}_3(\xic_1;\xic_2)$ are defined in
(\ref{trick_add}) and (\ref{trick_add2}), respectively. The real
parts of functions (\ref{trick_add}) are shown in Fig.~3~(b). When
$\xic^{\al}_3(\xic_1;\xic_2)$ run along slowness surfaces the
functions have no  imaginary parts. Fig.~3~(b) shows that $\alpha=1$
always describes the $qSV$ mode, $\alpha=3$---the $qSH$ mode and
$\alpha=2$---the $qP$ or $qSV$ mode, depending on $\xic_1$. The
functions (\ref{trick_add2}) exhibit analogous behavior.
\begin{figure}[ht]
{\bf~a}$\quad\quad\quad\quad\quad\quad\quad\quad\quad\quad\quad\quad\quad\quad\quad\quad\quad\quad\quad${\bf~b}\hfil\break
\hfil
 \psfrag{x}{$x_1$} \psfrag{y}{$x_3$}
 \psfrag{a}{\bf a}
\includegraphics[height=50mm,width=60mm]{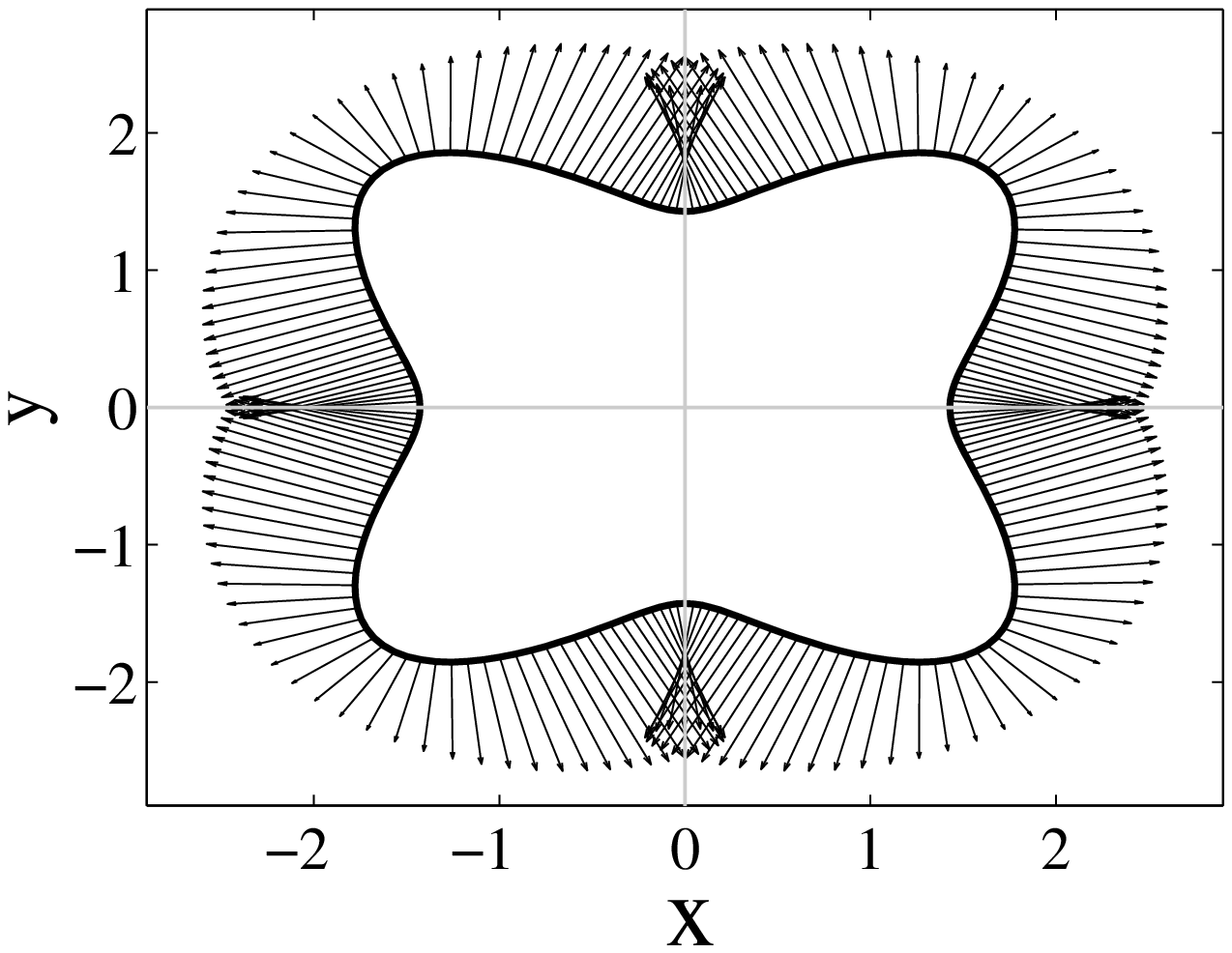}
\hfil
 \psfrag{k1}{$\kappa_1$} \psfrag{k2}{$\kappa_2$}
 \psfrag{k3}{$\kappa_3$}\psfrag{k4}{$\kappa_4$}
 \psfrag{xi1}{$\xic_1$} \psfrag{xi3}{$\xic_3$}
 \psfrag{b}{\bf b}
\includegraphics[height=50mm,width=60mm]{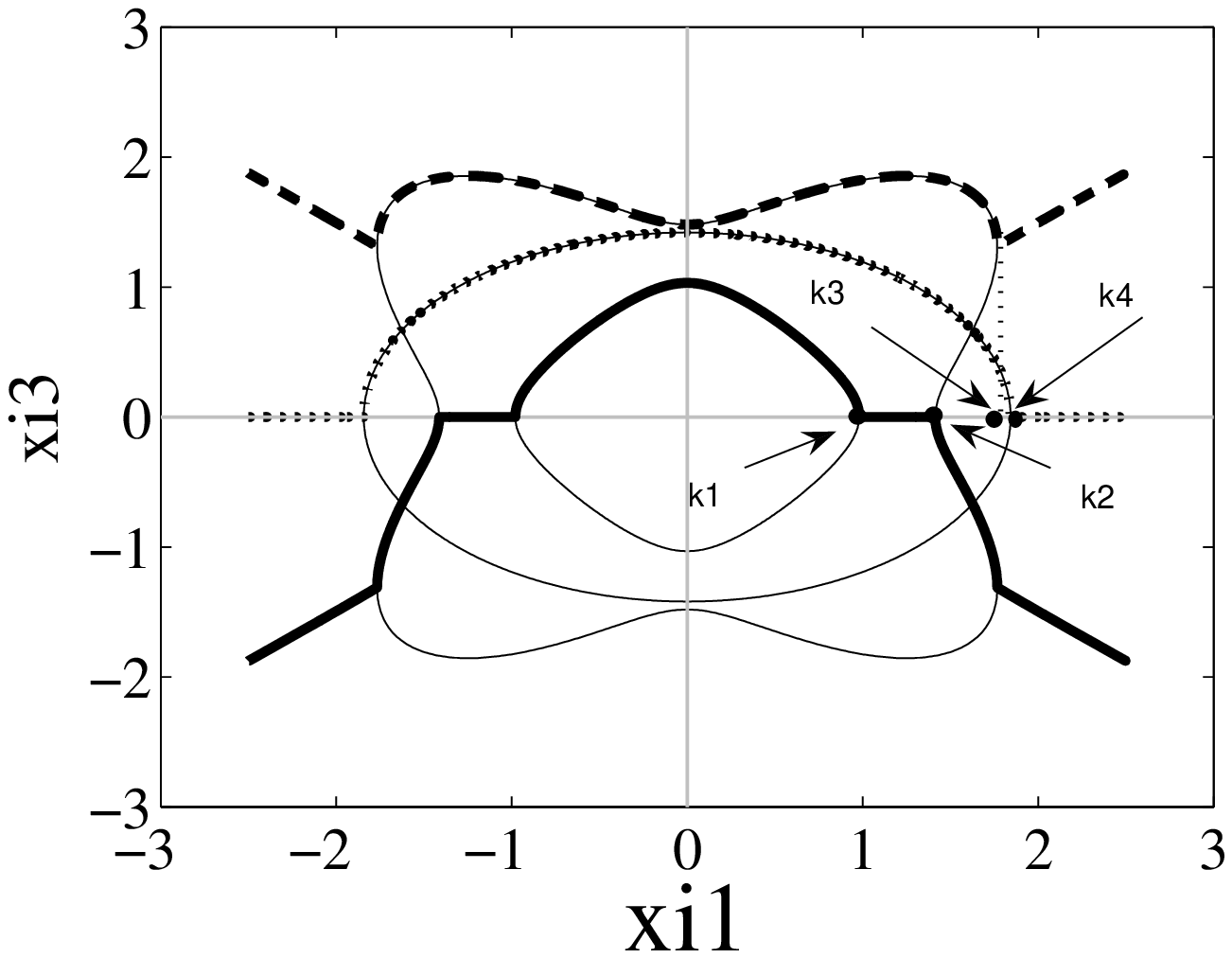}\hfil
\caption{(a) The $qSV$ energy flow (the mean Poynting) vectors $
{\bf P} =  -\frac{1}{2} {\rm Im~} [\omega \sigma^d \cdot
(\bf{u^d)}^* ]$, where the star denotes the complex conjugate.
(b)~The real parts of three roots $\xic^{(1)}_3(\xic_1;\xic_2)$
(dashed line), $\xic^{(2)}_3(\xic_1;\xic_2)$ (solid line) and
$\xic^{(3)}_3(\xic_1;\xic_2)$ (dotted line) imposed on the slowness
curves in Fig.~2~(a). When lying on the slowness curves the roots
are real. When sliding of these curves at the branch points
$\xic_1=\kappa_1, \kappa_2, \kappa_3$ or $\kappa_4$ the roots are
complex.}
\end{figure}

On adding a small negative imaginary part to all $c^\al$ (a small
positive part to the wave numbers), each real  pole moves away from
the real axis in the $\xic_3-$plane, with the sign of the resulting
$-\Im \xic^\alpm_3(\xi_1,c^\al)$ the same as the sign of the partial
derivative $\partial \xic^\alpm_3 (\xi_1,c^\al)/\partial c^\al$. In
other words, the resulting $-\Im \xic^\alpm_3(\xi_1,c^\al)$ has the
same sign as the vertical component of the gradient to the slowness
surface at $\xic_1$ (see Figs.~2~(a) and 3~(b)), that is  the sign
of the vertical component of the energy flux (see Fig.~3~(a)). Since
${\widehat {t}^G}(\bxic)$ in (\ref{def:teigen}) is a rational
function of $\xic_3$  we can now evaluate the integral (\ref{int})
using the Cauchy Residue Theorem to obtain \ba {\widetilde
t}^G(\xic_1,\xc_3;\xic_2)= \mp\i~ \frac { {\widehat \p}^\al (\bxic)
{\widehat \t}^\al(\bxic)} {\lambda^\al_{\xic_3}(\bxic)}\Big
|_{\xic_3=\xic^\alpm_3 (\xic_1;\xic_2);~\lambda^\al(\bxic)=1}
\ex^{-\i \xic^\alpm_3(\xic_1;\xic_2)\xc_3\,\,}, \label{tGres} \ea
where the subscript $\xic_3$ indicates differentiation with respect
to $\xic_3$ and the resulting plane waves satisfy the radiation
condition at infinity.  For the two cases considered in this paper
the denominators in (\ref{tGres}) are specified below in
(\ref{den_perp}) and (\ref{den_par}), respectively.
 We also note that ${\widetilde t}^G(\xic_1,0-;\xic_2)-{\widetilde
t}^G(\xic_1,0+;\xic_2)=I$, the identity matrix. This can be best
seen by using (\ref{eq:last}) to evaluate ${\widetilde
\U}^{sc}(\xic_1,0+;\xic_2)-{\widetilde \U}^{sc}(\xic_1,0-;\xic_2)$.

In order to utilize the boundary condition (\ref{bc}) in a
straightforward manner, we  apply the operator $-\i{
S}(\xic_1,\xic_2,\i{\acute \partial}_3)$ to Eq. (\ref{eq:last}) and
then write the resulting equation for the crack face $\xc_3=0$ to
obtain
\begin{eqnarray}
\overline \T^{sc} (\xic_1,0;\xic_2)=  -\widetilde
\tau^G(\xic_1,0;\xic_2) \overline{\Delta \U}^{sc} (\xic_1;\xic_2),
\label{eq:lastT}
\end{eqnarray}
where ${S}(\xic_1,\xic_2,\i{\acute \partial}_3)={ S}(\bxic)$, with
$\i\partial/\partial\xc_3$ substituted for $\xic_3$, and we have \ba
{\widetilde \tau}^G(\xic_1,\xc_3;\xic_2)= \pm\i~ \frac { {\widehat
\t}^\al (\bxic) {\widehat \t}^\al(\bxic)}
{\lambda^\al_{\xic_3}(\bxic)}\Big
|_{\xic_3=\xic^\alpm_3(\xic_1;\xic_2);~\lambda^\al(\bxic)=1}
\ex^{-\i \xic^\alpm_3(\xic_1;\xic_2)\xc_3\,\,}.\label{tau} \ea Note
that on the crack plane, $\T^{sc}$ must be a continuous function of
$\x$, so that $\overline {\T}^{sc} (\xic_1,0+;\xic_2)=\overline
{\T}^{sc} (\xic_1,0-;\xic_2)$ and therefore, ${\widetilde
\tau}^G(\xic_1,0;\xic_2)$ is well-defined.  This  can be verified
independently by expanding ${\widehat \tau}^G_{ij}(\bxic)$, the
inverse Fourier transform in $\xic_3$ of (\ref{tau}),  into the
Laurents series, to verify that for any crack orientation, its
components exhibit the following behavior at infinity:\ba {\widehat
\tau}^G_{ij}(\bxic)={\rm constant~ matrix}
+O(\frac{1}{\xic_3^2}),\,\, {\rm as~} \xic_3\rightarrow\infty.\ea
Then ${\widetilde \tau}^G(\xic_1,0;\xic_2)$ is well-defined via the
following consideration: \ba {\widetilde
\tau}^G(\xic_1,0+;\xic_2)-{\widetilde
\tau}^G(\xic_1,0-;\xic_2)=\lim_{R\rightarrow \infty}\oint {\widehat
\tau}^G(\bxic)~ d\xic_3=0,\ea where the integration contour is a
circle of radius $R$, circumscribed anticlockwise and centered at
the origin of coordinates. In view of the above, below we use the
simplified notations \ba \T^{sc} (\xc_1;\xic_2)=\T^{sc}
(\xc_1,0;\xic_2),\,\,\,\, {\widetilde
\tau}^G(\xic_1;\xic_2)={\widetilde \tau}^G(\xic_1,0;\xic_2),\ea

We can now make use of the boundary conditions (\ref{bc_scat}) and
apply the Fourier Transform to (\ref{tinc_def}) to put
 the functional equation (\ref{eq:lastT}) in the Wiener-Hopf form,
\begin{eqnarray}
\i\frac{{\widehat \t}^\be(\kbc^{in})}{\xic_1+\kc^{in}_1+i0}-
{\overline \T}^{sc-} (\xic_1;\xic_2)=\widetilde
\tau^G(\xic_1;\xic_2) \overline{\Delta \U}^{sc}(\xic_1;\xic_2),
\label{eq:functional}
\end{eqnarray}
where $\overline {\Delta \U}^{sc} (\xic_1;\xic_2)$ and ${\overline
\T}^{sc-}(\xic_1;\xic_2)$ are both unknown. Since $\Delta
\U^{sc}(\xc_1;\xc_2)={\bf 0}$ and $\T^{sc-}(\xc_1;\xc_2)={\bf 0}$
for $\xc_1\leq 0$ and $\xc_1\geq 0$,  respectively,
 their Fourier transforms $\overline{\Delta \U}^{sc}(\xic_1;\xic_2)$ and
 $\overline {\T}^{sc-}(\xic_1;\xic_2)$
 are analytic  in the upper and lower half of the complex
 $\xic_1$-plane, respectively.   This can be seen by studying the corresponding single Fourier integral.  Above, we also use the fact that we have
\begin{eqnarray}
\int_{0}^\infty \ex^{\i[\xic_1+\kc^{in} _1] \xc_1} d\xc_1 =
\frac{\i}{\xic_1+\kc^{in}_1 +i0}. \label{FTofH}
\end{eqnarray}

\section{Solution of the 3D functional equation when the symmetry axis is
perpendicular to the crack plane} \label{sec:case_per}
\subsection{The poles and branch points of ${\hat t}^G(\bxic)$}
\label{poles_perp} When the symmetry axis is perpendicular to the
crack plane, $\phi_0=\theta_0=0$, $Q=I$ and thus, all the
$\xi$-components coincide with the corresponding $\xic$-components.
As mentioned above, the functions $\pm
\xic^\al_3(\xic_1;\xic_2),\,\alpha=1,\,2$ and $\pm
\xic^{(3)}_3(\xic_1;\xic_2)$ which are the poles of the tensor $\hat
t^G(\bxic)$ are the respective roots of the quartic  and quadratic
expression in (\ref{eq:delta_quad}), with $\lambda^\al(\bxic)=1$, so
that we have \ba &&\xic^{(1)}_3(\xic_1;\xic_2)
=\xic_{31}(\xic_1;\xic_2)+\xic_{32}(\xic_1;\xic_2), \nn \\
&&\xic^{(2)}_3(\xic_1;\xic_2)=\xic_{31}(\xic_1;\xic_2)-\xic_{32}(\xic_1;\xic_2), \nn \\
&&\xic^{(3)}_3(\xic_1;\xic_2)=(B_1^{-1}B_3)^{1/2}\gamma_4(\xic_1;\xic_2),
\label{trick_add}\ea with \ba
&&\xic_{3n}(\xic_1;\xic_2)=\frac{\gamma_{6-3n}}{2(B_1A_{33})^{1/2}}\times\nn\\
&& \Bigl \{ \frac
{[A_{13}^2+2B_1A_{13}-A_{11}A_{33}]\xic_\bot^2+B_1+A_{33}- (-1)^n
2B_1(A_{11}A_{33})^{1/2}\gamma_1\gamma_2} { \gamma^2_{6-3n} }
\Bigr \}^{1/2}, \nn\\
&&\,\,\,\,\,\,\,\,\,\,\,\,\,\,\,\,\,\,\,\,\,\,\,\,\,\,\,\,\,\,\,\,\,\,\,\,\,\,
\,\,\,\,\,\,\,\,\,\,\,\,\,\,\,\,\,\,\,\,\,\,\,\,\,\,\,\,\,\,\,\,\,\,\,\,\,\,\,
\,\,\,\,\,\,\,\,\,\,\,\,\,\,\,\,\,\,\,\,\,\,\,\,\,\,\,\,\,\,\,\,\,\,\,\,\,\,\,
\,\,\,\,\,\,\,\,\,\,\,\,\,\,\,\,\,\,\,\,\,\,\,\,\,\,\,\,\,\,\,\,\,\,\,\,\,\,\,\,
\,\,\,\,\,\,\,\,\,\,\,\,\,\,\,\,\,\,\,\,\,\,
n=1,2,\label{trick_add_d}
 \ea $\{\}^{1/2}$ denoting the principal value, \ba
&&\gamma_0=\gamma_0(\xic_1;\xic_2)=(\kappa_0^2+\xic_1^2)^{1/2} \nn \\
&&\gamma_\ell=\gamma_\ell(\xic_1;\xic_2)=[(\kappa_\ell+\i0)^2-\xic_1
^2]^{1/2}, \,\,\,\,\,\,\ell=1,2,3,4, \ea with the branch points $
\pm\i\kappa_0$ and $\pm\kappa_\ell,\,\ell=1,2,3,4$, where we have
\ba
&&\kappa_0=(C_0^2+\xic_2^2)^{1/2}, \nn \\
&&\kappa_1= (A_{11}^{-1}-\xic_2^2)^{1/2}, \nn \\
&&\kappa_2= (B_1^{-1}-\xic_2^2)^{1/2}, \nn \\
&&\kappa_3=(C_3^2-\xic_2^2)^{1/2}, \nn \\
&&\kappa_4= (B_3^{-1}-\xic_2^2)^{1/2} \ea and \ba
&&C_\ell=D^{-1}\{2(A_{33}B_1)^{1/2}(B_1+A_{13})[A_{13}^2-
A_{11}A_{33}+B_1(A_{11}+2A_{13}+A_{33})]^{1/2}-
\nn\\
&&(-1)^\ell[A_{13}(2B_1+A_{13})(B_1+A_{33})+A_{33}(2B_1^2+A_{11}B_1-
A_{11}A_{33})]\}^{1/2},
\,\,\,\ell=0,3, \nn\\
&&D=\{(A_{11}A_{33}-A_{13}^2)[(2B_1+A_{13})^2-A_{11}A_{33}]\}^{1/2}.
\label{CD} \ea It can be checked, e.g.~using MATHEMATICA, that when
$n=2$ ($n=1$), $\pm\i\kappa_0$  ($\pm\kappa_3$) is a zero of the
corresponding numerator in the curly brackets in
(\ref{trick_add_d}). Therefore, the square roots of the expressions
in the curly brackets never vanish. Note that all branch points are
either real or purely imaginary and for the austenitic steels that
are described e.g.~in \cite{grid} and are considered in this
article, direct evaluation shows that we have the following
inequalities: \ba
&&\Re \kappa_1 \le \Re \kappa_2 \le \Re \kappa_3 \le \Re \kappa_4,\nn\\
&&\Im \kappa_4 \le \Im \kappa_3 \le \Im \kappa_2 \le \Im \kappa_1
\le \kappa_0. \label{kappa_con} \ea Therefore, the phases of the
principal values of $\xic_{32}$ and $\xic_{31}$ are determined by
the position of $\xic_1$ on the complex $\xic_1$-plane relative to
$\kappa_2$ and $\kappa_3$.  For a fixed values of $\xic_2$, the real
parts of functions $\xic^\al_3, \alpha=1,2,3$ described by equations
(\ref{trick_add}) are represented in Fig.~3~(b).

\subsection{Decoupling  the normal COD from the tangential COD}    In the case
under consideration, $t'_i=\sigma'_{i3}$ and   the crack and medium
coordinates coincide.  Therefore, in view of (\ref{d_tt})
 the transfer operator $S(\bxic)$ is
 \ba {S}(\bxic)= \left[
\begin{array}{ccc}
B_1\xic_3 & 0 & B_1\xic_1 \\
0 & B_1\xic_3 &  B_1\xic_2 \\
A_{13}\xic_1 &  A_{13}\xic_2 & A_{33}\xic_3
\end{array}
\right]. \ea It can be checked, e.g.~using  MATHEMATICA, that we can
write
\begin{eqnarray}
{\widetilde \tau}^G(\xic_1;\xic_2) = \left[
\begin{array}{ccc}
\widetilde\tau^{red}(\xic_1;\xic_2) & 0 \\
0 & \mu_3(\xic_1;\xic_2)
\end{array}
\right], \label{taured}
\end{eqnarray}
where we have
\begin{eqnarray}
\mu_3(\xic_1;\xic_2)=
\Bigl(\frac{A_{33}}{A_{11}}\Bigr)^{1/2}\frac{\gamma_2(\xic_1;\xic_2)}{\gamma_1(\xic_1;\xic_2)}
\mu_1(\xic_1;\xic_2),\nn\\\label{mus}
\end{eqnarray}
and
\begin{eqnarray}
\widetilde\tau^{red}(\xic_1;\xic_2)=\AA^{-1}(\xic_1;\xic_2)\DD(\xic_1;\xic_2)\AA(\xic_1;\xic_2),
\end{eqnarray}
with \ba \AA(\xic_1;\xic_2)=\left[
\begin{array}{ccc}
\xic_1 & \xic_2 \\
\xic_2 & -\xic_1
\end{array}
\right], \label{def:A}\ea

\ba \DD(\xic_1;\xic_2) = \left[
\begin{array}{ccc}
\mu_1(\xic_1;\xic_2) & 0 \\
0 & \mu_2(\xic_1;\xic_2)
\end{array}
\right], \label{def:D} \ea

\begin{eqnarray}
&&\mu_1(\xic_1;\xic_2)=\frac{\i R(\xic_1;\xic_2)}{4A_{33}\xic_{31}(\xic_1;\xic_2)},\nn\\
&&\mu_2(\xic_1;\xic_2)=\frac{1}{2}\Bigl (-B_1B_3 \Bigr
)^{1/2}\gamma_4(\xic_1;\xic_2),
\end{eqnarray}
and the Rayleigh function is defined by
\ba
R(\xic_1;\xic_2)=(A_{13}^2-A_{11}A_{33})(\xic_1^2+\xic_2^2)+A_{33}+
(A_{11}A_{33})^{1/2}
\frac{\gamma_1(\xic_1;\xic_2)}{\gamma_2(\xic_1;\xic_2)}. \ea
 Then substituting
(\ref{taured}) into the 3D vector functional equation
(\ref{eq:functional}), the latter  decouples into  the scalar
equation describing the COD component normal to the crack face and
2D vector functional equation describing the tangential COD
components,
\begin{eqnarray}
\mu_3(\xic_1;\xic_2) \overline{\Delta U}^{sc}_3(\xic_1;\xic_2) =
\i\frac{{\widehat t}^\be_3(\kbc^{in})}{\xic_1+\kc^{in}_1+\i0} -
\overline T_3^{sc-}(\xic_1;\xic_2) \label{eq:functionalsc1}
\end{eqnarray}
and
\begin{eqnarray}
\DD(\xic_1;\xic_2)\overline{\Delta \UU}^{sc}(\xic_1;\xic_2) =
\i\frac{\tt^\be(\kbc^{in})}{\xic_1+\kc^{in}_1+i0}-\overline
\TT^{sc-}(\xic_1;\xic_2), \label{eq:functionalsc2}
\end{eqnarray}
 where the calligraphic script is used to denote two dimensional vectors  expanded in the eigenvectors of
 $\widetilde \tau^{red}$, so that  for any vector $\AAb=\overline{\Delta \UU}^{sc},\,\overline
\TT^{sc-}$ or $\tt^\be$  we have \ba &&{\cal
A}_j(\xic_1;\xic_2)=\AA_{jk}(\xic_1;\xic_2)A_k(\xic_1;\xic_2),
\,\,\,\,j,k=1,2, \ea where we remind the reader that the repeated
index summation is applied to $k=1,\,2.$
\subsection {Solving the scalar functional equation for the normal COD}
First, let us consider the scalar  equation
(\ref{eq:functionalsc1}). It can be solved using the Wiener-Hopf
technique by introducing the following factorization of the
tangential traction $\mu_3 $:
\begin{eqnarray}
\mu_3 (\xic_1;\xic_2) =
\mu^+_3(\xic_1;\xic_2)\mu^-_3(\xic_1;\xic_2),
\end{eqnarray}
where $+(-)$ means that the function is analytic in the upper
(lower) half of the complex $\xic_1$-plane. In order to effect this
factorization, we first note that there exist values
$\xic_1=\pm\kappa_R,\,\kappa_R=(k_R^2-\xic_2^2)^{1/2}$ such that \ba
R(\pm\kappa_R;\xic_2)=0, \ea which are called the Rayleigh poles.
Above,  $k_R=c_0/c_R,$ where $c_R$ is the speed of the surface
Rayleigh wave. Only the pole $-\kappa_R$ corresponds to the outgoing
wave. Then the considerations and notations introduced in Appendices
A and B allow us to define $\mu^\pm_1(\xic_1;\xic_2)$, \ba
\mu^\pm_1(\xic_1;\xic_2)=\Bigl(\frac{\i l_0}{4l_1
A_{33}}\Bigr)^{1/2} \frac{K_0^ \pm(\xic_1;\xic_2)}{\gamma_3^
\pm(\xic_1;\xic_2)K_1^\pm(\xic_1;\xic_2)}(\kappa_R\pm \xic_1), \ea
and therefore, $\mu^\pm_3(\xic_1;\xic_2)$, \ba
\mu^\pm_3(\xic_1;\xic_2)=
\Bigl(\frac{A_{33}}{A_{11}}\Bigr)^{1/4}\frac{\gamma^\pm_2(\xic_1;\xic_2)}{\gamma^\pm_1(\xic_1;\xic_2)}
\mu^\pm_1(\xic_1;\xic_2), \ea where throughout we use the notations
\ba &&\gamma^\pm_0(\xic_1;\xic_2)=({\i\kappa_0\pm\xic_1})^{1/2},\nn\\
&&\gamma^\pm_\ell(\xic_1;\xic_2)=({\kappa_\ell\pm\xic_1})^{1/2},\,\,\ell=1,...,4.\label{gammapm}
\ea

To continue, as mentioned at the end of Section 5, $\overline{\Delta
\U}^{sc} (\xic_1;\xic_2)$ is analytic in the upper half of the
$\xic_1$-plane. Hence Eq.~(\ref{eq:functionalsc1}) can be rewritten
as
\begin{eqnarray}
-\mu_3^+(\xic_1;\xic_2) \overline{\Delta U}^{sc}_3(\xic_1;\xic_2)
+\i\frac{ {\widehat
t}^\be_3(\kbc^{in})}{(\xic_1+\kc^{in}_1+i0)\mu_3^-(-\kc^{in}_1;\xic_2)}
= \overline T_3^{mod-}, \label{eq:funct1}
\end{eqnarray}
where the superscript ${}^{mod}$ stands for modified and we have
\begin{eqnarray}
\overline T_3^{mod-} = \frac{\overline
T_3^{sc-}(\xic_1;\xic_2)}{\mu_3^-(\xic_1;\xic_2)} + \i\bigl[
\frac{1}{\mu_3^-(-\kc^{in}_1;\xic_2)}-\frac{1}{\mu_3^-(\xic_1;\xic_2)}
\bigr] \frac{ {\widehat t}^\be_3(\kbc^{in})}{\xic_1+\kc^{in}_1+i0}.
\end{eqnarray}
Note that the modification leading to $\overline T_3^{mod-}$ has
been introduced to assure that the left-hand (right-hand) side of
(\ref{eq:funct1}) is analytic  in the upper (lower) half of the
complex $\xic_1-$plane.  This can be true only if both sides are one
and the same entire function. Furthermore it is easy to check that
at infinity we have
\begin{eqnarray}
\mu_3^+ (\xic_1;\xic_2) = O(\xic_1^{1/2}).
\end{eqnarray}
Also, the tip condition (\ref{tip_cond}) implies that ${\bf \Delta}
\U^{sc}(\xc_1,\xc_2)=\O(\xc_1^{1/2})$, which in its turn implies \ba
\overline{\Delta \U}^{sc} (\xic_1;\xic_2)= \O (\xic_1^{-3/2}). \ea
It follows that as $\xic_1\to\infty$, the left-hand side of Eq.
(\ref{eq:funct1}) and therefore, its right-hand side are both
functions of order $O([\xic_1]^{-1})$, obviously  bounded. According
to the Liouville's Theorem, any such entire function is in fact
zero. Therefore, Eq. (\ref{eq:funct1}) implies
\begin{eqnarray}
\overline{\Delta  U}^{sc}_3(\xic_1;\xic_2) = \i\frac{{\widehat
t}^\be_3(\kbc^{in})}{(\xic_1+\kc^{in}_1+i0)\mu^+_3(\xic_1;\xic_2)\mu^-
_3(-\kc^{in}_1;\xic_2)}. \label{def:usc}
\end{eqnarray}

\subsection {Solving the 2D vector functional equation for the tangential COD components}
Let us now turn to the 2D vector functional equation
(\ref{eq:functionalsc2}). The matrix $M$ can be factorized, \ba
M(\xic_1;\xic_2)=M^+(\xic_1;\xic_2)M^-(\xic_1;\xic_2), \ea so that
the matrix \ba M^+(\xic_1;\xic_2)= \left[
\begin{array}{ccc}
\mu_1^+(\xic_1;\xic_2) & 0 \\
0 & \mu_2^+(\xic_1;\xic_2)
\end{array}
\right] \ea is analytic in the upper half of the complex
$\xic_1-$plane, and matrix \ba M^-(\xic_1;\xic_2)= \left[
\begin{array}{ccc}
\mu_1^-(\xic_1;\xic_2) & 0 \\
0 & \mu_2^-(\xic_1;\xic_2)
\end{array}
\right] \ea is analytic and has no zeros in the lower half of this
plane.  The factorization of $\mu_1(\xic_1;\xic_2)$ has been
described in the previous section, and $\mu_2$ can be readily
factorized into the factors \ba \mu^\pm_2(\xic_1;\xic_2)=\Bigl
(-\frac{B_1B_3}{4} \Bigr )^{1/4}\gamma_4^\pm(\xic_1;\xic_2). \ea
Below we also use the matrices $N^\pm$ which are respective inverses
of $M^\pm.$ Now, multiplying (\ref{eq:functionalsc2}) by matrix
$\EEm(\xic_1;\xic_2)$ gives us
\begin{eqnarray}\label{eq:2dfunc-step1}
\DD^+(\xic_1;\xic_2)\overline{\Delta
\UU}^{sc}(\xic_1;\xic_2)=\i\EEm(\xic_1;\xic_2)\frac{
\tt^\be(\kbc^{in})}{\xic_1+\kc^{in}_1+i0}-\EEm(\xic_1;\xic_2)\overline
\TT^{sc-}(\xic_1;\xic_2).\nn\\
\end{eqnarray}
It can be transformed into
\begin{eqnarray}
-\DD^+(\xic_1;\xic_2)\overline{\Delta
\UU}^{sc}(\xic_1;\xic_2)+\i\EEm(-\kc^{in}_1;\xic_2)\frac{
\tt^\be(\kbc^{in})}{\xic_1+\kc^{in}_1+i0}=\overline
\TT^{mod-}(\xic_1;\xic_2),\label{eq:2dfunc-step2}
\end{eqnarray}
where we have
\begin{eqnarray}\label{def:tmod}
&&\overline \TT^{mod-}(\xic_1;\xic_2) =
\nn\\
&&\EEm(\xic_1;\xic_2)\overline \TT^{sc-}(\xic_1;\xic_2) + \i\bigl[
\EEm(-\kc^{in}_1;\xic_2)-\EEm(\xic_1;\xic_2) \bigr] \frac{
\tt^\be(\kbc^{in})}{\xic_1+\kc^{in}_1+i0},\nn\\
\end{eqnarray}
so that the left-hand (right-hand) side of (\ref{eq:2dfunc-step2})
is analytic in the upper (lower) half of the complex $\xic_1-$plane.
This means that both are one and the same entire function. It is
easy to see that as $\xic_1 \rightarrow \infty$, this function is
$O(1)$ and therefore, bounded. According to Liouville's Theorem, any
such entire function is in fact a constant.  Let us call it $\g$.
Then multiplying both sides of (\ref{eq:2dfunc-step2}) first by
$(\DD^+)^{-1}(\xic_1;\xic_2)=\EEp(\xic_1;\xic_2)$ and then by
$\AA^{-1}(\xic_1;\xic_2)=\AA(\xic_1;\xic_2)/(\xi_1^2+\xi_2^2)$ we
obtain
\begin{eqnarray}
\overline{\Delta U}^{sc}_i(\xic_1;\xic_2)&=&-\Big[
\AA^{-1}(\xic_1;\xic_2)\EEp(\xic_1;\xic_2)\cdot\nn\\
&&\Big ( \frac{-\i\EEm(-\kc^{in}_1;\xic_2)\AA(\xic_1;\xic_2)\tt
^\be(\kbc^{in})}{\xic_1+\kc^{in}_1+\i0}+\g \Big )\Big]_i
,\nn\\
\overline T^{sc-}_i(\xic_1;\xic_2)_i&=
&\Big[\AA^{-1}(\xic_1;\xic_2)\DD^-(\xic_1;\xic_2) \cdot\nn\\
&&\Big ( \frac{-\i[\EEm(-\kc^{in}_1;\xic_2)-\EEm(\xic_1;\xic_2) ]
\AA(\xic_1;\xic_2)\tt^\be(\kbc^{in})}{\xic_1+\kc^{in}_1+i0}+\g\Big
)\Big]_i\,\,\,\,i=1,2, \nn\\ \label{sys_for_g}
\end{eqnarray}
where we have employed the formula (\ref{t_t}). In order to find
$\g$, we note that in view of (\ref{def:A}), the plus function
$\overline{\Delta \U}^{sc}(\xic_1;\xic_2)$ has an apparent pole at
$\xic_1=\i \xic_2$; similarly, the minus function $\overline
\T^{sc-}(\xic_1;\xic_2)$ has an apparent pole at $-i\xic_2$. Since
this is impossible, both  corresponding residues must vanish. This
gives us the following linear system for $\g$:
\begin{eqnarray}
&&\AA(\i\xic_2;\xic_2)\EEp(\i\xic_2;\xic_2)\Big (
\frac{-\i\EEm(-\kc^{in}_1;\xic_2)\AA(\i\xic_2;\xic_2)\tt^\be(\kbc^{in})
}{\i\xic_2+\kc^{in}_1+\i0}+\g \Big ) ={\bf 0},\nn
\\
&&\AA(-\i\xic_2;\xic_2)\DD^-(-\i\xic_2;\xic_2)\cdot\nn\\
&& \Big ( \frac{-\i[\EEm(-\kc^{in}_1;\xic_2)-\EEm(-\i\xic_2;\xic_2)
] \AA(-\i\xic_2;\xic_2)\tt^\be
(\kbc^{in})}{-\i\xic_2+\kc^{in}_1+i0}+\g\Big )={\bf 0}.
\end{eqnarray}
Since $|\AA(\i \xic_2;\xic_2)|=|\AA(-\i \xic_2;\xic_2)|=0$, only two
of the four equations above are linearly independent and thus, $\g$
is well defined.

\section{Solution of the 3D functional equation
when the symmetry axis lies in the crack plane perpendicularly to
the crack edge} \label{sec:case_par}
\subsection{The poles and branch points of ${\hat t}^G(\bxic)$}

When the symmetry axis lies in the crack plane perpendicularly to
the crack edge we have $\phi_0=\theta_0=\pi/2$,
 \ba Q=\left[
\begin{array}{ccc}
0 & 0 & 1 \\
-1& 0 &  0 \\
0 & -1 & 0
\end{array}
\right], \label{perQ-matrix} \ea and therefore, \ba
\xi_1=-\xic_2,\,\xi_2=-\xic_3,\, \xi_3=\xic_1. \label{2dcoordtrans}
\ea The case of a normal incidence,  $\xic_2\equiv 0$, lends itself
to an easy analytical treatment. Indeed, the poles of $\tilde
t^G(\xic_1;0)$, are respectively, the roots of the quartic equation
\ba (B_1\xic_1^2 + A_{11}\xic_3^2 -1 ) (A_{33}\xic_1^2 + B_1
\xic_3^2 - 1 ) -(A_{13} + B_1)^2 \xic_1^2 \xic_3^2=0 \label{biquad2}
\ea and the quadratic equation \ba B_1 \xic_1^2-1 + B_3 \xic_3^2 =0,
\label{quad2}\ea which correspond, respectively, to the first and
second lines in (\ref{eq:delta_quad}), with the corresponding
$\lambda^\al(\bxic)=1$, both rewritten in terms of the crack
coordinates. Moreover, Eq. (\ref{biquad2}) may be obtained from the
first equation in (\ref{eq:delta_quad}) simply by putting $\xic_2$
to zero and allowing $A_{11}$ and $A_{33}$ to exchange places. It
follows that we can follow the form of solution presented in Section
\ref{poles_perp} and write \ba &&\xic^{(1)}_3(\xic_1;0)
=\xic_{31}(\xic_1;0)+\xic_{32}(\xic_1;0), \nn \\
&&\xic^{(2)}_3(\xic_1;0)=\xic_{31}(\xic_1;0)-\xic_{32}(\xic_1;0), \nn \\
&&\xic^{(3)}_3(\xic_1;0)=(B_3^{-1}B_1)^{1/2}\gamma_2(\xic_1;0),
\label{trick_add2}\ea with \ba
&&\xic_{3n}(\xic_1;0)=\frac{\gamma_{6-3n}}{2(B_1A_{11})^{1/2}}\times\nn\\
&& \Bigl \{ \frac
{[A_{13}^2+2B_1A_{13}-A_{11}A_{33}]\xic_1^2+B_1+A_{11}- (-1)^m
2B_1(A_{11}A_{33})^{1/2}\gamma_1\gamma_2} { \gamma^2_{6-3n} }
\Bigr \}^{1/2}, \nn\\
&&\,\,\,\,\,\,\,\,\,\,\,\,\,\,\,\,\,\,\,\,\,\,\,\,\,\,\,\,\,\,\,\,\,\,\,\,\,\,
\,\,\,\,\,\,\,\,\,\,\,\,\,\,\,\,\,\,\,\,\,\,\,\,\,\,\,\,\,\,\,\,\,\,\,\,\,\,\,
\,\,\,\,\,\,\,\,\,\,\,\,\,\,\,\,\,\,\,\,\,\,\,\,\,\,\,\,\,\,\,\,\,\,\,\,\,\,\,
\,\,\,\,\,\,\,\,\,\,\,\,\,\,\,\,\,\,\,\,\,\,\,\,\,\,\,\,\,\,\,\,\,\,\,\,\,\,\,\,
\,\,\,\,\,\,\,\,\,\,\,\,\,\,\,\,\,\, n=1,2,
 \ea and \ba
&&\gamma_0=\gamma_0(\xic_1;0)=(\kappa_0^2+\xic_1^2)^{1/2},\nn\\
&&\gamma_\ell=\gamma_\ell(\xic_1;0)=(\kappa_\ell^2-\xic_1^2)^{1/2},\,\ell=1,2,3,
\ea with the branch points $\pm \i \kappa_0$ and $\pm
\kappa_\ell,\,\ell=1,2,3,$ where we have \ba \kappa_0^2=C_0^2,\quad
\kappa_1^2= A_{33}^{-1},\quad\kappa_2^2=
B_{1}^{-1},\quad\kappa_3^2=C_3^2 \ea and \ba
&&C_\ell=\nn\\
&&D^{-1}\{2(A_{11}B_1)^{1/2}(B_1+A_{13})[A_{13}^2-
A_{11}A_{33}+B_1(A_{11}+2A_{13}+A_{33})]^{1/2}-
\nn\\
&&(-1)^\ell[A_{13}(2B_1+A_{13})(B_1+A_{11})+A_{11}(2B_1^2+A_{33}B_1-
A_{11}A_{33})]\}^{1/2},\nn\\
&&\,\,\,\,\,\,\,\,\,\,\,\,\,\,\,\,\,\,\,\,\,\,\,\,\,\,\,\,\,\,\,\,\,\,\,\,\,\,
\,\,\,\,\,\,\,\,\,\,\,\,\,\,\,\,\,\,\,\,\,\,\,\,\,\,\,\,\,\,\,\,\,\,\,\,\,\,\,
\,\,\,\,\,\,\,\,\,\,\,\,\,\,\,\,\,\,\,\,\,\,\,\,\,\,\,\,\,\,\,\,\,\,\,\,\,\,\,\,\,\,\,\,\,\,\,\,\,
\,\,\,\,\,\,\,\,\,\,\,\,\,\,\,\,\,\,\,\,\,\,\,\,\,\,\,
\,\,\,\,\,\,\,\,\,\,\,\,\,\,\,\,\,\,\,\,\,\,\ell=0,3, \nn\\
&&D=\{(A_{11}A_{33}-A_{13}^2)[(2B_1+A_{13})^2-A_{11}A_{33}]\}^{1/2}.
\label{CD2} \ea

The factorization described in Appendices A and B is applicable
provided we have \ba \kappa_1<\kappa_2<\kappa_3. \ea This condition
is satisfied by the austenitic steel under study.

\subsection{Decoupling the vector functional equations into the scalar functional equations}

In this case, $t'_i=\sigma'_{i2}$ and the components of the transfer
operator $S_m(\nabla)$ are given by
\begin{eqnarray}
(S_m)_{ik}(\nabla)= \Sigma^{(2)}_{ik}(\nabla).
\end{eqnarray}
Therefore, in view of (\ref{d_tt}), the transfer operator
${S}(\bxic)$ at $\xic_2=0$ is \ba {S}(\xic_1,0,\xic_3)= \left[
\begin{array}{ccc}
-B_3\xic_3 & 0 & 0 \\
0 & -A_{11}\xic_3 &  A_{13}\xic_1 \\
0 &  B_{1}\xic_1 & -B_{1}\xic_3
\end{array}
\right]. \ea It is easy to check that if  $\xic_2=0$ and $A_{11}$
and $A_{33}$ exchange  places the Fourier transform of the
elastodynamic equation is the same as in the previous case.  This
means that ${\widetilde \tau}^G(\xic_1;0)$ has the following
diagonal form
\begin{eqnarray}
{\widetilde \tau}^G(\xic_1;0) = \left[
\begin{array}{ccc}
\mu_1(\xic_1;0) & 0 & 0 \\
0 & \mu_2(\xic_1;0) & 0 \\
0 & 0 & \mu_3(\xic_1;0)
\end{array}
\right], \label{tau_par}
\end{eqnarray}
with the eigenvalues \ba
&&\mu_1(\xic_1;0)=\frac{1}2(-B_1B_3)^{1/2}\gamma_2(\xic_1;0),\nn\\
&&\mu_2(\xic_1;0)=\Bigl(\frac{A_{11}}{A_{33}}\Bigr)^{1/2}\frac{\gamma_2(\xic_1;0)}{\gamma_1(\xic_1;0)}\mu_3(\xic_1;0),\\
&&\mu_3(\xic_1;0)=\frac{\i
R(\xic_1;0)}{4A_{11}\xic_{31}(\xic_1;0)},\nn \ea and the Rayleigh
function \ba R(\xic_1;0)=(A_{13}^2-A_{11}A_{33})\xic_1^2+A_{11}+
(A_{11}A_{33})^{1/2} \frac{\gamma_1(\xic_1;0)}{\gamma_2(\xic_1;0)}.
\ea Substituting (\ref{tau_par}) into the 3D vector functional
equation (\ref{eq:functional}), the latter decouples into three
scalar equations,
\begin{eqnarray}
\mu_i(\xic_1;0) \overline{\Delta U}^{sc}_i(\xic_1;0) =
\i\frac{{\widehat
t}^\be_i(\kc^{in}_1,\kc^{in}_2,0)}{\xic_1+\kc^{in}_1+i0}-\overline
T_i^{sc-}(\xic_1;0), \,\,i=1,2,3. \ea

The equations have the same structure as (\ref{eq:functionalsc1})
and therefore, can all be solved using the Wiener-Hopf technique by
introducing the factorization
\begin{eqnarray}
\mu_i (\xic_1;0) =\mu^+_i (\xic_1;0)\mu^-_i(\xic_1;0),\,\,i=1,2,3.
\end{eqnarray}
On allowing  $A_{11}$ and $A_{33}$ to  exchange places, the
tangential traction $\mu_3(\xic_1;0)$  is the same as the tangential
traction $\mu_1(\xic_1;0)$  in Section \ref{sec:case_per}.
Therefore, it can be factored in  the same manner, with
$l_0,\,l_1,\,K_0$ and $K_1$ the same as in the Appendices A and B
but with $A_{11}$ standing in place of $A_{33}$ and vice versa.

It follows that the  considerations  and notations introduced in
Appendices A and B allow us to define $\mu^\pm_i (\xic_1;0)$ as \ba
&&\mu^\pm_1(\xic_1;0)=\Bigl(-\frac{B_1B_3}{4}\Bigr)^{1/4}\gamma_2^\pm(\xic_1;0),\nn\\
&&\mu^\pm_2(\xic_1;0)=\Bigl (\frac{A_{11}}{A_{33}}\Bigr
)^{1/4}\frac{\gamma_2^\pm(\xic_1;0)}{\gamma_1^\pm(\xic_1;0)}
\mu^\pm_3(\xic_1;0),\nn\\
&&\mu^\pm_3(\xic_1;0)=\Bigl(-\frac{l_0}{4l_1 A_{11}}\Bigr)^{1/2}
\frac{K_0^\pm(\xic_1;0)}{\gamma_3^\pm(\xic_1;0)K_1^\pm(\xic_1;0)}(\kappa_R\pm
\xic_1), \ea where, similarly to (\ref{gammapm}), we have \ba
\gamma^\pm_\ell(\xic_1;0)=(\kappa_\ell\pm\xic_1)^{1/2},\,\,\ell=1,~2,~3.
\ea It is easy to see from their definition that the + (-) functions
above have no zeros in the upper (lower) half of the complex
$\xic_1-$plane.  It follows that we have
\begin{eqnarray}
\overline{\Delta  U}^{sc}_i(\xic_1;0) = \i\frac{{\widehat
t}^\be_i(\kbc^{in})}
{(\xic_1+\kc^{in}_1+i0)\mu^+_i(\xic_1;0)\mu^-_i(-\kc^{in}_1;0)},\,\,i=1,2,3.
\end{eqnarray}

\section{Diffraction coefficients}
\label{sec:dc} Let us consider the equation
\begin{eqnarray}
\U^{sc}(\xc_1,\xc_3;\xic_2) = -\frac{1}{2\pi} \int_{-\infty}^\infty
\widetilde t^G(\xic_1,\xc_3;\xic_2) \overline {\Delta
\U}^{sc}(\xic_1;\xic_2) \ex^{-\i\xic_1 \xc_1} d\xic_1,
\label{forsubstr}
\end{eqnarray}
which   follows from    Eq. (\ref{eq:last}) via the inverse Fourier
transform in $\xic_1$. Substituting (\ref{tGres}) into
(\ref{forsubstr}) leads us to \ba &&\U^{sc}(\xc_1,\xc_3;\xic_2) =
\pm\frac{\i}{2\pi} \cdot\nn\\ &&\int_{-\infty}^\infty \frac {
{\widehat \p}^\al (\bxic){\widehat \t}^\al(\bxic)}
{\lambda^\al_{\xic_3}(\bxic)}\Big
|_{\xic_3=\xic^\alpm_3(\xic_1;\xic_2);~\lambda^\al(\bxic)=1}
\overline {\Delta \U}^{sc}(\xic_1;\xic_2) \ex^{-\i[\xic_1 \xc_1+
\xic^\alpm_3(\xic_1;\xic_2)\xc_3]}d\xic_1, \nn\\ \label{Tsc} \ea
where summation over $\alpha$ is implied and the top (bottom) sign
is chosen when $\xc_3\geq0+$ ($\xc_3\leq0-$).

Let us now introduce the polar coordinates $\rr$ and $\thec$  such
that we have
\begin{eqnarray}
&&\xc_1 = \rr \,\cos\thec, \nn\\
&&\xc_3 = \rr\, \sin\thec,\,\,\thec\,\epsilon\,[0,2\pi].
\end{eqnarray}
When values of $\kc^{in} \rr$ are large, the main contributions to
(\ref{Tsc}) come from the stationary phase points and other critical
points that are described e.g.~in \cite{grid}. The method of the
uniform stationary phase (see e.g.~\cite{bor1}, \cite{blha}) allows
us to treat the situations when the critical points coalesce, that
is when an observation point lies in a transition zone between
geometrical regions, but here we treat only those observation points
for which the critical points are isolated, that is the observation
points that lie in geometrical zones. Thus, the formulas given below
are not applicable near the shadow boundaries, cuspidal edges and
conical points of the $qSV$ wave surface or points of tangential
contact of the $qSH$ and $qSV$ wave surfaces (see e.g.~\cite{grid}
and \cite{gf}) The contribution of each isolated phase stationary
point can be evaluated  using the standard
 stationary phase formula,
\begin{eqnarray}
\int_{-\infty}^\infty g(t) e^{irf(t)} dt \sim g(c)
\Bigl[\frac{2\pi\i}{r{\ddot f}(c)}\Bigr]^{1/2}\ex^{\i rf(c)},
\label{stphase}
\end{eqnarray}
where  the dot denotes the derivative with respect to the argument
(see e.g.~\cite{won89}); and $c$ is such that ${\dot f} (c)=0$. In
both symmetric cases considered above, the phase function in
(\ref{Tsc}) can be written as \ba -\xic_1
\xc_1+\xic^\al_3(\xic_1;\xic_2)|\xc_3|= -\rr[\xic_1\cos
\thec-\xic^\al_3(\xic_1;\xic_2)|\sin \thec|]. \label{phi} \ea
 Let us call each solution $\xic_1$ of the equation
\begin{eqnarray}
-\cos\thec+{\dot \xic^\al_3}(\xic_1;\xic_2)|\sin\thec|=0
\end{eqnarray}
the  stationary phase point $\xic^\al_1$. Then applying
(\ref{stphase}) to (\ref{Tsc}) and multiplying both sides of the
resulting formula by $\exp (\i \kc^{in}_2 \xc_2)$ the main
contributions to the scattered field in the geometrical zones are
given by the stationary points $\xic^\al_1$, so that for each fixed
$\alpha$ and $\beta$, we have the GTD approximation \ba
\u^{(\alpha)diff}(\xbc)&\sim&
 \frac{1}{ \rr ^{1/2}}
\D^\albe(\bzeta^{diff}) \ex^{-\i \bzeta^{diff}\cdot \xbc},
\label{GTD}\ea where using (\ref{xic2}), the dimensionless
diffracted wave vector is \ba \bzeta^{diff}=\Bigl(\xic^\al_1,
-\kc^{in}_2,
-\sgn(\sin\thec)\xic^\al_3(\xic^\al_1;-\kc^{in}_2)\Bigr);
\label{diffvector} \ea and the vector diffraction coefficients
$\D^{\albe}$ are such that we can write \ba \D^\albe(\bzeta^{diff})
&=& -\sgn(\sin\thec)
\frac{1} {\left[2\pi\i{ |\sin\thec|\ddot
\xic^\al_3}(\kc^{diff}_1;-\kc^{in}_2) \right]^{1/2}} \cdot\nn\\&&
\frac { {\widehat \p}^\al(\bzeta^{diff}) {\widehat
\t}^\al(\bzeta^{diff})}{\lambda^\al_{\xic_3}(\bzeta^{diff})}\Big
|_{\lambda^\al(\bzeta^{diff})=1} \widehat {\Delta
\U}^{sc}(\kc^{diff}_1;-\kc^{in}_2), \label{diffmatrix} \ea with no
summation over $\alpha$. We remind the reader that the first
superscript $\alpha=1,2$ or $3$ describes a diffracted wave and the
second superscript $\beta=1,2$ or $3$,  an incident wave. In the
case of the symmetry axis perpendicular to the crack plane we have
\ba &&\lambda^\al_{\xic_3}(\bxic)=-2\xic_3
\frac{(A_{13}^2+2B_1A_{13}-A_{11}A_{33})\xic_\perp^2-2B_1A_{33}\xic_3^2+(B_1+A_{33})\lambda^\al(\bxic)}
{(B_1+A_{11})\xic_\perp^2+(B_1+A_{33})\xic_3^2-2\lambda^\al(\bxic)},\nn\\
&&\,\,\,\,\,\,\,\,\,\,\,\,\,\,\,\,\,\,\,\,\,\,\,\,\,\,\,\,\,\,\,\,\,\,
\,\,\,\,\,\,\,\,\,\,\,\,\,\,\,\,\,\,\,\,\,\,\,\,\,\,\,\,\,\,\,\,\,\,
\,\,\,\,\,\,\,\,\,\,\,\,\,\,\,\,\,\,\,\,\,\,\,\,\,\,\,\,\,\,\,\,\,\,
,\,\,\,\,\,\,\,\,\,\,\,\,\,\,\,\,\,\,\,\,\,\,\,\,\,\,\,\,\,\,\,\,\,
\,\,\,\,\,\,\,\,\,\,\,\,\,\,\,\,\,\,\,\,\,\,\,\,\,\,\,\,\,\,\,\,\,\,
\alpha=1,2,\nn\\
&&\lambda^{(3)}_{\xic_3}(\bxic)=2B_1\xic_3, \label{den_perp}\ea
where when $|\xic_3|=|\xic^\al_3(\xic_1;-\kc^{in}_2)|$,
$\lambda^\al(\bxic)=1$. In the case of the symmetry axis lying in
the crack plane perpendicularly to the crack edge the above formulas
apply for $\xic_2=0$ if $A_{11}$ and $A_{33}$ exchange places, so
that we have \ba &&\lambda^\al_{\xic_3}(\bxic)=-2\xic_3
\frac{(A_{13}^2+2B_1A_{13}-A_{11}A_{33})\xic_1^2-2B_1A_{11}\xic_3^2+(B_1+A_{11})\lambda^\al(\bxic)}
{(B_1+A_{33})\xic_1^2+(B_1+A_{11})\xic_3^2-2\lambda^\al(\bxic)},\nn\\
&&\,\,\,\,\,\,\,\,\,\,\,\,\,\,\,\,\,\,\,\,\,\,\,\,\,\,\,\,\,\,\,\,\,
\,\,\,\,\,\,\,\,\,\,\,\,\,\,\,\,\,\,\,\,\,\,\,\,\,\,\,\,\,\,\,\,\,\,
\,\,\,\,\,\,\,\,\,\,\,\,\,\,\,\,\,\,\,\,\,\,\,\,\,\,\,\,\,\,\,\,\,\,
,\,\,\,\,\,\,\,\,\,\,\,\,\,\,\,\,\,\,\,\,\,\,\,\,\,\,\,\,\,\,\,\,\,
\,\,\,\,\,\,\,\,\,\,\,\,\,\,\,\,\,\,\,\,\,\,\,\,\,\,\,\,\,\,\,\,\,\,
\alpha=1,2,\nn\\
&&\lambda^{(3)}_{\xic_3}(\bxic)=2B_3\xic_3. \label{den_par}\ea

Let us now introduce the spherical polar angles
$\varphic$ and $\varthec$ associated with the
crack coordinate system.  Then each incident unit
wave vector can be expressed in terms of its
medium coordinates as \ba
\n^{in}=\left(\sin\varphic^{in}\cos\varthec^{in},\,\,\cos\varphic^{in},\,\,\sin\varphic^{in}\sin\varthec^{in}\right).
\label{numexpnprime} \ea  When the symmetry axis
lies in the crack plane and is perpendicular to
the crack edge we only consider the incident
vectors that are  perpendicular to the crack edge,
i.e. we assume $\varphic^{in}=90^0$. Using the
transpose of (\ref{perQ-matrix}), this yields the
following medium coordinates of the incident unit
wave vector $\n^{in}$: \ba
\n^{in}=\left(0,\,\,-\sin\varthec^{in},\,\,
\cos\varthec^{in}\right). \ea The  incident wave
number $\kc^\be$ and parameter $\kc^{in}_2$ can be
found by using the formulae (\ref{kbe}) and
(\ref{xic2}), respectively. In each case, the
polarization of the incident wave ${\widehat
\p}^\be(\kbc^{in})$ is given by (\ref{ue}) and
(\ref{p_conv}).

Thus, each incident wave vector of type $\beta = 1$, $2$ or $3$
produces three families of diffracted wave vectors of type $\alpha=
1$, $2$ or $3$, each covering {\it a portion} of the cone surface,
$qP$, $qSV$ or $qSH$, respectively, with each cone having a
cross-section in the shape of the respective slowness surface (see
Fig. 3 b). Each wave vector can be expressed as \ba
\bzeta^{diff}=-(\varrhoc^\al(\varthec;-\kc^{in}_2)~\cos\varthec,\,\,\kc^{in}_2,\,\,\varrhoc^\al(\varthec;-\kc^{in}_2)~
\sin\varthec),\,\,\,\varthec\in[0,2\pi], \label{diffvector1} \ea
where in the case of the symmetry axis perpendicular to the crack
plane we have \ba
&&\varrhoc^{(1)}(\varthec;-\kc^{in}_2)=\left[\frac{-D_2+(D_2^2-4 D_1
D_3)^{1/2}}{2 D_1}\right]^{1/2},\nn\\
&&\varrhoc^{(2)}(\varthec;-\kc^{in}_2)=\left[\frac{-D_2-(D_2^2-4 D_1
D_3)^{1/2}}{2 D_1}\right]^{1/2},\nn\\
&&\varrhoc^{(3)}(\varthec;-\kc^{in}_2)=\left[\frac{1-B_3(\kc^{in}_2)^2}{B_3~\cos\varthec^2+B_1~\sin\varthec^2}\right]^{1/2},
\ea with \ba
&&D_1=B_1A_{33}~{\rm sin}^4\varthec-B_4~{\rm cos}^2\varthec~{\rm sin}^2\varthec+A_{11}B_1~{\rm cos}^4\varthec,\nn\\
&&D_2=(2A_{11}[\kc^{in}_2]^2B_1-B_1-A_{11})~{\rm cos}^2\varthec-(B_4[\kc^{in}_2]^2+A_{33}+B_1)~{\rm sin}^2\varthec,\nn\\
&&D_3=(A_{11}[\kc^{in}_2]^2-1)(B_1[\kc^{in}_2]^2-1),\quad
B_4=A_{13}^2+2A_{13}B_1-A_{11}A_{33}, \ea and in the case of the
symmetry axis lying in the crack plane perpendicularly to the crack
edge we can write \ba
&&\varrhoc^{(1)}(\varthec;0)=\left[\frac{-D_2+(D_2^2-4 D_1
)^{1/2}}{2 D_1}\right]^{1/2},\nn\\
&&\varrhoc^{(2)}(\varthec;0)=\left[\frac{-D_2-(D_2^2-4 D_1
)^{1/2}}{2 D_1}\right]^{1/2},\nn\\
&&\varrhoc^{(3)}(\varthec;0)=\left[\frac{1}{B_1~{{\rm
cos}}^2\varthec+B_3~{\rm sin}^2\varthec}\right]^{1/2}. \ea with \ba
&&D_1=B_1A_{11}~{\rm sin}^4\varthec-B_4~{{\rm cos}}^2\varthec~{\rm sin}^2\varthec+A_{33}B_1~{{\rm cos}}^4\varthec,\nn\\
&&D_2=-(B_1+A_{33})~{{\rm cos}}^2\varthec-(A_{11}+B_1)~{\rm
sin}^2\varthec. \ea Using the fact that the wave front is a polar
reciprocal of the slowness surface,  it can be shown that in all
cases, the polar angles $\thec$ and $\varthec$ of the ray and wave
vector, respectively,  are related by the following formula \ba
\thec=\varthec-{\rm tan}^{-1}\Bigl(\frac{1}{\varrhoc}\frac{
\partial \varrhoc}{\partial\varthec}\Bigr). \ea

Similarly to \cite{gau2}, in Appendices C and D we plot for
 various
incident polar angles $\varphic^{in}$ and $\varthec^{in}$, the
magnitudes of the diffracted coefficients $D^{\gamma,\delta}$, where
$\gamma,\,\delta=qP$, $qSV$ or $qSH$ versus the diffracted polar
angle $\varthec\in[0,2\pi]$, thus tracing their variation over the
corresponding slowness curves. As expected, all the graphs for the
magnitudes $D^{(qSH,qSH)}$ are similar to the respective graphs for
the magnitudes  $D^{SH,SH}$ given in \cite{gau2}. Two representative
graphs of the magnitudes of the diffracted coefficients
$D^{(qP,qP)}$ versus the physical polar angle $\thec\in[0,2\pi]$,
that is tracing their variation over the corresponding wave curves
are given in Fig.~4.
\begin{figure}[ht]
{\bf~a}$\quad\quad\quad\quad\quad\quad\quad\quad\quad\quad\quad\quad\quad\quad\quad\quad\quad\quad\quad${\bf~b}\hfil\break
\psfrag{x}{$\thec,\,{}^{\rm o}$}
\psfrag{y}{$|D^{(qP,qP)}|$}
 \hfil
  \includegraphics[height=50mm,width=60mm]{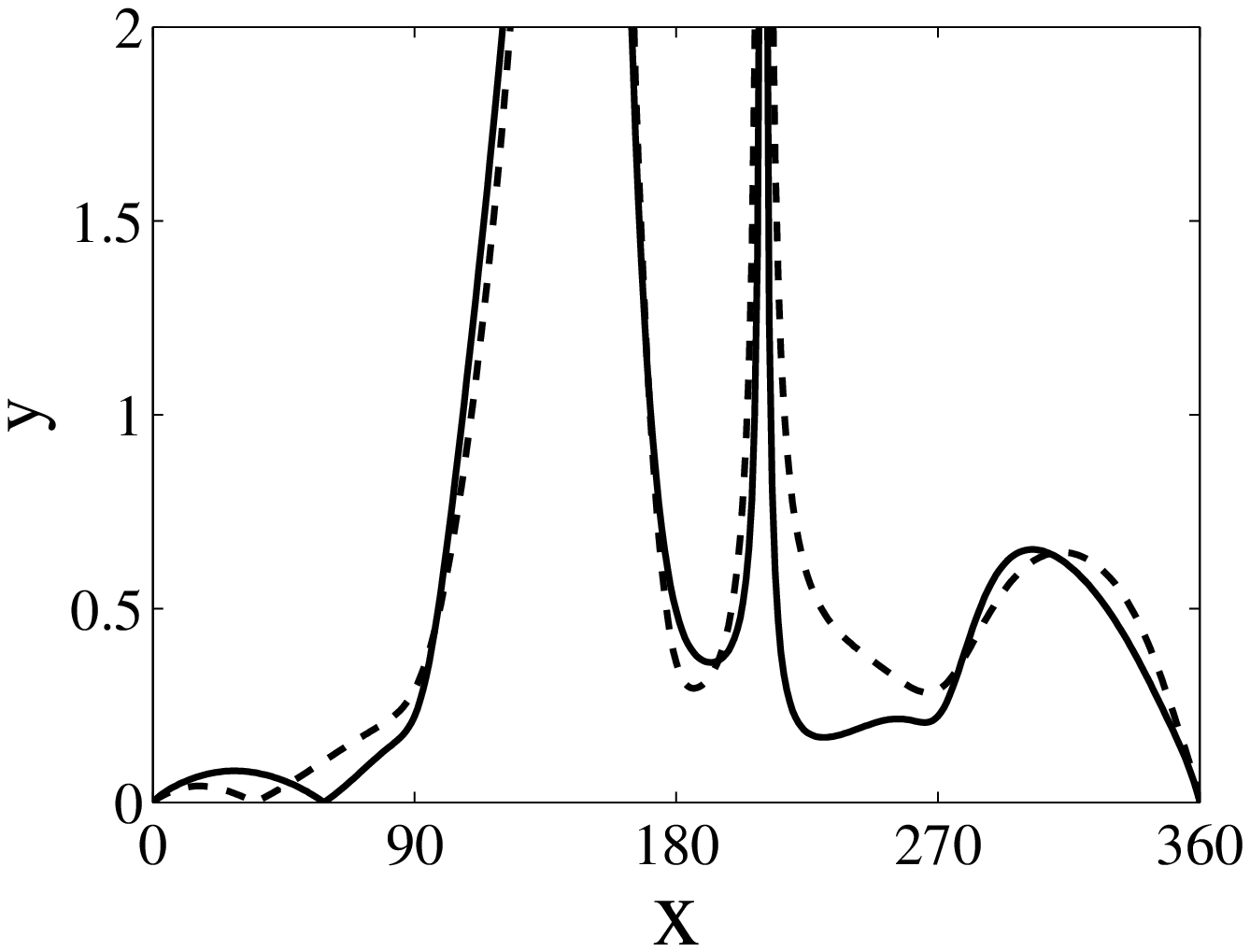}
\hfil
  \includegraphics[height=50mm,width=60mm]{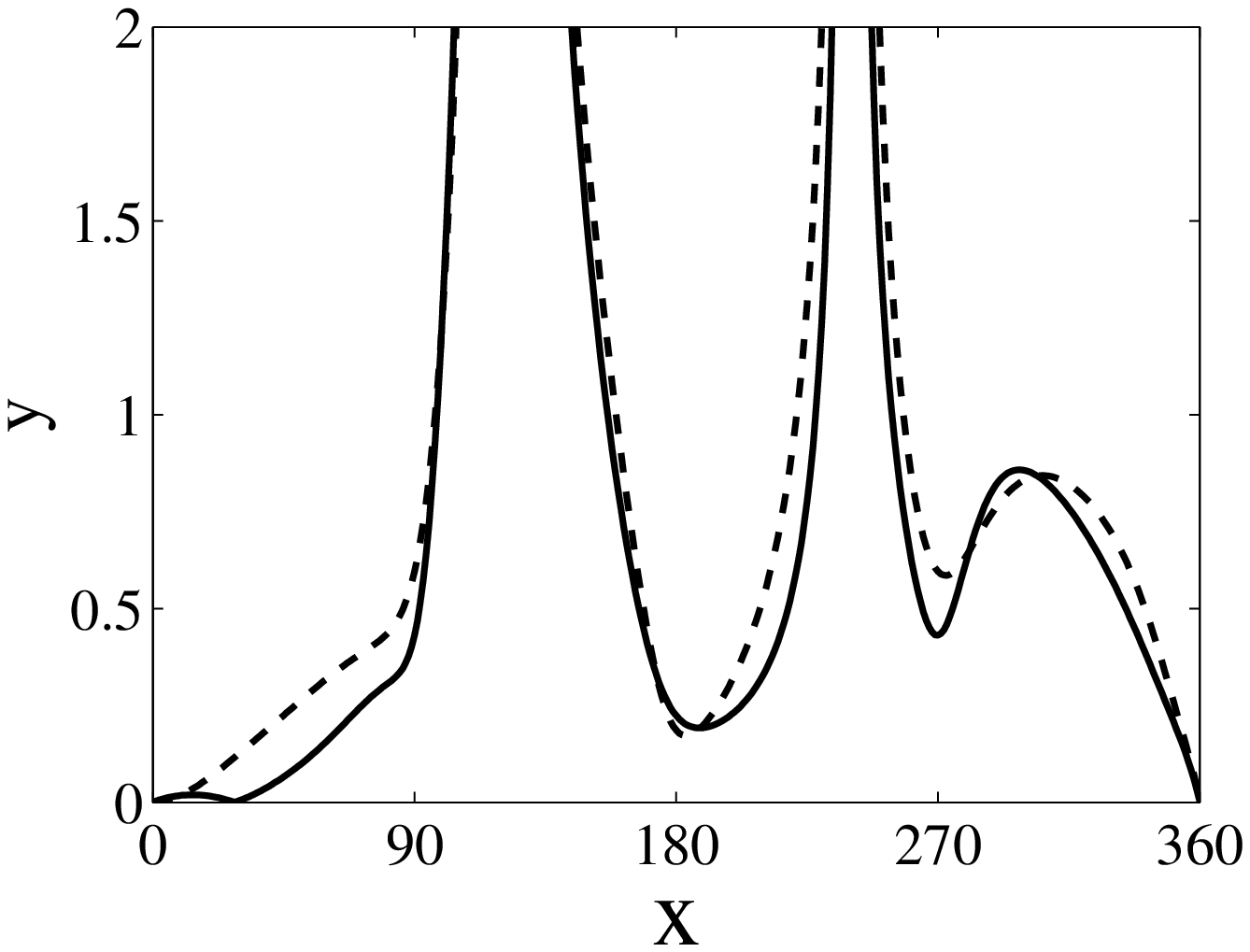}
\hfil
  \caption{The magnitude of the
 diffraction coefficient $D^{(qP,qP)}$ versus the physical polar angle
 $\thec$ for the symmetry axis perpendicular to the crack plane (solid line) and the symmetry axis lying in the crack plane perpendicularly to
 the crack edge (dashed line).  The incident angles are ${\acute \phi}^{in}=90^{\rm o}$  and (a)~${\thec}^{in}=120^{\rm
 o}$, (b)  ${\thec}^{in}=150^{\rm o}$.}
\end{figure}

Finally, we note that the GTD approximation (\ref{GTD}) breaks down
in the so-called transition regions between various geometrical
zones when the denominator in (\ref{diffmatrix}) is small or
vanishes. The regions center on the light-shadow boundaries,
cuspidal edges and conical points where, respectively,
$\kc^{diff}_1+\kc^{in}_1$, ${\ddot
\xic^\al_3}(\kc^{diff}_1;-\kc^{in}_2)$ and
$\lambda^\al_{\xic_3}(\bzeta^{diff})$ turn to zero. When the
singularities are few and far between, like in the isotropic case
considered in \cite{gau2} or in the cases presented in our
Appendices C and D, there is no practical need for developing the
Uniform GTD that provides a good approximation of the diffracted
field in transition zones as well as the geometrical ones.  However,
when cuspidal edges and conical points are present  the formula
(\ref{diffmatrix}) ceases to be of much practical use and the
Uniform GTD description should be used instead of (\ref{GTD}) (see
e.g. \cite {grid}). We illustrate the situation in Appendix E but do
not plot the corresponding "diffraction coefficients" in Appendices
C and D.

\section{\bf Conclusions}

We have considered a semi-infinite crack embedded in a transversely
isotropic medium and studied two  special cases, one, in which the
axis of symmetry is normal to the crack face and the wave incidence
is   arbitrary and another, in which the axis lies in the crack
plane   normal to the edge and the incident wave vector is also
normal  to the edge.  The problem is of interest in  NDE, because
austenitic steels that are found in claddings and other welds in the
nuclear reactors can often be modeled as transversely isotropic. In
both cases, we have expressed the scattered field in   a closed form
and computed the corresponding diffraction coefficients.

A simple case of normal incidence in a TI material that supports
three convex slowness surfaces   has been considered before
\cite{nor84}. This article and its published version
\cite{TIDiffGZF} addresses an extra complication which arises when a
slowness surface contains inflections and in the first of the above
cases, the incidence is allowed to be
 oblique.
 In future, we plan to cross-validate our code with other numerical codes, such as the one reported
 in \cite{lew96},
and  also validate it against experimental data---whenever the
latter become available.

\section{\bf Acknowledgement}
 Partial support of this work has been provided by CEA LIST under the
CIVA 2012 Carnot project. We are grateful to Professor Borovikov for
many useful comments.

\appendix
\vfill\eject

\section {Factorization of the numerator of $\mu_1(\xic_1;-\kc^{in}_2)$}

In order to factorize the numerator we introduce \ba
K_0(\xic_1;\xic_2)=
K_0^*(\xic_1,\frac{\gamma_1(\xic_1;\xic_2)}{\gamma_2(\xic_1;\xic_2)})=\frac{R(\xic_1;\xic_2)}
{l_0
 (\kappa_R^2-\xic_1^2)},
\ea where $\kappa_R=(k_R^2-\xic_2^2)^{1/2}$ and \ba
l_0=\lim_{\xic_1\to\infty}
\frac{R(\xic_1;\xic_2)}{\kappa_R^2-\xic_1^2}=A_{11}A_{33}-A_{13}^2.
\ea

The function $ K_0(\xic_1;\xic_2)$  thus defined tends to unity at
infinity, is never zero and involves  the ratio $\gamma_1/\gamma_2$.
The branch cuts of $R(\xic_1;\xic_2)$ and therefore
$K_0(\xic_1;\xic_2)$ that lie in the lower half plane are presented
in Fig.~A.1.
\begin{figure}[ht]
 \psfrag{a}{\bf a}
  \psfrag{k2}{$-\kappa_2-i\epsilon$}
\psfrag{k1}{$-\kappa_1-i\epsilon$} \psfrag{xi1p}{$(\xic_1)$}
\psfrag{rxi}{$\Re\xic_1$} \psfrag{ixi}{$\Im\xic_1$}
  \includegraphics[height=40mm,width=35mm]{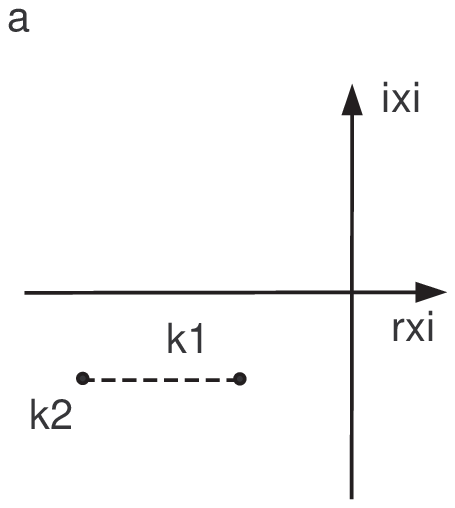}
   \psfrag{b}{\bf b}
\psfrag{k2}{$-\kappa_2-i\epsilon$} \psfrag{k1}{$-|\kappa_1|$}
\psfrag{xi1p}{$(\xic_1)$} \psfrag{rxi}{$\Re\xic_1$}
\psfrag{ixi}{$\Im\xic_1$} \hfil
  \includegraphics[height=40mm,width=35mm]{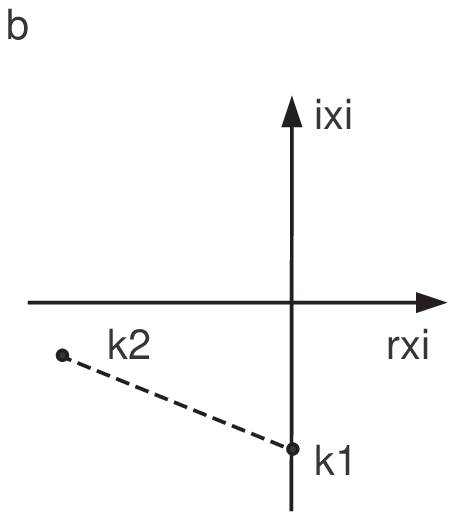}
   \psfrag{c}{\bf c}
\psfrag{k2}{$-|\kappa_2|$} \psfrag{k1}{$-|\kappa_1|$}
\psfrag{xi1p}{$(\xic_1)$} \psfrag{rxi}{$\Re\xic_1$}
\psfrag{ixi}{$\Im\xic_1$} \hfil
  \includegraphics[height=40mm,width=35mm]{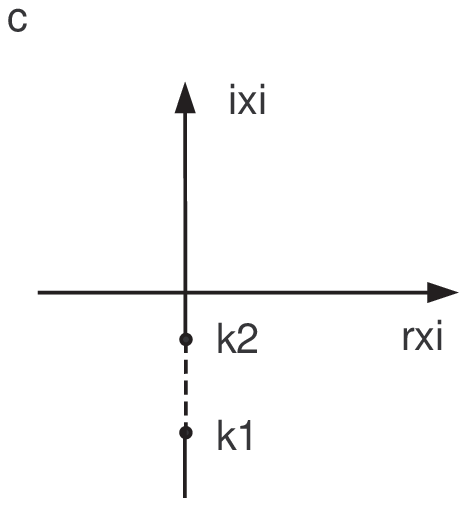}
  \caption{The branch cuts of function $K_0(\xic_1;\xic_2)$ for a)
$\kappa_1$ imaginary and $\kappa_2$ real;  b) $\kappa_1$ and
$\kappa_2$ real; c) $\kappa_1$ and $\kappa_2$ imaginary.}
\end{figure}

Closing the integration contour in the lower half-space, we can
write \ba &&\ln K^+_0(\xic_1;\xic_2)=\frac{1}{2
\pi\i}\int_{-\infty}^\infty
\frac{\ln K_0(t;\xic_2)}{t-\xic_1}dt=\nn\\
&&\frac{1}{2\pi\i}\int_{-\kappa_2}^{-\kappa_1}\frac {\Delta \ln
K_0(t;\xic_2)}
{t-\xic_1}\,dt,\nn\\
\label{K01} \ea where $\ln K^+_0(\xic_1;\xic_2)$ is  analytic in the
upper half of the $\xic_1$ plane, the jump over the cut is \ba
\Delta \ln K_0(t;\xic_2)=
 \ln K^*_0(t,\frac{\gamma_1(t;\xic_2)}{\gamma_2(t;\xic_2)})
-\ln K^*_0(t,-\frac{\gamma_1(t;\xic_2)}{\gamma_2(t;\xic_2)}) \ea and
we have \ba K_0(\xic_1;\xic_2)=
K^+_0(\xic_1;\xic_2)K^-_0(\xic_1;\xic_2), \ea with \ba
K^-_0(\xic_1;\xic_2)= K^+_0(-\xic_1;\xic_2) \ea analytic in the
lower half of the $\xic_1-$plane. By changing $t$ to $-t$, the
formula (\ref{K01}) can be simplified to \ba \ln
K^+_0(\xic_1;\xic_2)&=&-\frac{1}{2\pi\i}
\int_{\kappa_1}^{\kappa_2}\frac {\Delta \ln
K_0(t;\xic_2)}{t+\xic_1}\,dt. \,\,\label{K02} \ea

When  $-\xic_1$ lies  on the branch cut $[\kappa_2,\,\kappa_1]$,
integration of the singular integrand can be avoided by using the
formula \ba K^+_0(\xic_1;\xic_2) =\frac{1}{K^+_0(-\xic_1;\xic_2)}
K_0(\xic_1;\xic_2). \ea

\section {Factorization of the denominator of
$\mu_1(\xic_1;\xic_2)$}

In order to factorize the denominator of $\mu_1(\xi_1)$  we
introduce the function \ba
K_1(\xic_1;\xic_2)=K_1^*(\xic_1,\frac{\gamma_1(\xic_1;\xic_2)}{\gamma_2(\xic_1;\xic_2)})=\frac{\xic_{31}(\xic_1;\xic_2)}{l_1\gamma_3(\xic_1;\xic_2)},
\ea where we have \ba l_1=\lim_{\xic_1\to\infty}
\frac{\xic_{31}(\xic_1;\xic_2)} {\gamma_3(\xic_1;\xic_2)} = \Bigl[
\frac {2B_1(A_{11}A_{33})^{1/2}-A_{13}^2-2B_1A_{13}+A_{11}A_{33}}
{4B_1A_{33}} \Bigr]^{1/2}. \ea The function tends to unity at
infinity, is never zero and involves  the ratio $\gamma_1/\gamma_2$.
Using the same considerations as above, leads us to \ba \ln
K_1^+(\xic_1;\xic_2)&=&-\frac{1}{2 \pi\i}\int_{\kappa_1}^{\kappa_2}
\frac{\Delta \ln K_1(t;\xic_2)}{t+\xic_1}\,dt,\label{K1} \ea where
the jump over the cut is \ba \Delta \ln K_1(t;\xic_2)=
 \ln K^*_1(t,\frac{\gamma_1(t;\xic_2)}{\gamma_2(t;\xic_2)})
-\ln K^*_1(t,-\frac{\gamma_1(t;\xic_2)}{\gamma_2(t;\xic_2)}).
\label{F1} \ea We note that \ba K_1(\xic_1;\xic_2)=
K^+_1(-\xic_1;\xic_2)K^-_1(\xic_1;\xic_2), \ea with the function \ba
K^-_1(\xic_1;\xic_2)= K^+_1(-\xic_1;\xic_2) \ea analytic in the
lower half of the $\xic_1-$plane. It follows that when $-\xic_1$
lies on  the branch cut $[\kappa_2,\,\kappa_1]$, integration of the
singular integrand can  be avoided by using the formula \ba
K^+_1(\xic_1;\xic_2) =\frac{1}{K^+_1(-\xic_1;\xic_2)}
K_1(\xic_1;\xic_2). \ea

\vfill\eject
\section {Magnitudes of various diffraction coefficients for the axis of symmetry perpendicular to the crack}

\noindent In Fig. C.4 below, when the mode of the incident wave is
$qSV$ and the corresponding diffracted wave vectors are complex, we
artificially set the diffraction coefficients to zero.

\begin{figure}[ht]
{\bf~a}$\quad\quad\quad\quad\quad\quad\quad\quad\quad\quad\quad\quad\quad\quad\quad\quad\quad\quad\quad${\bf~d}\hfil\break
\psfrag{x}{$\varthec,\,{}^{\rm o}$} \psfrag{y}{$|D|$}
\includegraphics[height=49mm,width=65mm]{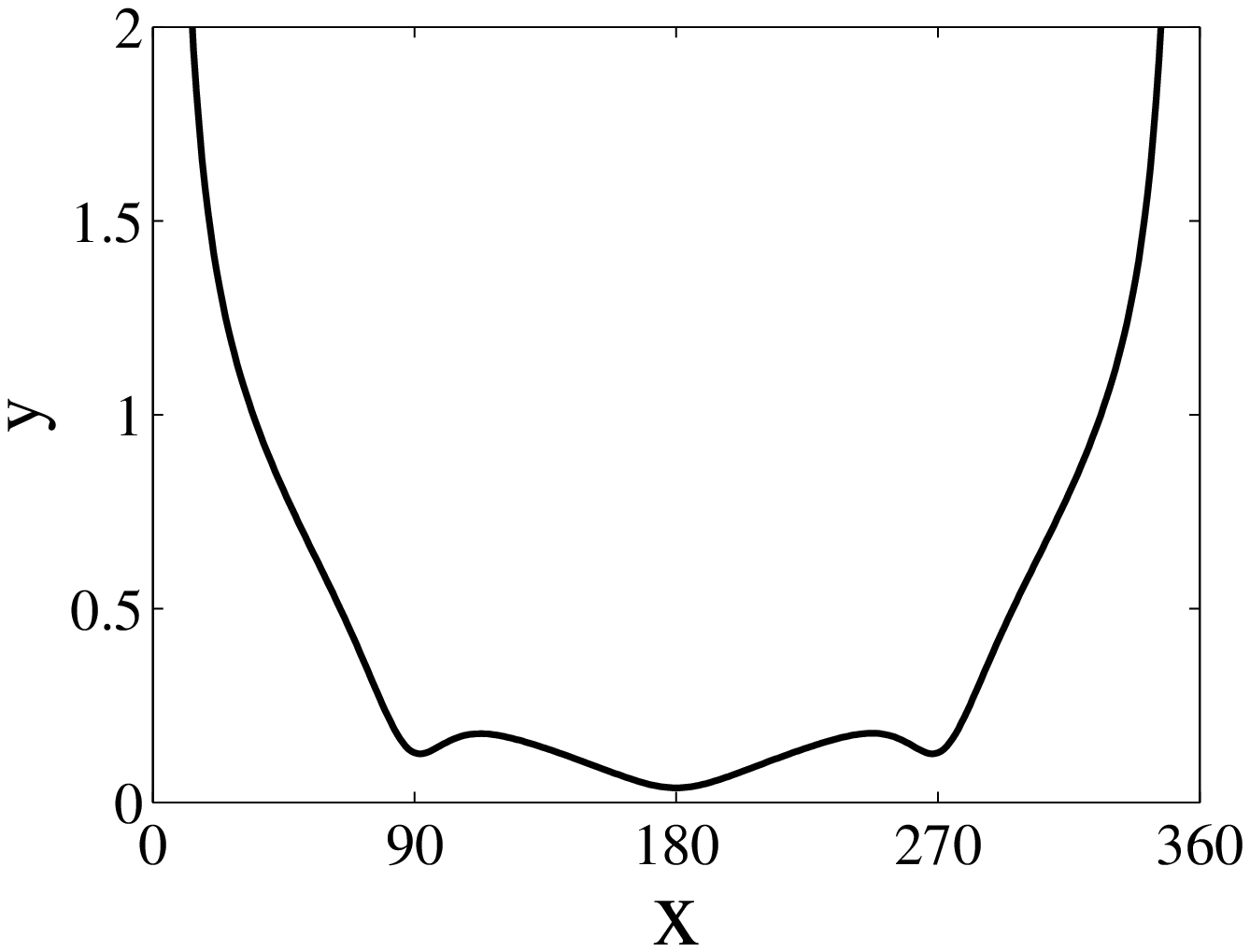}
  \includegraphics[height=49mm,width=65mm]{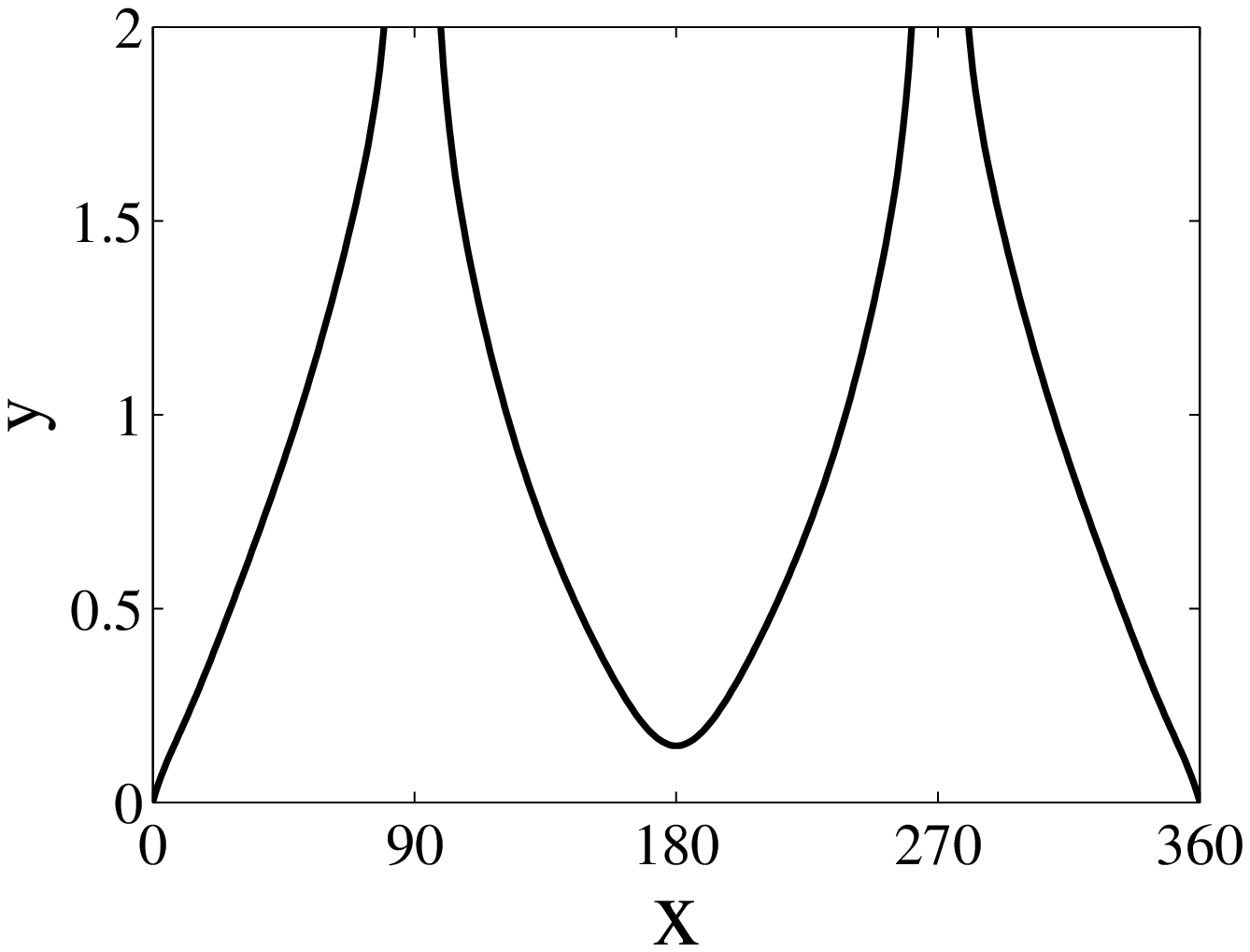}
  {\bf~b}$\quad\quad\quad\quad\quad\quad\quad\quad\quad\quad\quad\quad\quad\quad\quad\quad\quad\quad\quad${\bf~e}\hfil\break
  \includegraphics[height=49mm,width=65mm]{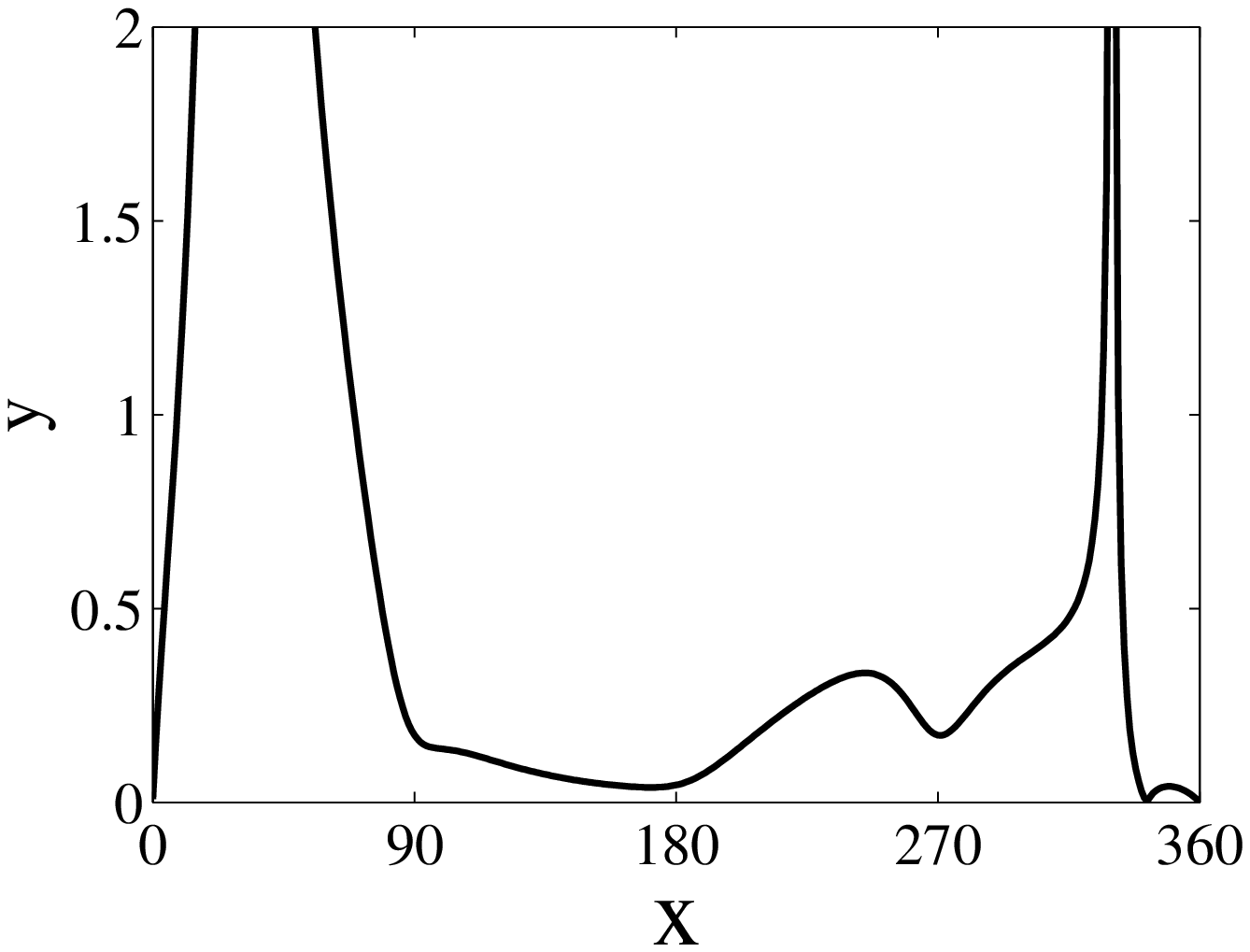}
  \includegraphics[height=49mm,width=65mm]{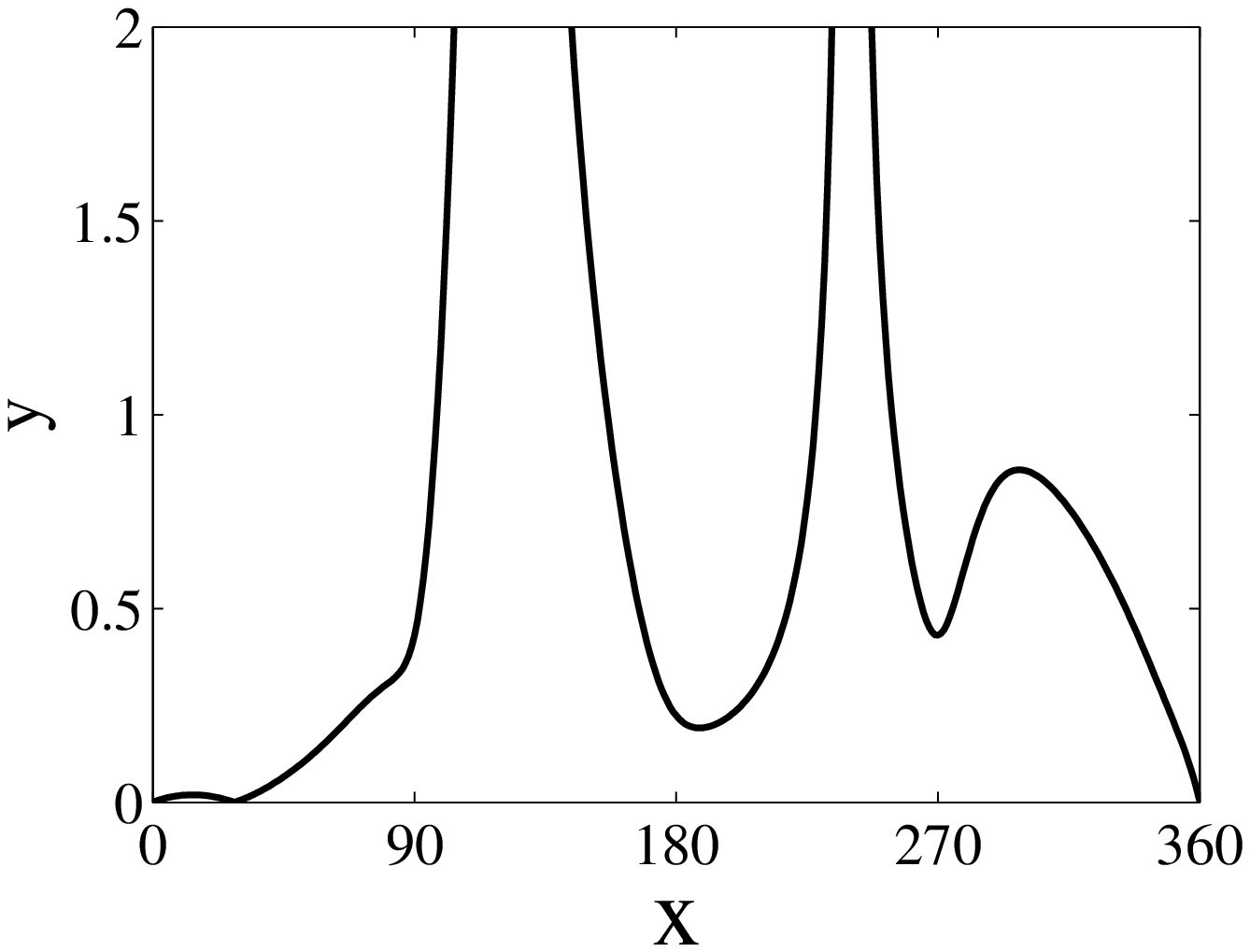}
  {\bf~c}$\quad\quad\quad\quad\quad\quad\quad\quad\quad\quad\quad\quad\quad\quad\quad\quad\quad\quad\quad${\bf~f}\hfil\break
  \includegraphics[height=49mm,width=65mm]{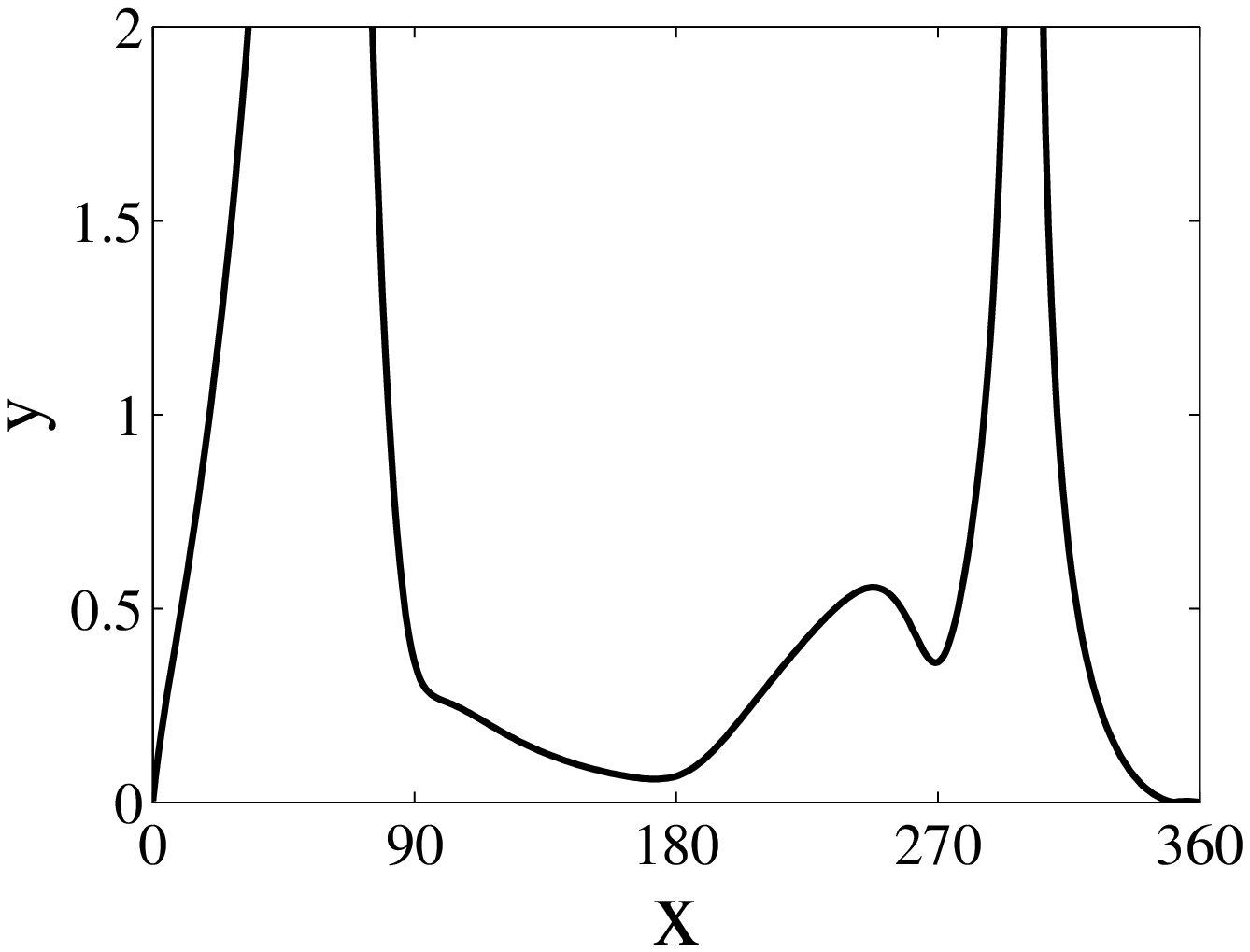}
  \hfil\hfil  \hfil\hfil   \hfil\hfil  \hfil\hfil
  \hfil\hfil  \hfil\hfil   \hfil\hfil  \hfil\hfil
  \hfil\hfil  \hfil\hfil   \hfil\hfil  \hfil\hfil
  \hfil\hfil  \hfil\hfil   \hfil\hfil  \hfil\hfil
  \includegraphics[height=49mm,width=65mm]{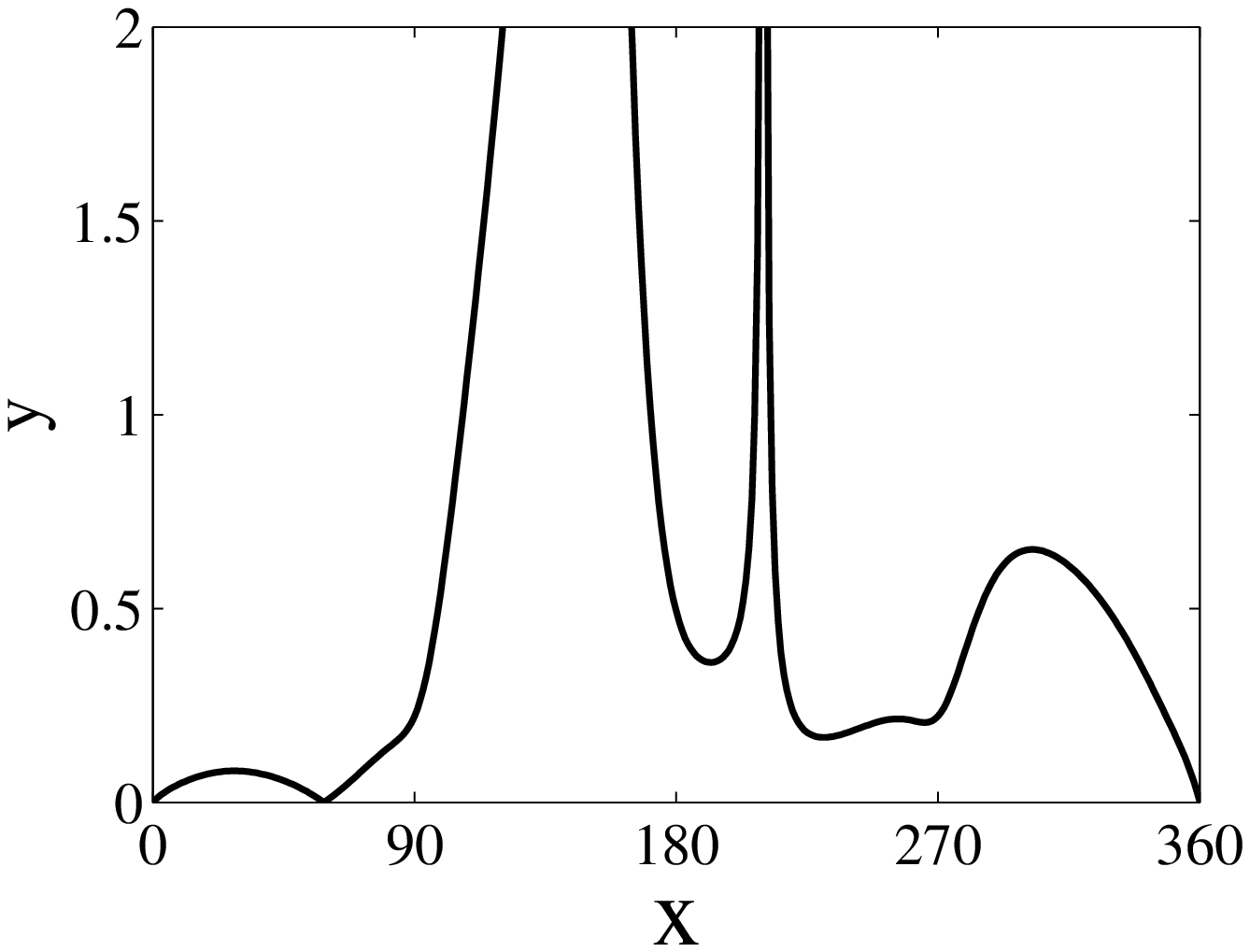}
  \caption{Magnitude $D^{(qP,qP)}$ versus $\varthec$,  $\varphic^{inc}=90^0$ and $\varthec^{inc}=0^0$(a), $30^0$(b),
$60^0$(c),  $90^0$(d), $120^0$(e), $150^0$(f).}
\end{figure}

\vfill\eject

\begin{figure}[ht]
{\bf~a}$\quad\quad\quad\quad\quad\quad\quad\quad\quad\quad\quad\quad\quad\quad\quad\quad\quad\quad\quad${\bf~d}\hfil\break
\psfrag{x}{$\varthec,\,{}^{\rm o}$} \psfrag{y}{$|D|$}
\includegraphics[height=51mm,width=65mm]{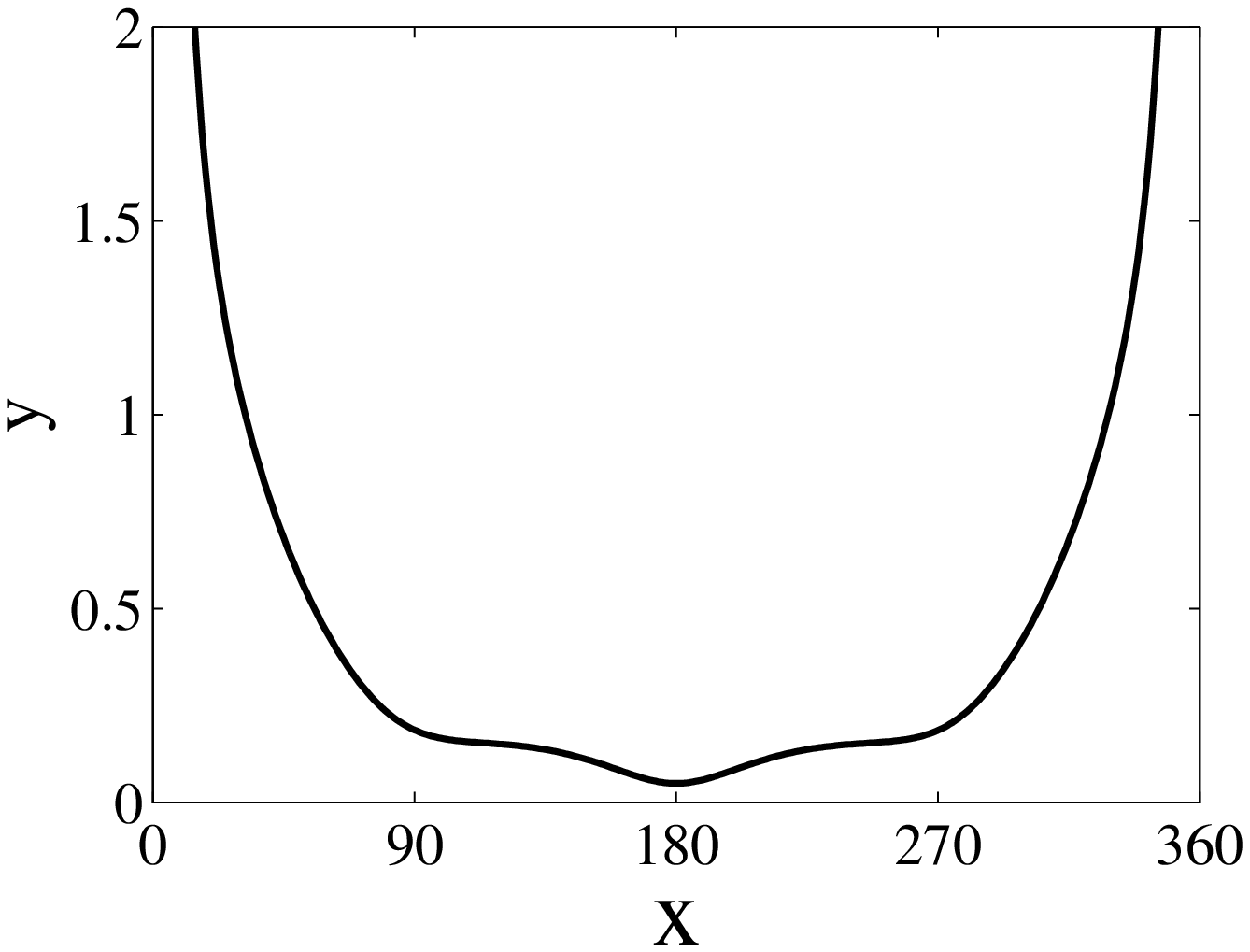}
  \includegraphics[height=51mm,width=65mm]{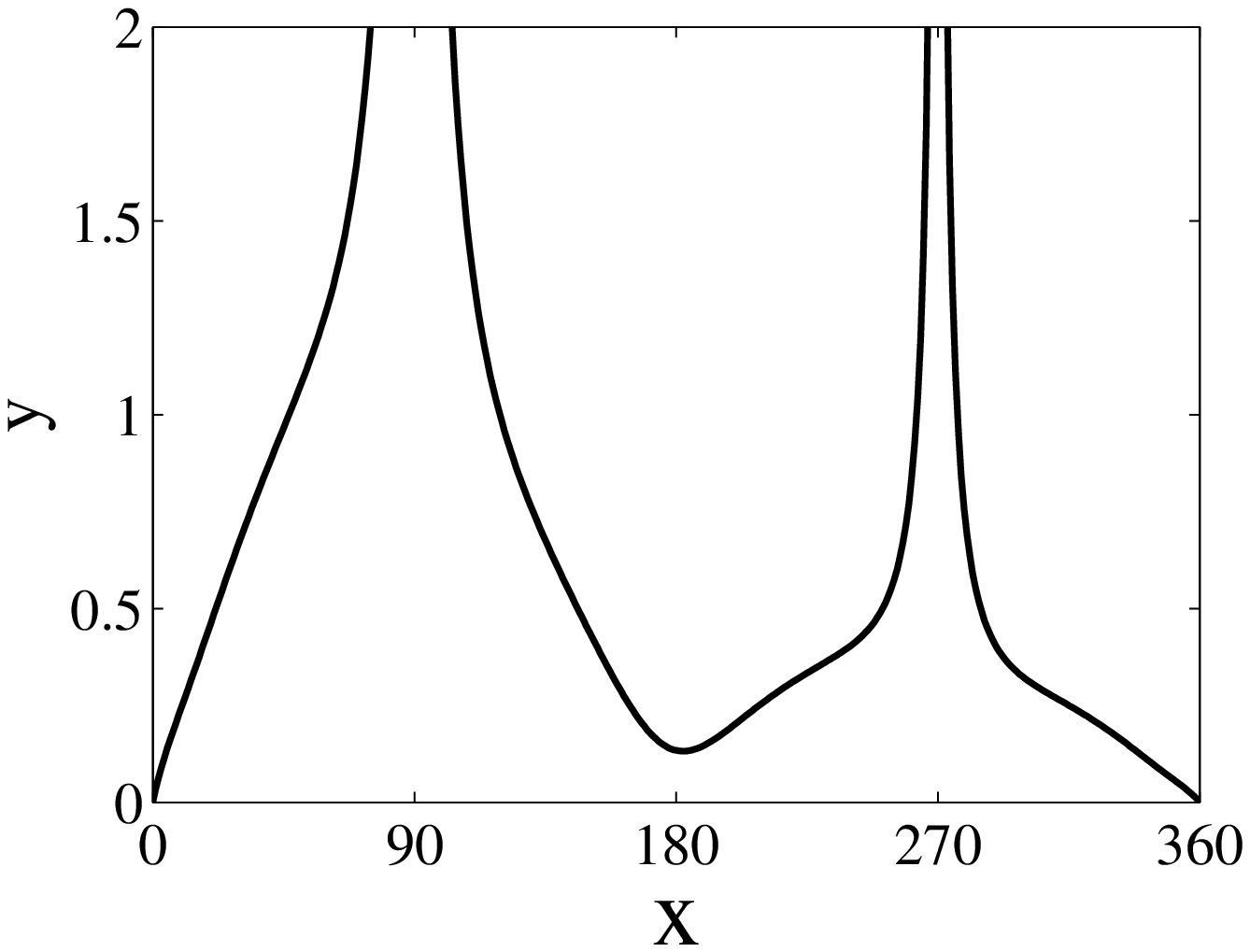}
  {\bf~b}$\quad\quad\quad\quad\quad\quad\quad\quad\quad\quad\quad\quad\quad\quad\quad\quad\quad\quad\quad${\bf~e}\hfil\break
  \includegraphics[height=51mm,width=65mm]{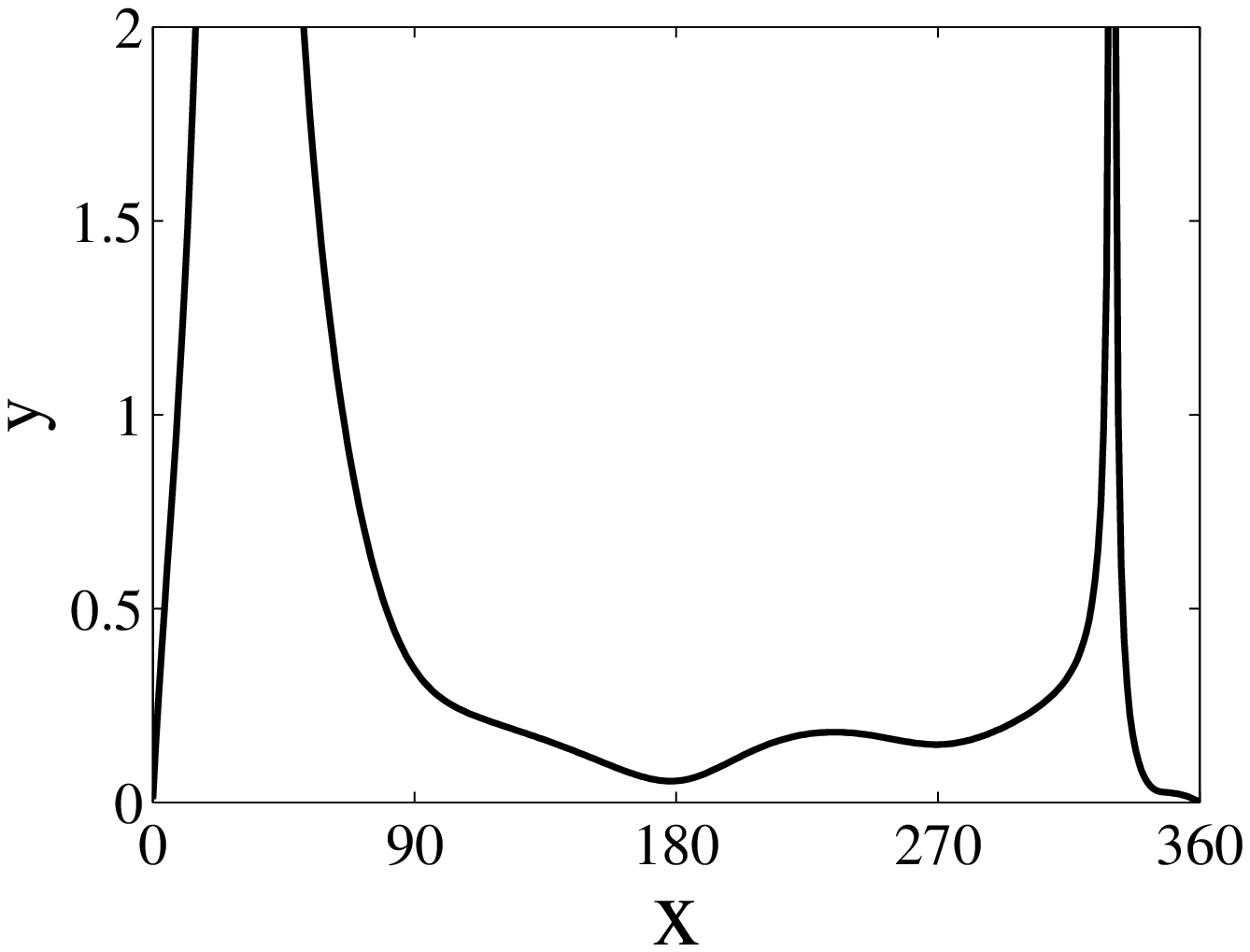}
  \includegraphics[height=51mm,width=65mm]{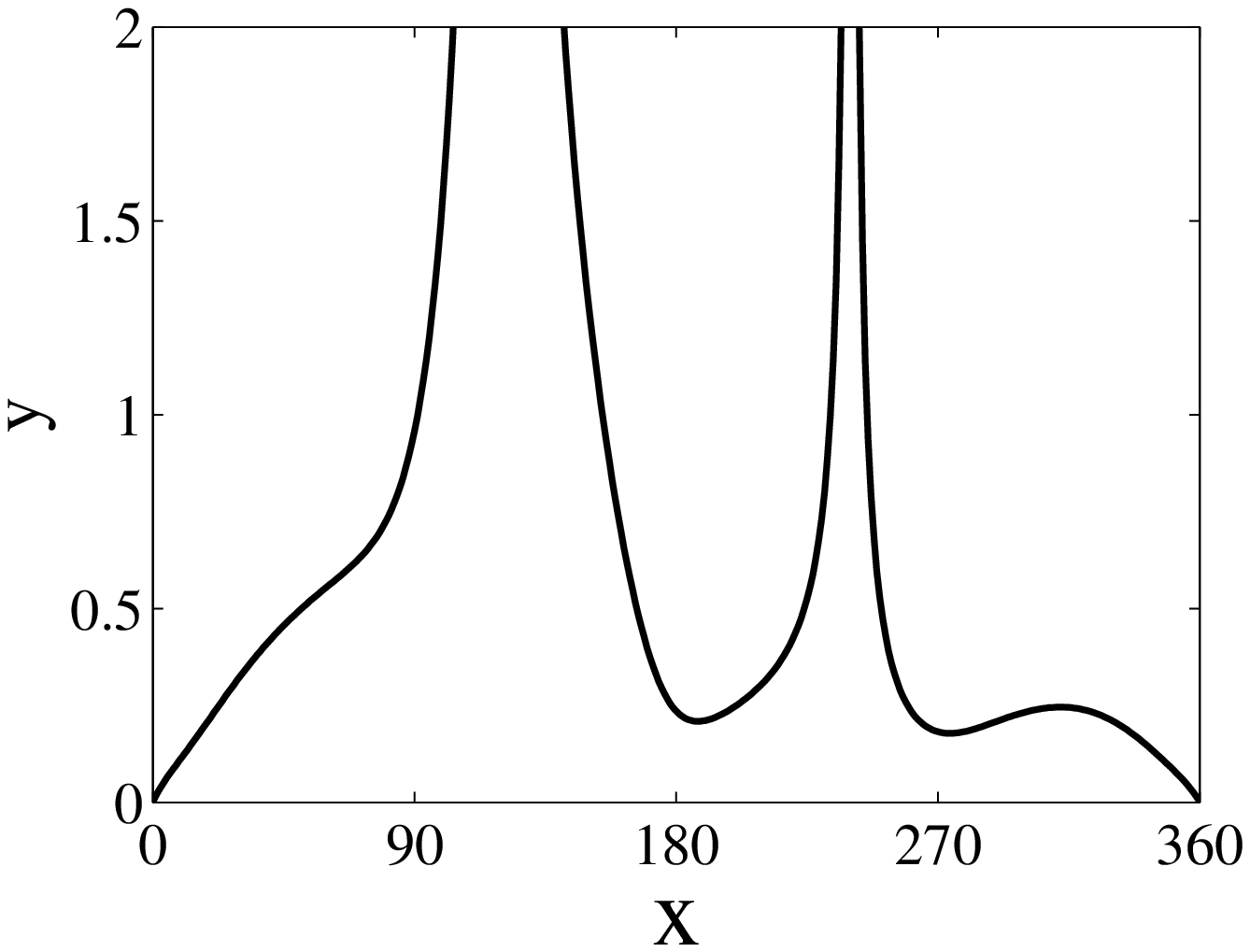}
  {\bf~c}$\quad\quad\quad\quad\quad\quad\quad\quad\quad\quad\quad\quad\quad\quad\quad\quad\quad\quad\quad${\bf~f}\hfil\break
  \includegraphics[height=51mm,width=65mm]{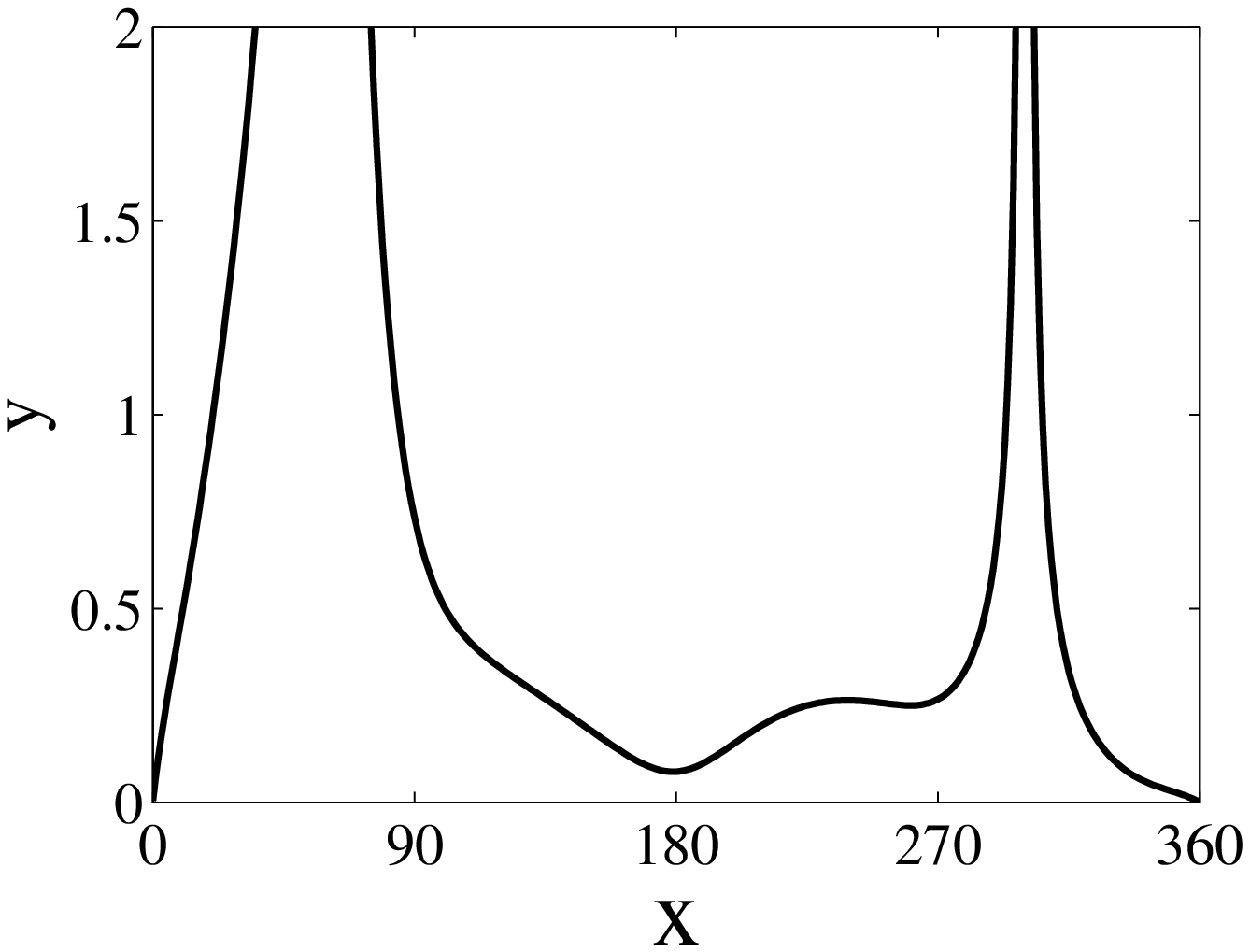}
  \hfil\hfil  \hfil\hfil   \hfil\hfil  \hfil\hfil
  \hfil\hfil  \hfil\hfil   \hfil\hfil  \hfil\hfil
  \hfil\hfil  \hfil\hfil   \hfil\hfil  \hfil\hfil
  \hfil\hfil  \hfil\hfil   \hfil\hfil  \hfil\hfil
  \includegraphics[height=51mm,width=65mm]{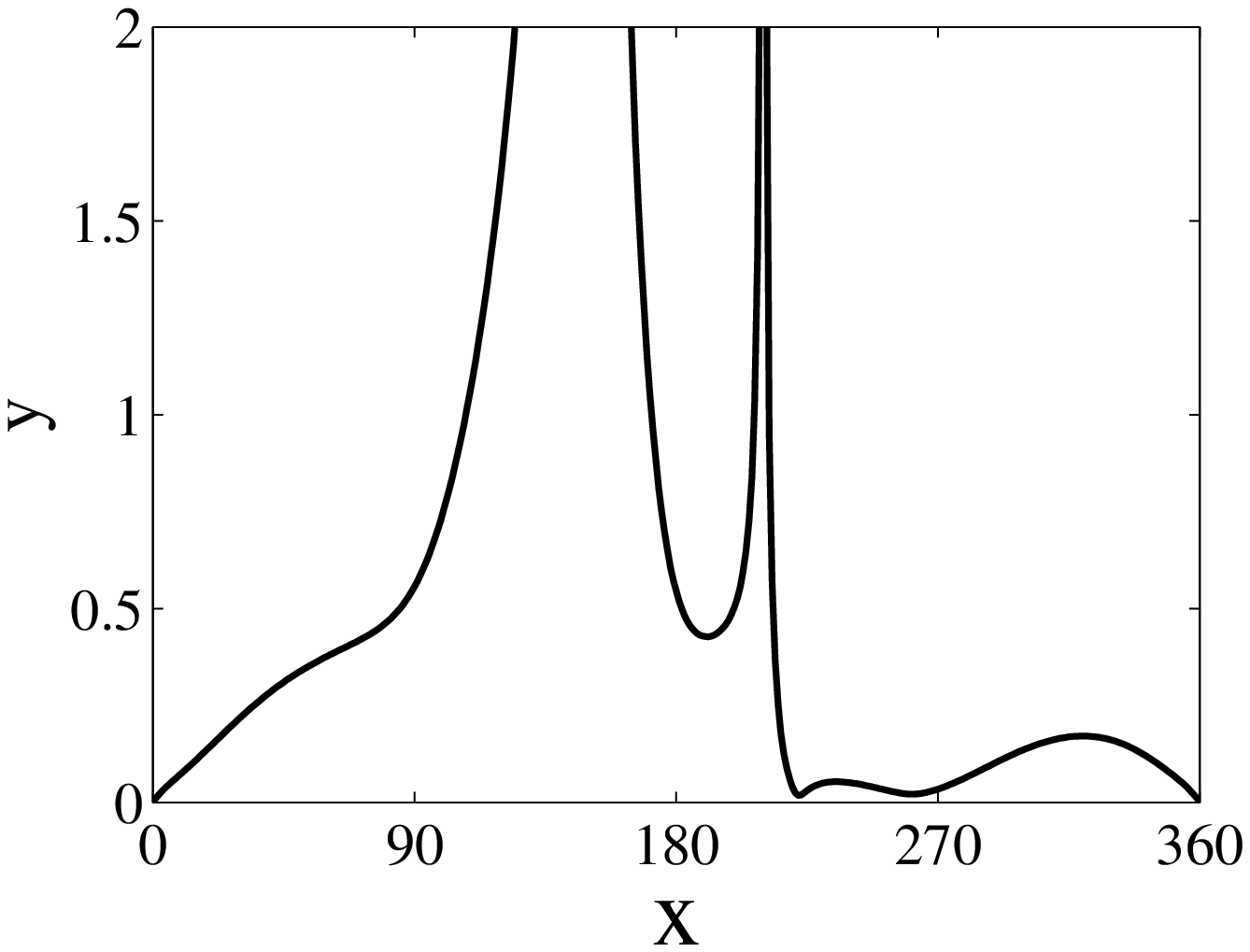}
 \caption{Magnitude $D^{(qP,qP)}$ versus $\varthec$,  $\varphic^{inc}=60^0$ and $\varthec^{inc}=0^0$(a), $30^0$(b),
$60^0$(c),  $90^0$(d), $120^0$(e), $150^0$(f).}
\end{figure}
\vfill\eject

\begin{figure}[ht]
{\bf~a}$\quad\quad\quad\quad\quad\quad\quad\quad\quad\quad\quad\quad\quad\quad\quad\quad\quad\quad\quad${\bf~d}\hfil\break
\psfrag{x}{$\varthec,\,{}^{\rm o}$} \psfrag{y}{$|D|$}
\includegraphics[height=51mm,width=65mm]{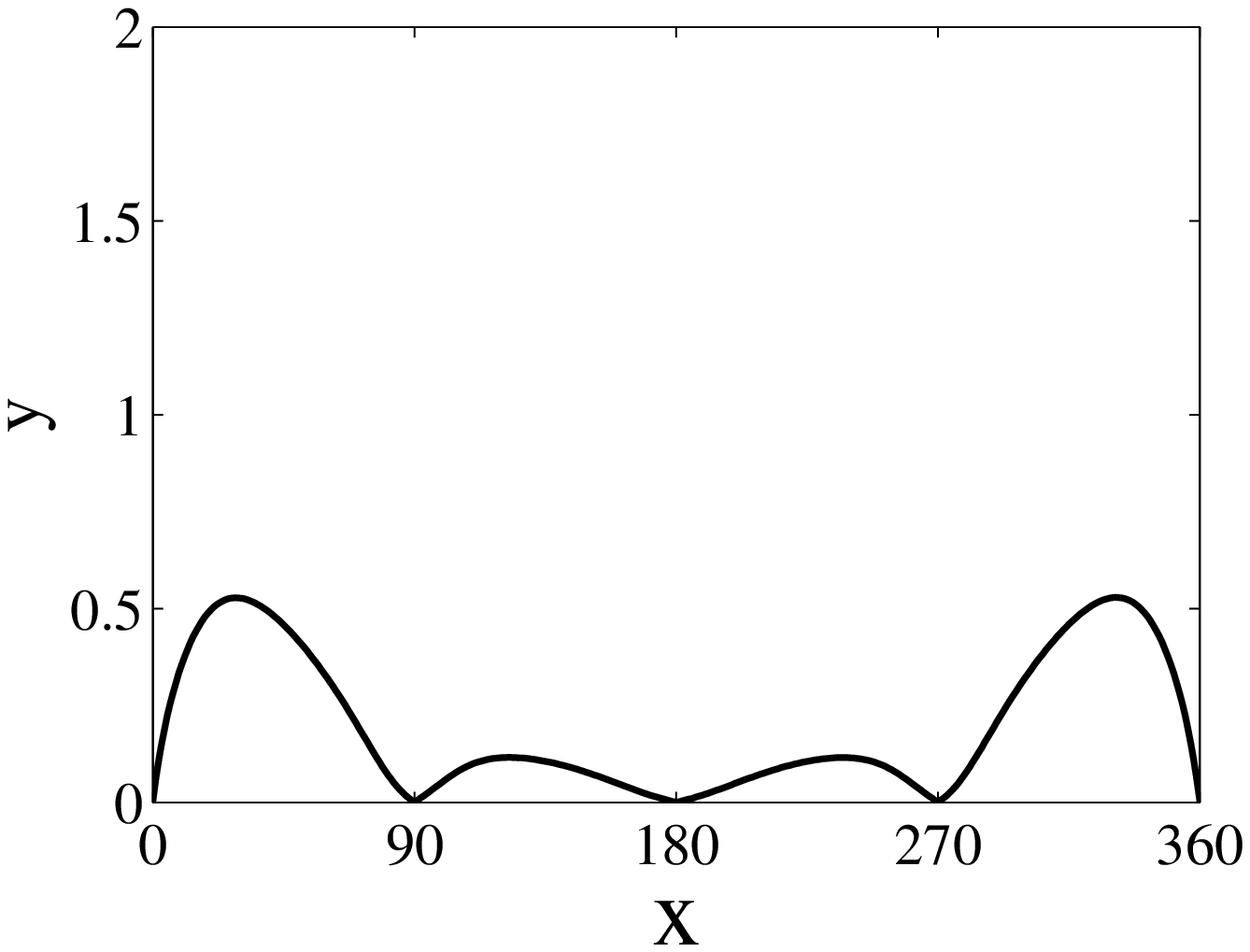}
  \includegraphics[height=51mm,width=65mm]{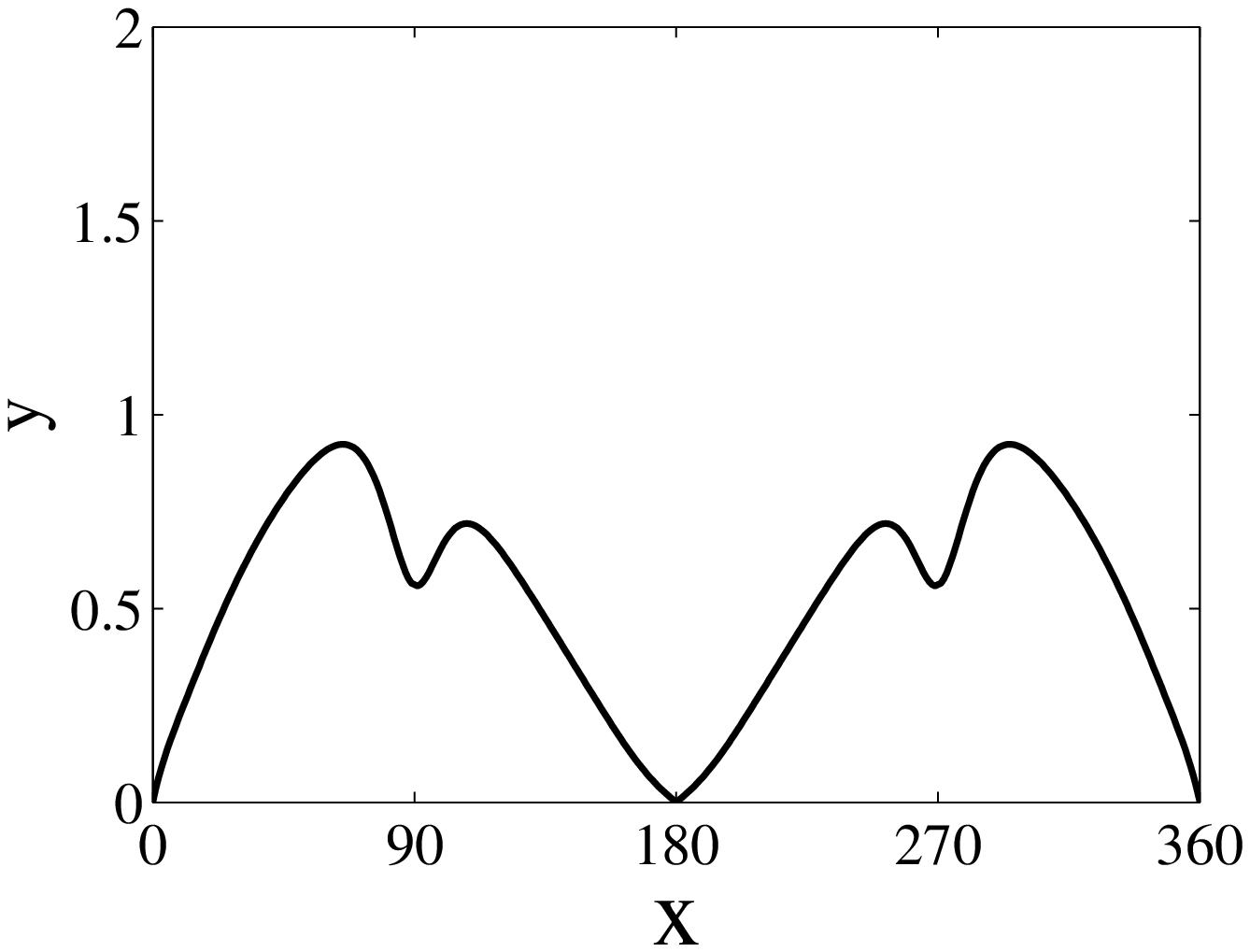}
  {\bf~b}$\quad\quad\quad\quad\quad\quad\quad\quad\quad\quad\quad\quad\quad\quad\quad\quad\quad\quad\quad${\bf~e}\hfil\break
  \includegraphics[height=51mm,width=65mm]{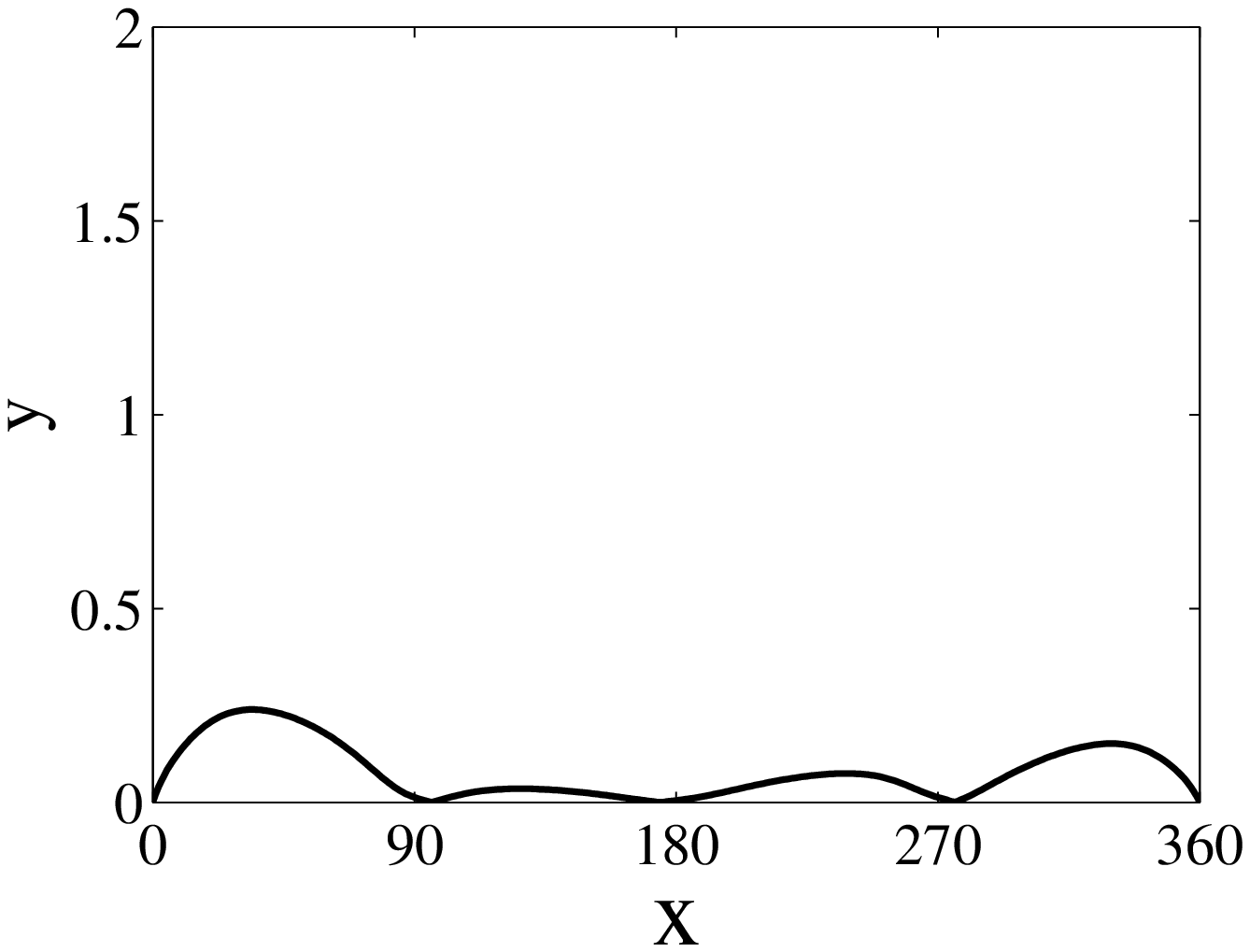}
  \includegraphics[height=51mm,width=65mm]{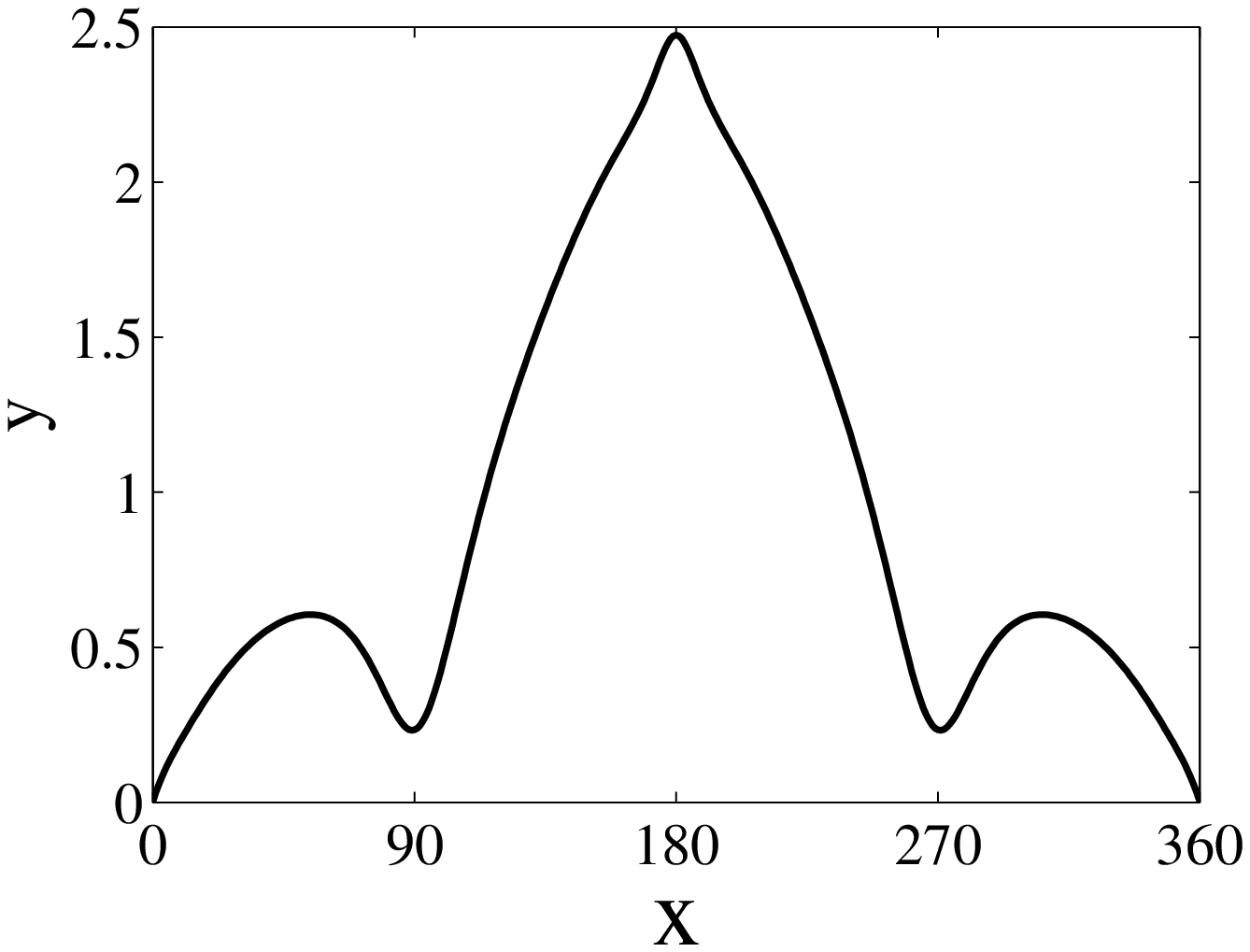}
  {\bf~c}$\quad\quad\quad\quad\quad\quad\quad\quad\quad\quad\quad\quad\quad\quad\quad\quad\quad\quad\quad${\bf~f}\hfil\break
  \includegraphics[height=51mm,width=65mm]{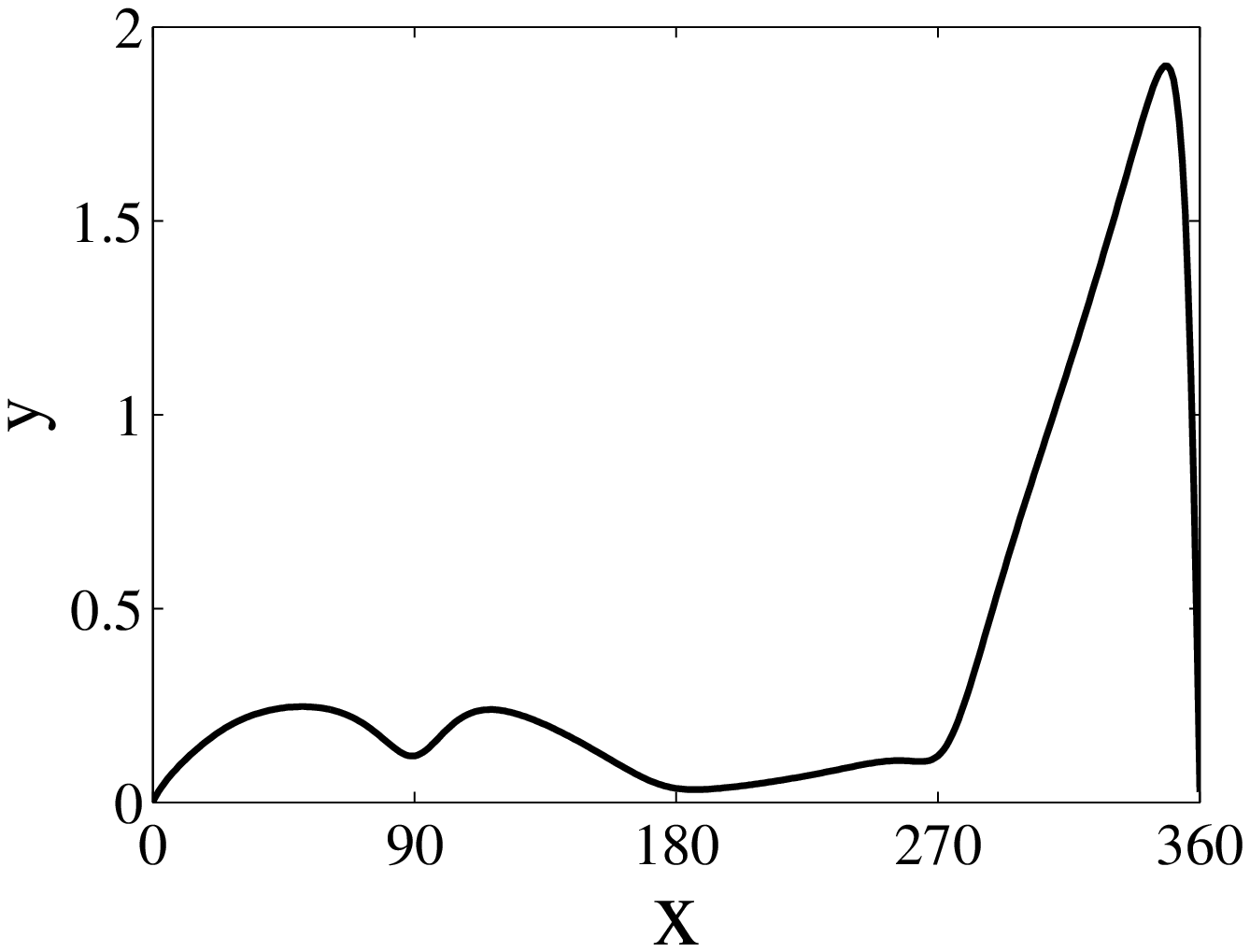}
  \hfil\hfil  \hfil\hfil   \hfil\hfil  \hfil\hfil
  \hfil\hfil  \hfil\hfil   \hfil\hfil  \hfil\hfil
  \hfil\hfil  \hfil\hfil   \hfil\hfil  \hfil\hfil
  \hfil\hfil  \hfil\hfil   \hfil\hfil  \hfil\hfil
  \includegraphics[height=51mm,width=65mm]{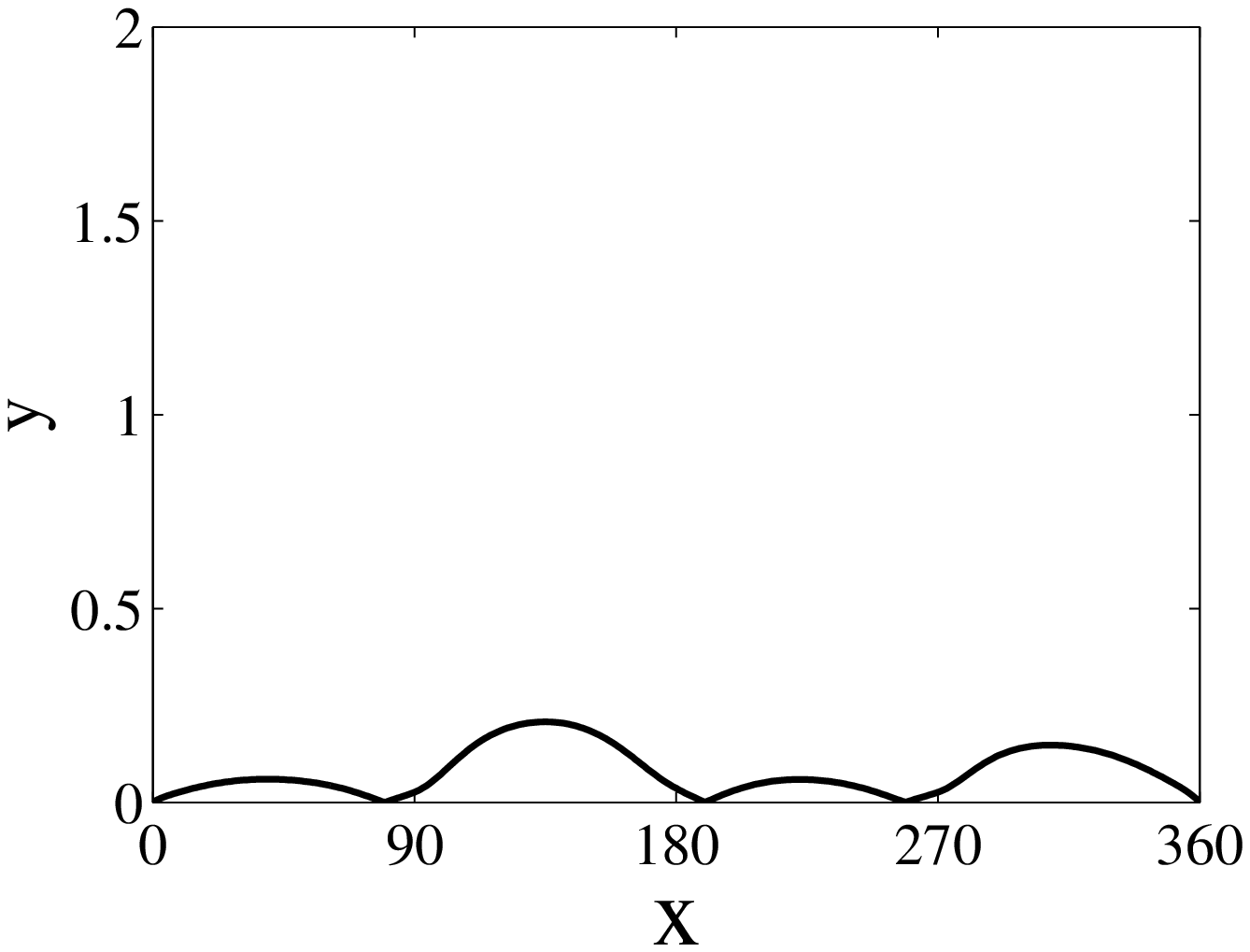}
  \caption{Magnitude $D^{(qP,qSV)}$ versus $\varthec$,  $\varphic^{inc}=90^0$ and $\varthec^{inc}=0^0$(a), $30^0$(b),
$60^0$(c),  $90^0$(d), $120^0$(e), $150^0$(f).}
\end{figure}

\vfill\eject

\begin{figure}[ht]
{\bf~a}$\quad\quad\quad\quad\quad\quad\quad\quad\quad\quad\quad\quad\quad\quad\quad\quad\quad\quad\quad${\bf~d}\hfil\break
\psfrag{x}{$\varthec,\,{}^{\rm o}$} \psfrag{y}{$|D|$}
\includegraphics[height=51mm,width=65mm]{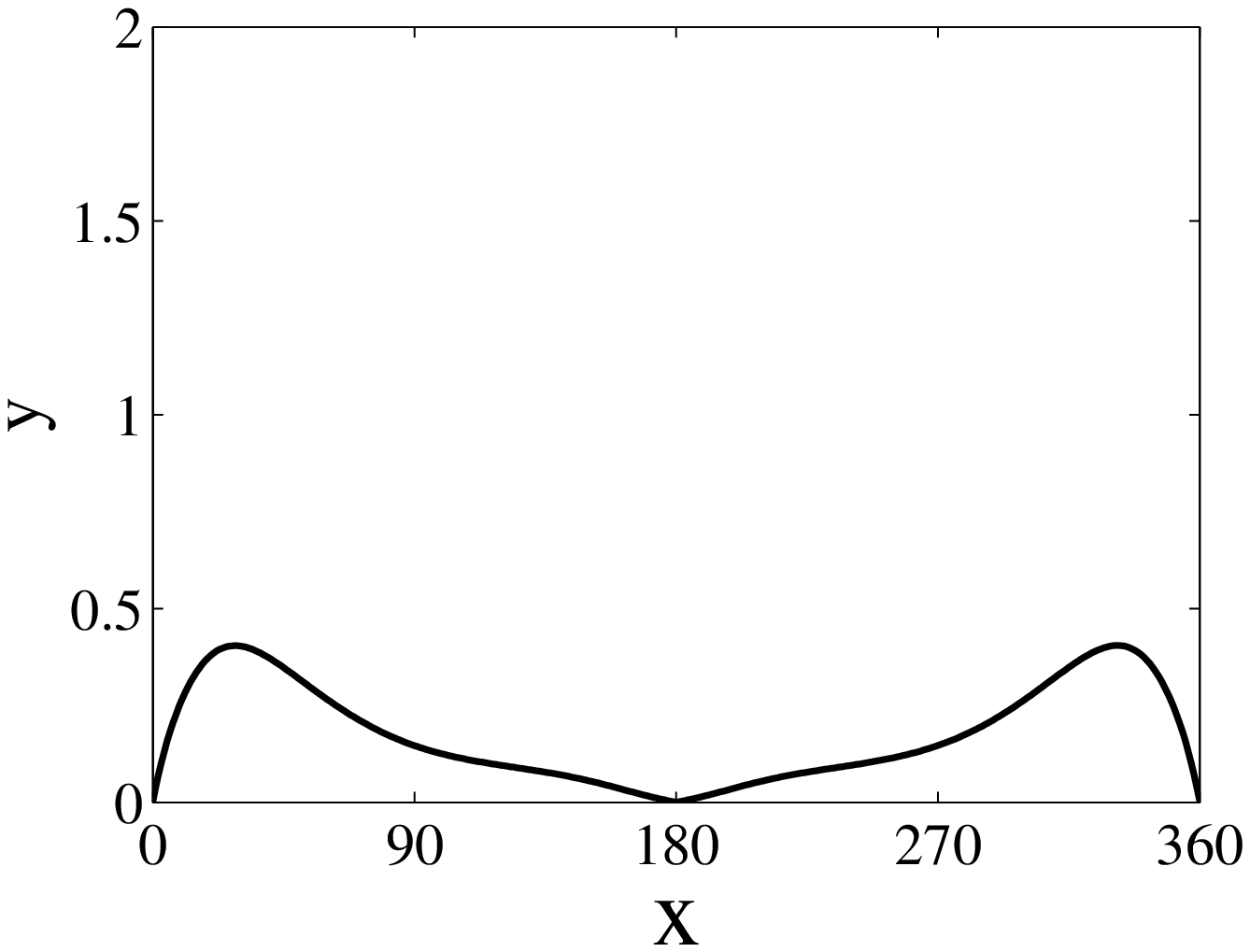}
  \includegraphics[height=51mm,width=65mm]{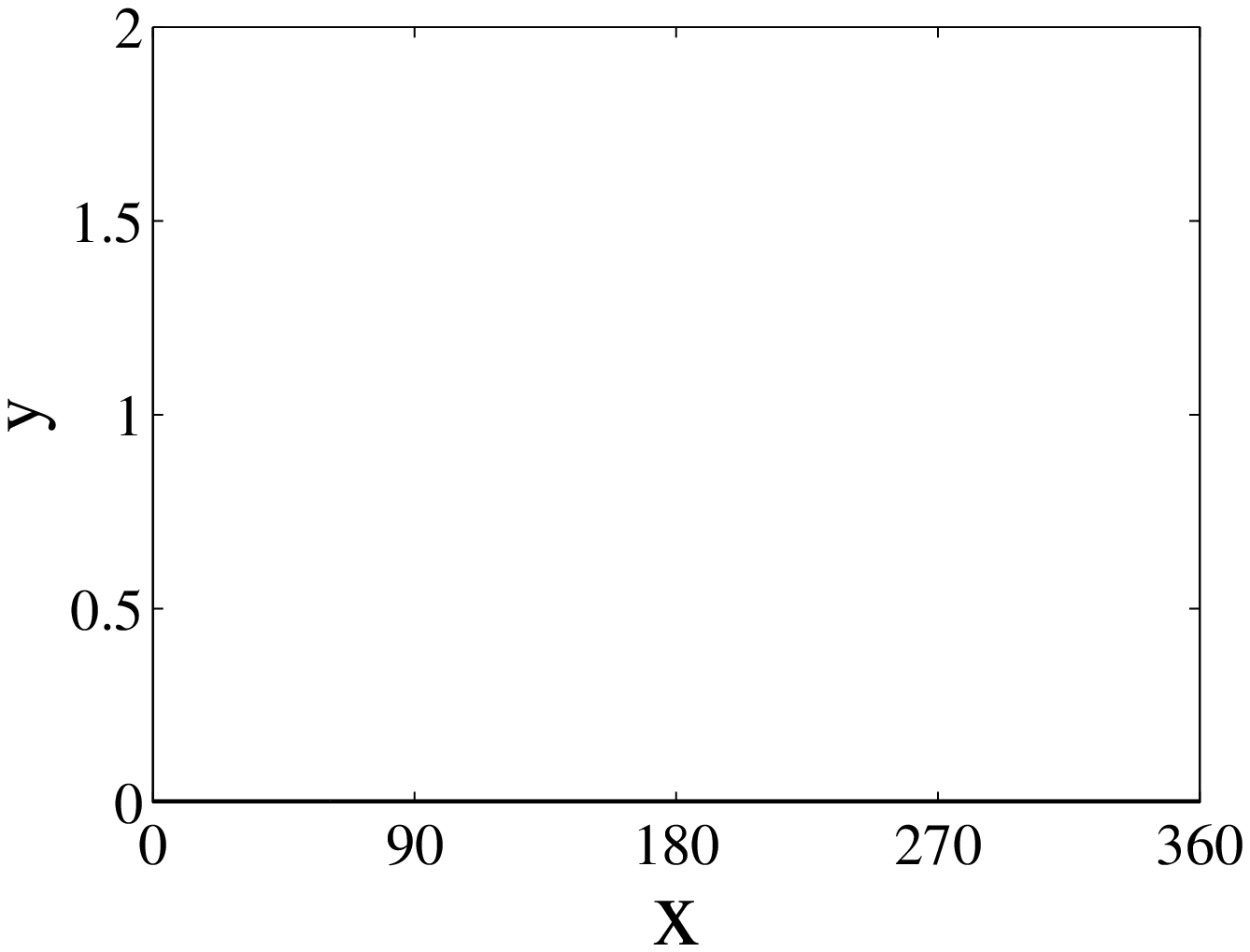}
  {\bf~b}$\quad\quad\quad\quad\quad\quad\quad\quad\quad\quad\quad\quad\quad\quad\quad\quad\quad\quad\quad${\bf~e}\hfil\break
  \includegraphics[height=51mm,width=65mm]{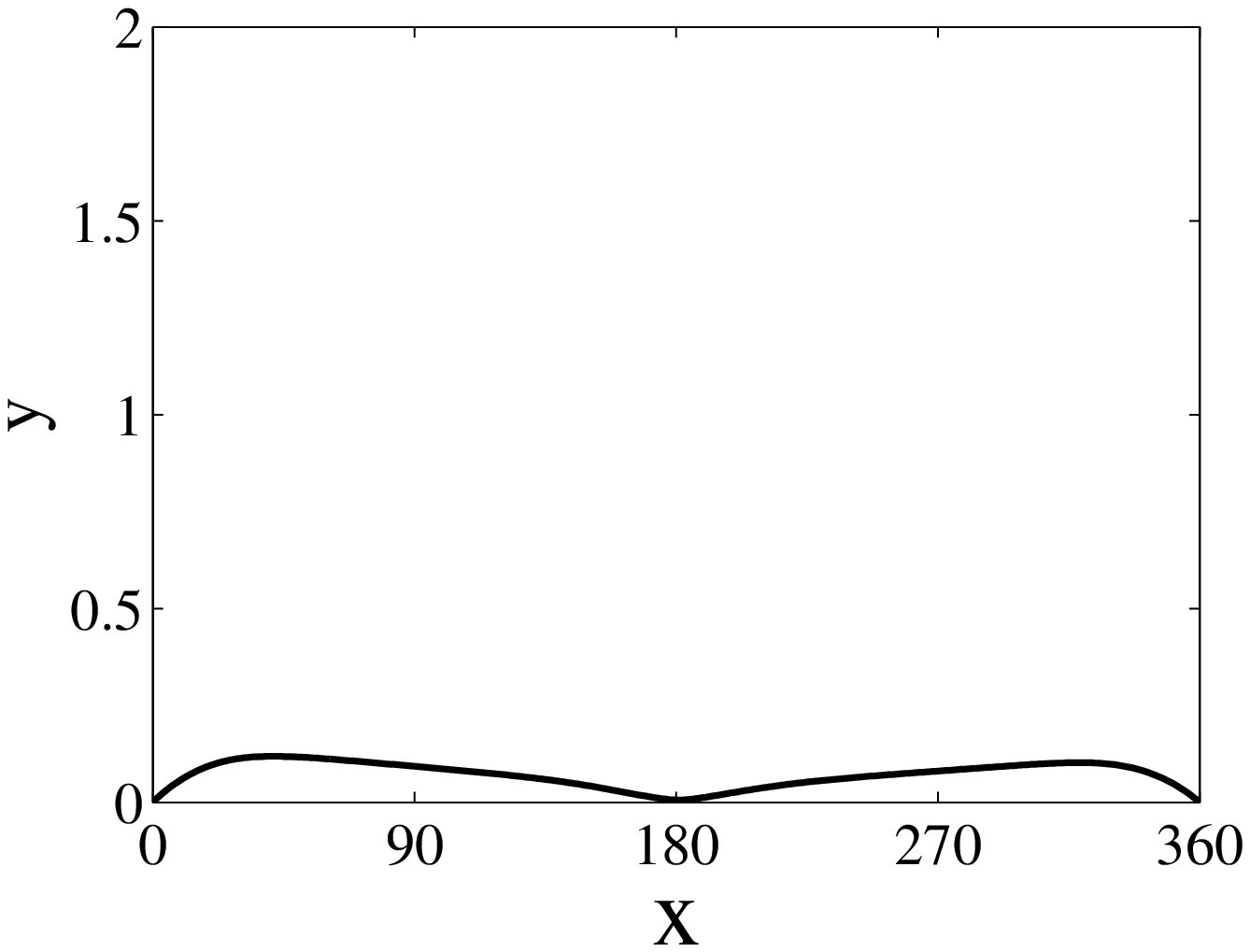}
  \includegraphics[height=51mm,width=65mm]{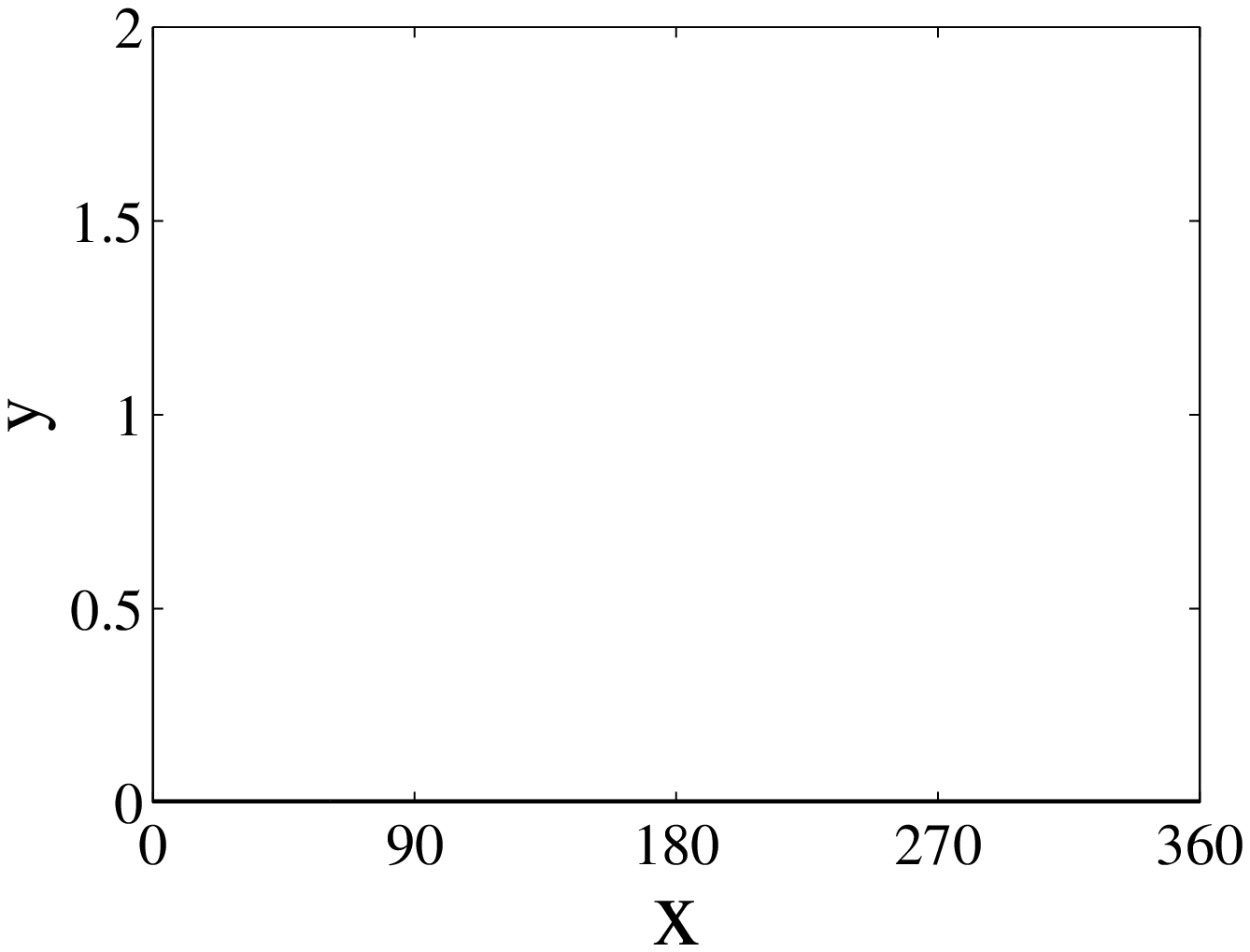}
  {\bf~c}$\quad\quad\quad\quad\quad\quad\quad\quad\quad\quad\quad\quad\quad\quad\quad\quad\quad\quad\quad${\bf~f}\hfil\break
  \includegraphics[height=51mm,width=65mm]{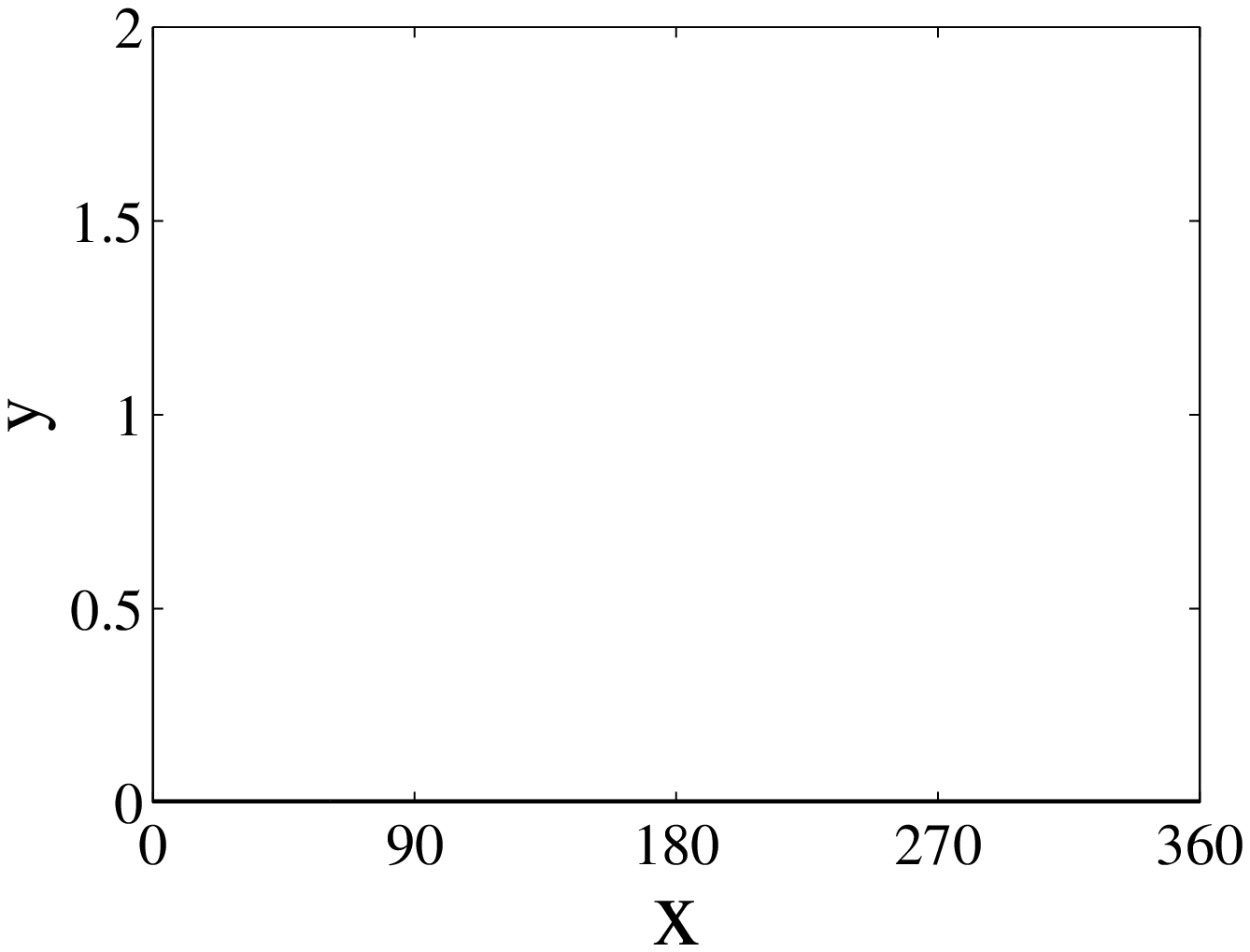}
  \hfil\hfil  \hfil\hfil   \hfil\hfil  \hfil\hfil
  \hfil\hfil  \hfil\hfil   \hfil\hfil  \hfil\hfil
  \hfil\hfil  \hfil\hfil   \hfil\hfil  \hfil\hfil
  \hfil\hfil  \hfil\hfil   \hfil\hfil  \hfil\hfil
  \includegraphics[height=51mm,width=65mm]{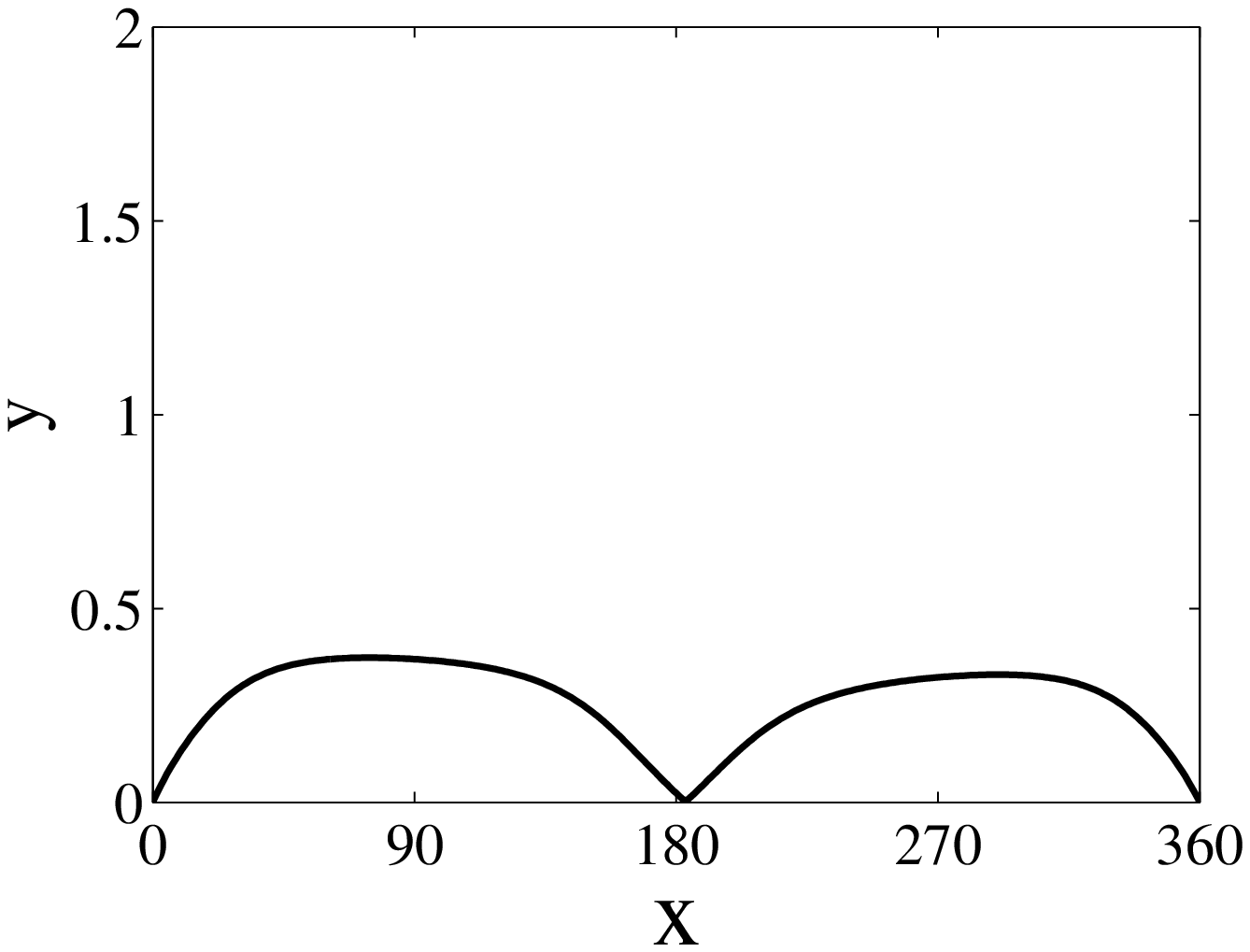}
  \caption{Magnitude $D^{(qP,qSV)}$ versus $\varthec$,  $\varphic^{inc}=60^0$ and $\varthec^{inc}=0^0$(a), $30^0$(b),
$60^0$(c),  $90^0$(d), $120^0$(e), $150^0$(f).}
\end{figure}

\vfill\eject

\begin{figure}[ht]
{\bf~a}$\quad\quad\quad\quad\quad\quad\quad\quad\quad\quad\quad\quad\quad\quad\quad\quad\quad\quad\quad${\bf~d}\hfil\break
\psfrag{x}{$\varthec,\,{}^{\rm o}$} \psfrag{y}{$|D|$}
\includegraphics[height=51mm,width=65mm]{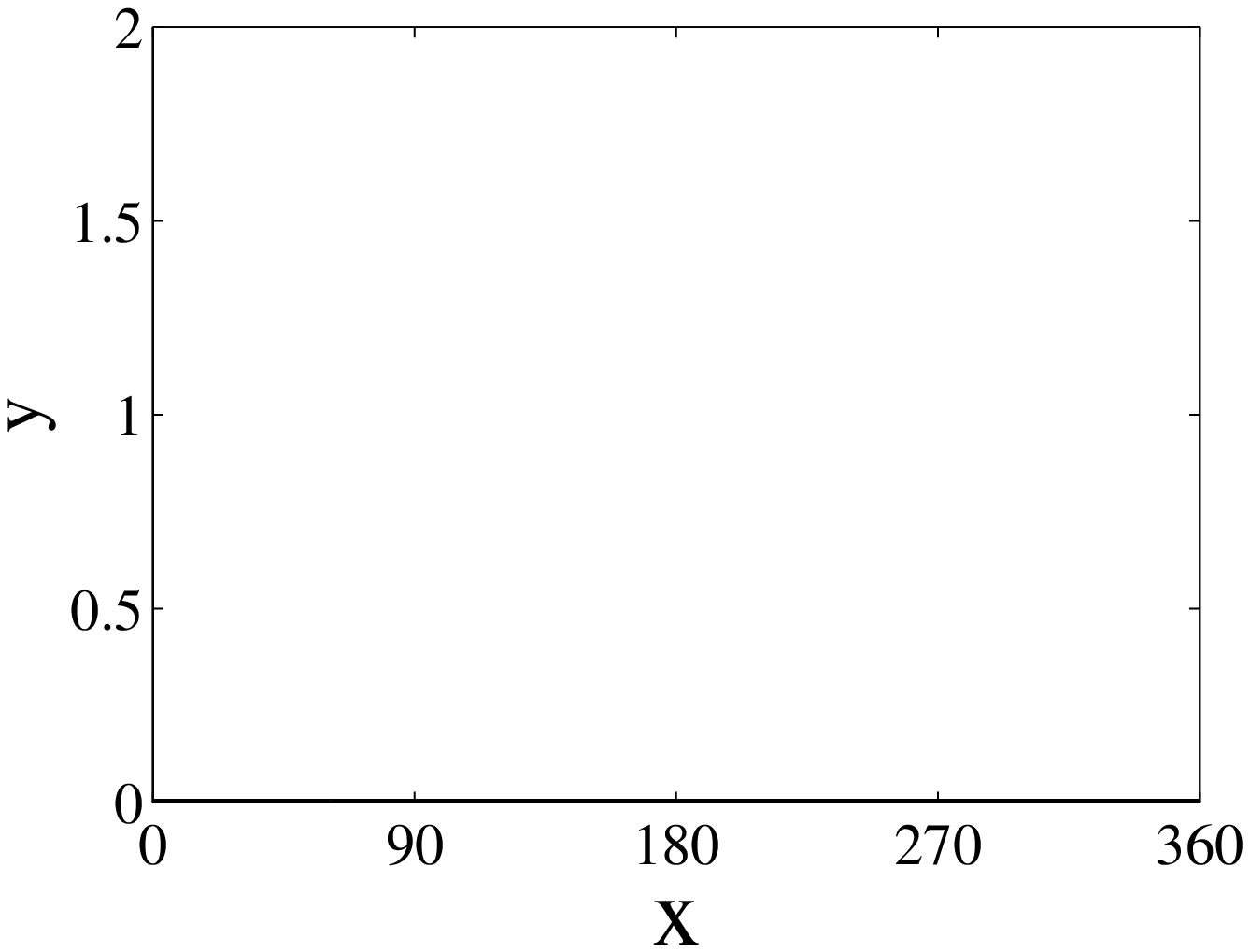}
  \includegraphics[height=51mm,width=65mm]{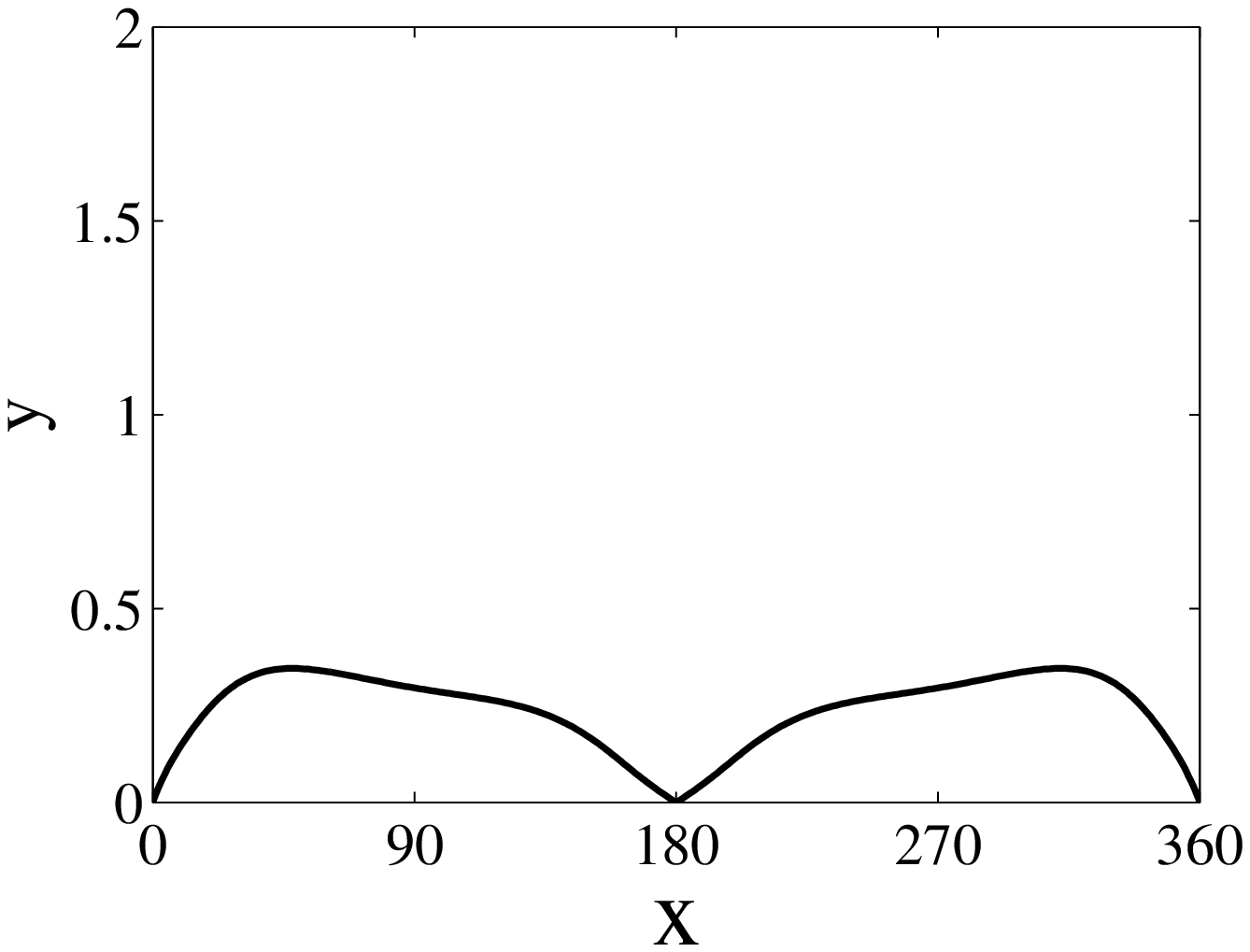}
  {\bf~b}$\quad\quad\quad\quad\quad\quad\quad\quad\quad\quad\quad\quad\quad\quad\quad\quad\quad\quad\quad${\bf~e}\hfil\break
  \includegraphics[height=51mm,width=65mm]{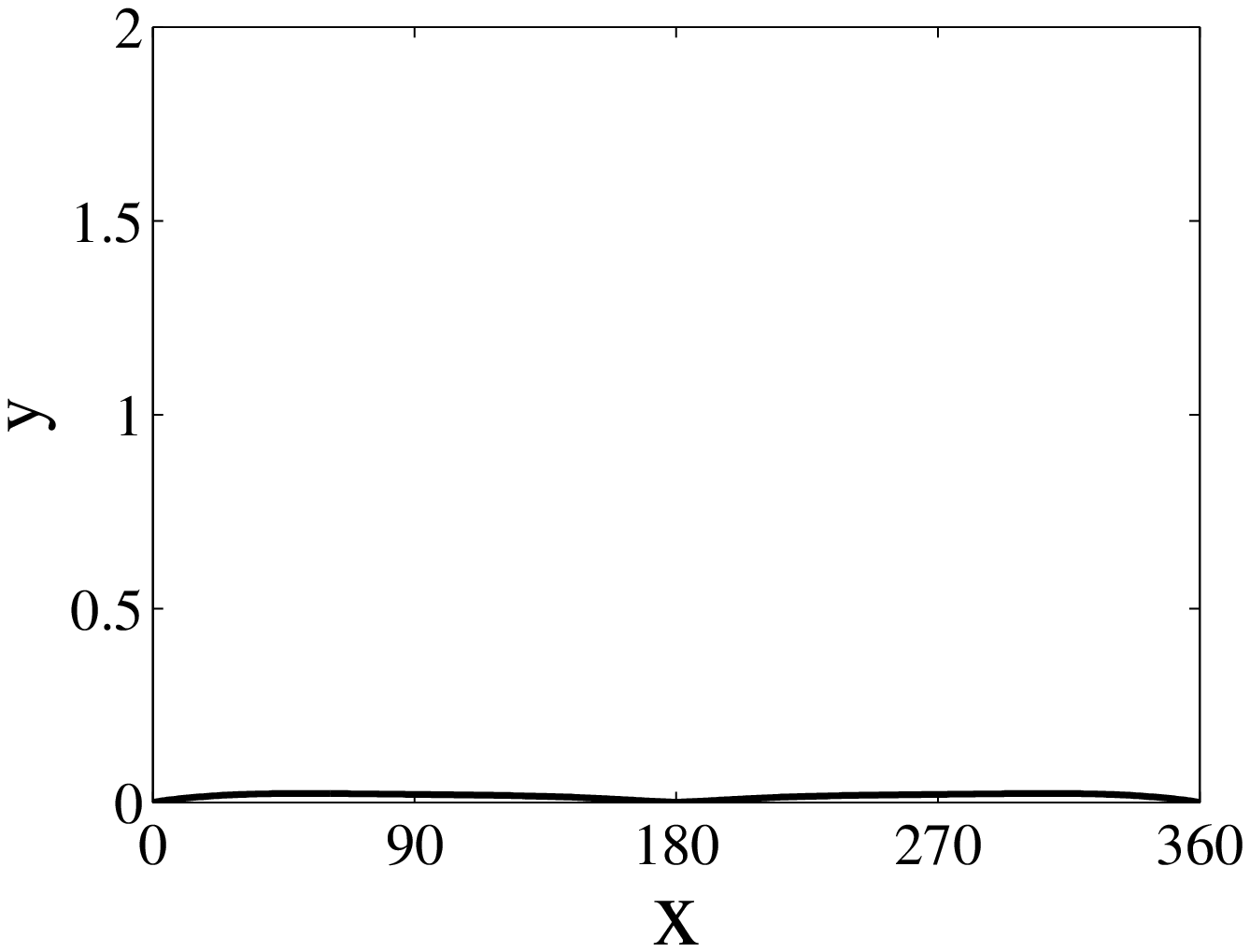}
  \includegraphics[height=51mm,width=65mm]{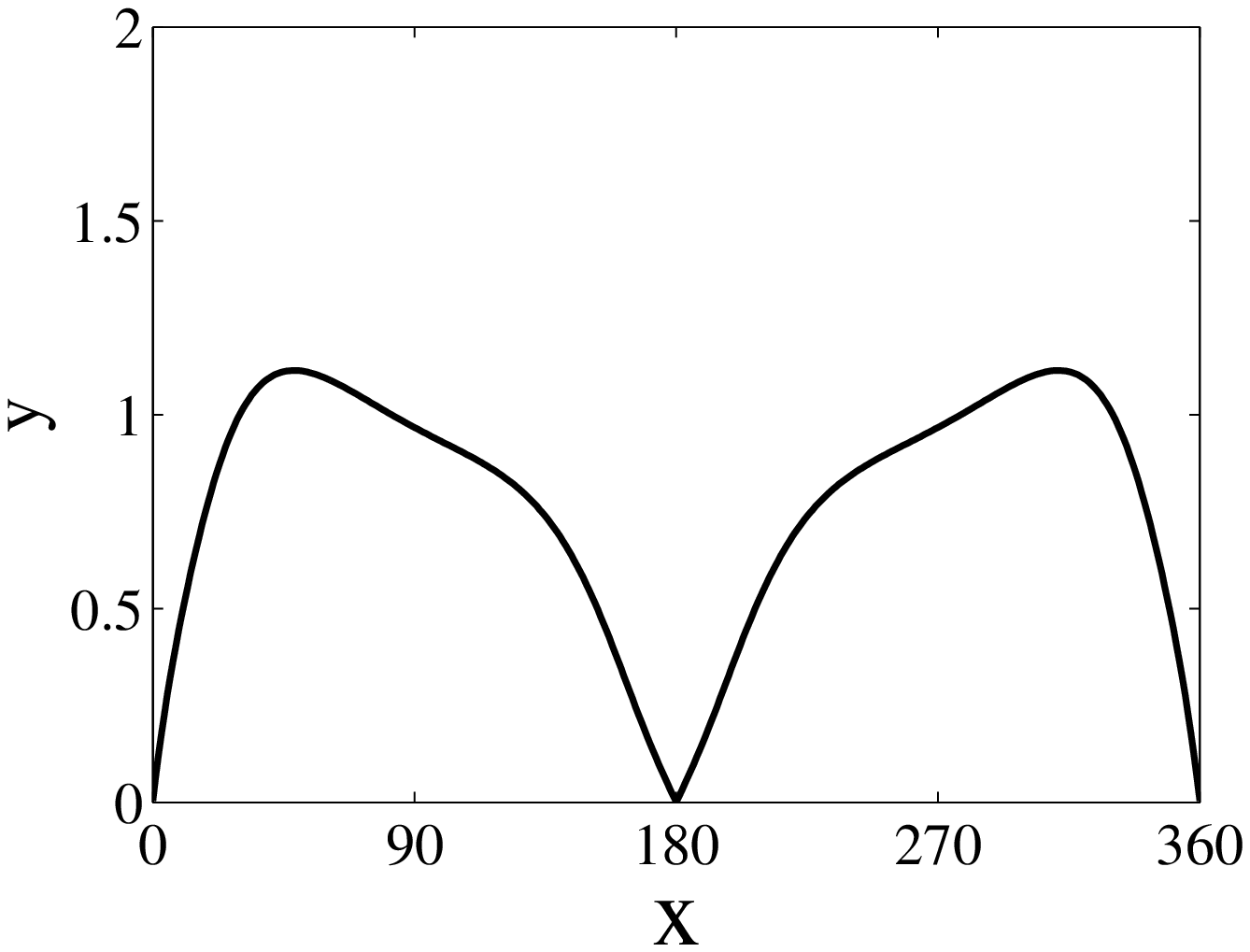}
  {\bf~c}$\quad\quad\quad\quad\quad\quad\quad\quad\quad\quad\quad\quad\quad\quad\quad\quad\quad\quad\quad${\bf~f}\hfil\break
  \includegraphics[height=51mm,width=65mm]{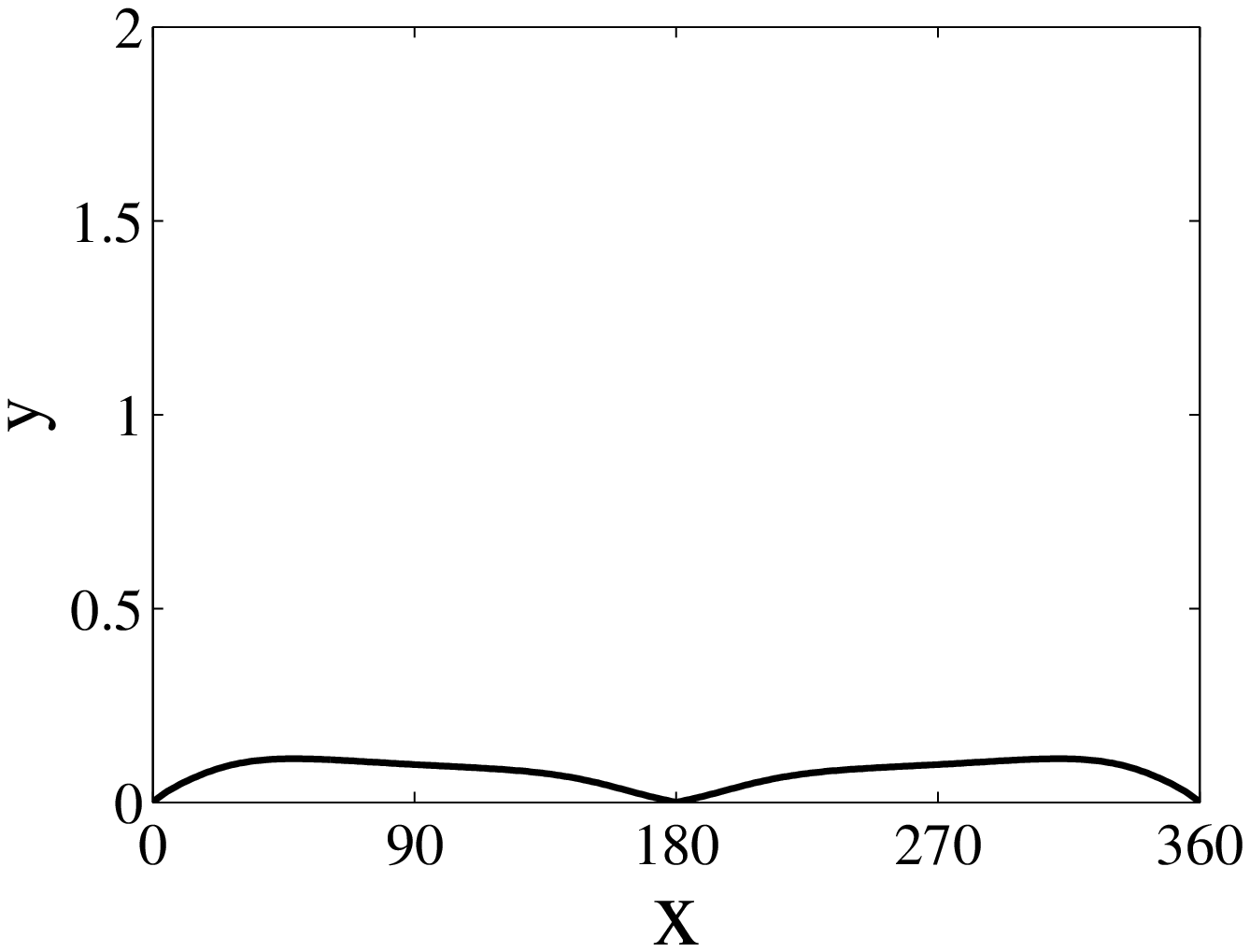}
  \hfil\hfil  \hfil\hfil   \hfil\hfil  \hfil\hfil
  \hfil\hfil  \hfil\hfil   \hfil\hfil  \hfil\hfil
  \hfil\hfil  \hfil\hfil   \hfil\hfil  \hfil\hfil
  \hfil\hfil  \hfil\hfil   \hfil\hfil  \hfil\hfil
  \includegraphics[height=51mm,width=65mm]{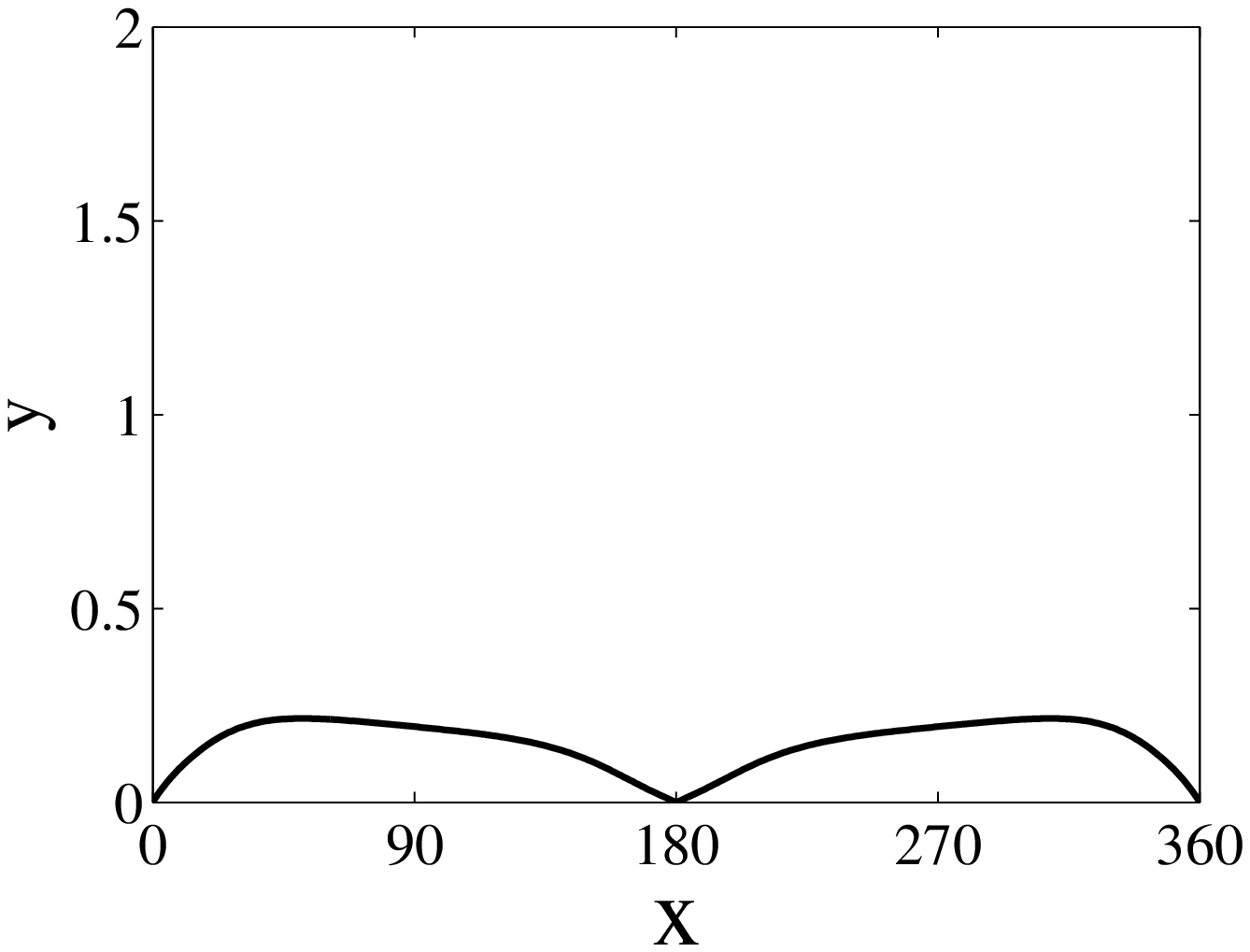}
   \caption{Magnitude $D^{(qP,qSH)}$ versus $\varthec$,  $\varphic^{inc}=60^0$ and $\varthec^{inc}=0^0$(a), $30^0$(b),
$60^0$(c),  $90^0$(d), $120^0$(e), $150^0$(f).}
\end{figure}

\vfill\eject

\begin{figure}[ht]
{\bf~a}$\quad\quad\quad\quad\quad\quad\quad\quad\quad\quad\quad\quad\quad\quad\quad\quad\quad\quad\quad${\bf~d}\hfil\break
\psfrag{x}{$\varthec,\,{}^{\rm o}$} \psfrag{y}{$|D|$}
\includegraphics[height=51mm,width=65mm]{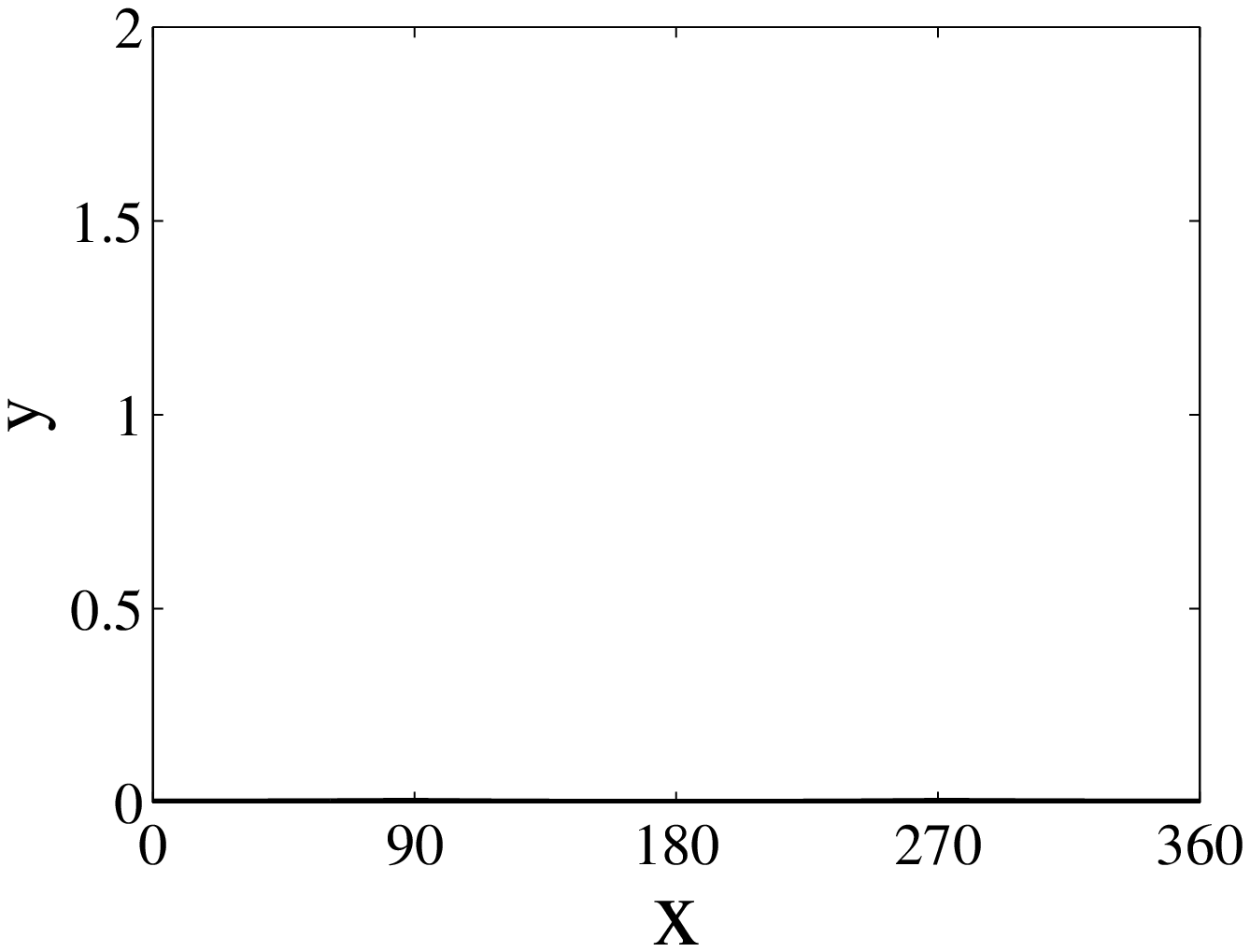}
  \includegraphics[height=51mm,width=65mm]{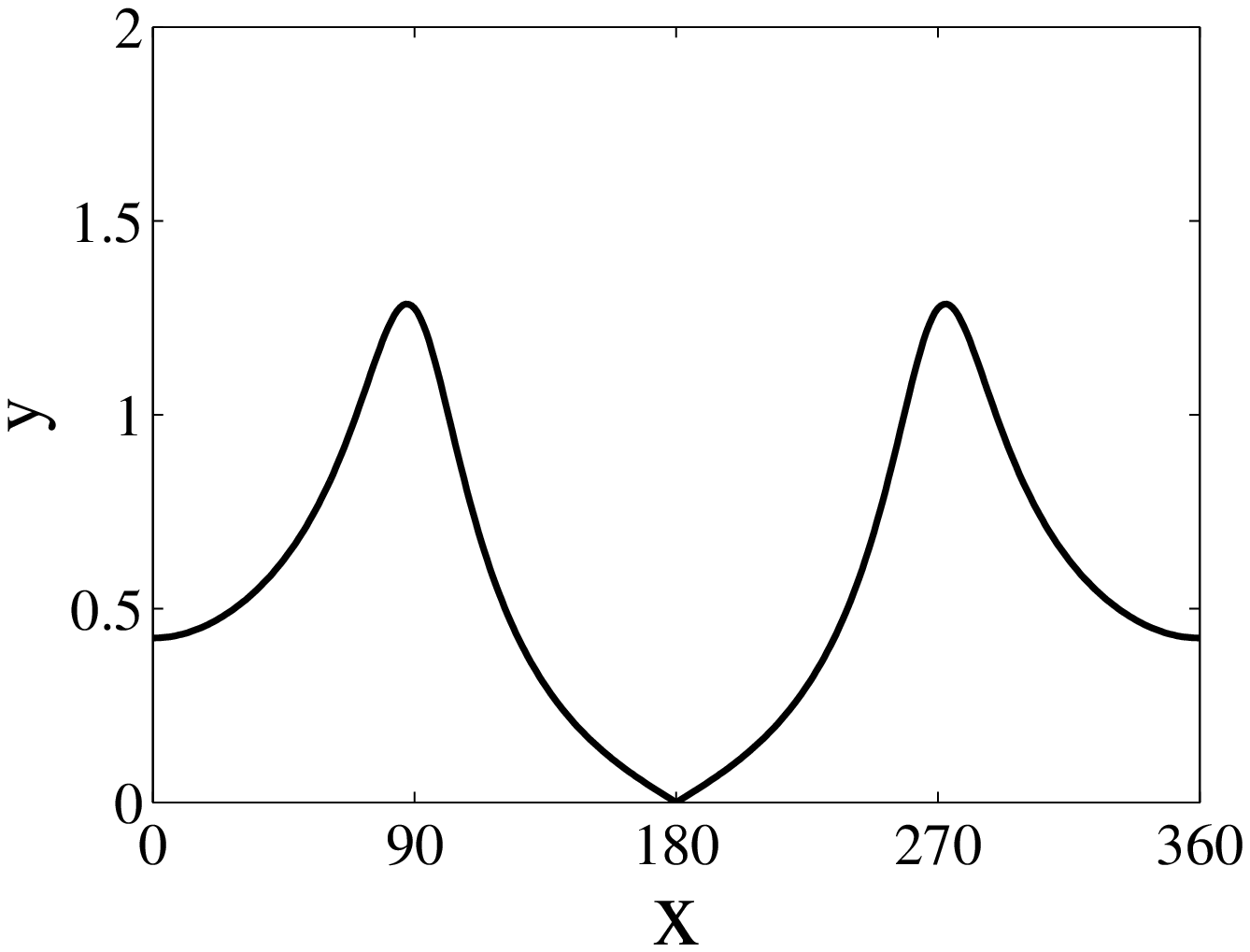}
  {\bf~b}$\quad\quad\quad\quad\quad\quad\quad\quad\quad\quad\quad\quad\quad\quad\quad\quad\quad\quad\quad${\bf~e}\hfil\break
  \includegraphics[height=51mm,width=65mm]{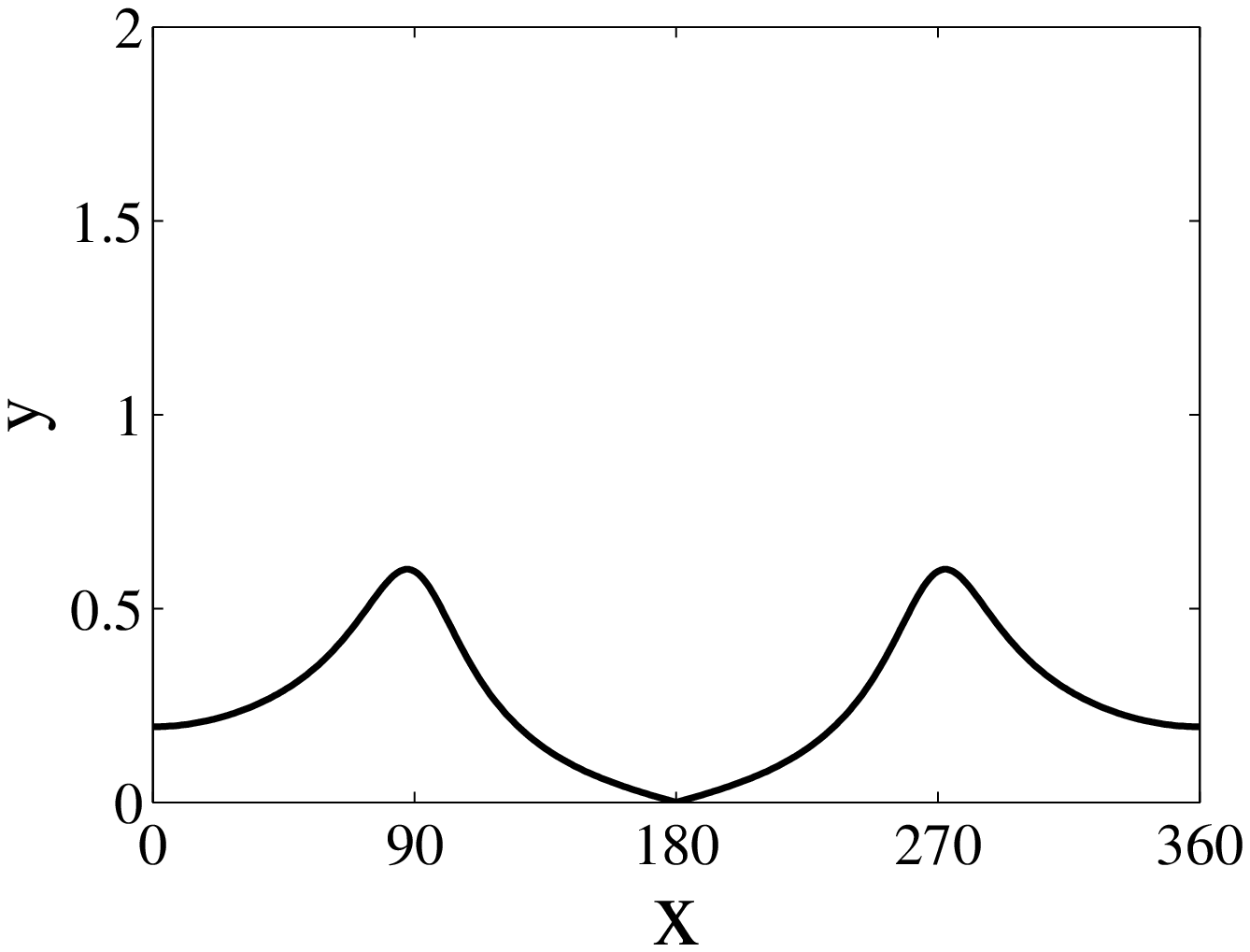}
  \includegraphics[height=51mm,width=65mm]{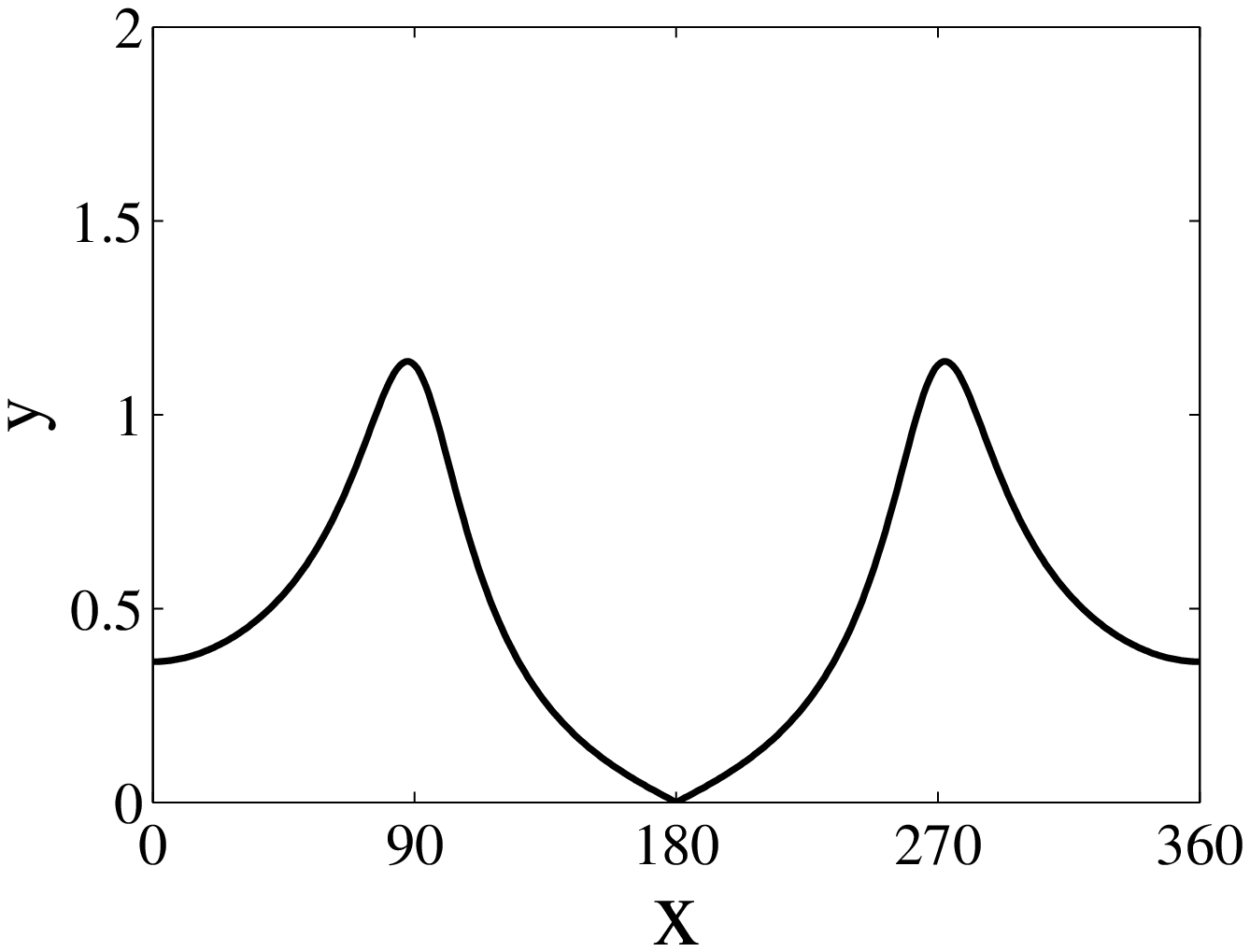}
  {\bf~c}$\quad\quad\quad\quad\quad\quad\quad\quad\quad\quad\quad\quad\quad\quad\quad\quad\quad\quad\quad${\bf~f}\hfil\break
  \includegraphics[height=51mm,width=65mm]{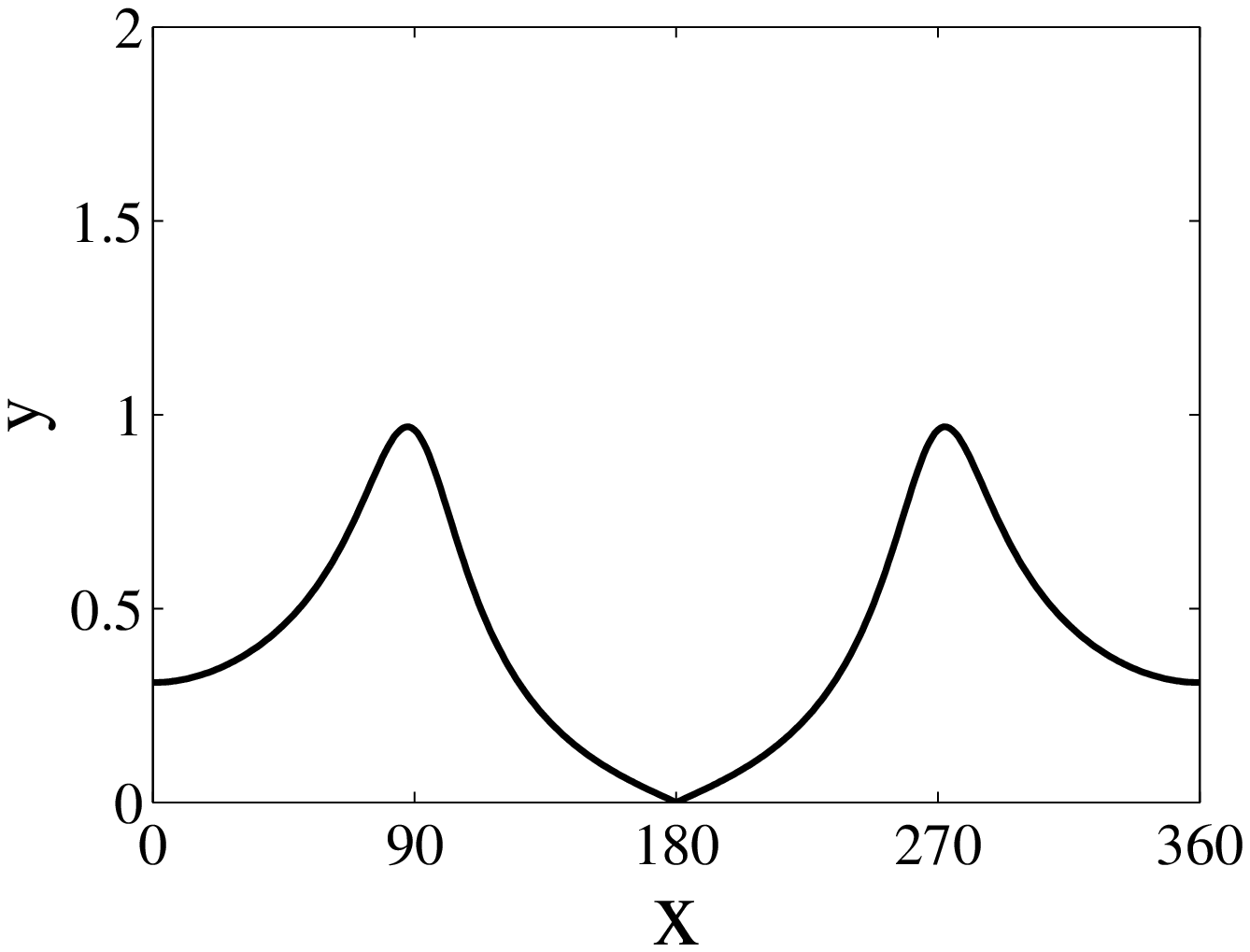}
  \hfil\hfil  \hfil\hfil   \hfil\hfil  \hfil\hfil
  \hfil\hfil  \hfil\hfil   \hfil\hfil  \hfil\hfil
  \hfil\hfil  \hfil\hfil   \hfil\hfil  \hfil\hfil
  \hfil\hfil  \hfil\hfil   \hfil\hfil  \hfil\hfil
  \includegraphics[height=51mm,width=65mm]{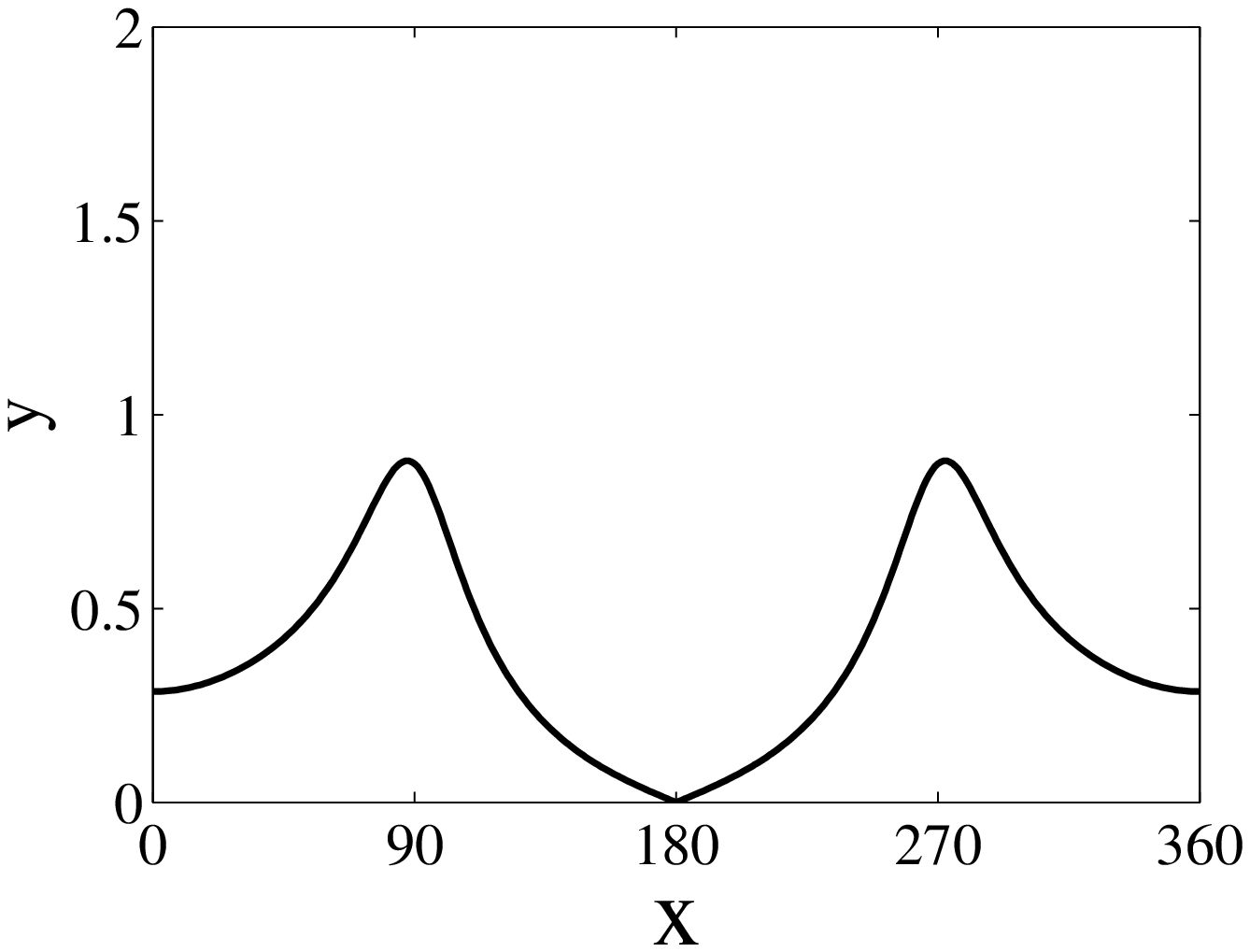}
  \caption{Magnitude $D^{(qSH,qP)}$ versus $\varthec$,  $\varphic^{inc}=60^0$ and $\varthec^{inc}=0^0$(a), $30^0$(b),
$60^0$(c),  $90^0$(d), $120^0$(e), $150^0$(f).}
\end{figure}

\vfill\eject

\begin{figure}[ht]
{\bf~a}$\quad\quad\quad\quad\quad\quad\quad\quad\quad\quad\quad\quad\quad\quad\quad\quad\quad\quad\quad${\bf~d}\hfil\break
\psfrag{x}{$\varthec,\,{}^{\rm o}$} \psfrag{y}{$|D|$}
\includegraphics[height=51mm,width=65mm]{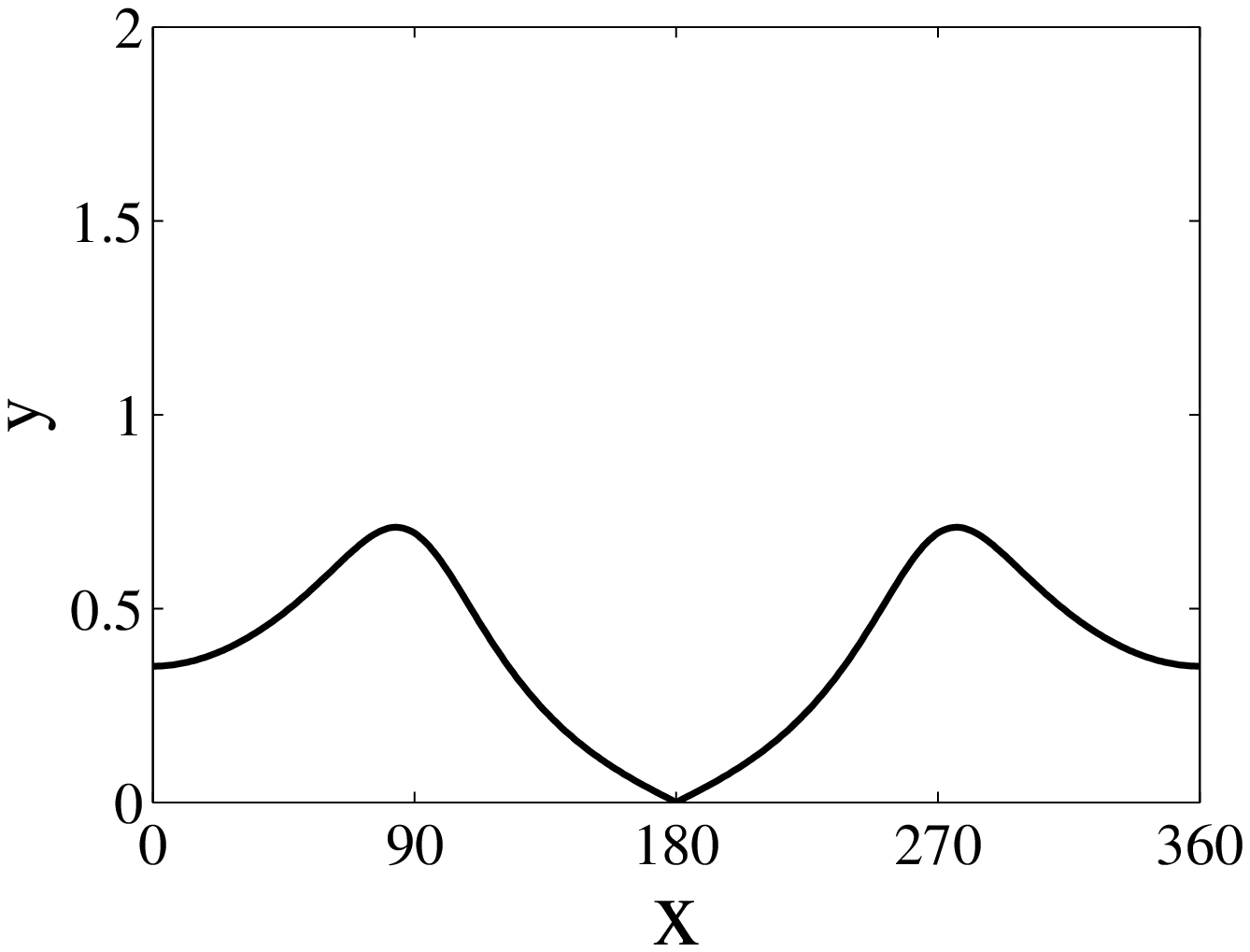}
  \includegraphics[height=51mm,width=65mm]{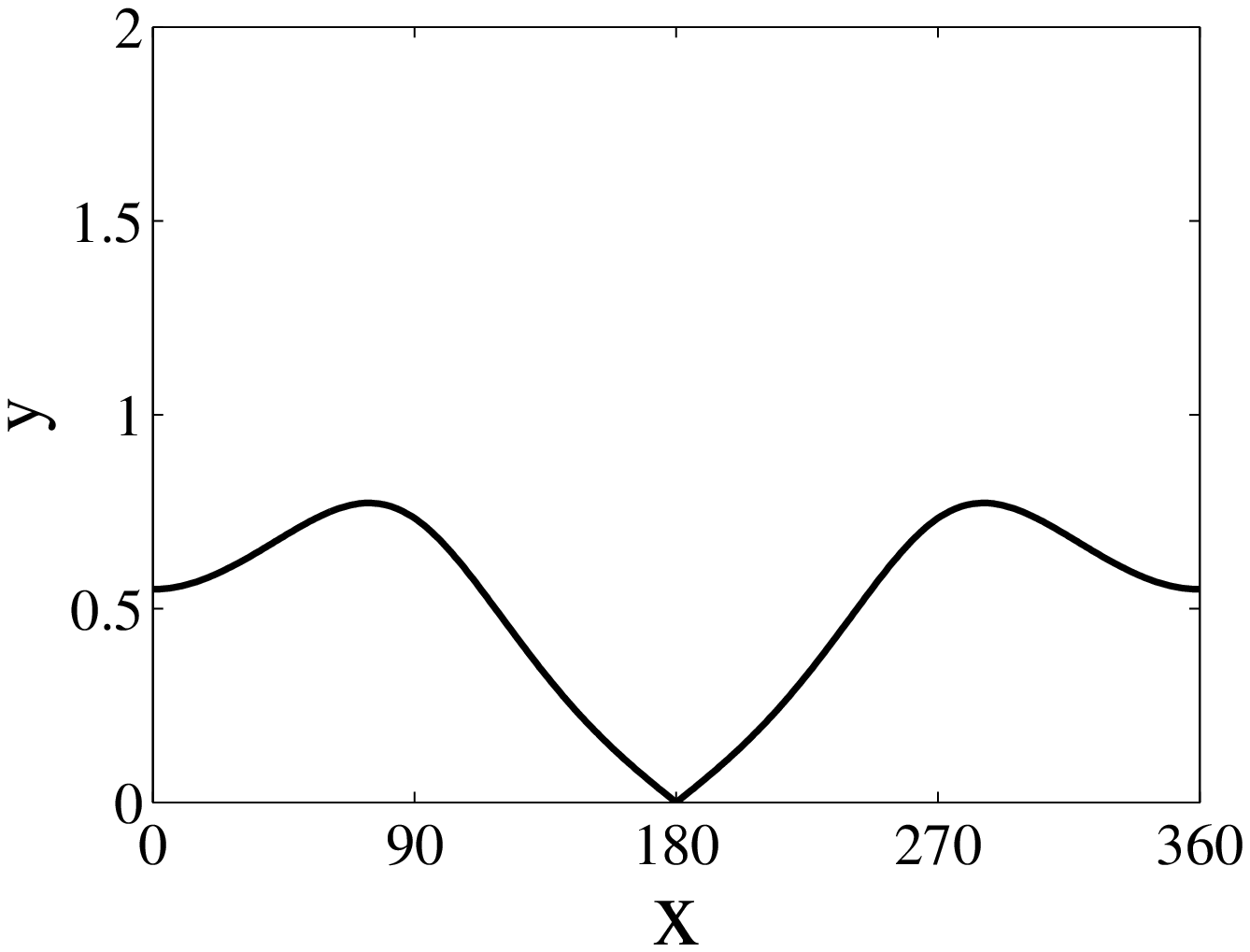}
  {\bf~b}$\quad\quad\quad\quad\quad\quad\quad\quad\quad\quad\quad\quad\quad\quad\quad\quad\quad\quad\quad${\bf~e}\hfil\break
  \includegraphics[height=51mm,width=65mm]{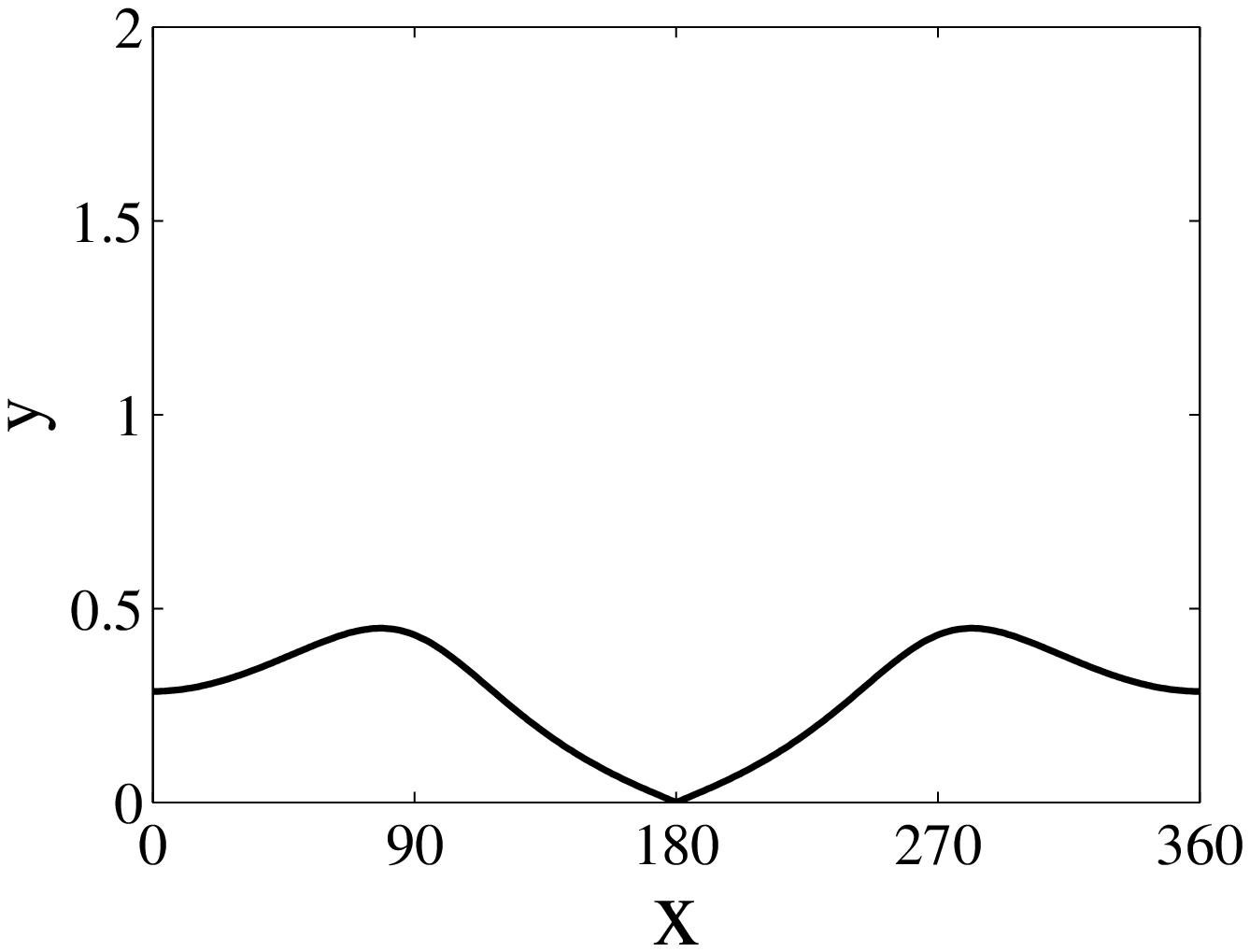}
  \includegraphics[height=51mm,width=65mm]{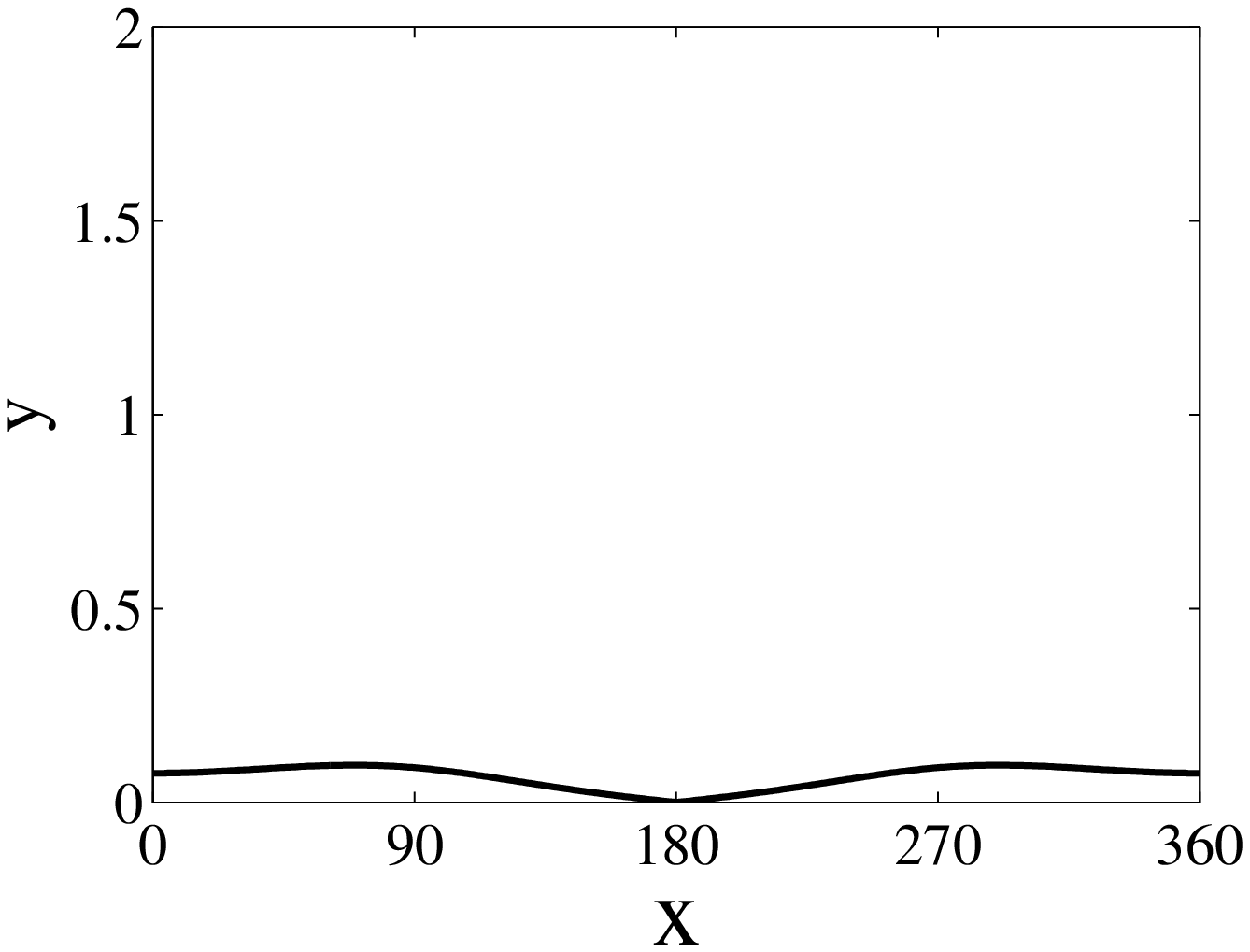}
  {\bf~c}$\quad\quad\quad\quad\quad\quad\quad\quad\quad\quad\quad\quad\quad\quad\quad\quad\quad\quad\quad${\bf~f}\hfil\break
  \includegraphics[height=51mm,width=65mm]{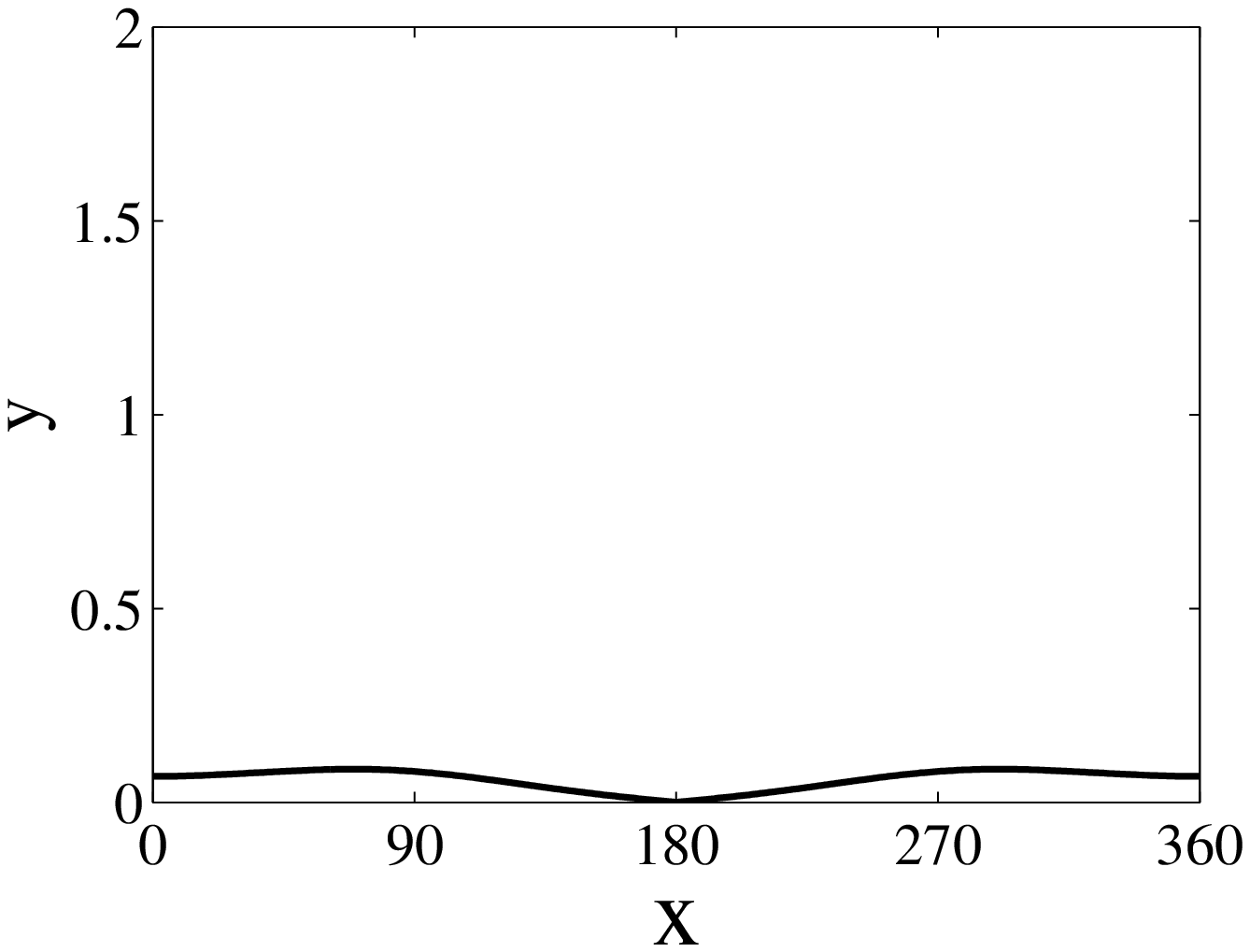}
  \hfil\hfil  \hfil\hfil   \hfil\hfil  \hfil\hfil
  \hfil\hfil  \hfil\hfil   \hfil\hfil  \hfil\hfil
  \hfil\hfil  \hfil\hfil   \hfil\hfil  \hfil\hfil
  \hfil\hfil  \hfil\hfil   \hfil\hfil  \hfil\hfil
  \includegraphics[height=51mm,width=65mm]{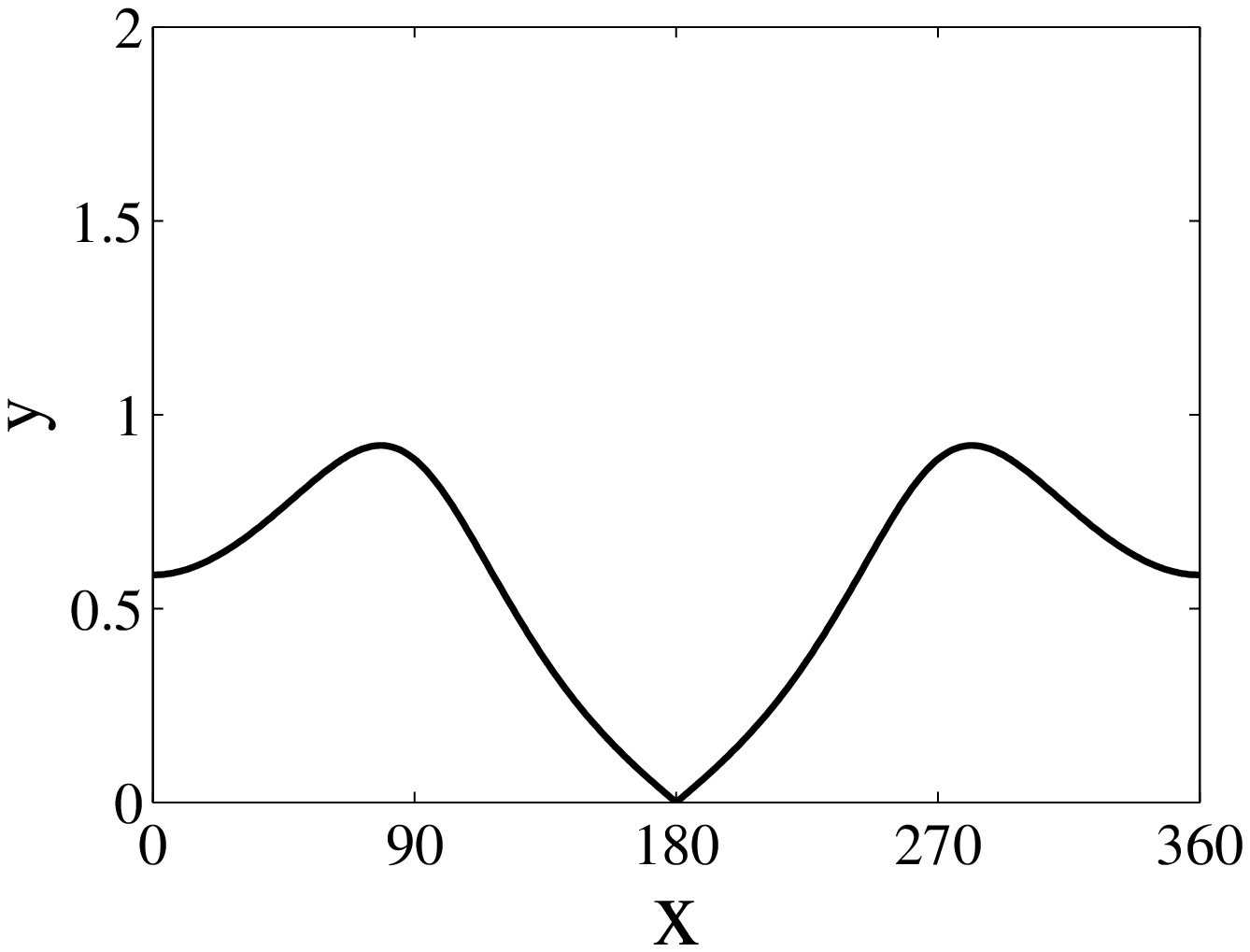}
  \caption{Magnitude $D^{(qSH,qSV)}$ versus $\varthec$,  $\varphic^{inc}=60^0$ and $\varthec^{inc}=0^0$(a), $30^0$(b),
$60^0$(c),  $90^0$(d), $120^0$(e), $150^0$(f).}
\end{figure}

\vfill\eject

\begin{figure}[ht]
{\bf~a}$\quad\quad\quad\quad\quad\quad\quad\quad\quad\quad\quad\quad\quad\quad\quad\quad\quad\quad\quad${\bf~d}\hfil\break
\psfrag{x}{$\varthec,\,{}^{\rm o}$}
\psfrag{y}{$|D|$}
\includegraphics[height=51mm,width=65mm]{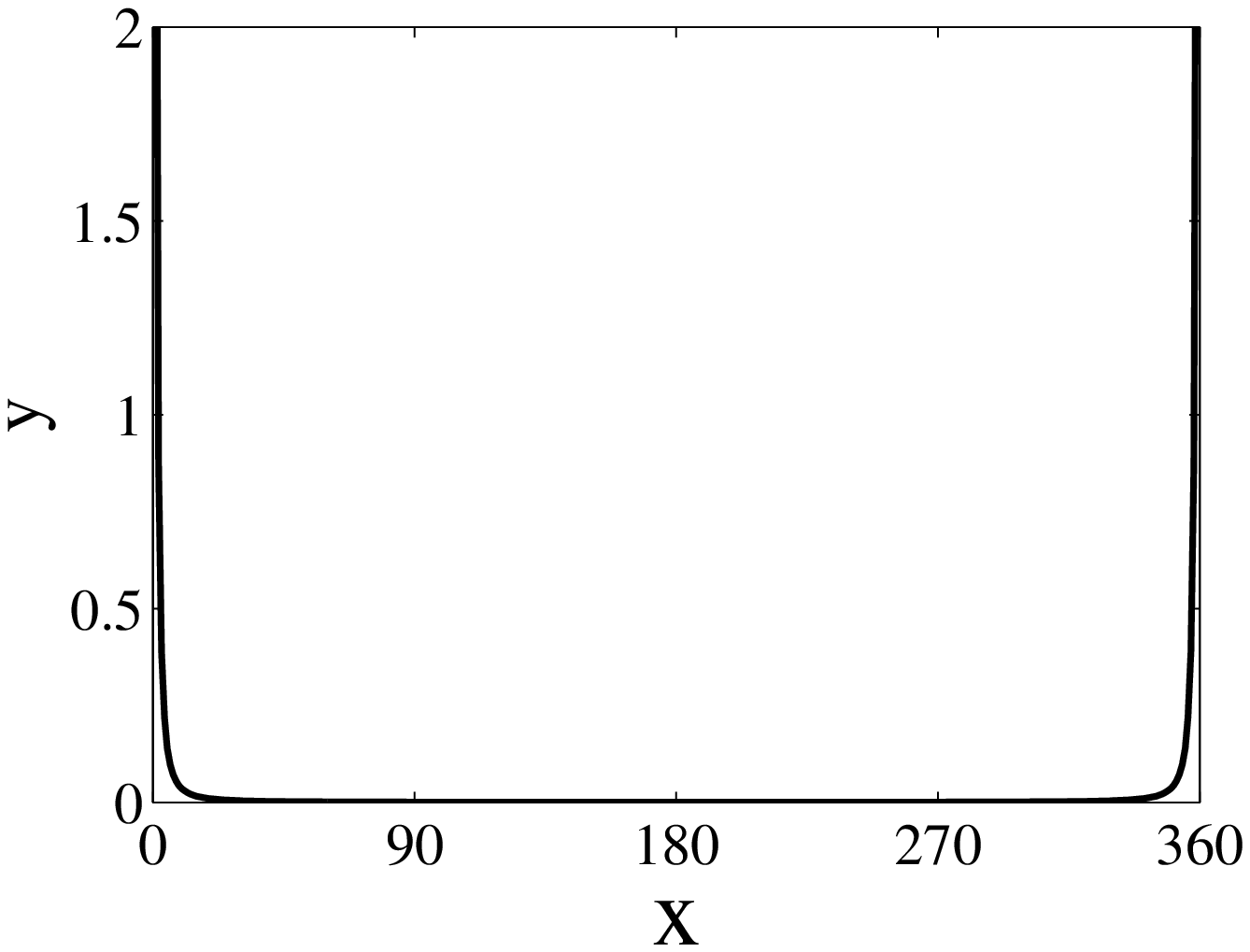}
\includegraphics[height=51mm,width=65mm]{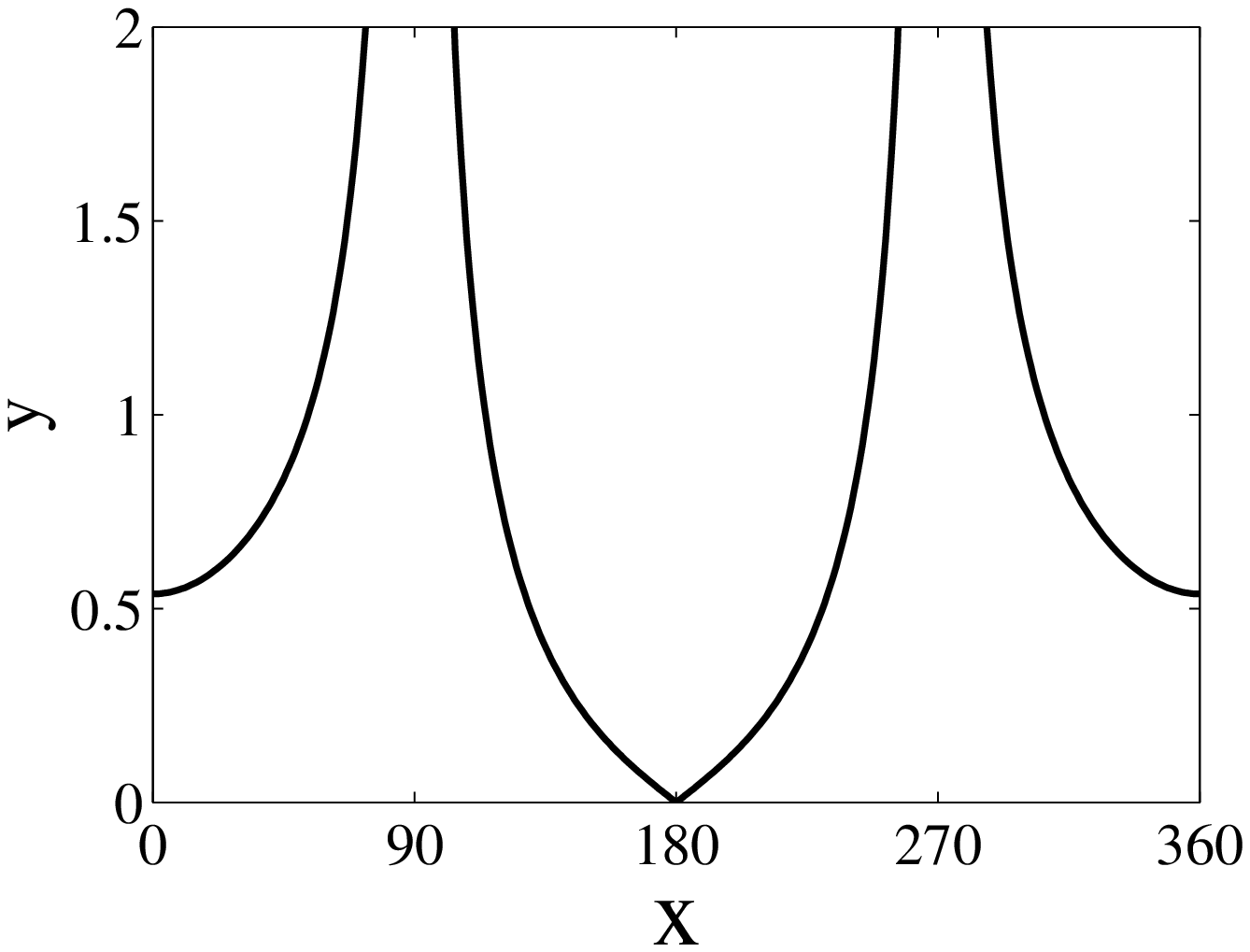}
{\bf~b}$\quad\quad\quad\quad\quad\quad\quad\quad\quad\quad\quad\quad\quad\quad\quad\quad\quad\quad\quad${\bf~e}\hfil\break
\includegraphics[height=51mm,width=65mm]{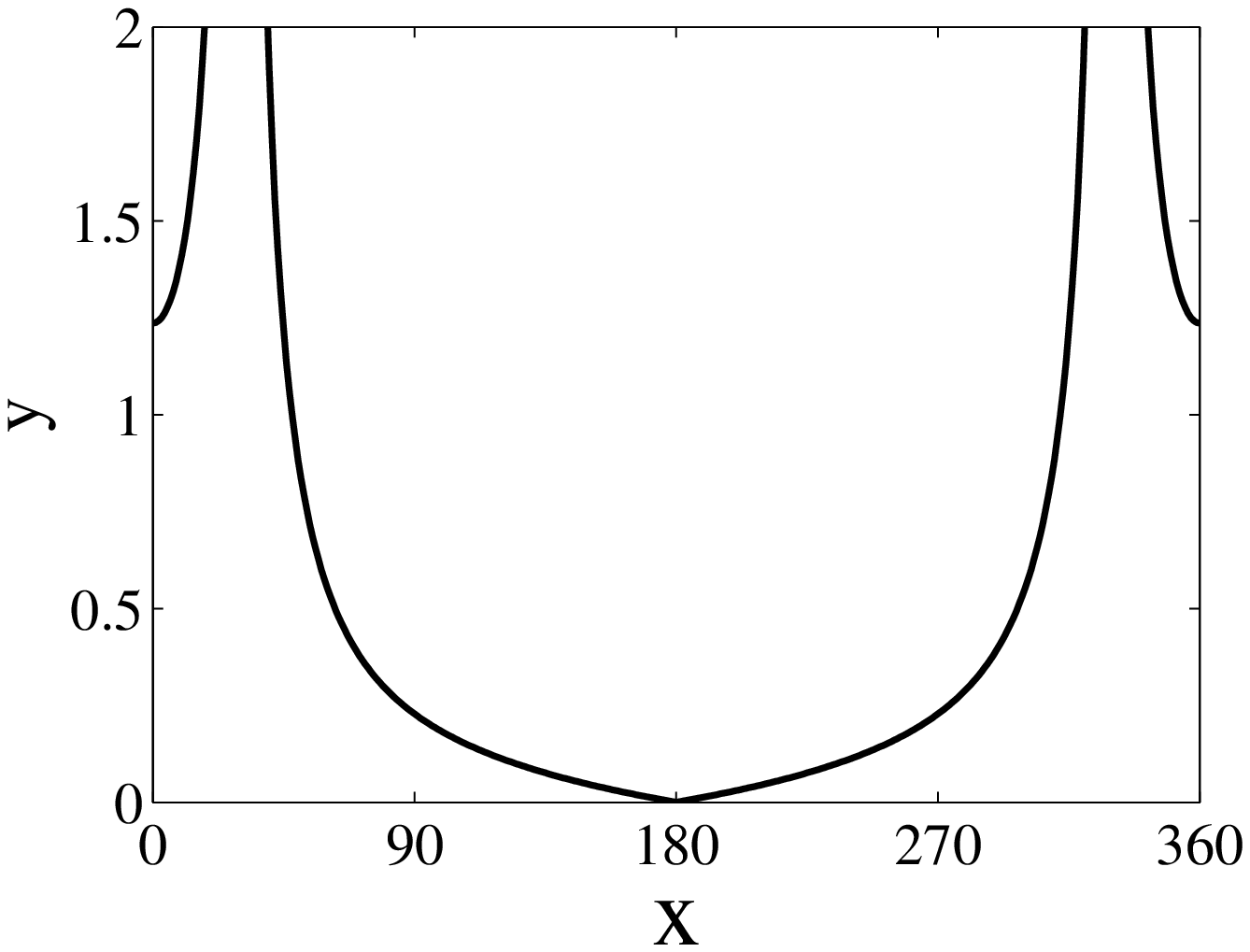}
\includegraphics[height=51mm,width=65mm]{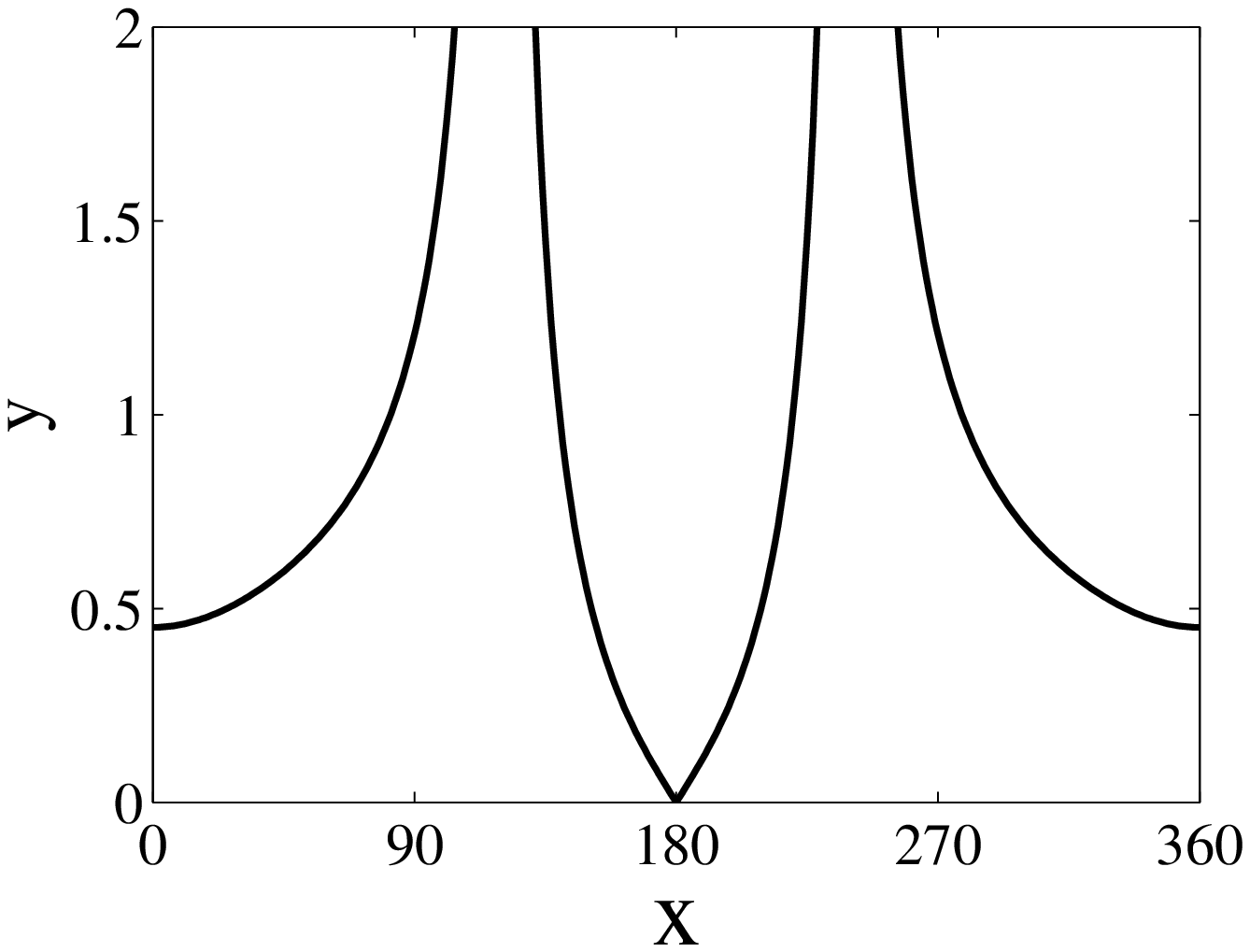}
{\bf~c}$\quad\quad\quad\quad\quad\quad\quad\quad\quad\quad\quad\quad\quad\quad\quad\quad\quad\quad\quad${\bf~f}\hfil\break
\includegraphics[height=51mm,width=65mm]{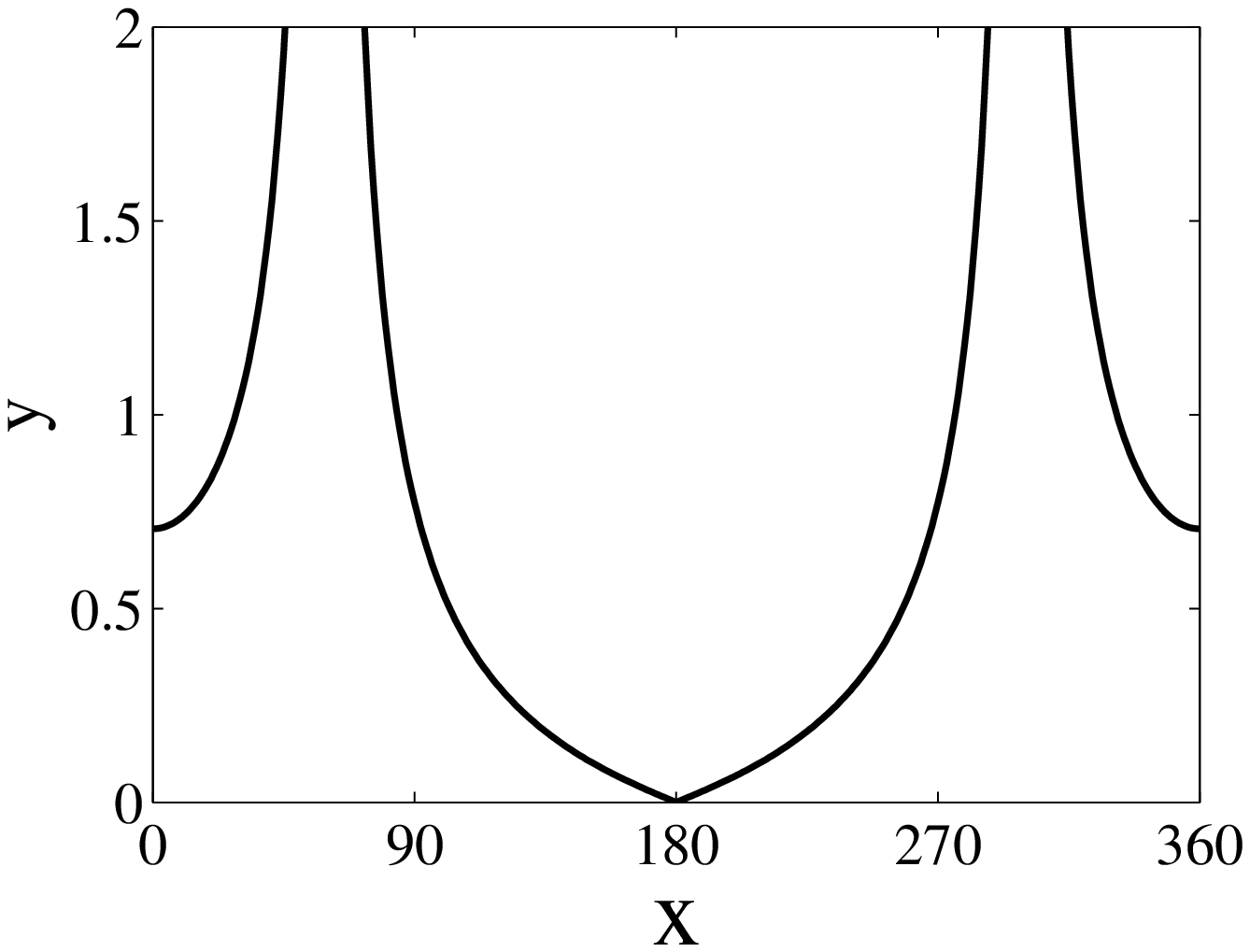}
  \hfil\hfil  \hfil\hfil   \hfil\hfil  \hfil\hfil
  \hfil\hfil  \hfil\hfil   \hfil\hfil  \hfil\hfil
  \hfil\hfil  \hfil\hfil   \hfil\hfil  \hfil\hfil
  \hfil\hfil  \hfil\hfil   \hfil\hfil  \hfil\hfil
\includegraphics[height=51mm,width=65mm]{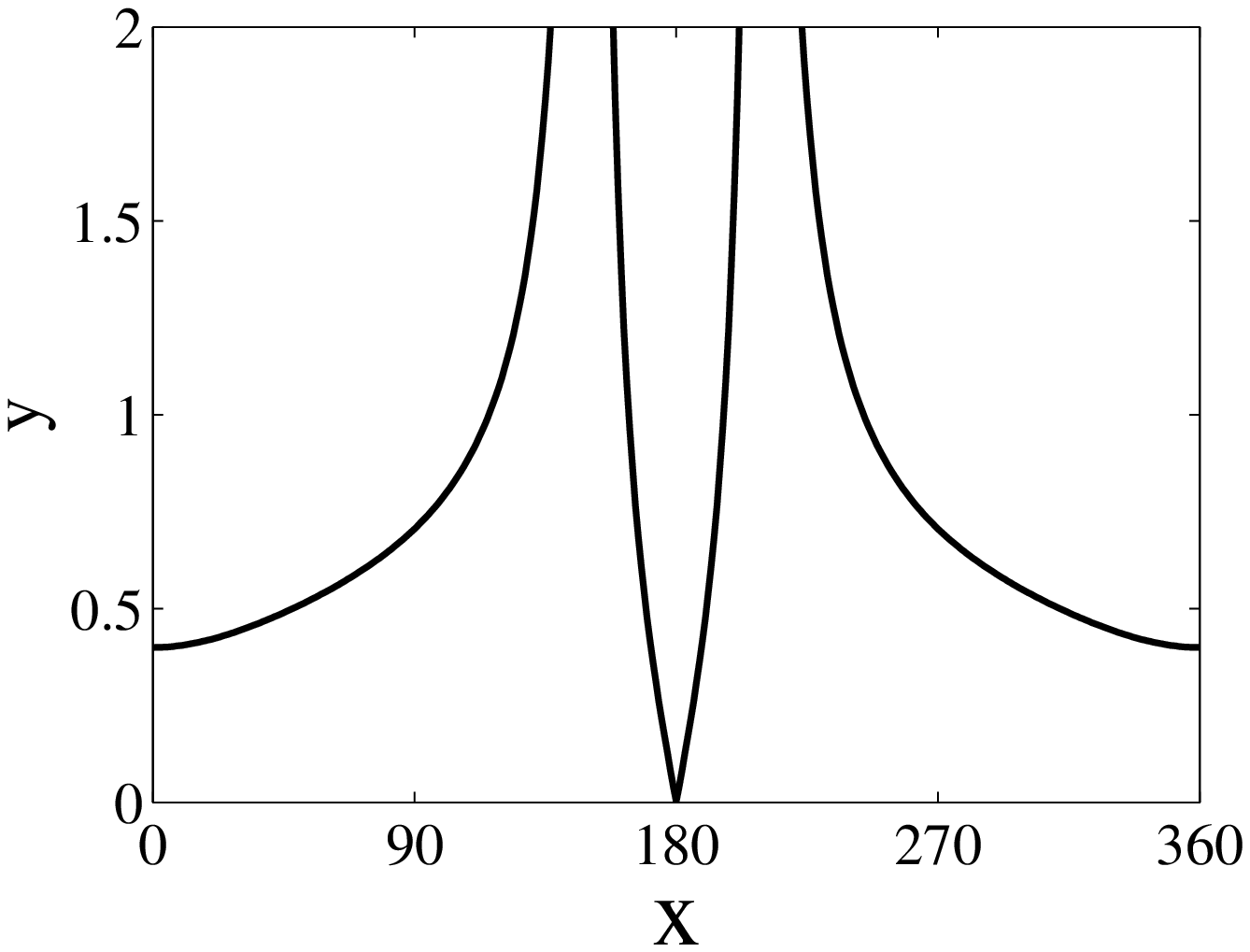}
  \caption{Magnitude $D^{(qSH,qSH)}$ versus $\varthec$,  $\varphic^{inc}=90^0$ and $\varthec^{inc}=0^0$(a), $30^0$(b), $60^0$(c),  $90^0$(d), $120^0$(e), $150^0$(f).}
  \label{3DQSHQSH90}
\end{figure}

\vfill\eject

\begin{figure}[ht]
{\bf~a}$\quad\quad\quad\quad\quad\quad\quad\quad\quad\quad\quad\quad\quad\quad\quad\quad\quad\quad\quad${\bf~d}\hfil\break
\psfrag{x}{$\varthec,\,{}^{\rm o}$}
\psfrag{y}{$|D|$}
\includegraphics[height=51mm,width=65mm]{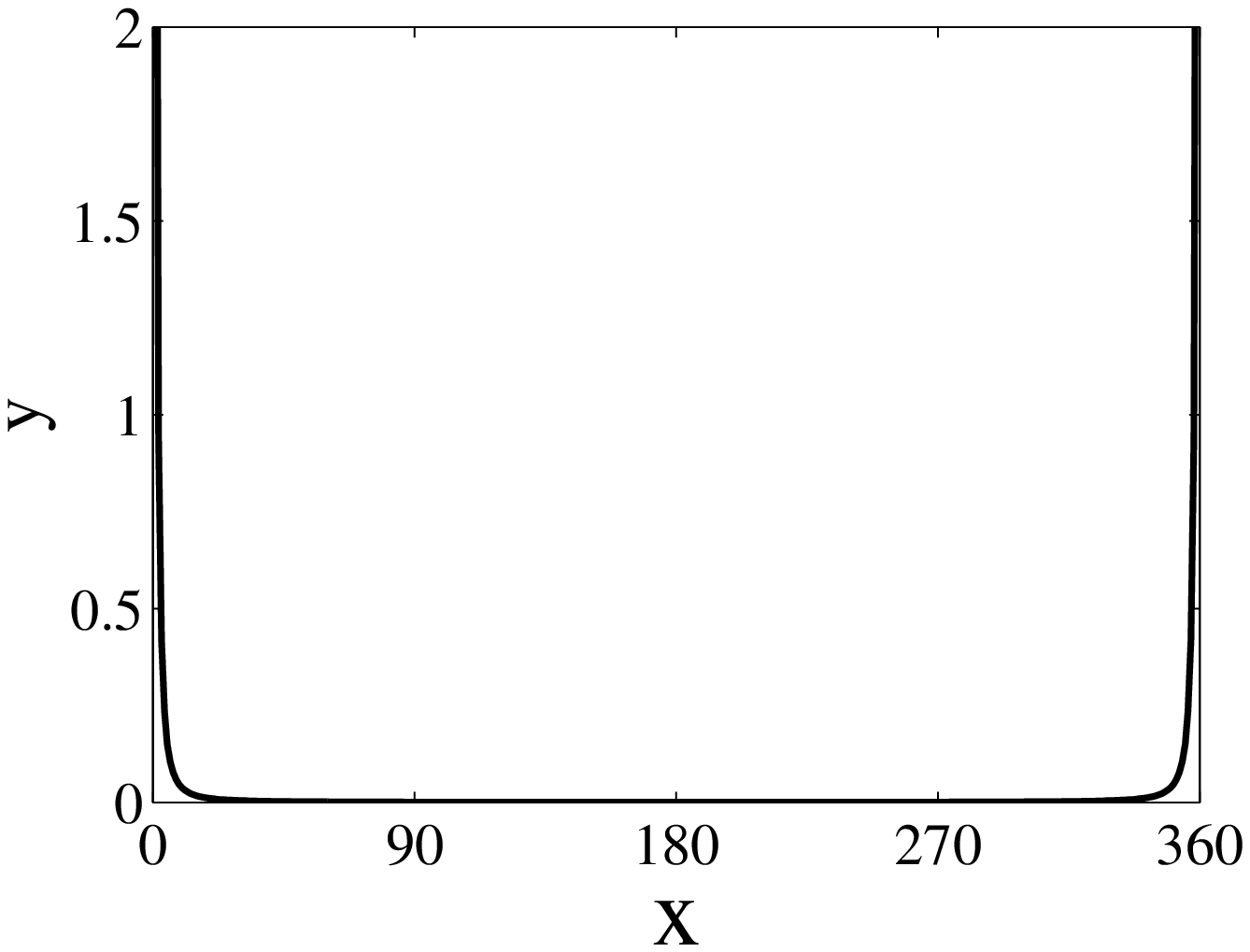}
  \includegraphics[height=51mm,width=65mm]{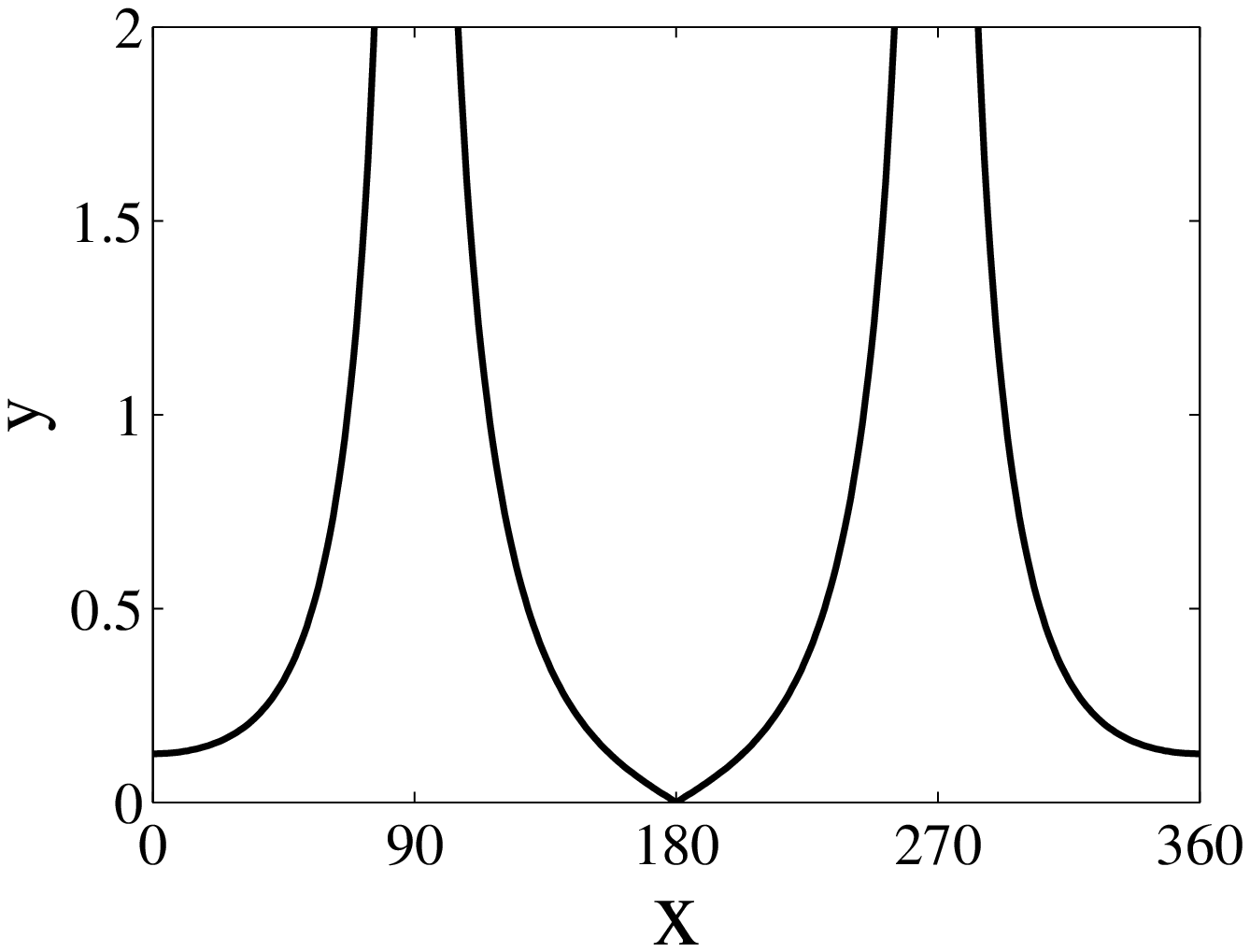}
  {\bf~b}$\quad\quad\quad\quad\quad\quad\quad\quad\quad\quad\quad\quad\quad\quad\quad\quad\quad\quad\quad${\bf~e}\hfil\break
  \includegraphics[height=51mm,width=65mm]{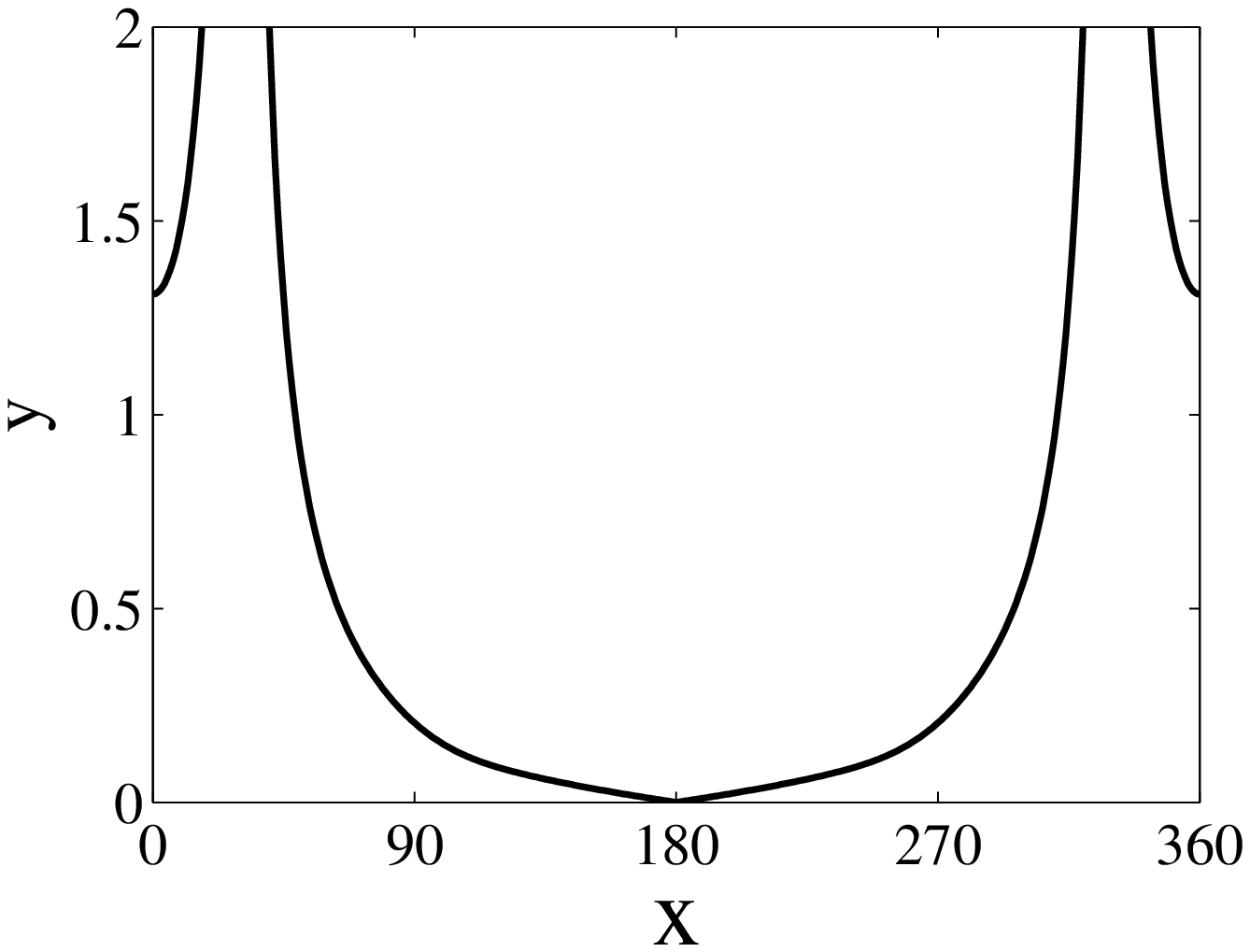}
  \includegraphics[height=51mm,width=65mm]{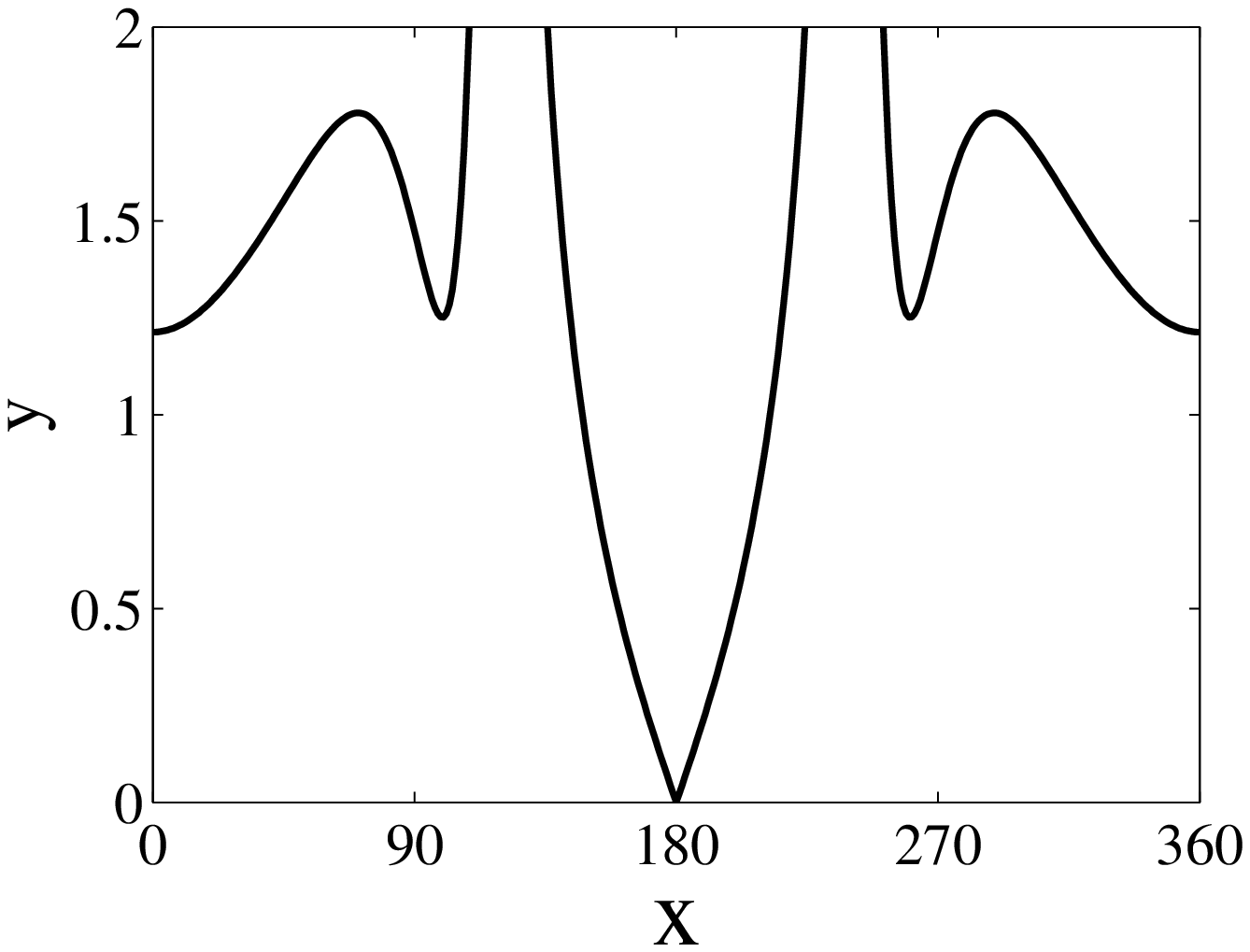}
  {\bf~c}$\quad\quad\quad\quad\quad\quad\quad\quad\quad\quad\quad\quad\quad\quad\quad\quad\quad\quad\quad${\bf~f}\hfil\break
  \includegraphics[height=51mm,width=65mm]{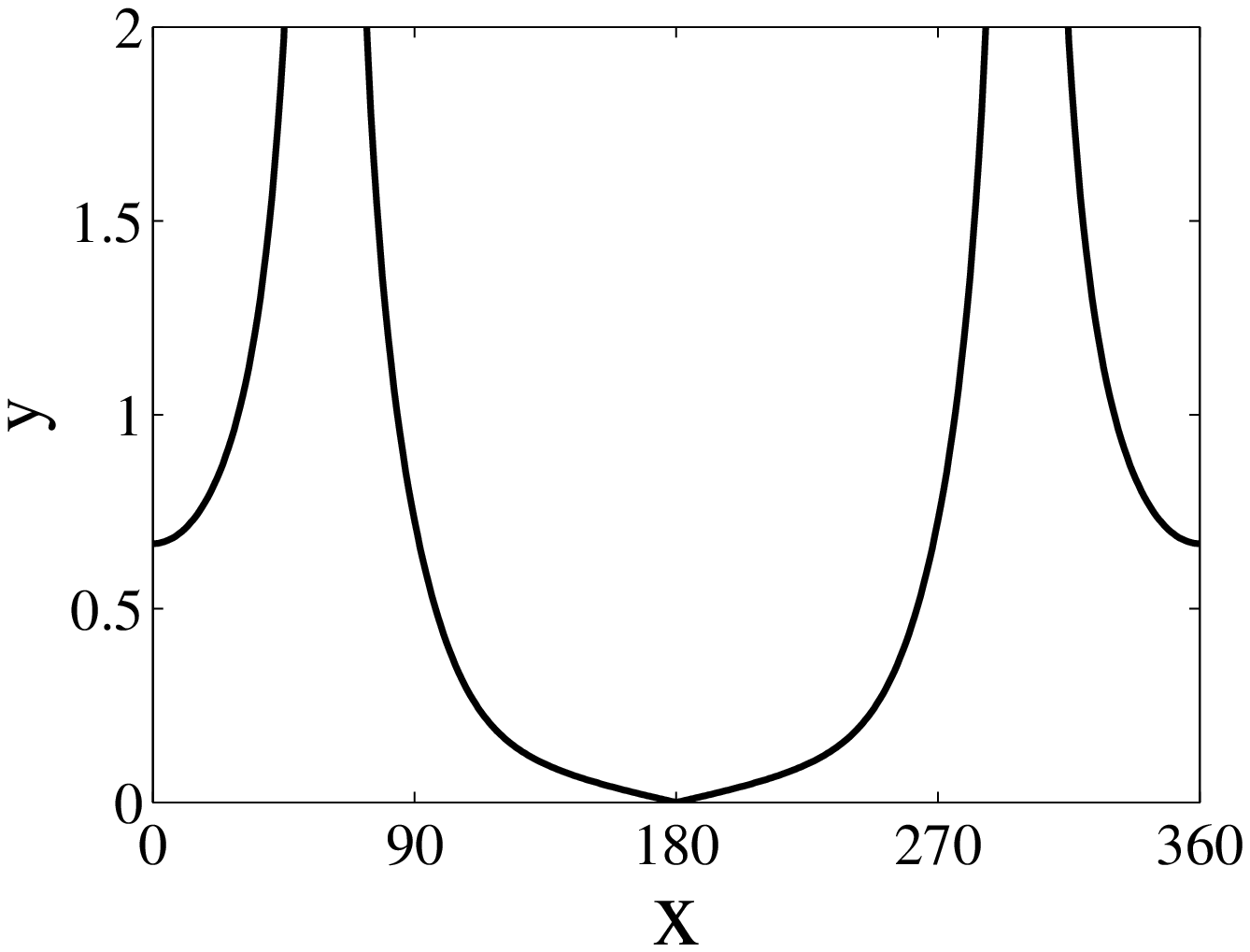}
  \hfil\hfil  \hfil\hfil   \hfil\hfil  \hfil\hfil
  \hfil\hfil  \hfil\hfil   \hfil\hfil  \hfil\hfil
  \hfil\hfil  \hfil\hfil   \hfil\hfil  \hfil\hfil
  \hfil\hfil  \hfil\hfil   \hfil\hfil  \hfil\hfil
  \includegraphics[height=51mm,width=65mm]{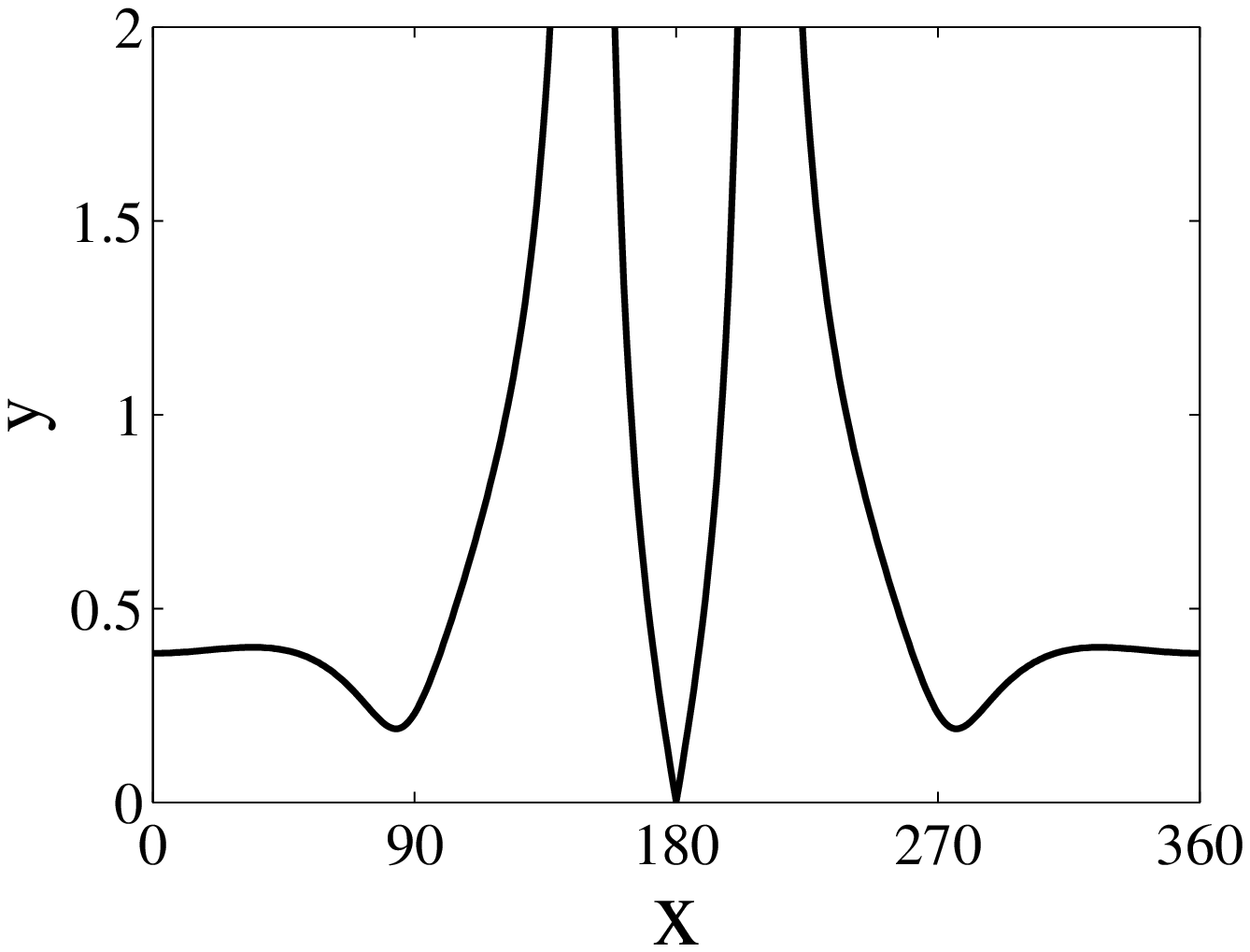}
  \caption{Magnitude $D^{(qSH,qSH)}$ versus $\varthec$,  $\varphic^{inc}=60^0$ and $\varthec^{inc}=0^0$(a), $30^0$(b),
$60^0$(c),  $90^0$(d), $120^0$(e), $150^0$(f).}
\end{figure}

\vfill\eject

\section {Magnitudes of various diffraction coefficients for the axis of symmetry lying in the crack plane perpendicularly to the crack edge}

\begin{figure}[ht]
{\bf~a}$\quad\quad\quad\quad\quad\quad\quad\quad\quad\quad\quad\quad\quad\quad\quad\quad\quad\quad\quad${\bf~d}\hfil\break
\psfrag{x}{$\varthec,\,{}^{\rm o}$}
\psfrag{y}{$|D|$}
\includegraphics[height=51mm,width=65mm]{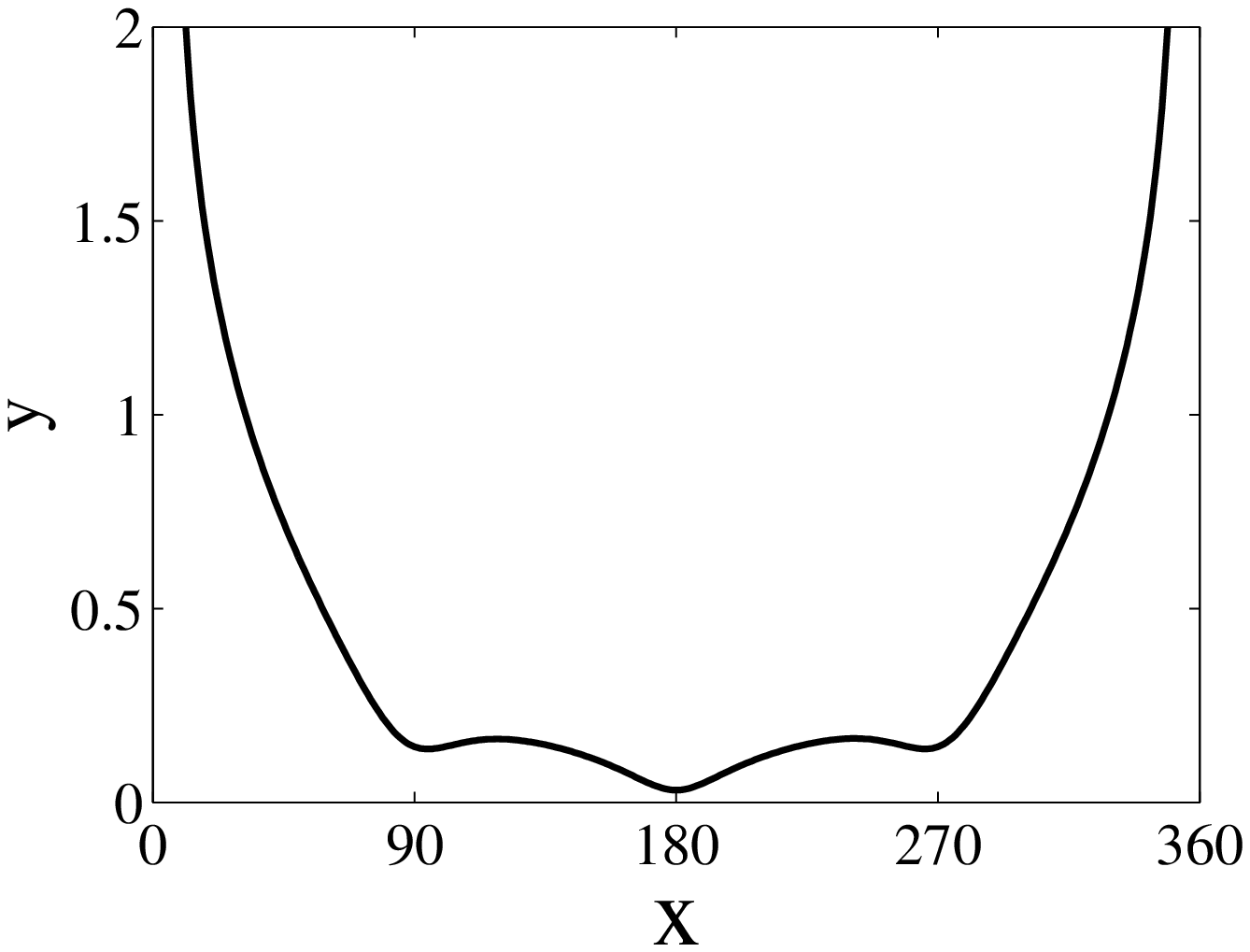}
  \includegraphics[height=51mm,width=65mm]{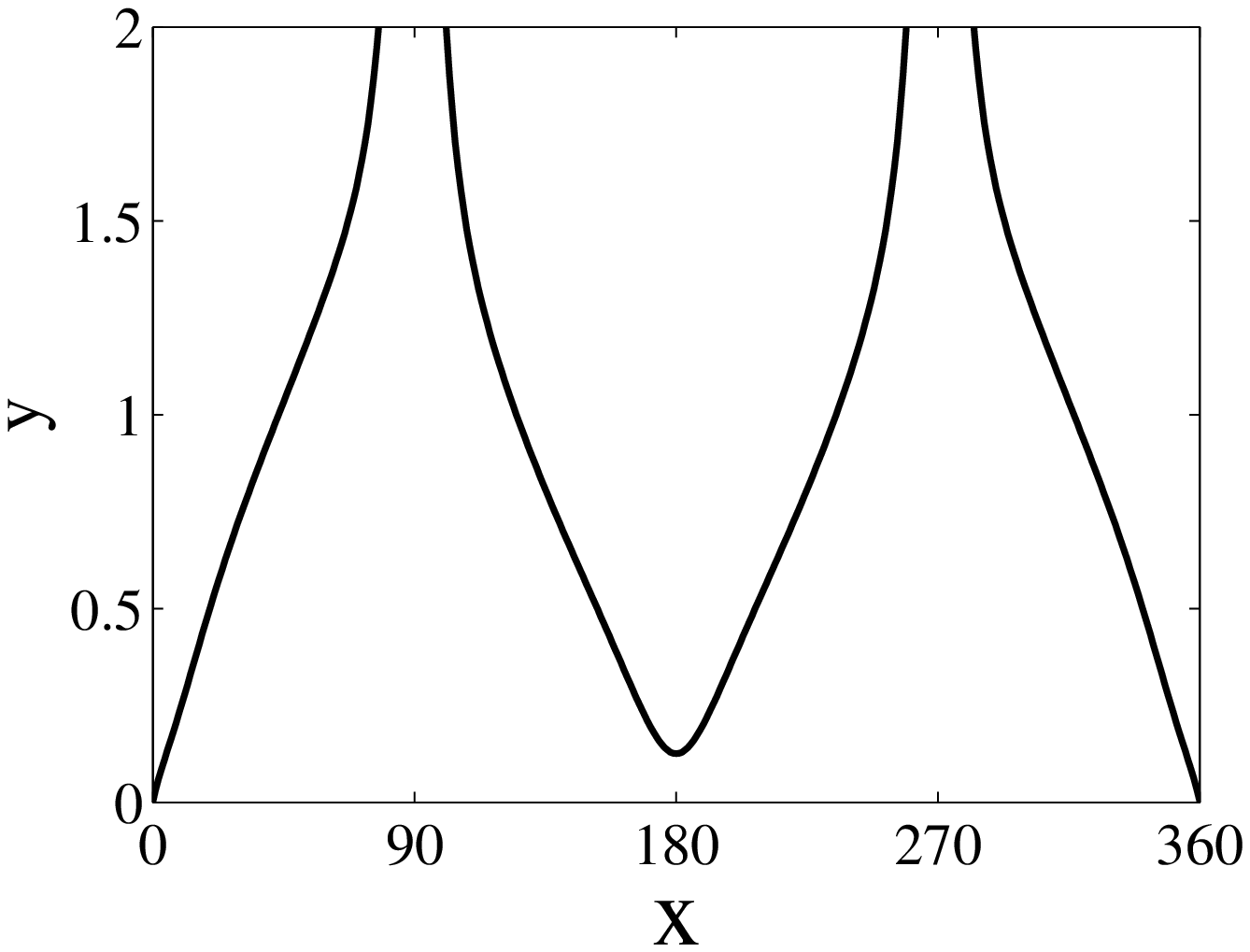}
  {\bf~b}$\quad\quad\quad\quad\quad\quad\quad\quad\quad\quad\quad\quad\quad\quad\quad\quad\quad\quad\quad${\bf~e}\hfil\break
  \includegraphics[height=51mm,width=65mm]{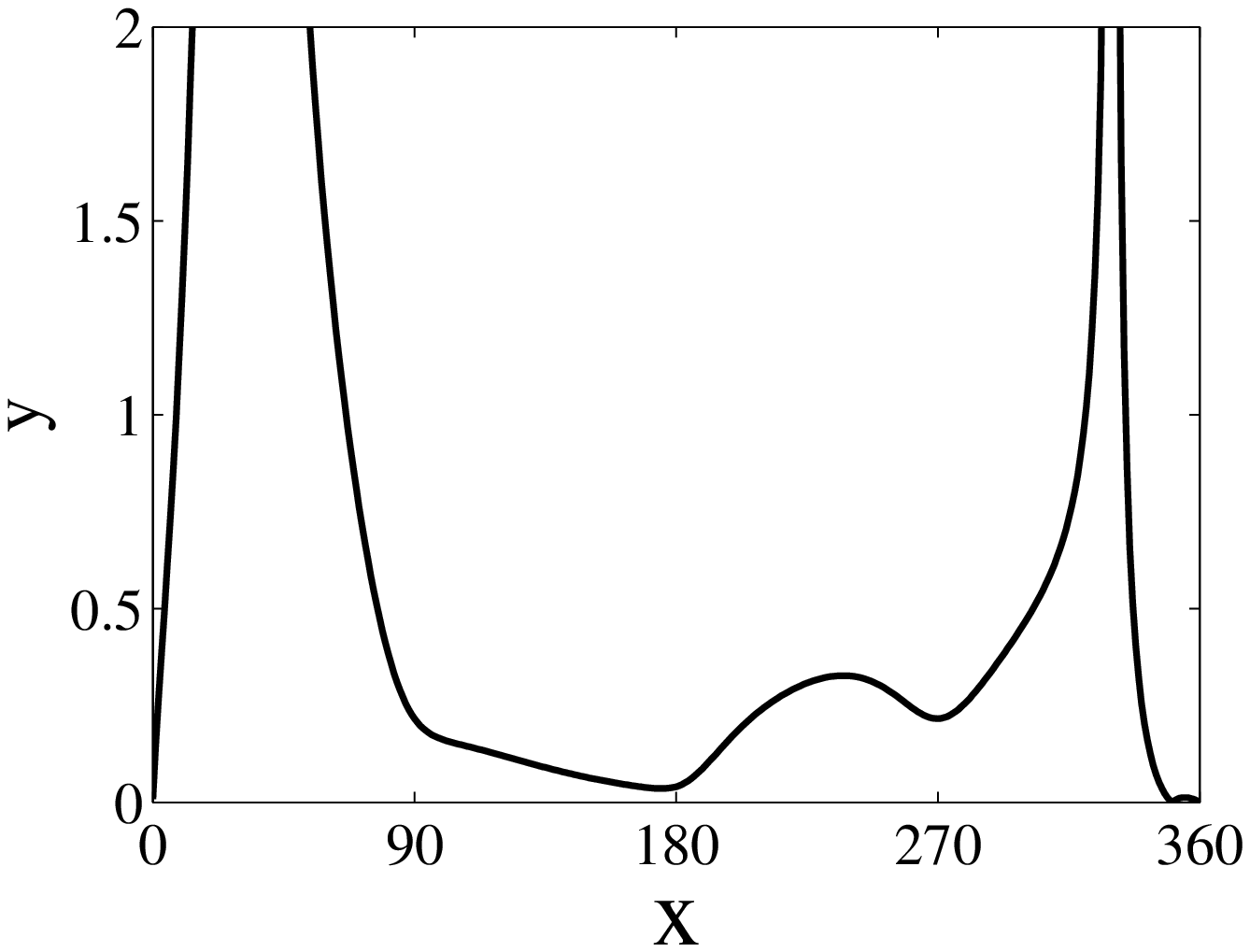}
  \includegraphics[height=51mm,width=65mm]{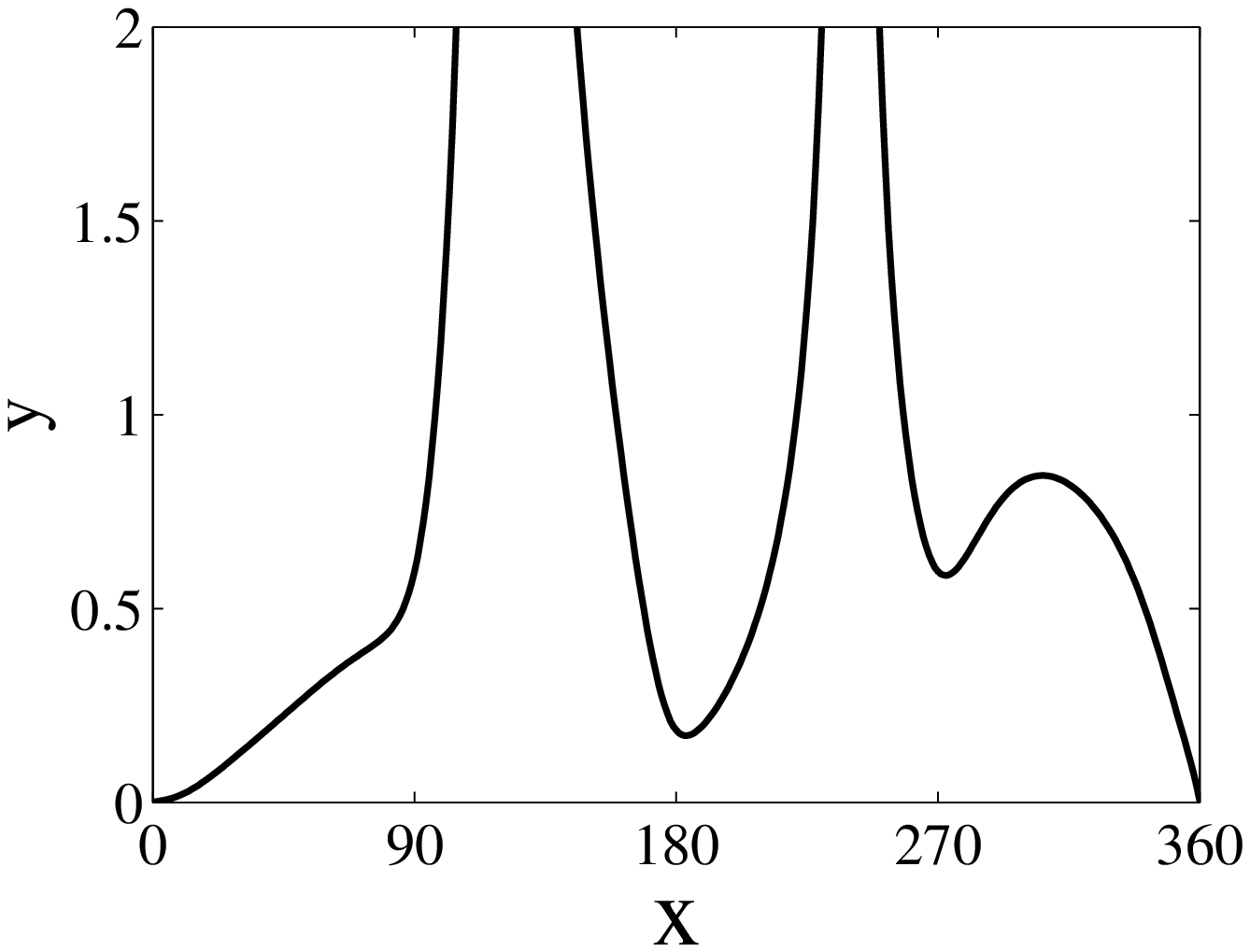}
  {\bf~c}$\quad\quad\quad\quad\quad\quad\quad\quad\quad\quad\quad\quad\quad\quad\quad\quad\quad\quad\quad${\bf~f}\hfil\break
  \includegraphics[height=51mm,width=65mm]{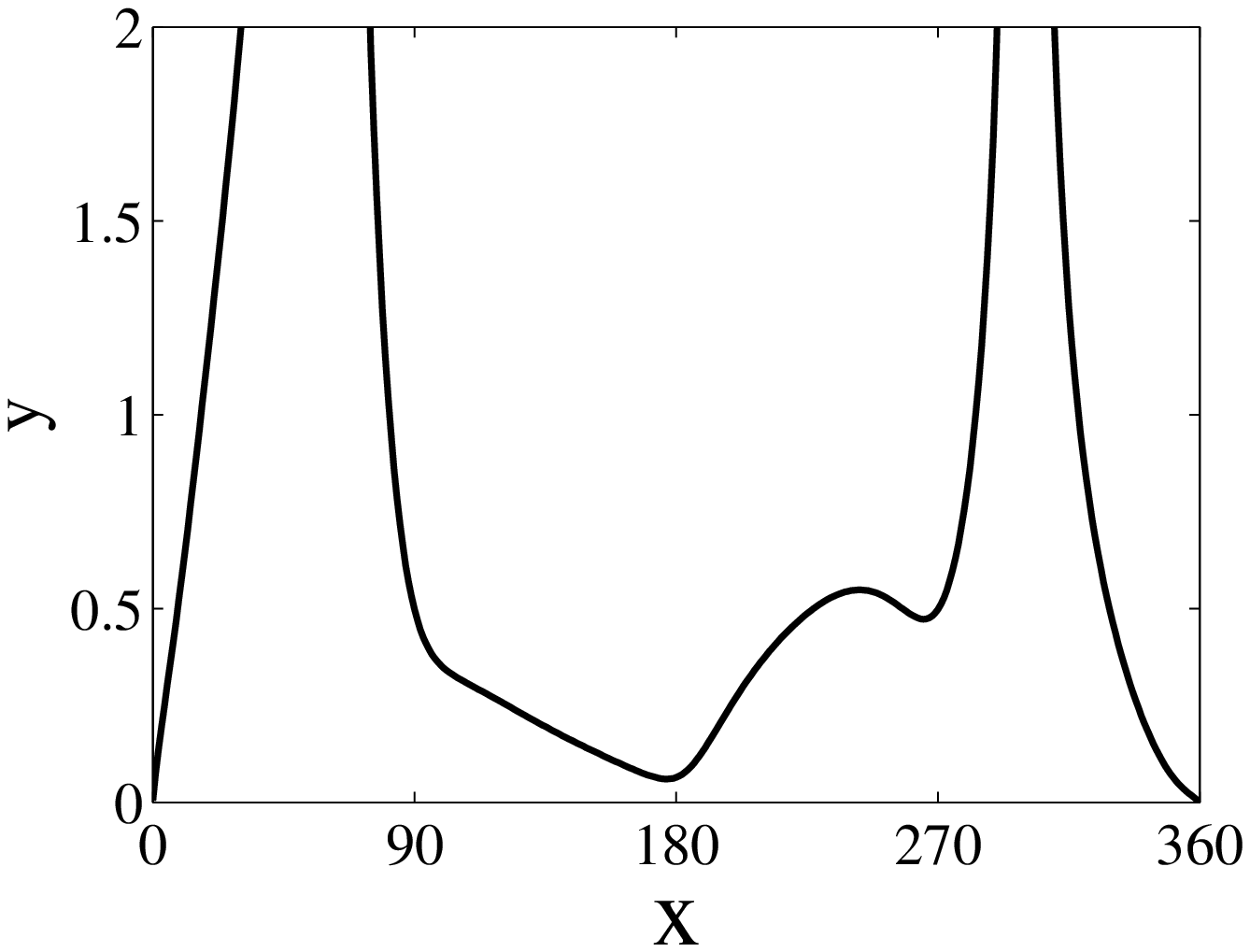}
  \hfil\hfil  \hfil\hfil   \hfil\hfil  \hfil\hfil
  \hfil\hfil  \hfil\hfil   \hfil\hfil  \hfil\hfil
  \hfil\hfil  \hfil\hfil   \hfil\hfil  \hfil\hfil
  \hfil\hfil  \hfil\hfil   \hfil\hfil  \hfil\hfil
  \includegraphics[height=51mm,width=65mm]{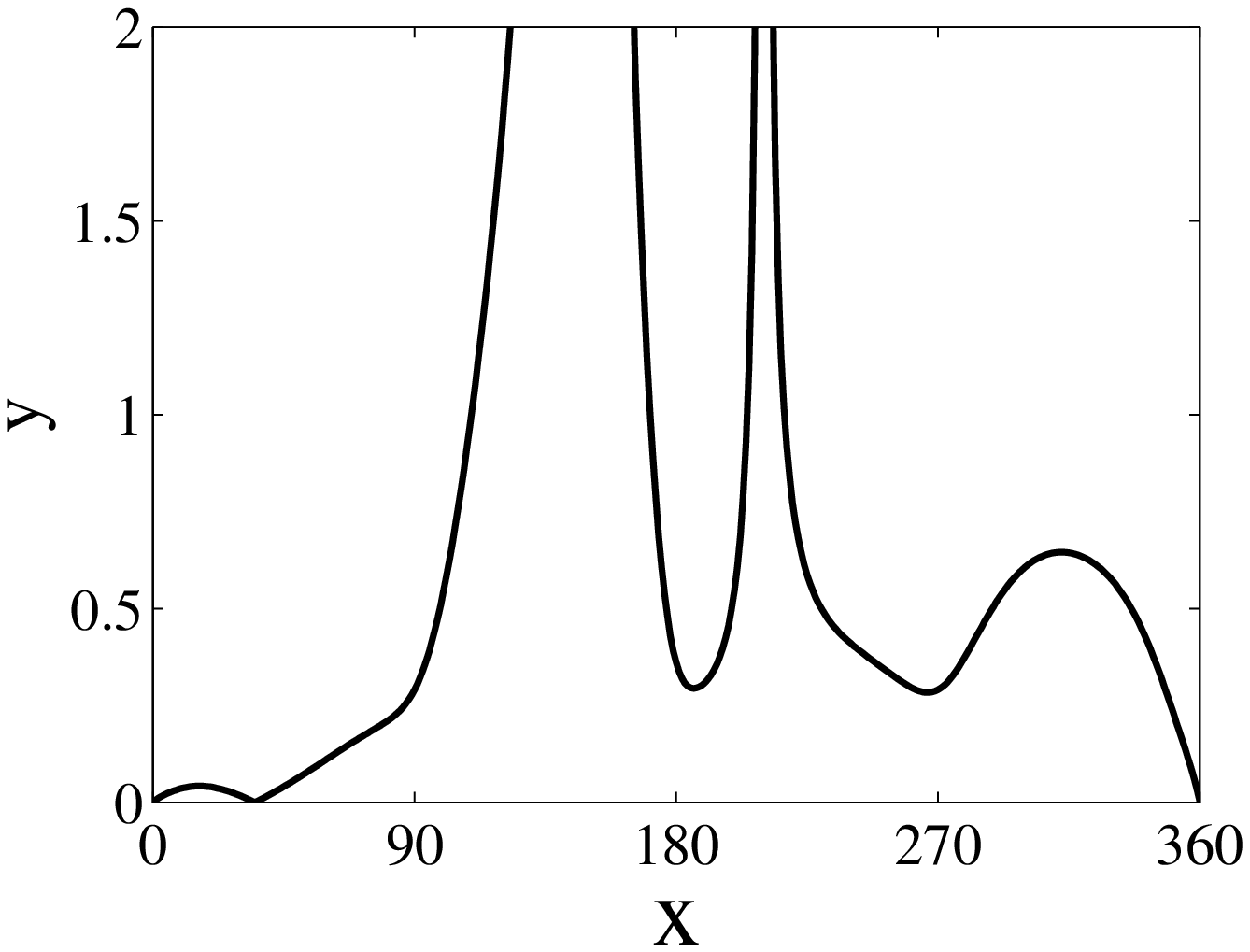}
   \caption{Magnitude $D^{(qP,qP)}$ versus $\varthec$,  $\varphic^{inc}=90^0$ and $\varthec^{inc}=0^0$(a), $30^0$(b),
$60^0$(c),  $90^0$(d), $120^0$(e), $150^0$(f).}
\end{figure}

\vfill\eject

\begin{figure}[ht]
{\bf~a}$\quad\quad\quad\quad\quad\quad\quad\quad\quad\quad\quad\quad\quad\quad\quad\quad\quad\quad\quad${\bf~d}\hfil\break
\psfrag{x}{$\varthec,\,{}^{\rm o}$}
\psfrag{y}{$|D|$}
\includegraphics[height=51mm,width=65mm]{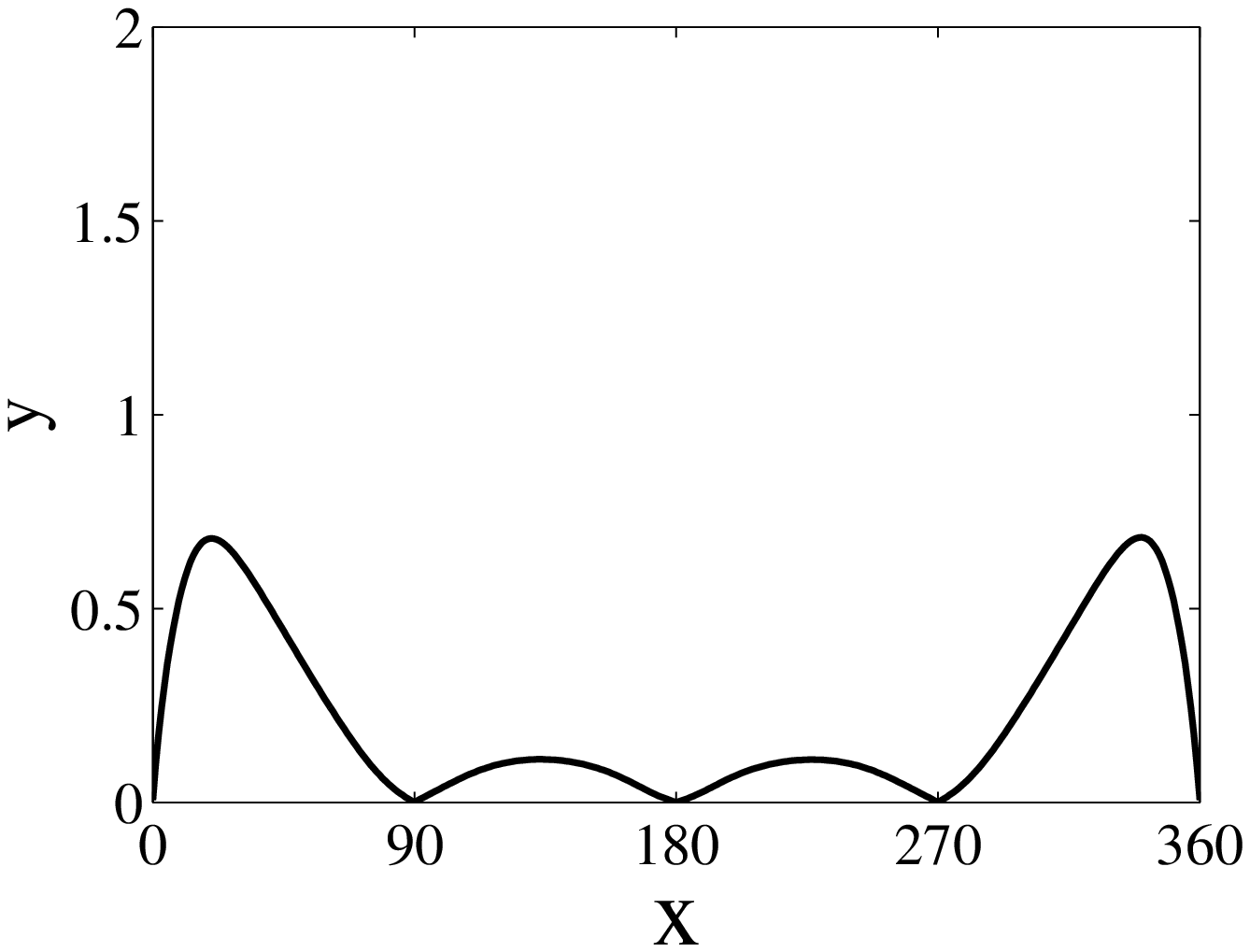}
  \includegraphics[height=51mm,width=65mm]{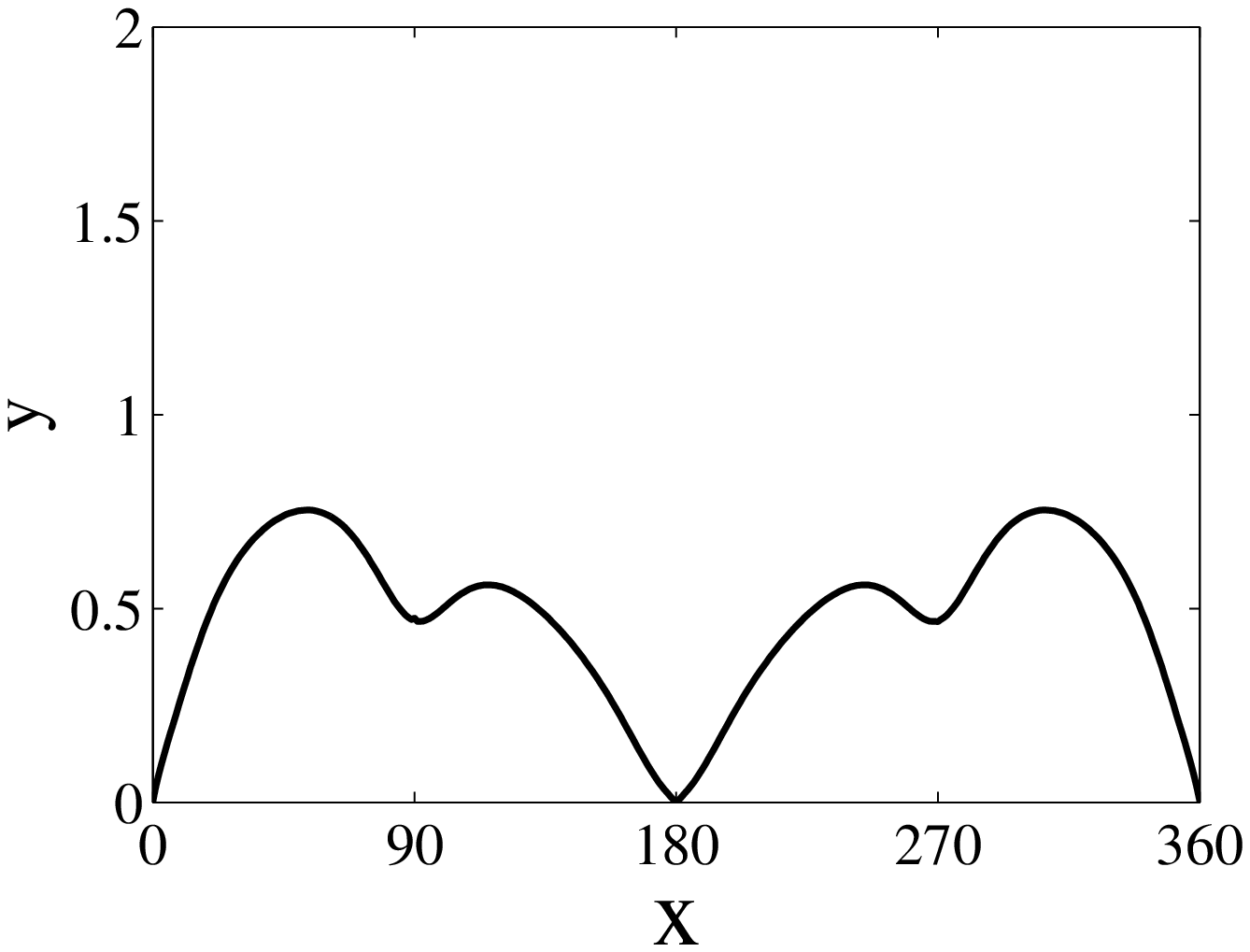}
  {\bf~b}$\quad\quad\quad\quad\quad\quad\quad\quad\quad\quad\quad\quad\quad\quad\quad\quad\quad\quad\quad${\bf~e}\hfil\break
  \includegraphics[height=51mm,width=65mm]{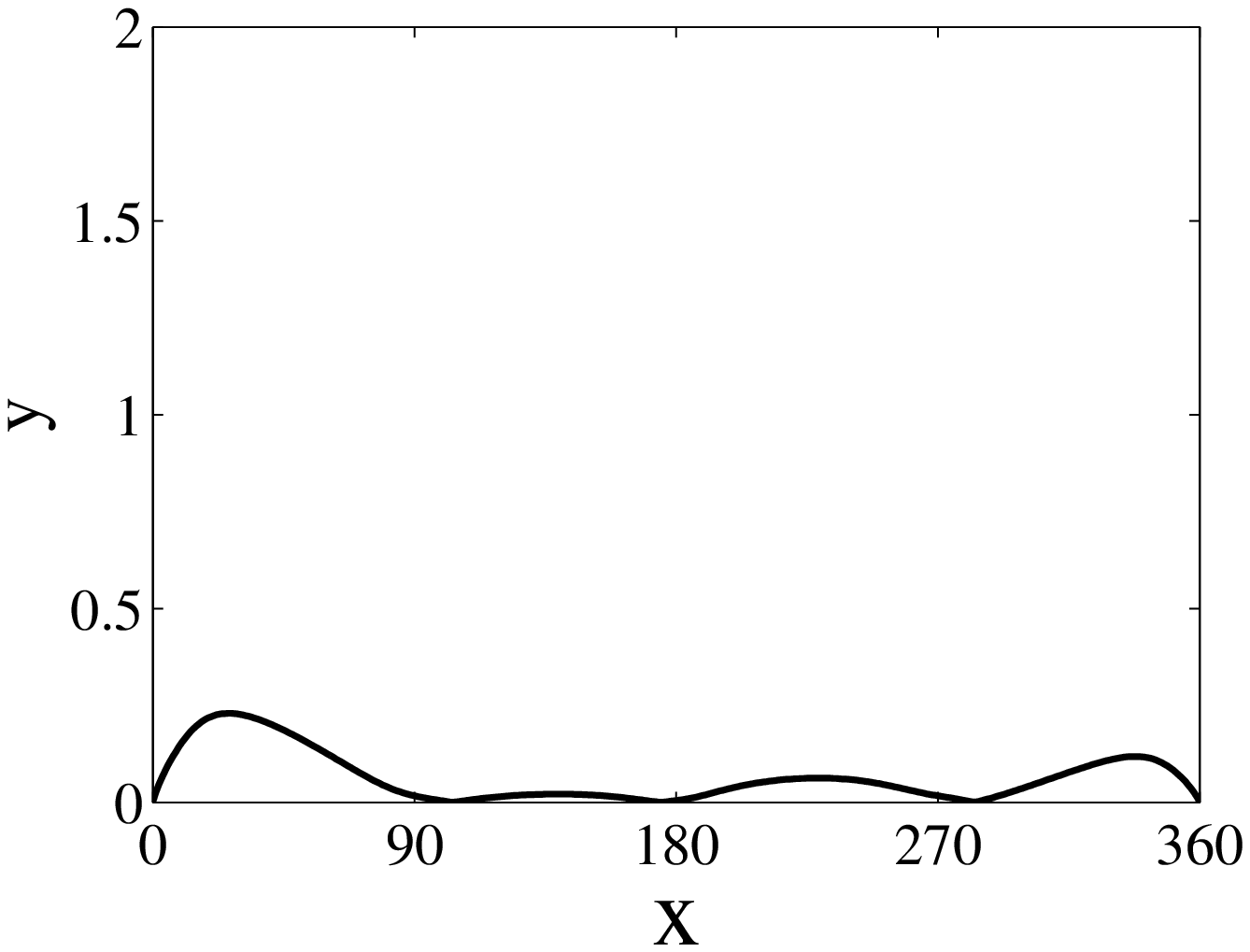}
  \includegraphics[height=51mm,width=65mm]{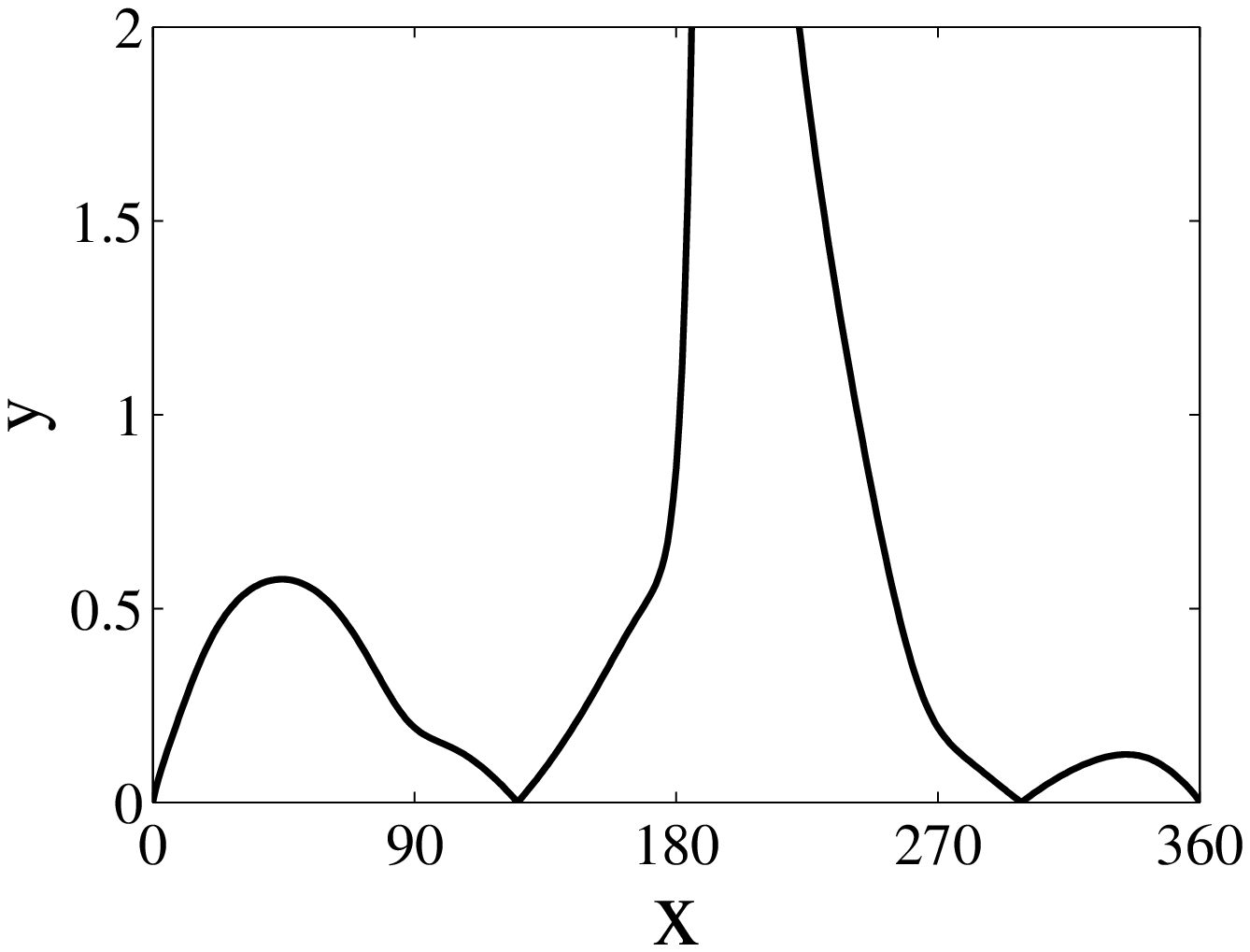}
  {\bf~c}$\quad\quad\quad\quad\quad\quad\quad\quad\quad\quad\quad\quad\quad\quad\quad\quad\quad\quad\quad${\bf~f}\hfil\break
  \includegraphics[height=51mm,width=65mm]{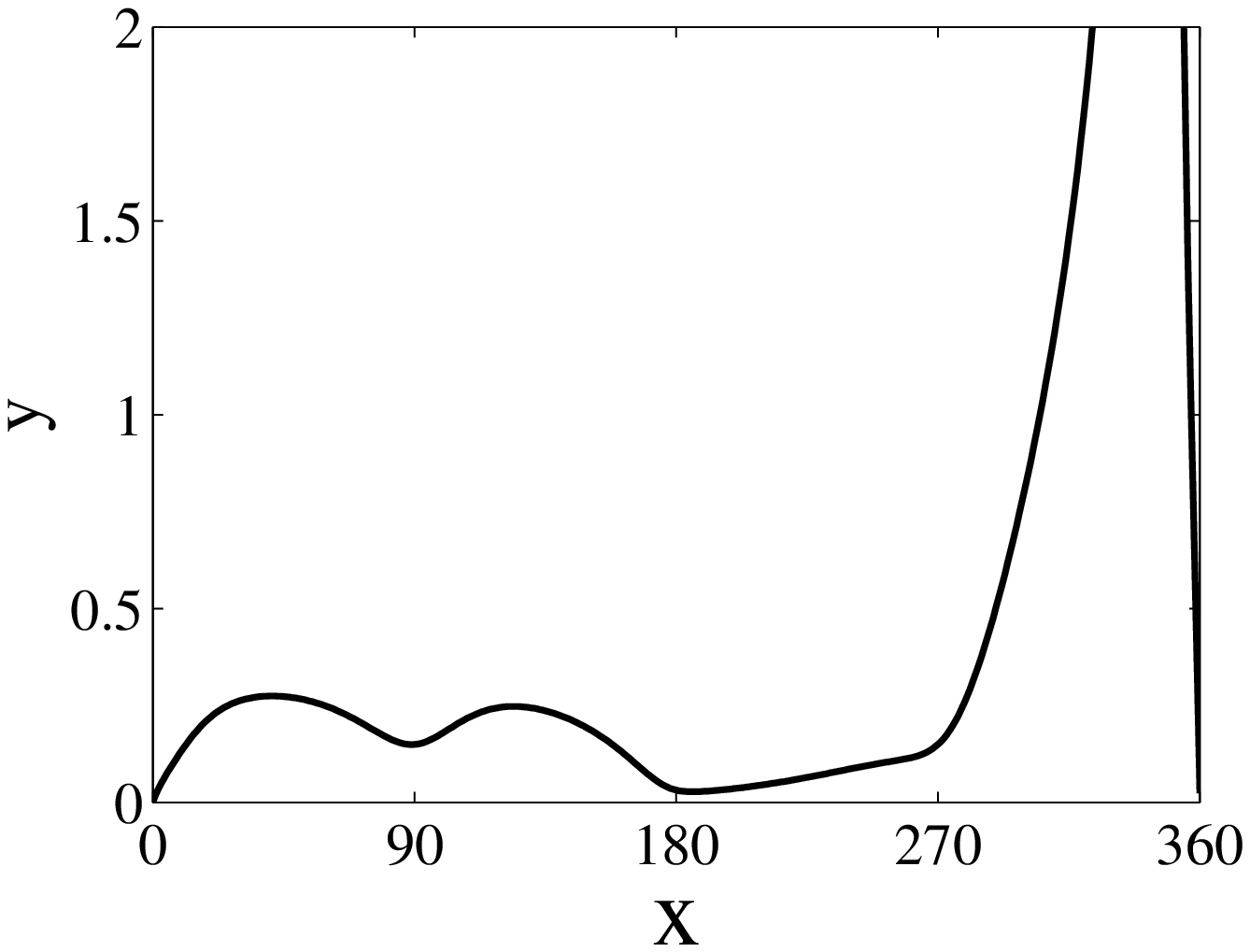}
  \hfil\hfil  \hfil\hfil   \hfil\hfil  \hfil\hfil
  \hfil\hfil  \hfil\hfil   \hfil\hfil  \hfil\hfil
  \hfil\hfil  \hfil\hfil   \hfil\hfil  \hfil\hfil
  \hfil\hfil  \hfil\hfil   \hfil\hfil  \hfil\hfil
  \includegraphics[height=51mm,width=65mm]{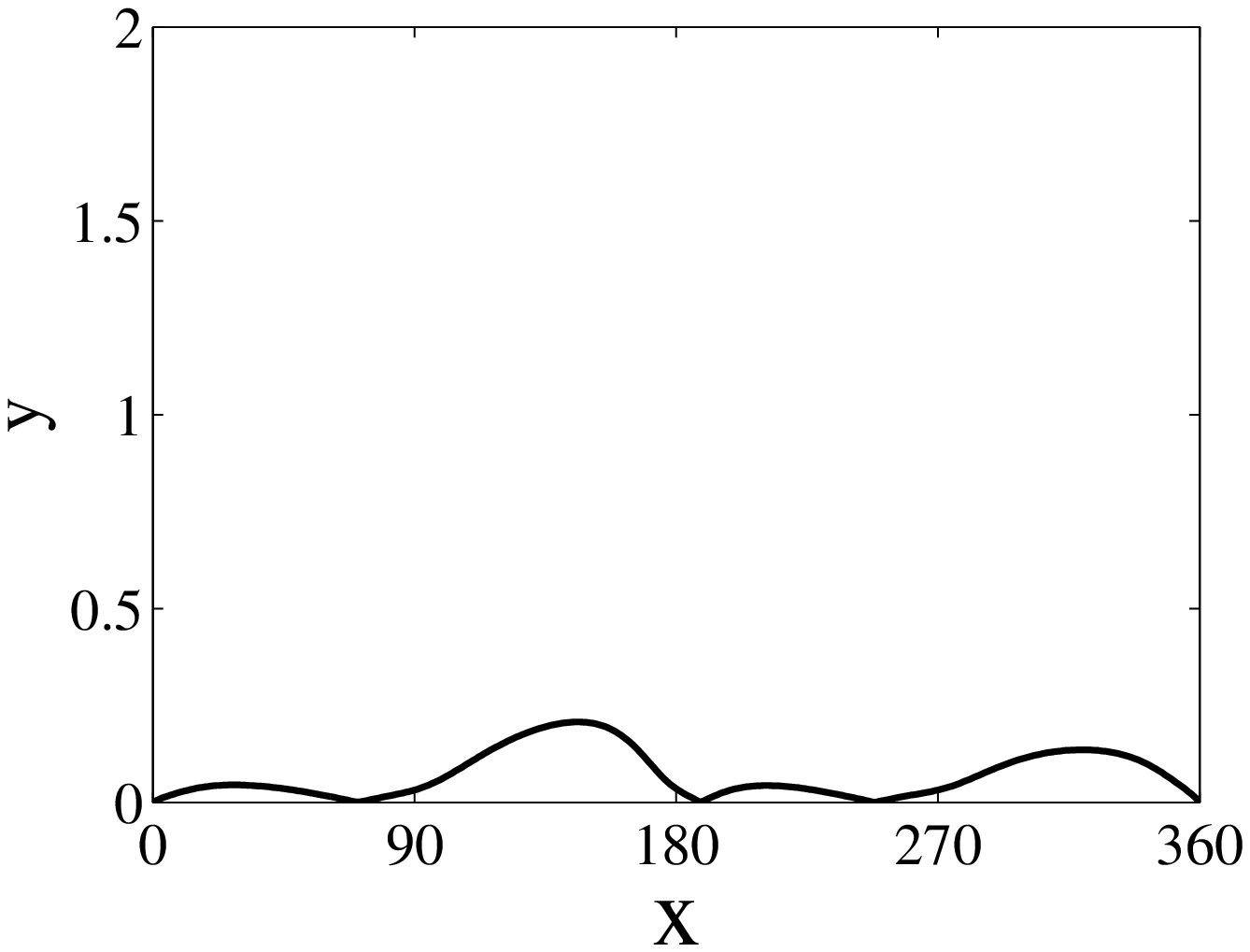}
   \caption{Magnitude $D^{(qP,qSV)}$ versus $\varthec$,  $\varphic^{inc}=90^0$ and $\varthec^{inc}=0^0$(a), $30^0$(b),
$60^0$(c),  $90^0$(d), $120^0$(e), $150^0$(f).}
   \label{2DQSVQP90}
\end{figure}

\vfill\eject

\begin{figure}[ht]
{\bf~a}$\quad\quad\quad\quad\quad\quad\quad\quad\quad\quad\quad\quad\quad\quad\quad\quad\quad\quad\quad${\bf~d}\hfil\break
\psfrag{x}{$\varthec,\,{}^{\rm o}$} \psfrag{y}{$|D|$}
\includegraphics[height=51mm,width=65mm]{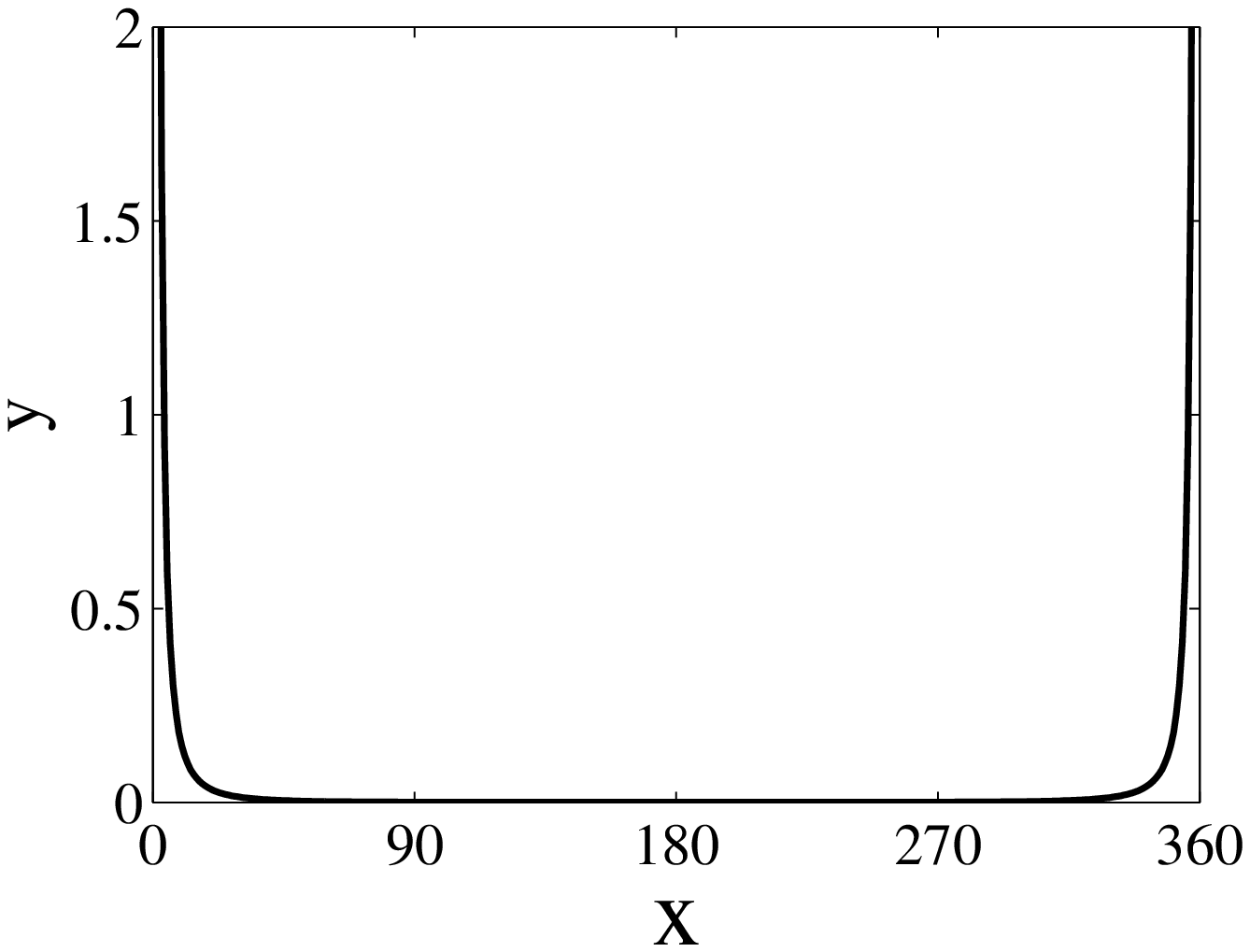}
  \includegraphics[height=51mm,width=65mm]{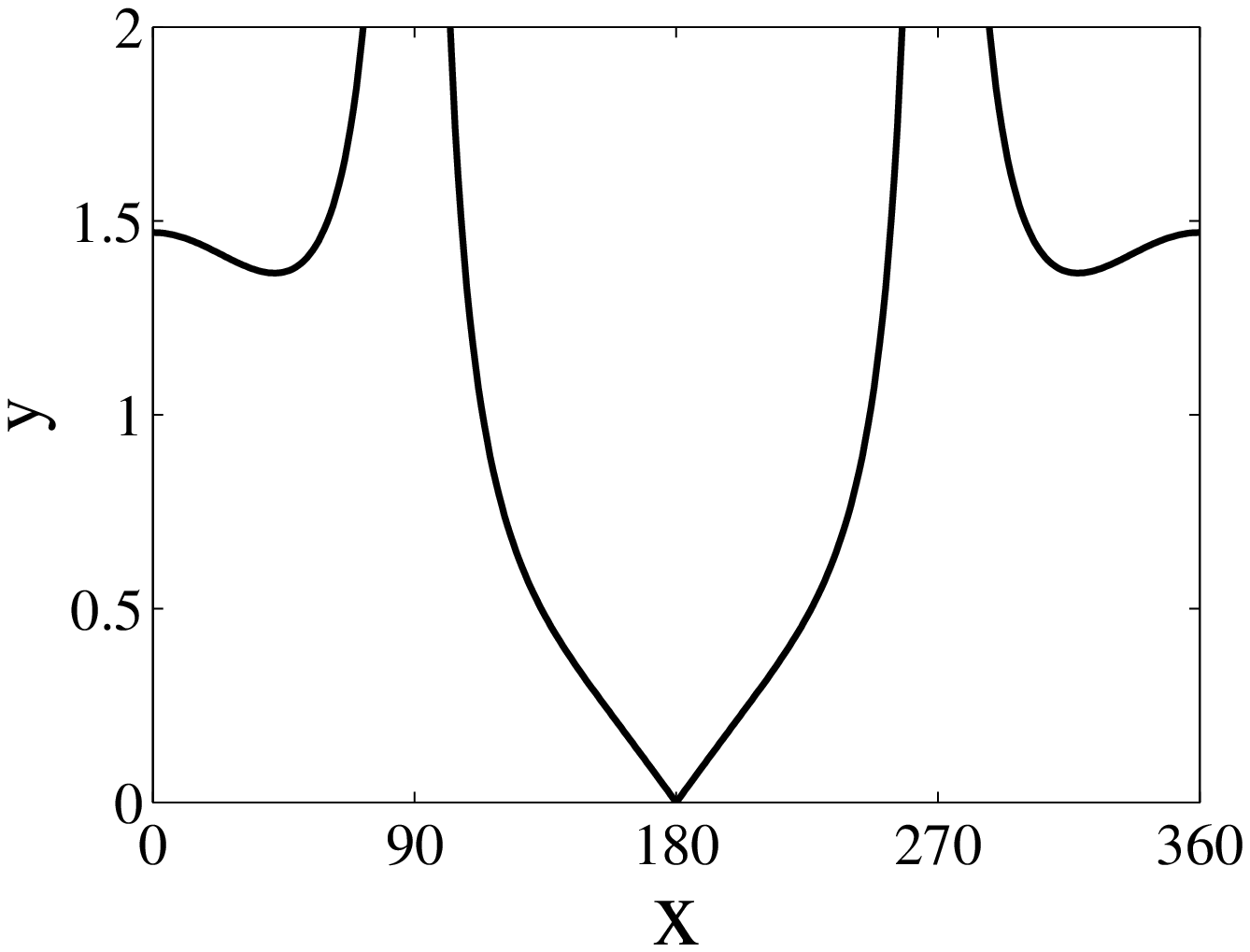}
  {\bf~b}$\quad\quad\quad\quad\quad\quad\quad\quad\quad\quad\quad\quad\quad\quad\quad\quad\quad\quad\quad${\bf~e}\hfil\break
  \includegraphics[height=51mm,width=65mm]{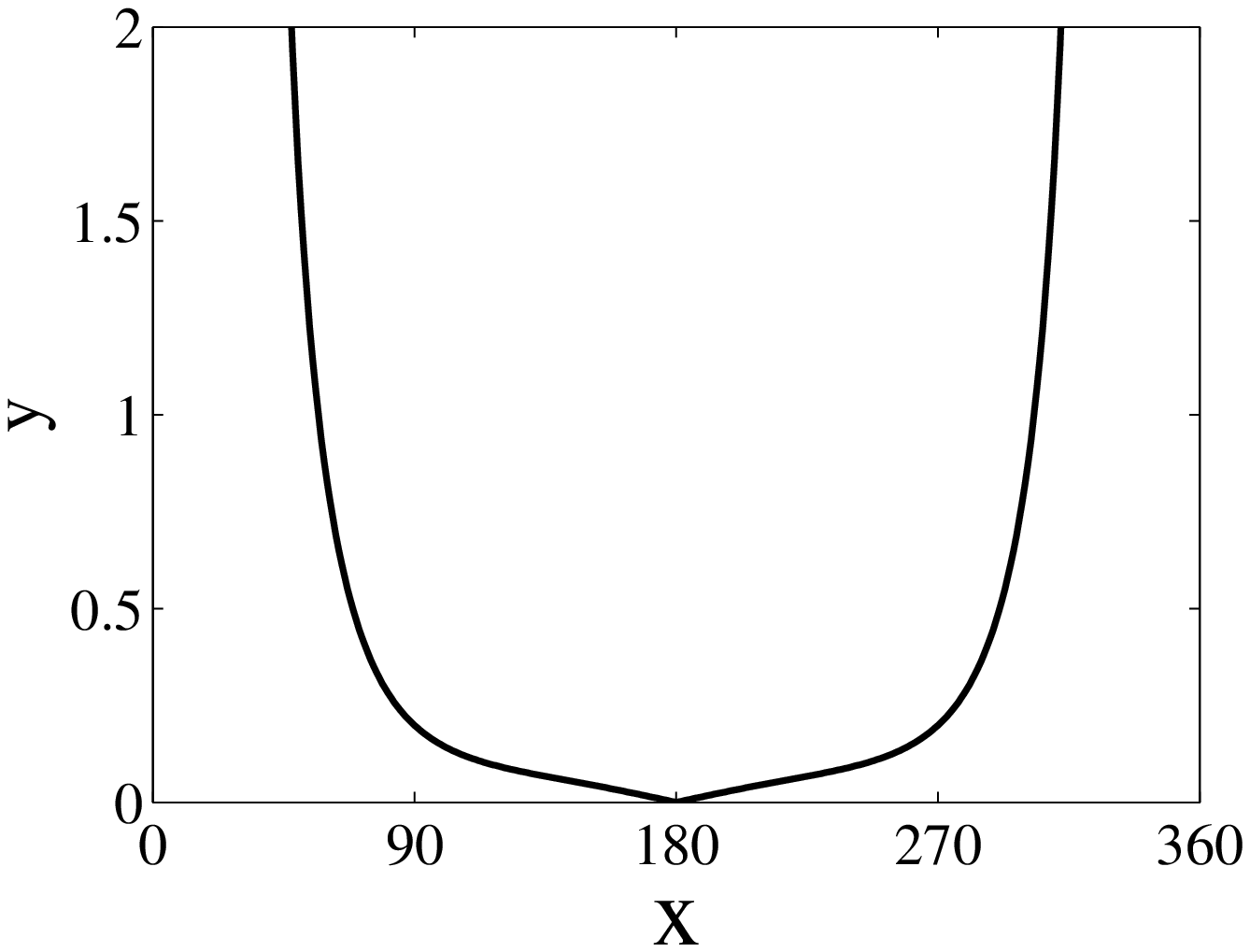}
  \includegraphics[height=51mm,width=65mm]{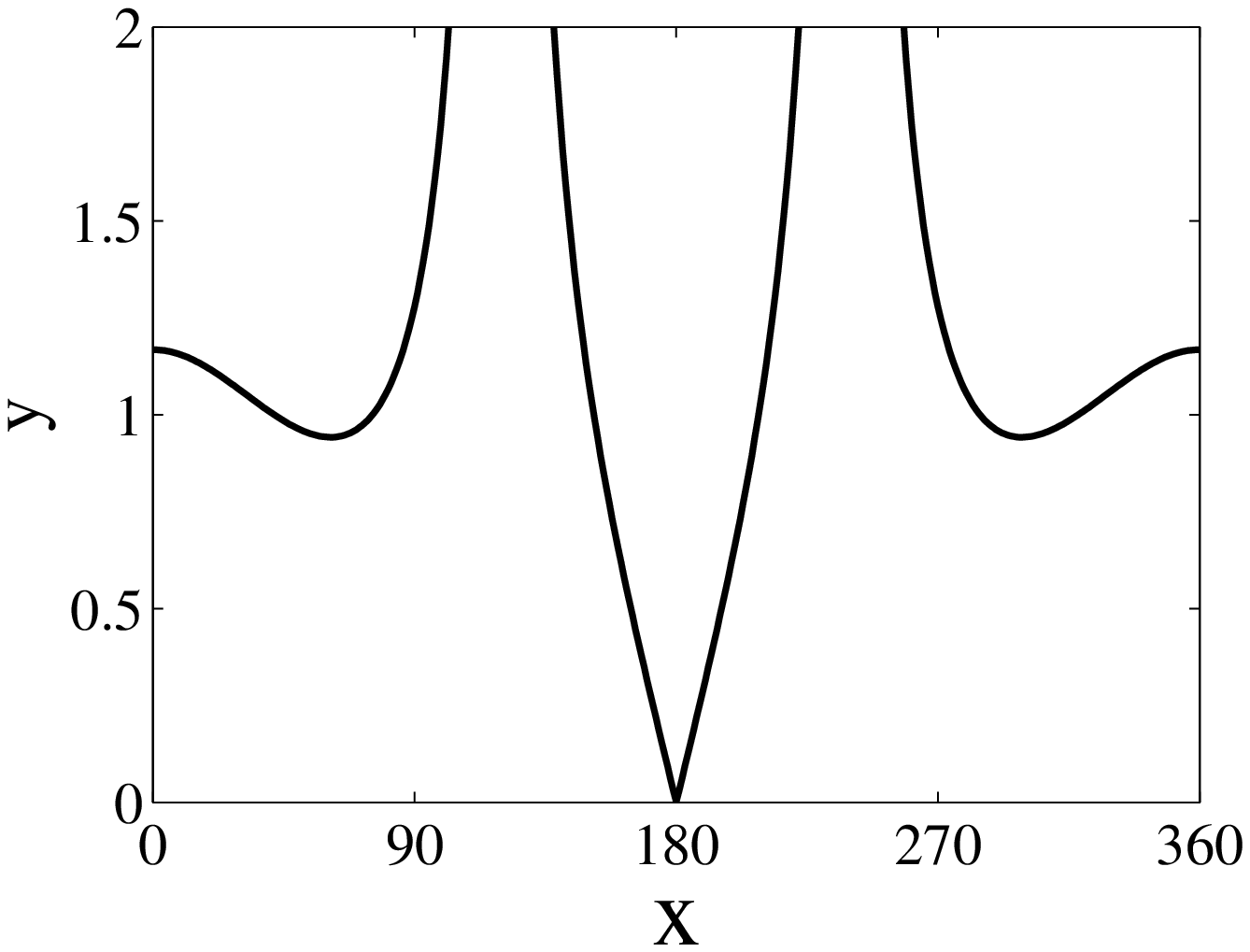}
  {\bf~c}$\quad\quad\quad\quad\quad\quad\quad\quad\quad\quad\quad\quad\quad\quad\quad\quad\quad\quad\quad${\bf~f}\hfil\break
  \includegraphics[height=51mm,width=65mm]{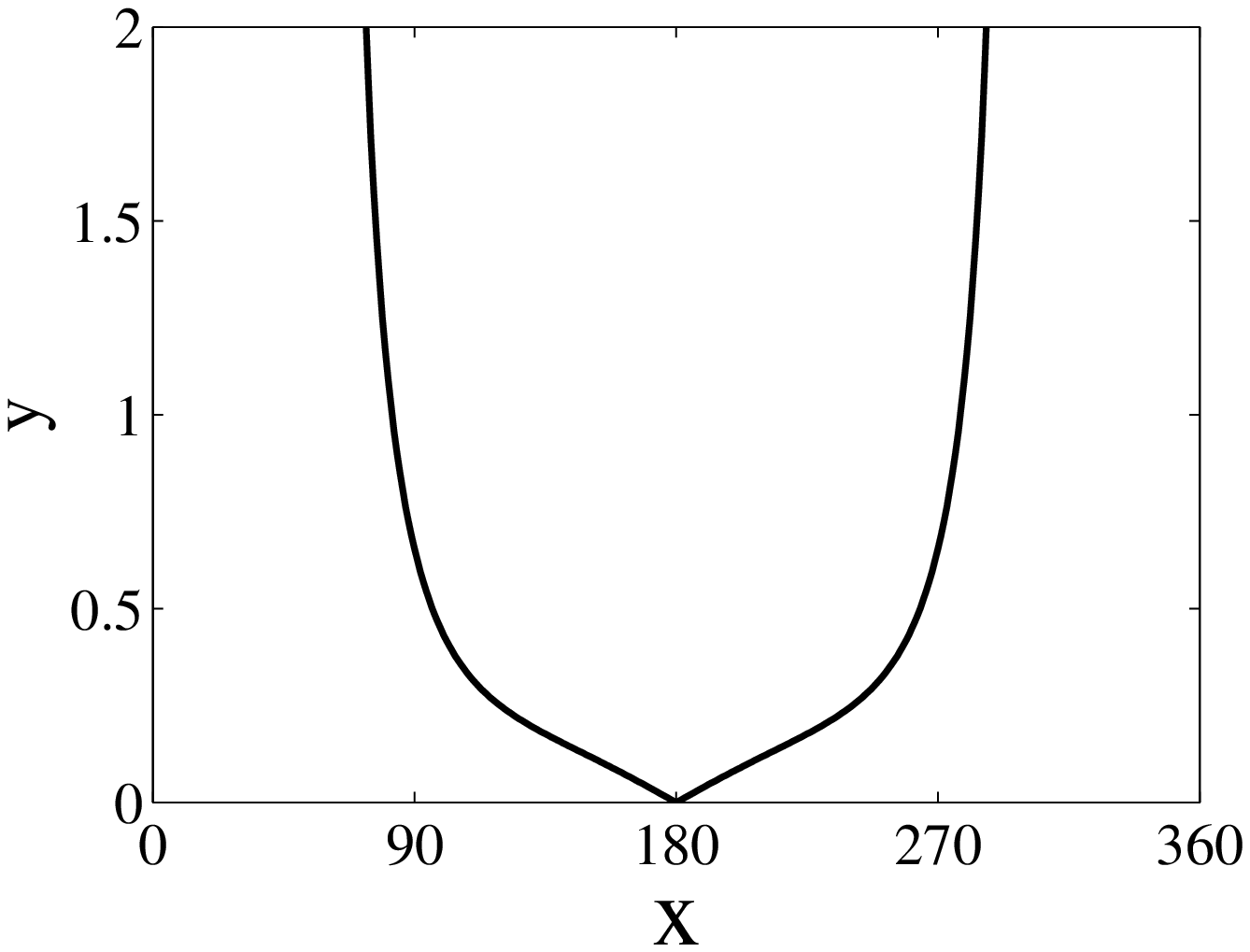}
  \hfil\hfil  \hfil\hfil   \hfil\hfil  \hfil\hfil
  \hfil\hfil  \hfil\hfil   \hfil\hfil  \hfil\hfil
  \hfil\hfil  \hfil\hfil   \hfil\hfil  \hfil\hfil
  \hfil\hfil  \hfil\hfil   \hfil\hfil  \hfil\hfil
  \includegraphics[height=51mm,width=65mm]{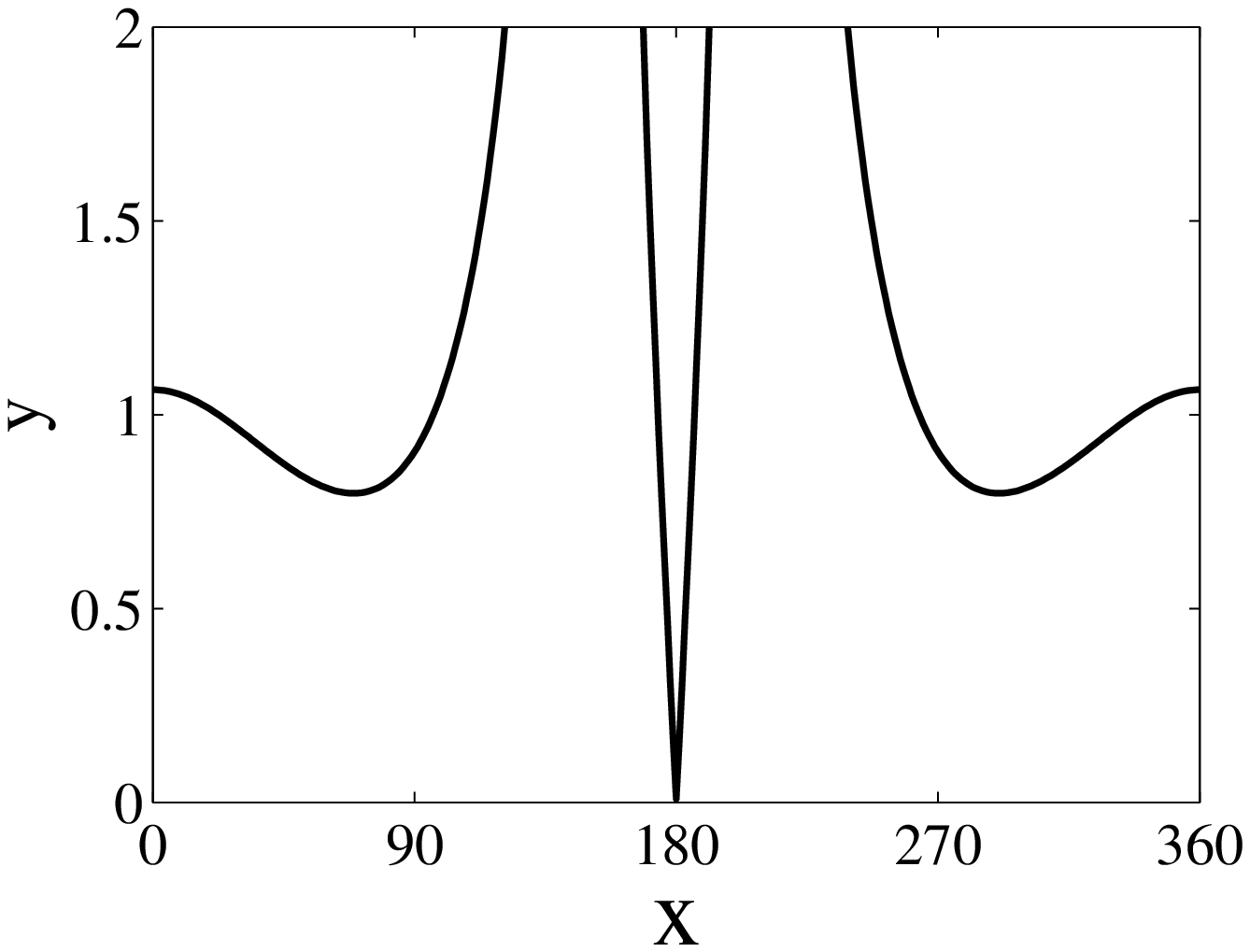}
   \caption{Magnitude $D^{(qSH,qSH)}$ versus $\varthec$,  $\varphic^{inc}=90^0$ and $\varthec^{inc}=0^0$(a), $30^0$(b), $60^0$(c),  $90^0$(d), $120^0$(e), $150^0$(f).}
\end{figure}

\vfill\eject

\section{The effect of cuspidal edges and conical points}

As mentioned at the end of Section 8, the cuspidal edges and conical
points on the $qSV$ wave surfaces lead to extra singularities in
(\ref{diffmatrix}). In Fig.~E.1 we plot the "diffraction
coefficients" as defined by (\ref{diffmatrix}) for the case of the
symmetry axis perpendicular to the
  crack. In Fig.~E.2 the
corresponding quantities are multiplied by $\sqrt{2\pi\i{
|\sin\thec|\ddot \xic^\al_3}(\kc^{diff}_1;-\kc^{in}_2) }$ to exclude
the influence of the cuspidal edges.  In Fig.~E.3 further
multiplication by $\lambda^\al_{\xic_3}(\bzeta^{diff})$ is carried
out to exclude the influence of the conical points too.  The
remaining singularities are due to shadow boundaries.

\begin{figure}[ht]
{\bf~a}$\quad\quad\quad\quad\quad\quad\quad\quad\quad\quad\quad\quad\quad\quad\quad\quad\quad\quad\quad${\bf~d}\hfil\break
\psfrag{x}{$\varthec,\,{}^{\rm o}$}
\psfrag{y}{$|D|$}
\includegraphics[height=43mm,width=65mm]{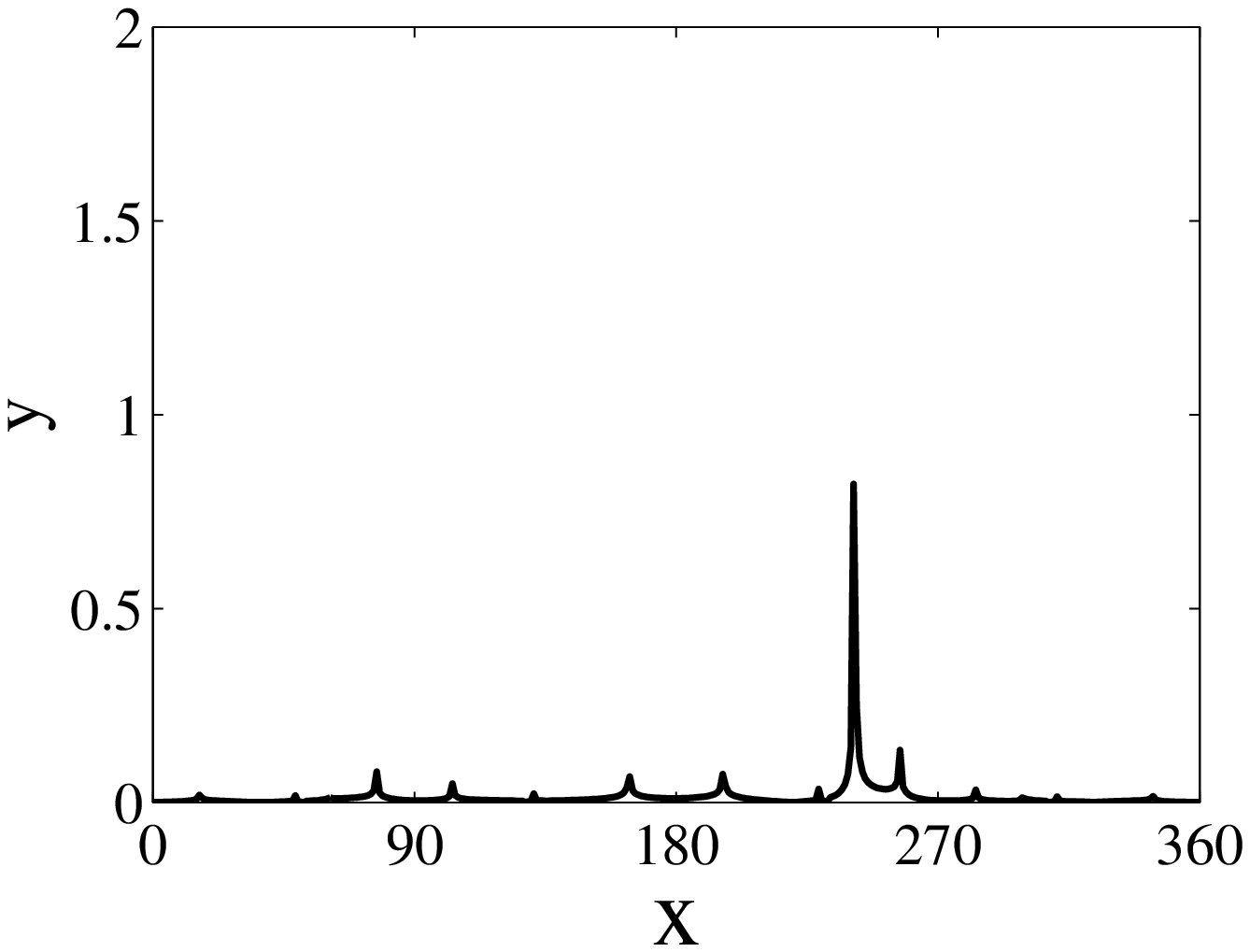}
\includegraphics[height=43mm,width=65mm]{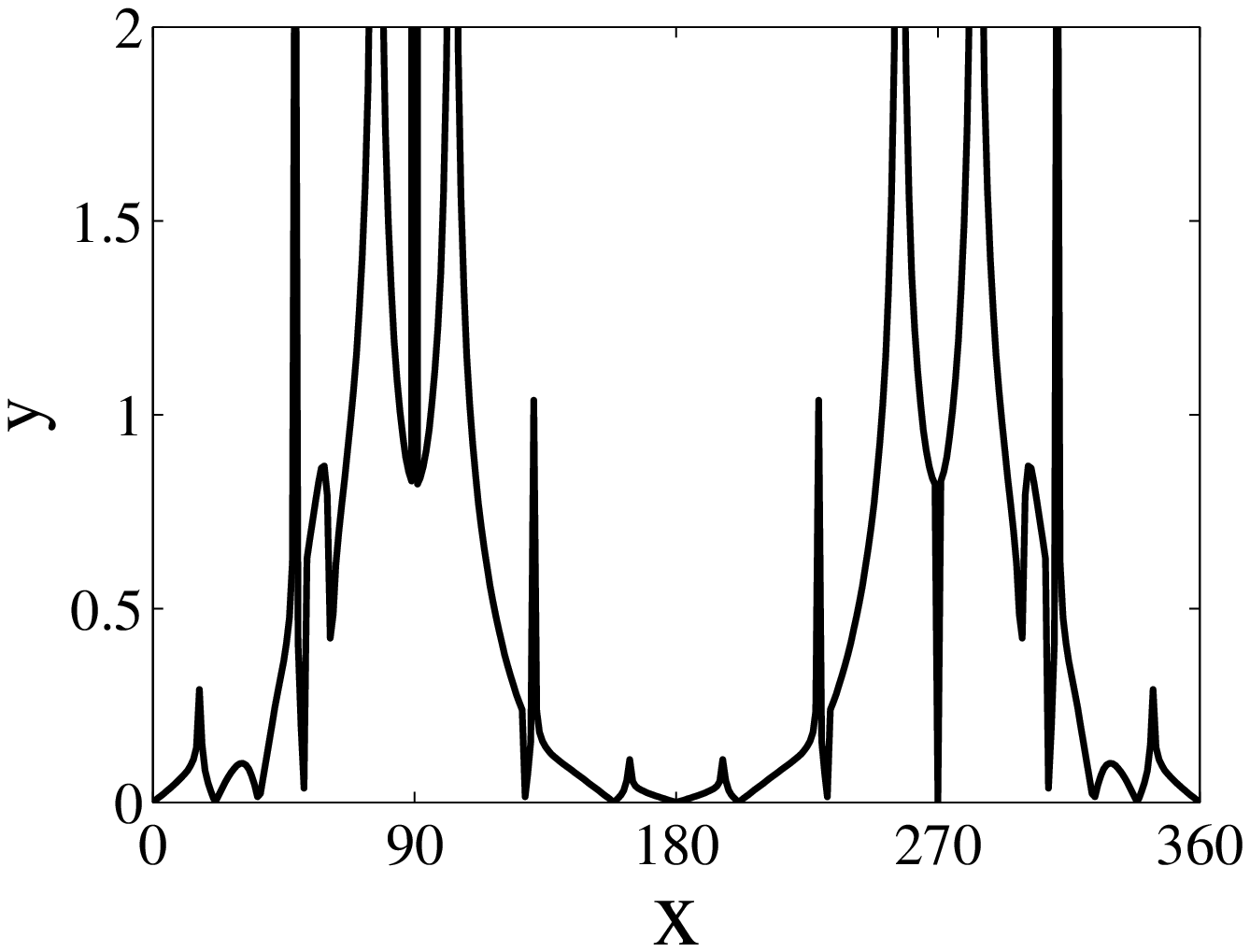}
{\bf~b}$\quad\quad\quad\quad\quad\quad\quad\quad\quad\quad\quad\quad\quad\quad\quad\quad\quad\quad\quad${\bf~e}\hfil\break
\includegraphics[height=43mm,width=65mm]{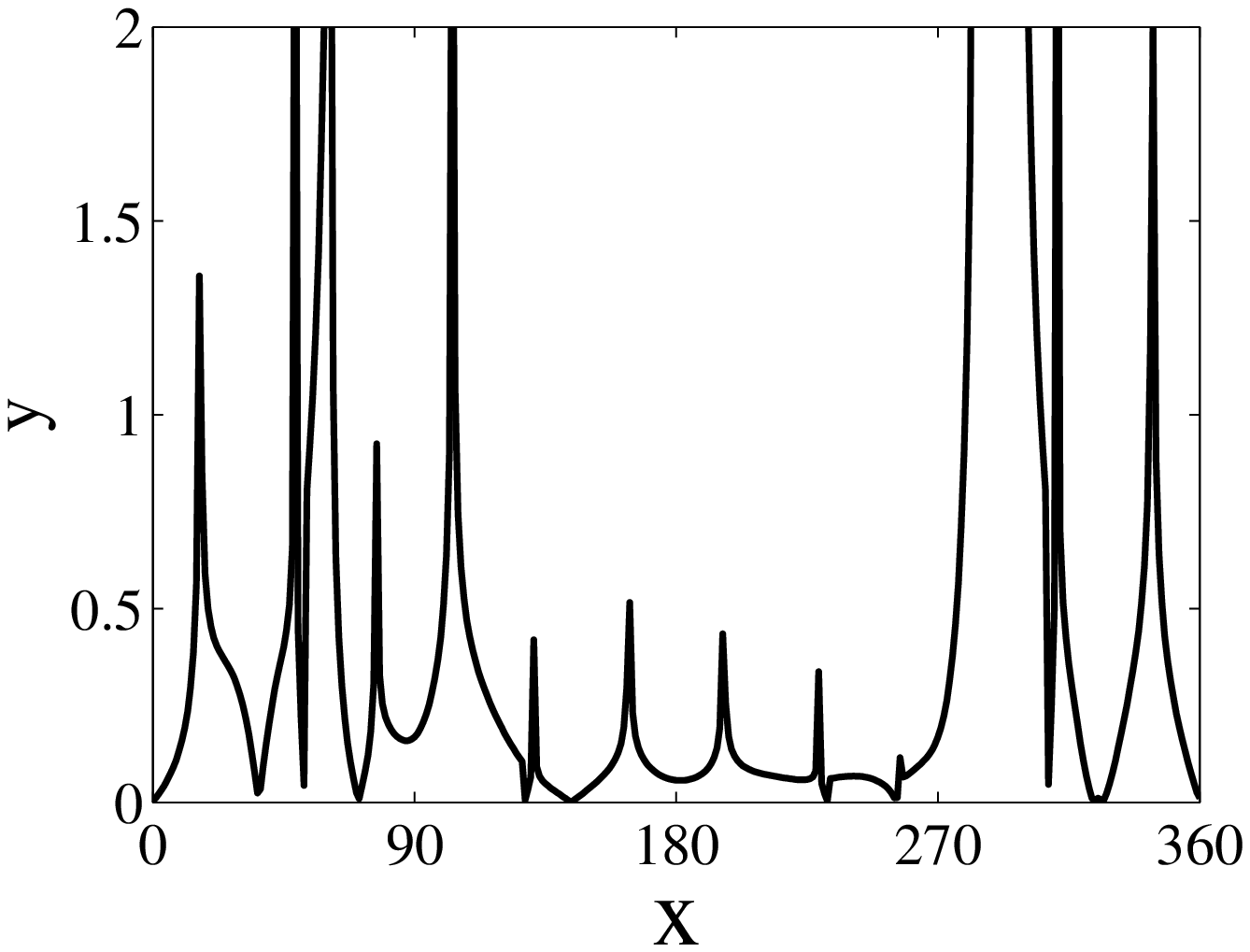}
\includegraphics[height=43mm,width=65mm]{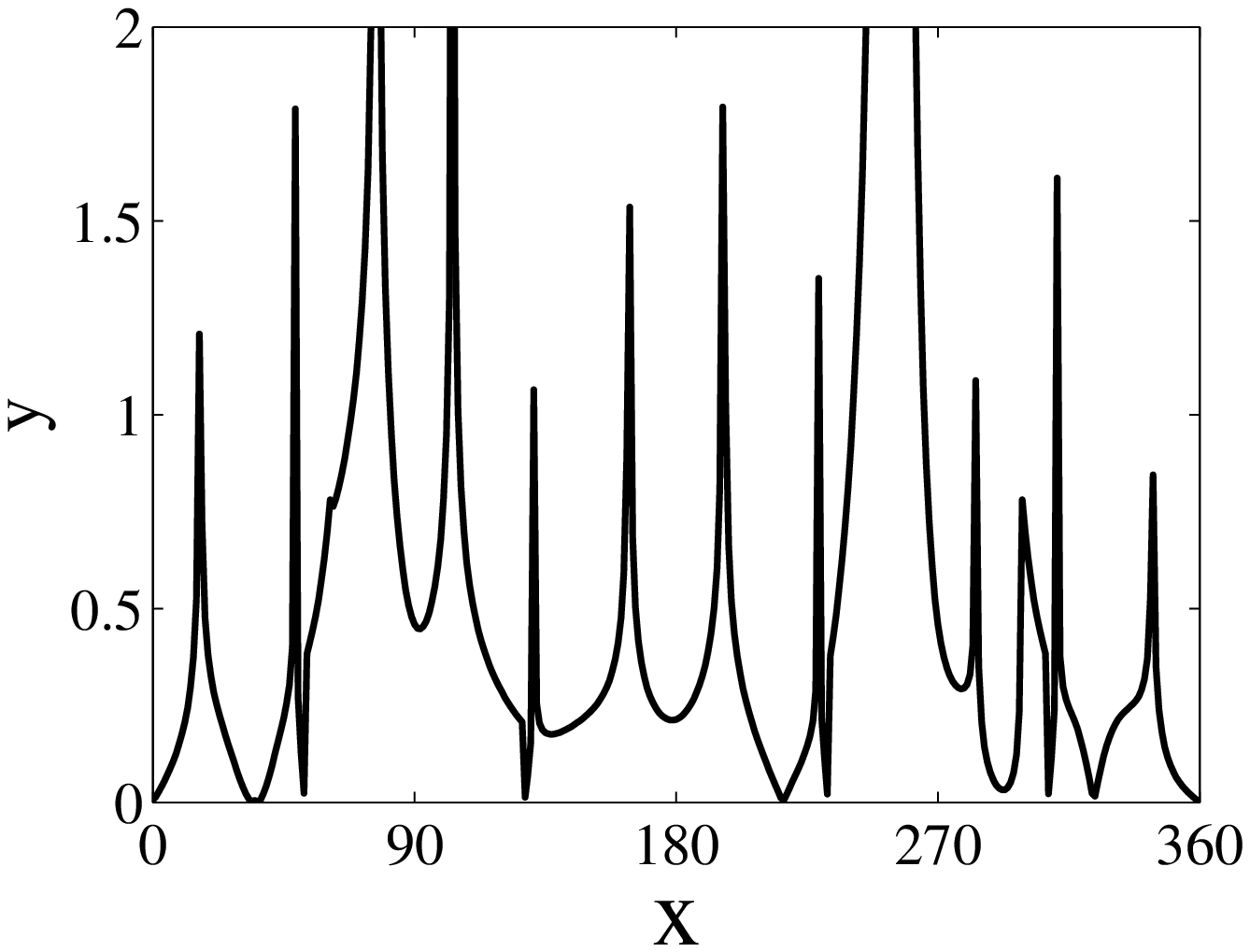}
{\bf~c}$\quad\quad\quad\quad\quad\quad\quad\quad\quad\quad\quad\quad\quad\quad\quad\quad\quad\quad\quad${\bf~f}\hfil\break
\includegraphics[height=43mm,width=65mm]{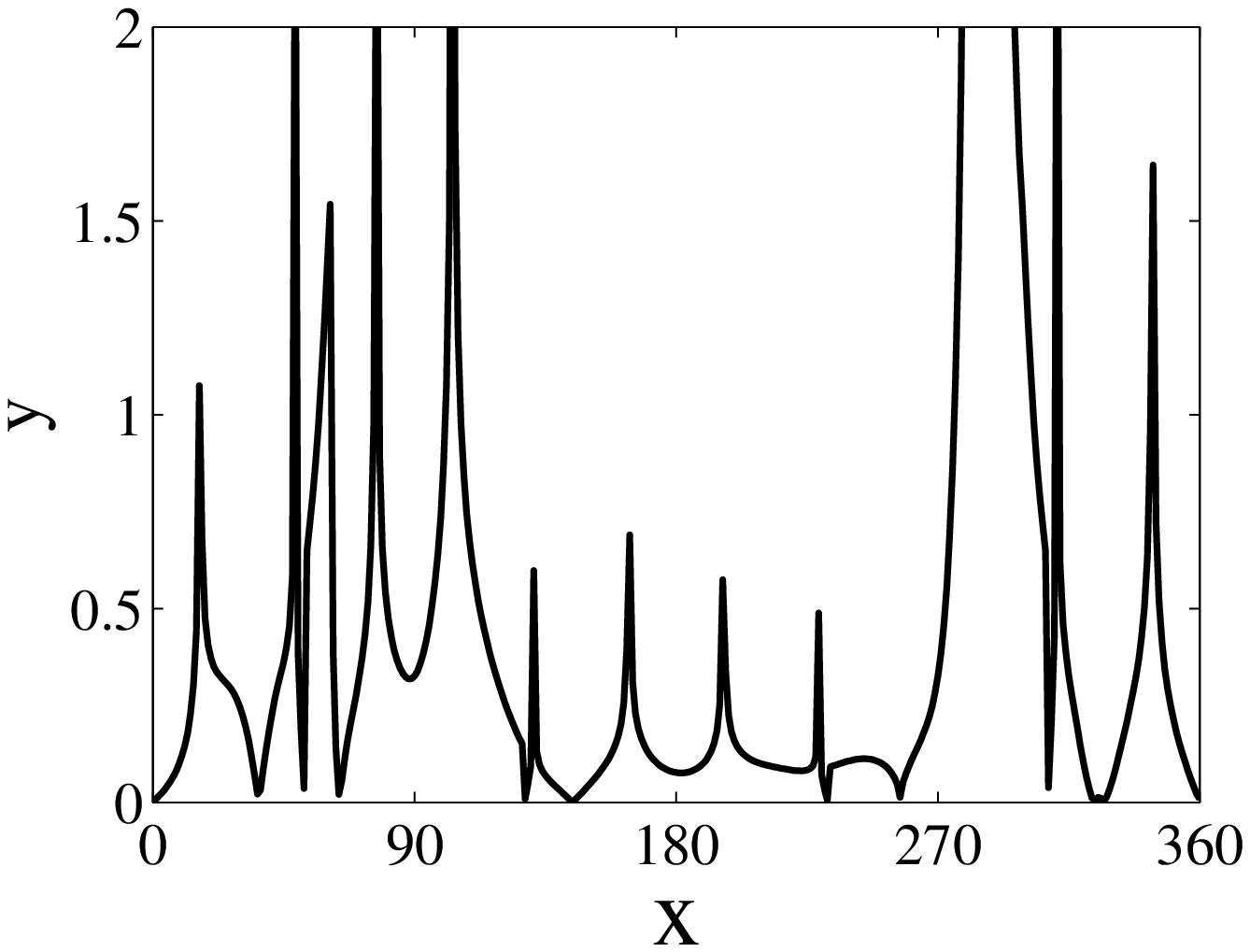}
  \hfil\hfil  \hfil\hfil   \hfil\hfil  \hfil\hfil
  \hfil\hfil  \hfil\hfil   \hfil\hfil  \hfil\hfil
  \hfil\hfil  \hfil\hfil   \hfil\hfil  \hfil\hfil
  \hfil\hfil  \hfil\hfil   \hfil\hfil  \hfil\hfil
\includegraphics[height=43mm,width=65mm]{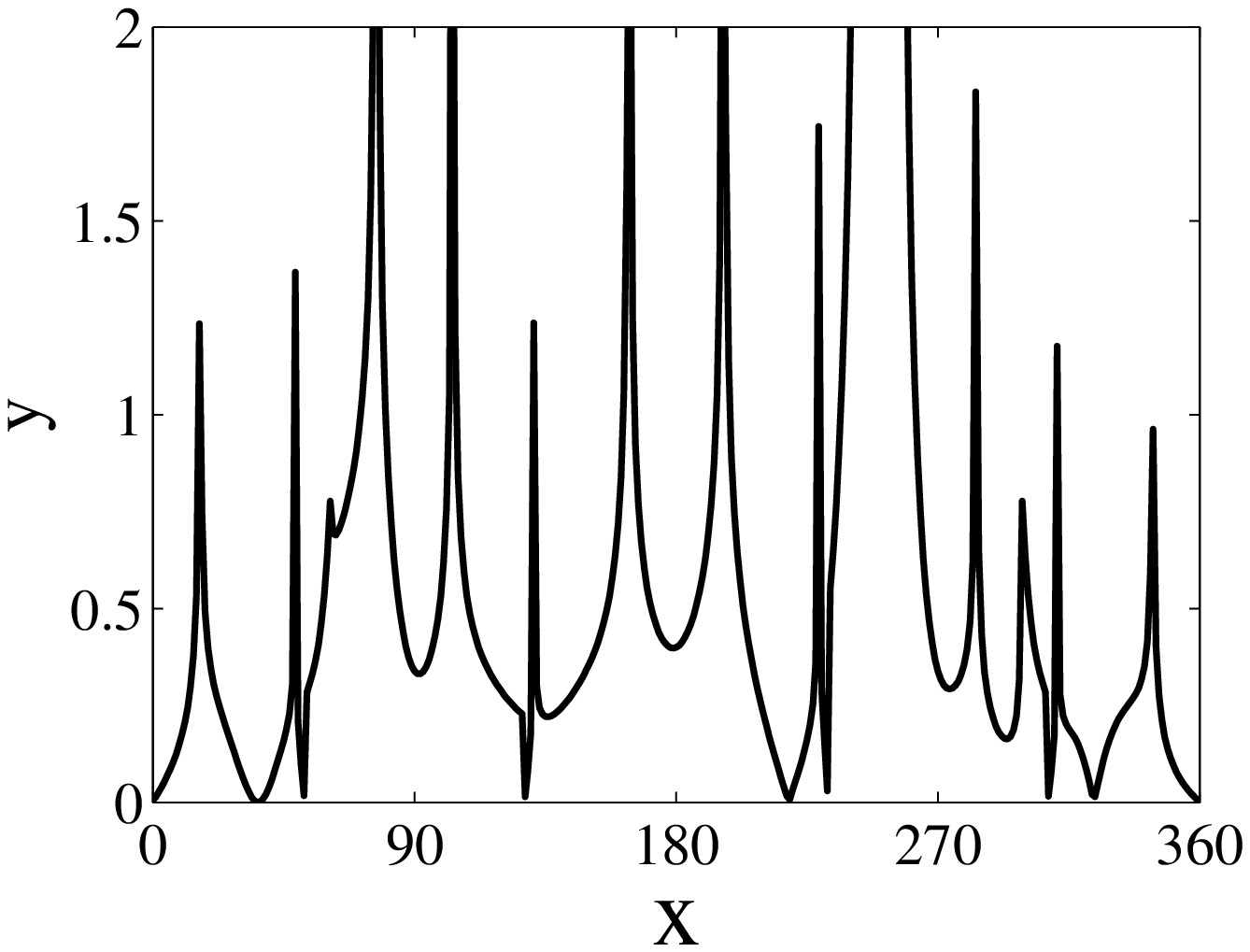}
  \caption{Magnitude $D^{(qSV,qP)}$ versus $\varthec$,  $\varphic^{inc}=90^0$ and $\varthec^{inc}=0^0$(a), $30^0$(b), $60^0$(c),  $90^0$(d), $120^0$(e), $150^0$(f).}
\end{figure}

\vfill\eject

\begin{figure}[ht]
{\bf~a}$\quad\quad\quad\quad\quad\quad\quad\quad\quad\quad\quad\quad\quad\quad\quad\quad\quad\quad\quad${\bf~d}\hfil\break
\psfrag{x}{$\varthec,\,{}^{\rm o}$}
\psfrag{y}{$|D|$}
\includegraphics[height=43mm,width=65mm]{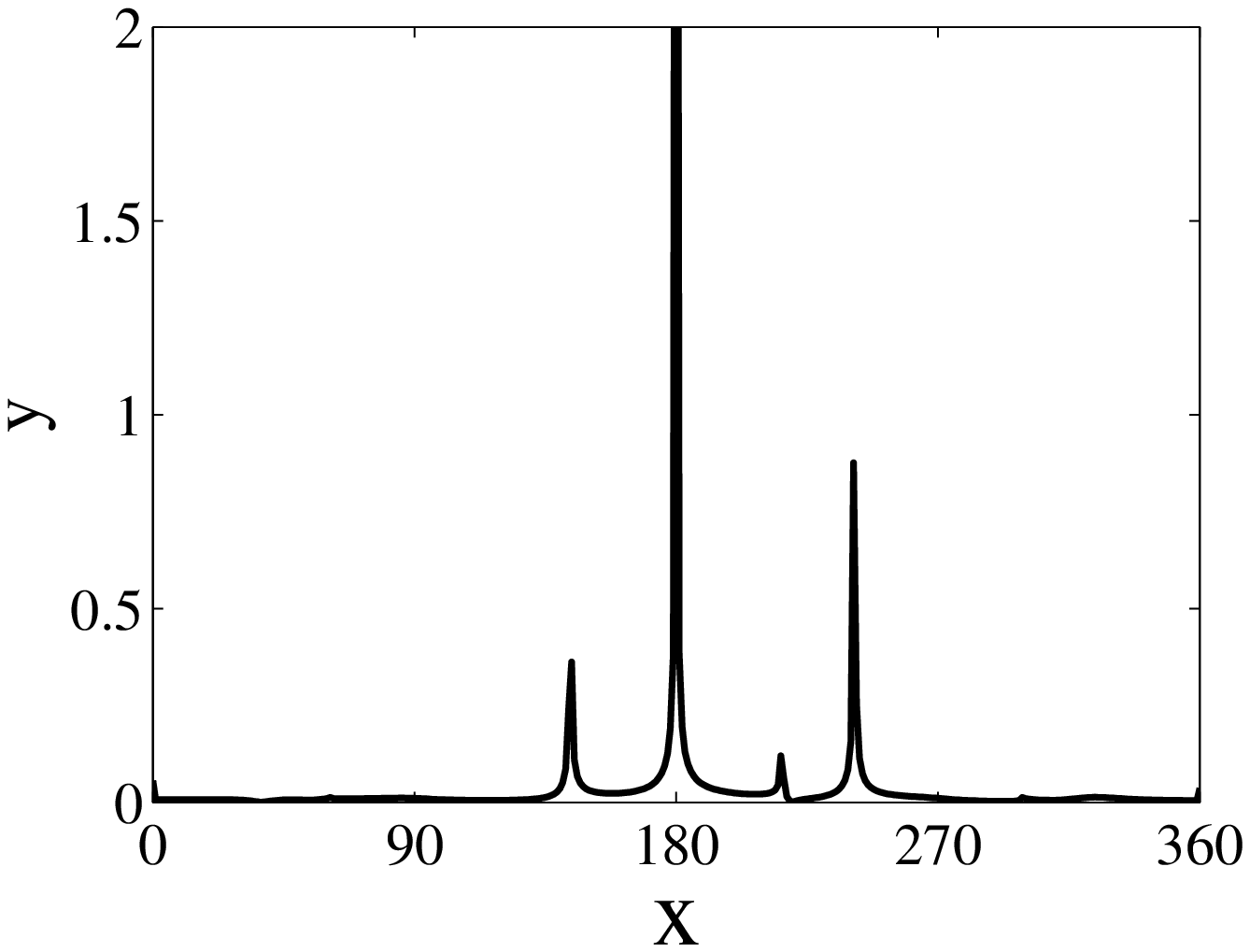}
\includegraphics[height=43mm,width=65mm]{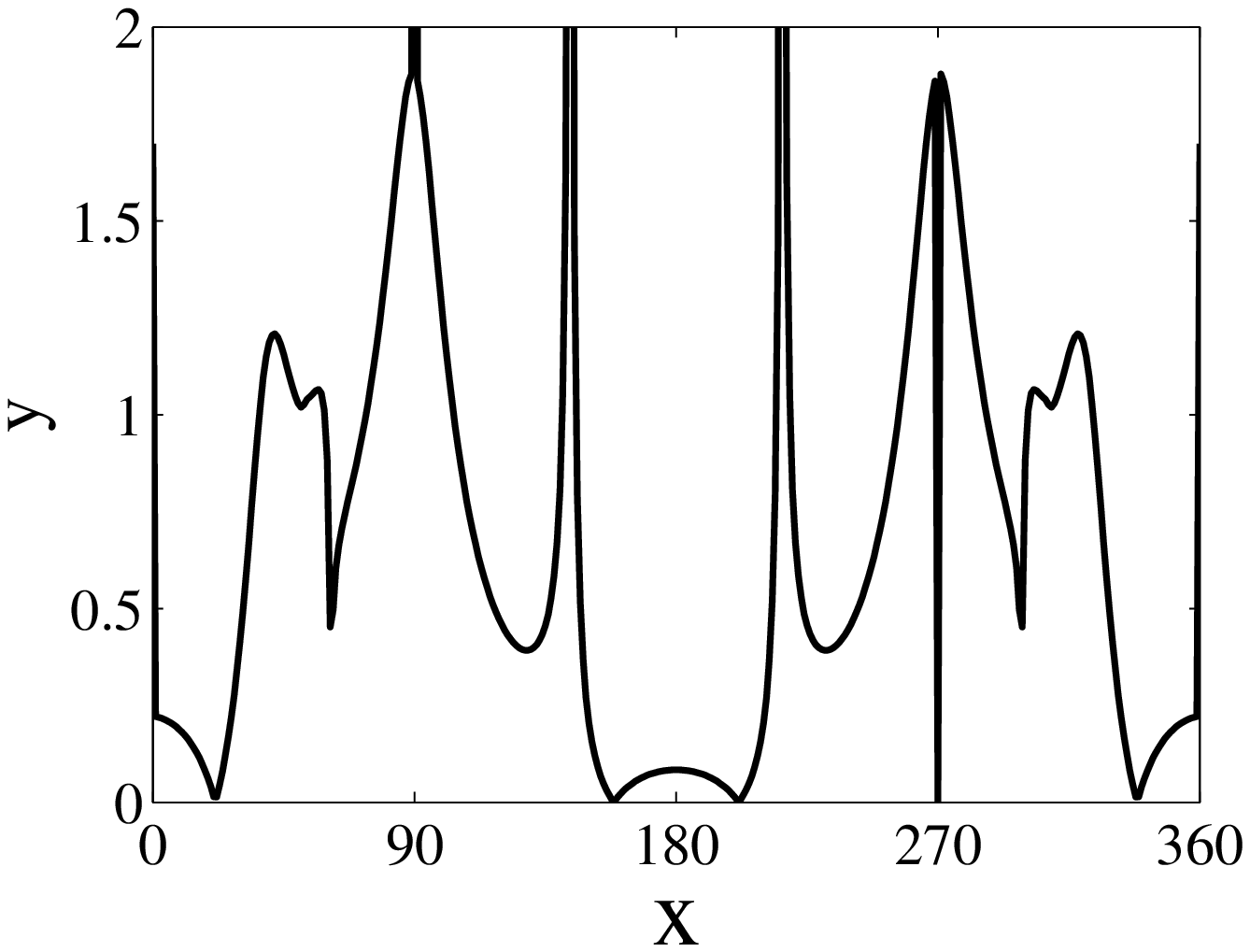}
{\bf~b}$\quad\quad\quad\quad\quad\quad\quad\quad\quad\quad\quad\quad\quad\quad\quad\quad\quad\quad\quad${\bf~e}\hfil\break
\includegraphics[height=43mm,width=65mm]{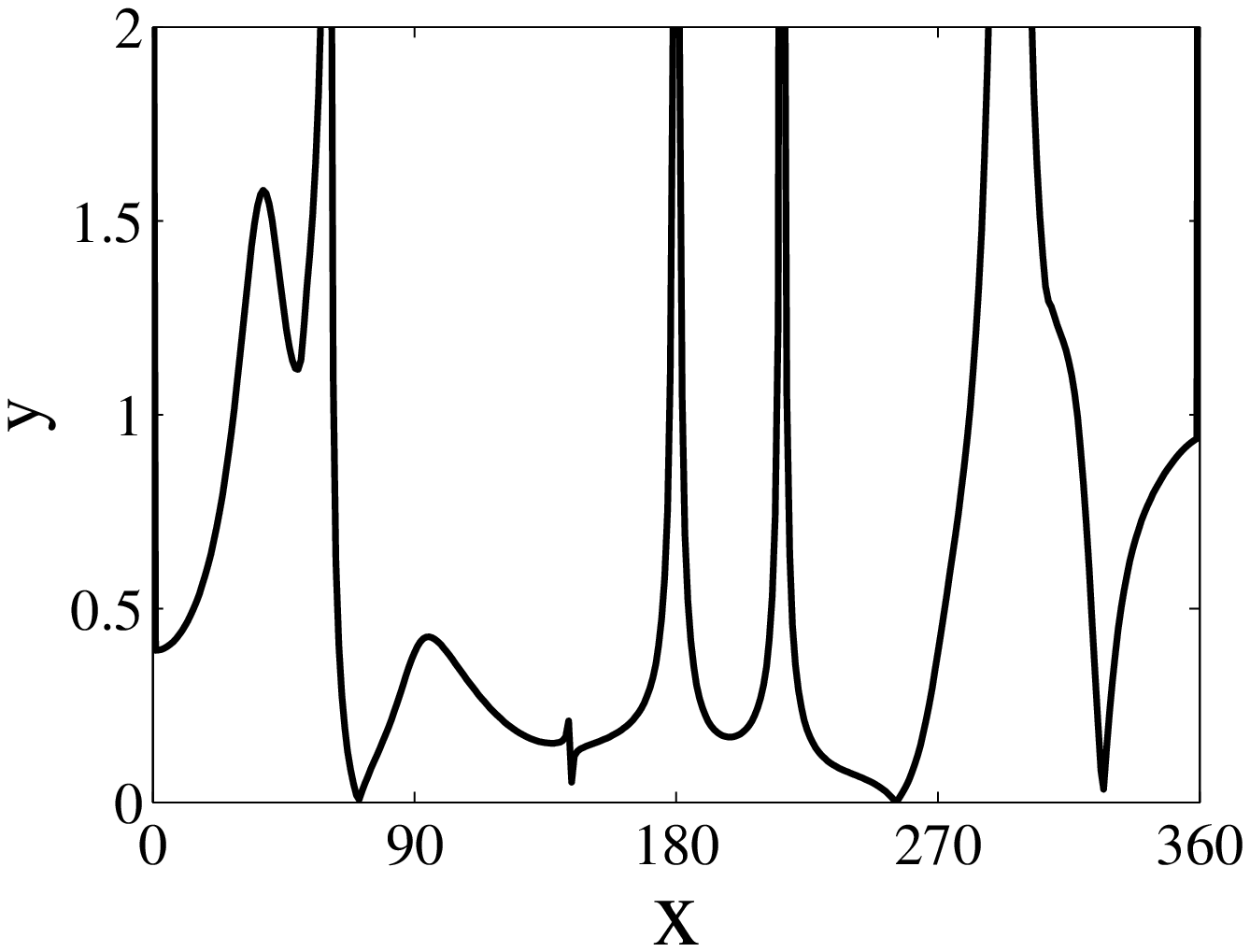}
\includegraphics[height=43mm,width=65mm]{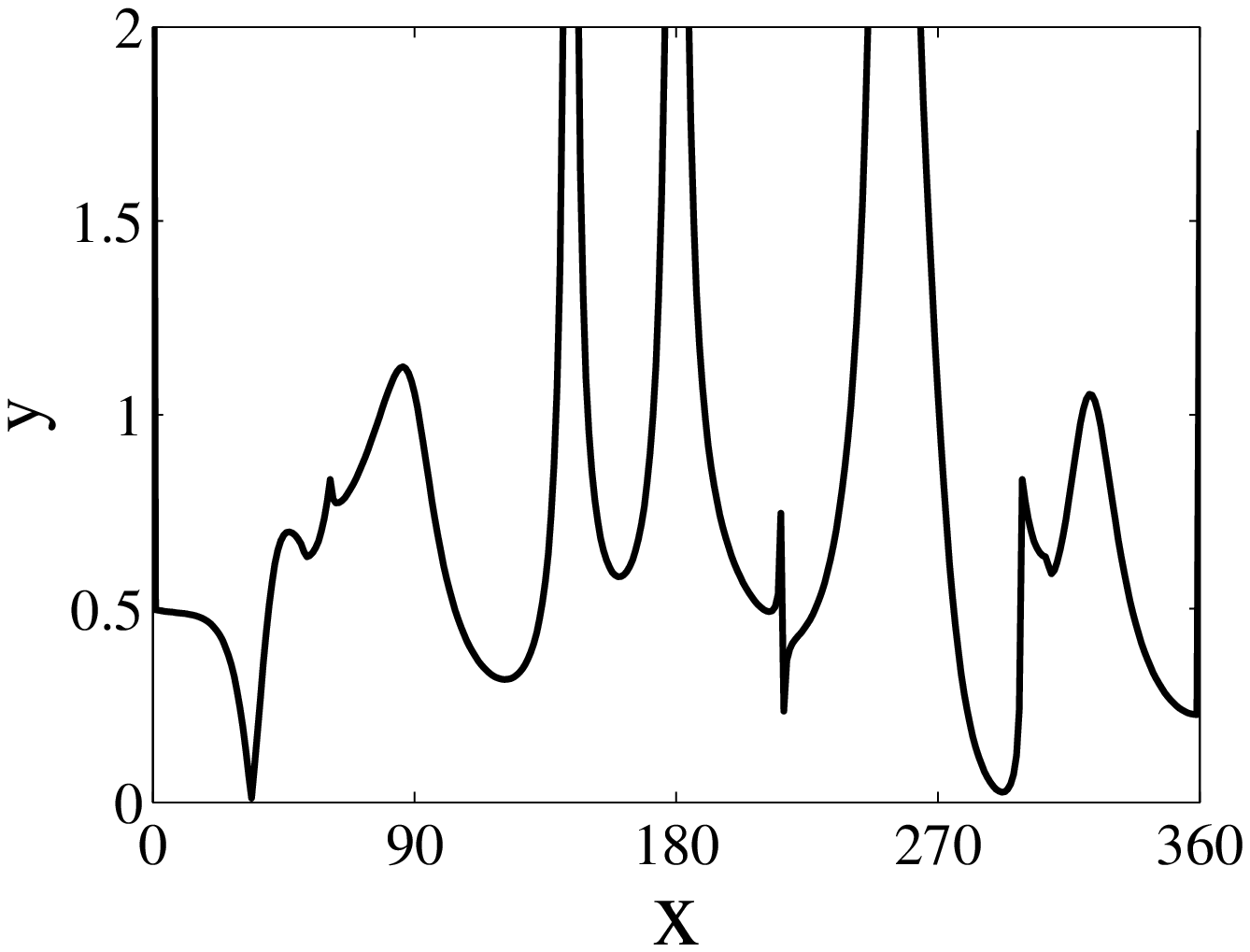}
{\bf~c}$\quad\quad\quad\quad\quad\quad\quad\quad\quad\quad\quad\quad\quad\quad\quad\quad\quad\quad\quad${\bf~f}\hfil\break
\includegraphics[height=43mm,width=65mm]{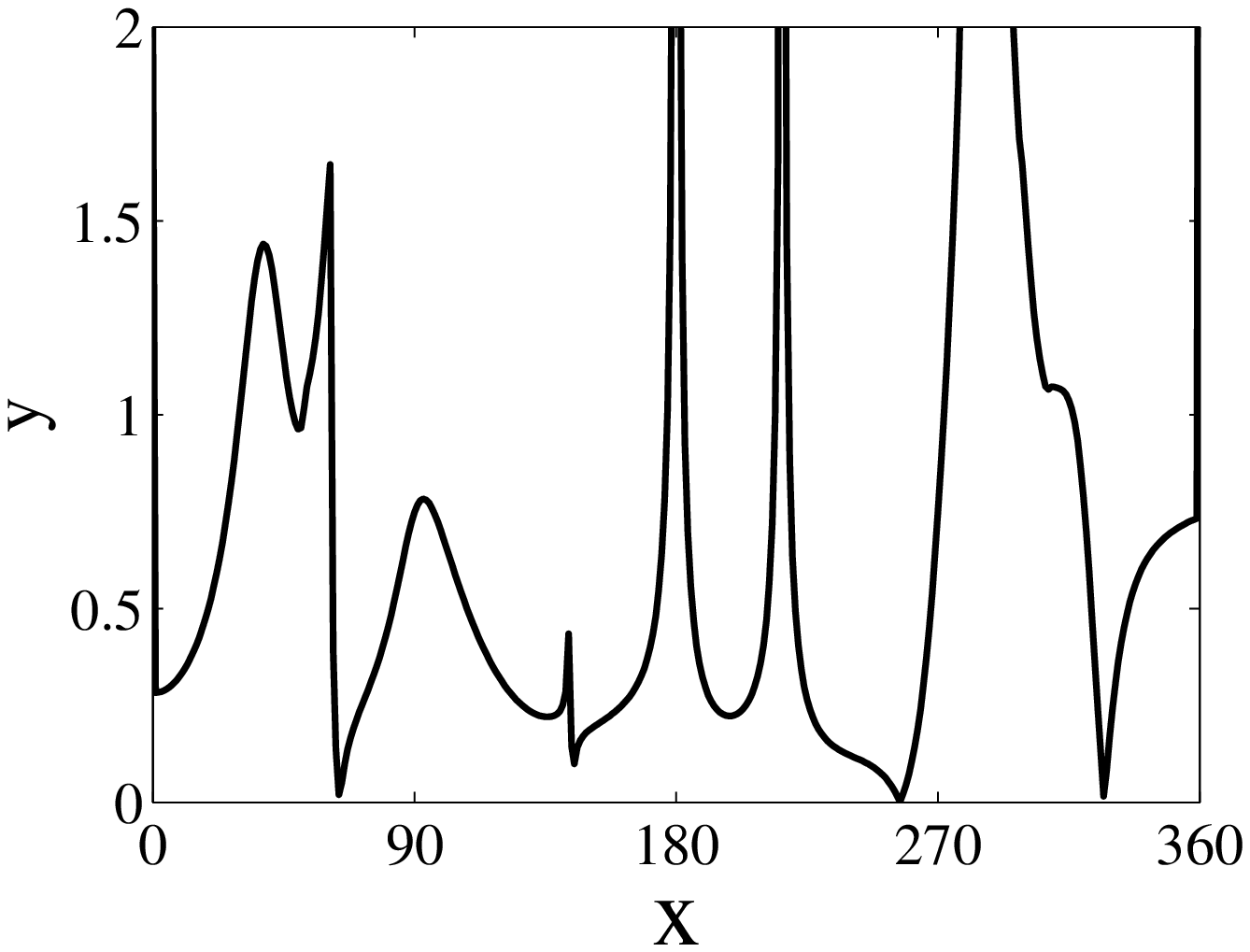}
  \hfil\hfil  \hfil\hfil   \hfil\hfil  \hfil\hfil
  \hfil\hfil  \hfil\hfil   \hfil\hfil  \hfil\hfil
  \hfil\hfil  \hfil\hfil   \hfil\hfil  \hfil\hfil
  \hfil\hfil  \hfil\hfil   \hfil\hfil  \hfil\hfil
\includegraphics[height=43mm,width=65mm]{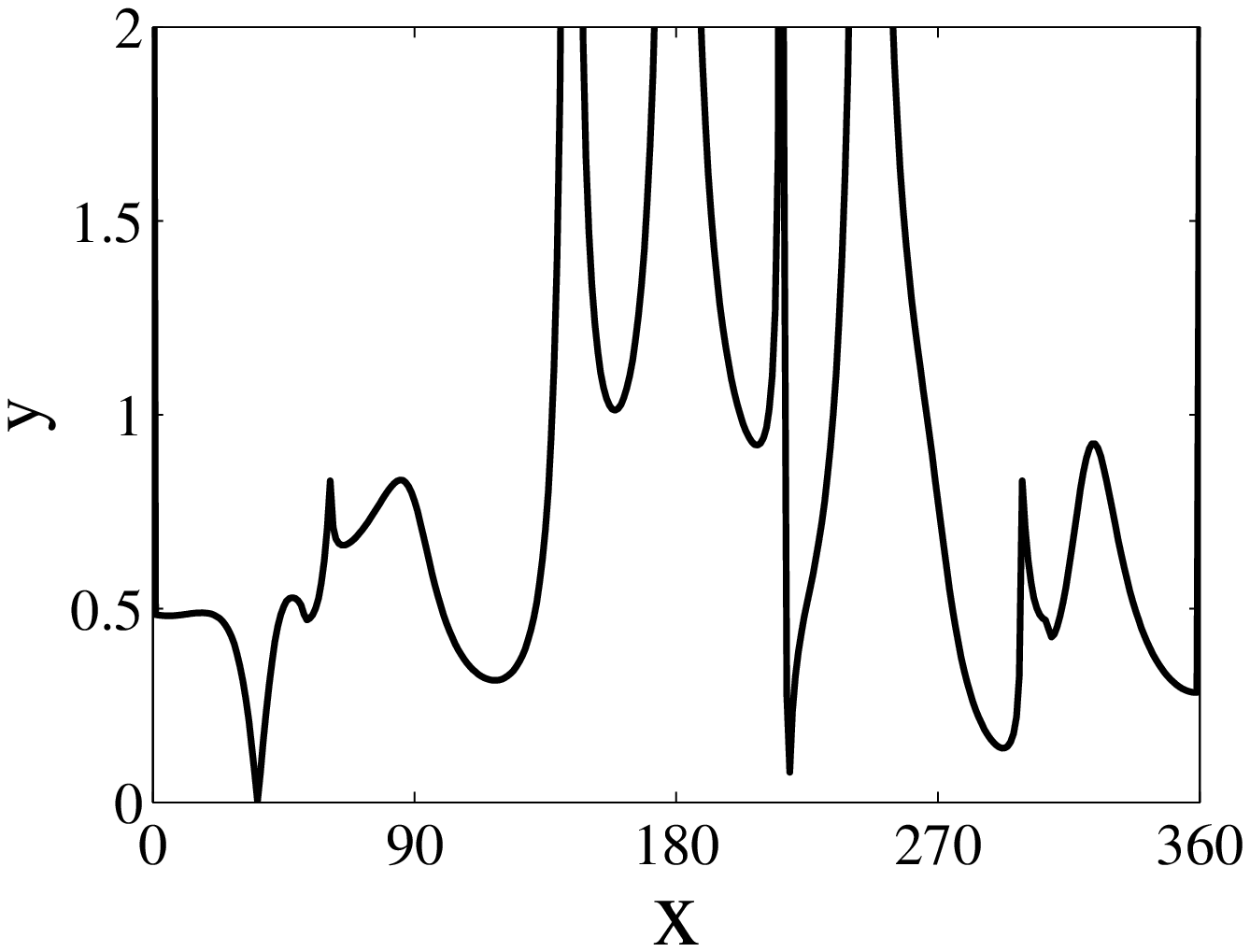}
  \caption{As in Fig.~E.1 but with the effect of cuspidal edges excluded.}
\end{figure}

\vfill\eject

\begin{figure}[ht]
{\bf~a}$\quad\quad\quad\quad\quad\quad\quad\quad\quad\quad\quad\quad\quad\quad\quad\quad\quad\quad\quad${\bf~d}\hfil\break
\psfrag{x}{$\varthec,\,{}^{\rm o}$}
\psfrag{y}{$|D|$}
\includegraphics[height=43mm,width=65mm]{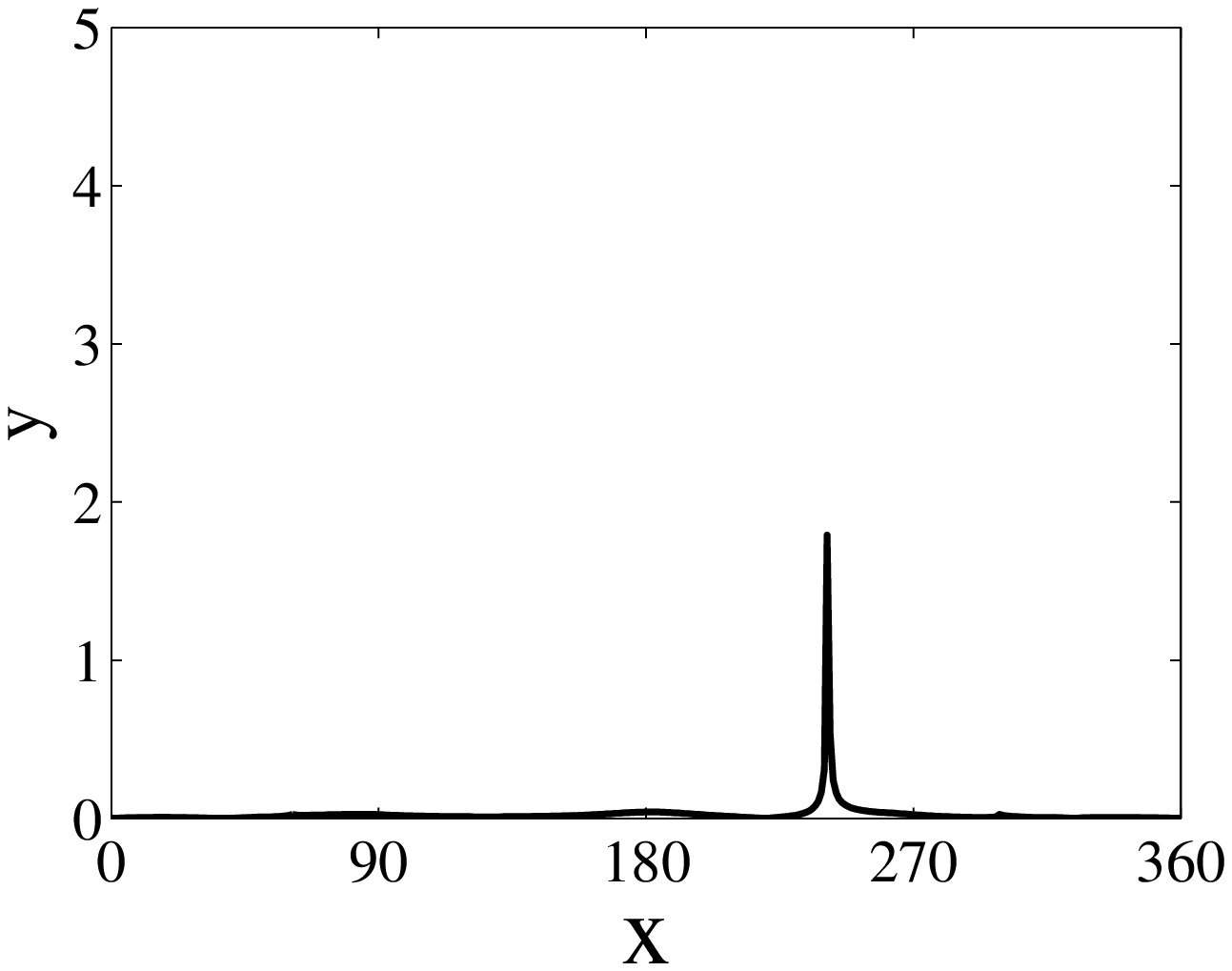}
\includegraphics[height=43mm,width=65mm]{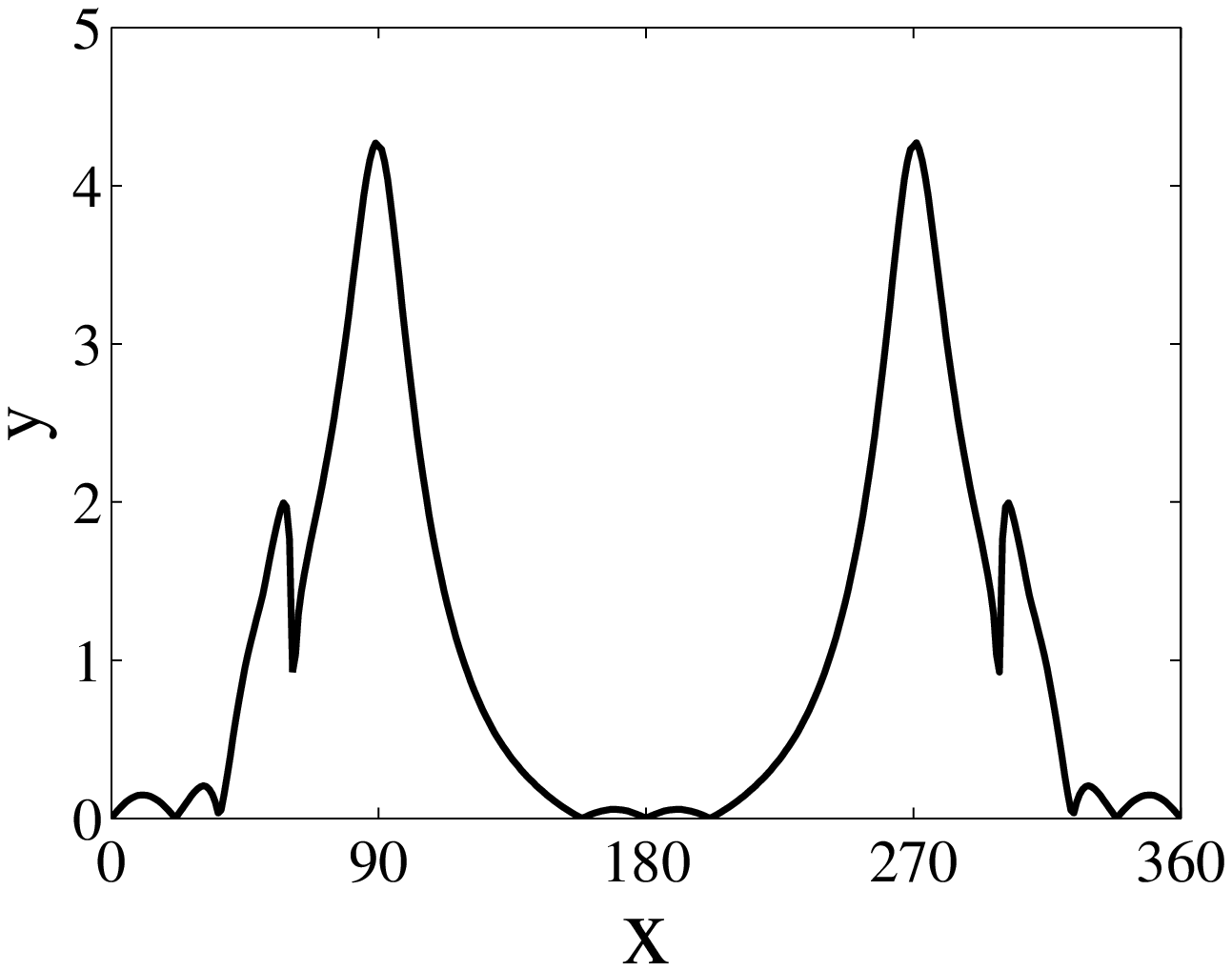}
{\bf~b}$\quad\quad\quad\quad\quad\quad\quad\quad\quad\quad\quad\quad\quad\quad\quad\quad\quad\quad\quad${\bf~e}\hfil\break
\includegraphics[height=43mm,width=65mm]{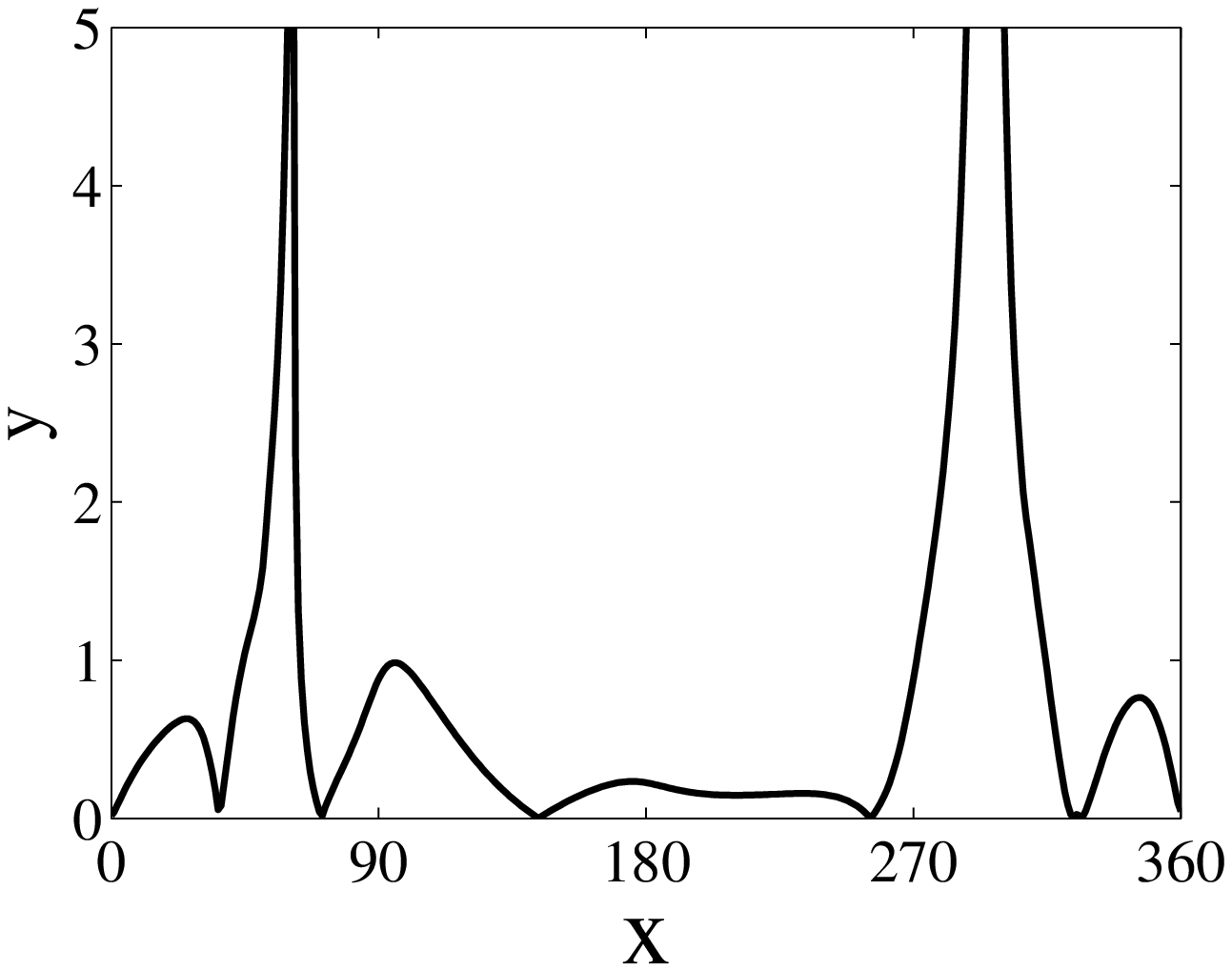}
\includegraphics[height=43mm,width=65mm]{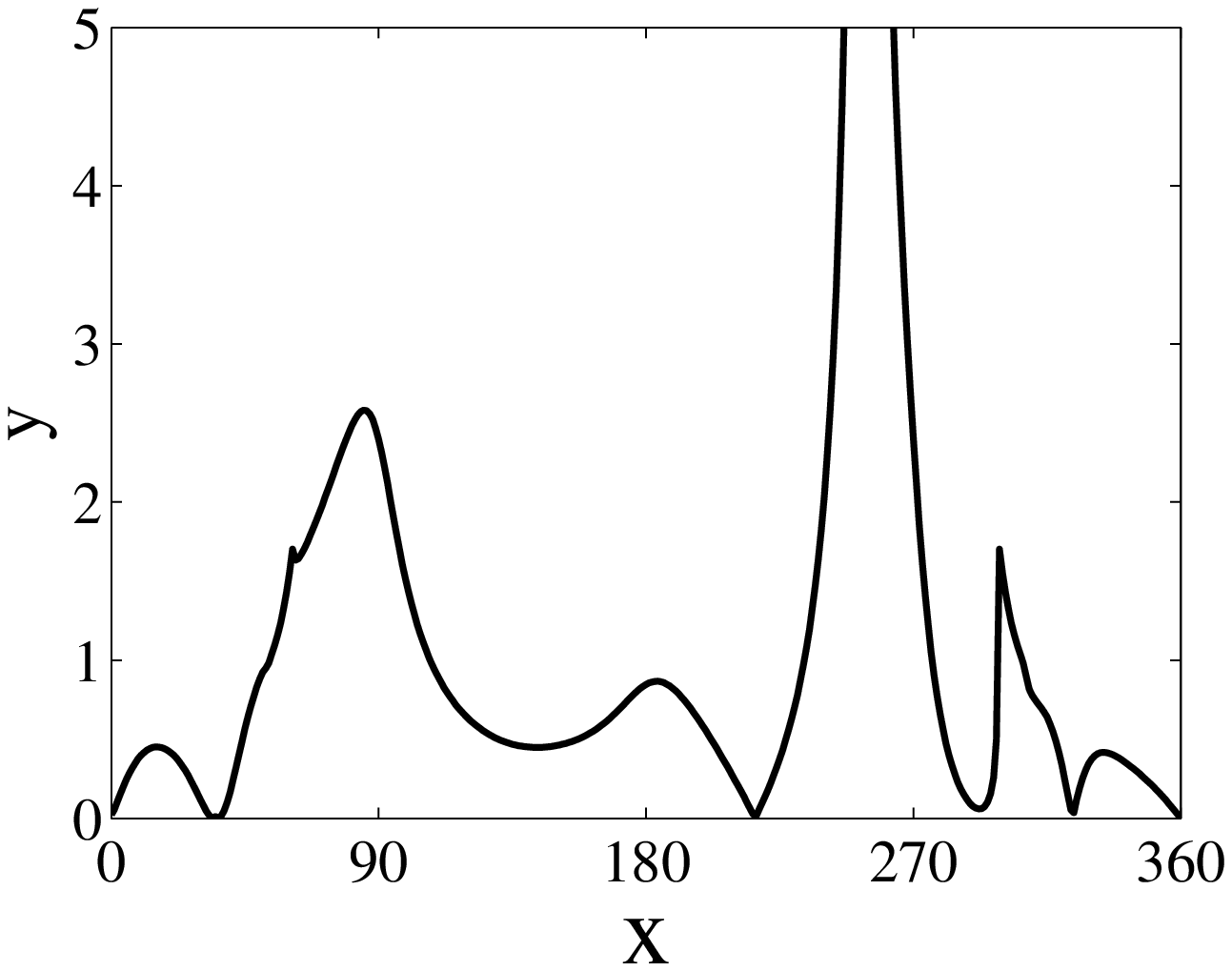}
{\bf~c}$\quad\quad\quad\quad\quad\quad\quad\quad\quad\quad\quad\quad\quad\quad\quad\quad\quad\quad\quad${\bf~f}\hfil\break
\includegraphics[height=43mm,width=65mm]{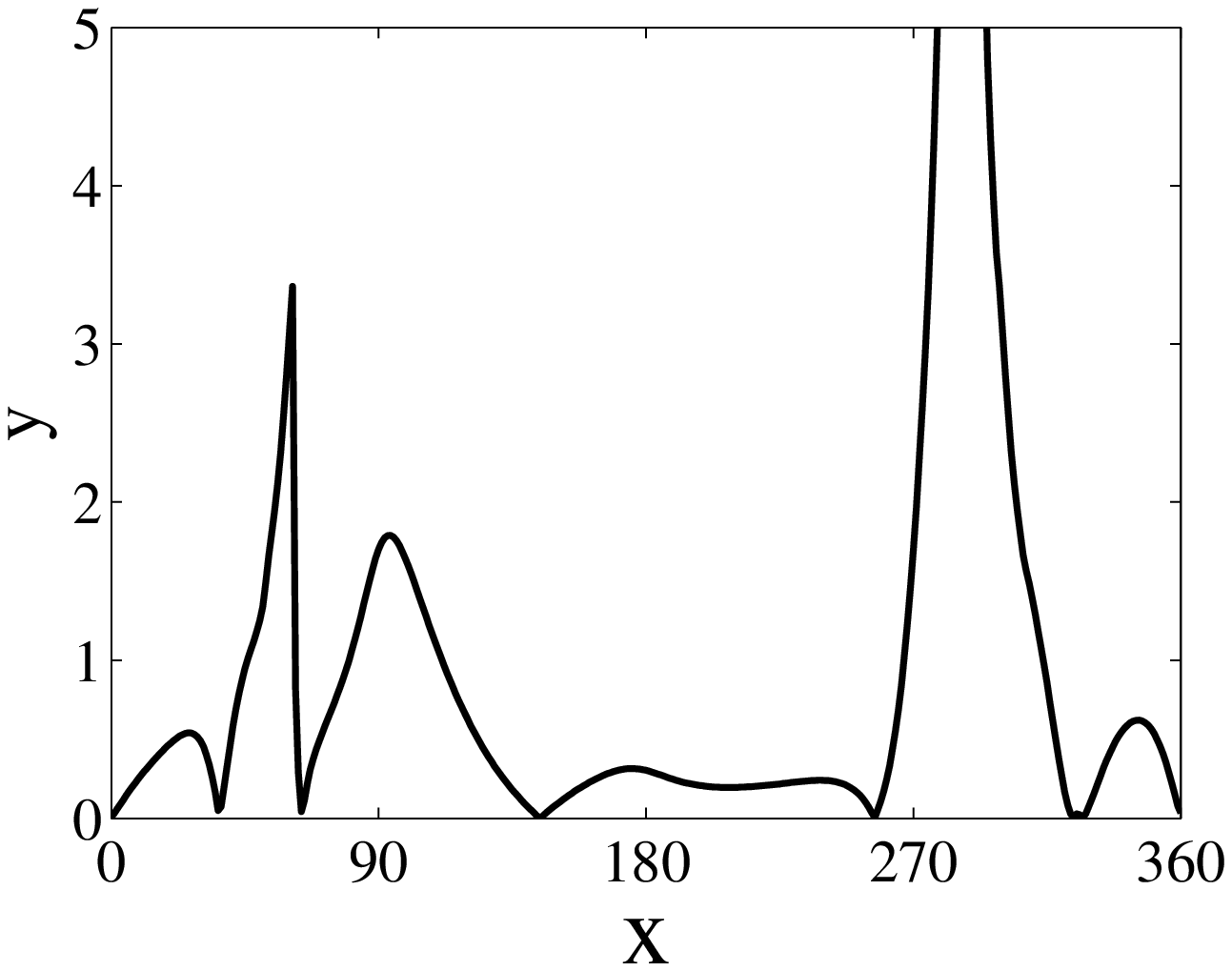}
  \hfil\hfil  \hfil\hfil   \hfil\hfil  \hfil\hfil
  \hfil\hfil  \hfil\hfil   \hfil\hfil  \hfil\hfil
  \hfil\hfil  \hfil\hfil   \hfil\hfil  \hfil\hfil
  \hfil\hfil  \hfil\hfil   \hfil\hfil  \hfil\hfil
\includegraphics[height=43mm,width=65mm]{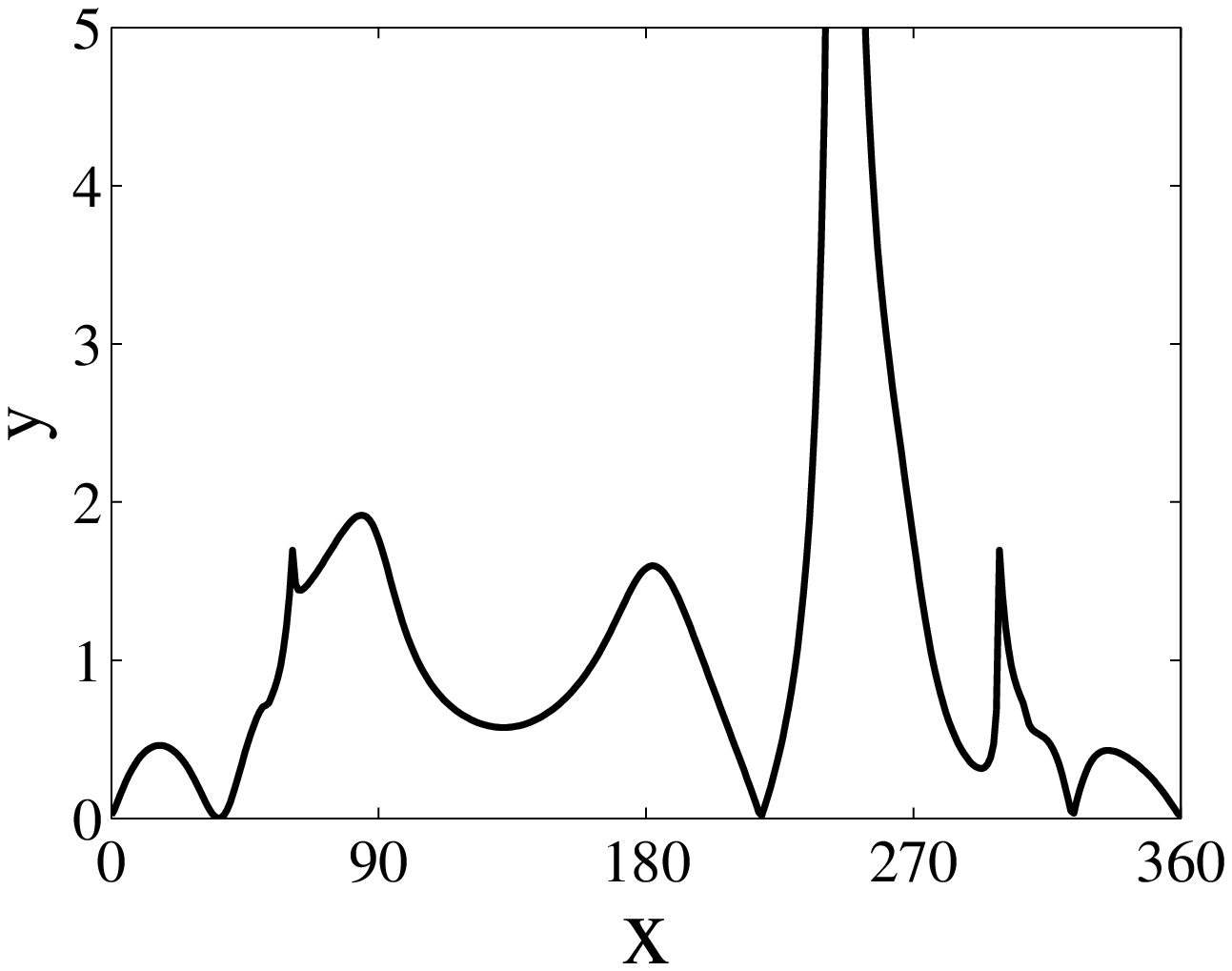}
  \caption{As in Fig.~E.2 but with the effect of conical points excluded.}
\end{figure}

\end{document}